\documentclass[prx,twocolumn,floatfix,superscriptaddress,longbibliography,nofootinbib]{revtex4-2}

\usepackage[latin9]{inputenc}
\usepackage{color}
\usepackage{amsmath}
\usepackage{amssymb}
\usepackage{stmaryrd}
\usepackage{graphicx}
\usepackage{romannum}
\usepackage{physics}
\usepackage[normalem]{ulem}
\usepackage{import}
\usepackage{victor}
\usepackage{chris}
\usepackage{laurens}
\usepackage{suffix}
\usepackage[unicode=true,bookmarks=true,bookmarksnumbered=false,bookmarksopen=false,breaklinks=false,backref=false,colorlinks=true]{hyperref}
\makeatletter
\@ifundefined{textcolor}{}
{%
 \definecolor{BLACK}{gray}{0}
 \definecolor{WHITE}{gray}{1}
 \definecolor{RED}{rgb}{1,0,0}
 \definecolor{GREEN}{rgb}{0,1,0}
 \definecolor{BLUE}{rgb}{0,0,1}
 \definecolor{CYAN}{cmyk}{1,0,0,0}
 \definecolor{MAGENTA}{cmyk}{0,1,0,0}
 \definecolor{YELLOW}{cmyk}{0,0,1,0}
}

\newcommand{\vva}[1]{{\color{blue} [VVA: #1]}}
\WithSuffix\newcommand\vva*[1]{{\color{blue} #1}}

\usepackage{amsfonts}\usepackage{tabularx}\usepackage{dcolumn}\usepackage{bm}\usepackage{graphicx}\usepackage{epstopdf}

\hypersetup{urlcolor=blue}

\usepackage{tensor}
\usepackage{braket}

\usepackage[capitalise,compress]{cleveref}
\makeatother

\makeatletter
\newsavebox{\@brx}
\newcommand{\llangle}[1][]{\savebox{\@brx}{\m@th{#1\langle}}%
  \mathopen{\copy\@brx\kern-0.5\wd\@brx\usebox{\@brx}}}
\newcommand{\rrangle}[1][]{\savebox{\@brx}{\m@th{#1\rangle}}%
  \mathclose{\copy\@brx\kern-0.5\wd\@brx\usebox{\@brx}}}
\makeatother

\begin{document}

\hypersetup{
 linkcolor=magenta, urlcolor=blue, citecolor=blue, unicode=true}
 
\title{Non-invertible symmetry-protected topological order in a group-based cluster state}

\author{ Christopher Fechisin}\thanks{These authors contributed equally to this work. Correspondence to fechisin@umd.edu.}
  \affiliation{Joint Quantum Institute, NIST/University of Maryland, College Park, MD, 20742, USA}
  \affiliation{Joint Center for Quantum Information and Computer Science, NIST/University of Maryland, College Park, MD, 20742, USA}

\author{ Nathanan Tantivasadakarn}
  \thanks{These authors contributed equally to this work. Correspondence to fechisin@umd.edu.}\affiliation{Walter Burke Institute for Theoretical Physics and Department of Physics, California Institute of Technology, Pasadena, CA 91125, USA}
  \affiliation{Microsoft Quantum, Station Q, Santa Barbara, CA 93106, USA}
  \affiliation{Department of Physics, Harvard University, Cambridge, MA 02138, USA}

 \author{ Victor V. Albert}
  \affiliation{Joint Center for Quantum Information and Computer Science, NIST/University of Maryland, College Park, MD, 20742, USA}

\date{\today}
\begin{abstract}

Despite growing interest in beyond-group symmetries in quantum condensed matter systems, there are relatively few microscopic lattice models explicitly realizing these symmetries, and many phenomena have yet to be studied at the microscopic level. We introduce a one-dimensional stabilizer Hamiltonian composed of group-based Pauli operators whose ground state is a $G\times \rep G$-symmetric state: the $G$ \textit{cluster state} introduced in [\href{http://doi.org/10.1088/1367-2630/17/2/023029}{Brell, New Journal of Physics \textbf{17}, 023029 (2015)}]. We show that this state lies in a symmetry-protected topological (SPT) phase protected by $G\times \rep G$ symmetry, distinct from the symmetric product state by a duality argument. We identify several signatures of SPT order, namely protected edge modes, string order parameters, and topological response. We discuss how $G$ cluster states may be used as a universal resource for measurement-based quantum computation, explicitly working out the case where $G$ is a semidirect product of abelian groups.

\end{abstract}
\maketitle

% disable subsections and subsubsections in the TOC
\makeatletter
\def\l@subsection#1#2{}
\def\l@subsubsection#1#2{}
\makeatother

\tableofcontents

\pagenumbering{arabic}

\section{Introduction}

Symmetry is a central organizing principle and computational tool in the study of condensed matter physics. Conventionally, we think of a family of symmetry operators which:
\begin{enumerate}
	\item act on the full system, e.g., as a tensor product ``on-site'' (internal) symmetry or by realizing a transformation, such as rotation or inversion, of the underlying lattice; and
	\item multiply according to the multiplication rule of some group.
\end{enumerate}
However, this is not the most general situation. Exotic phenomena can be described by symmetries which violate one or both of the above criteria. In loosening the definition of symmetry, we are able to unify a wider family of quantum phases within a single framework and explore novel phases which pose challenges of theoretical and practical interest (see e.g.~\cite{mcgreevy2023generalized,Cordova22}).

Rich physics has emerged from the study of two particular generalizations of global symmetry which act on only part of the system and therefore relax the first constraint above --- higher form symmetry~\cite{gaiotto_generalized_2015,KapustinThorngren2017,yoshida_topological_2016,tsui_lattice_2020} and subsystem symmetry~\cite{NewmanMoore1999,XuMoore2004,vijay_fracton_2016,Youetal2018,Devakuletal2018}. Both types of symmetry act on subsets of local degrees of freedom, such as qubits or qudits, that make up manifolds of nonzero codimension in the geometry of the many-body system.

Higher form symmetries act on homologically non-trivial manifolds and are deformable within their homology class via the Gauss law of a gauge theory \cite{Yoshida13,vijay_fracton_2016,Williamson2016}.
A notable $1$-form symmetry operator~\cite{gaiotto_generalized_2015} is the Wilson loop operator, which acts on a subsystem that forms a closed loop and which distinguishes the confined and deconfined phases of the $\mathbb{Z}_2$ gauge theory in 2+1D. Other higher form symmetries are important objects in the design of topological quantum computation schemes~\cite{Yoshida15,yoshida_topological_2016,yoshida_gapped_2017,barkeshli_codimension-2_2022,barkeshli2023highergroup,Zhu22,Zhu23,barkeshli2023higher}.

On the other hand, subsystem symmetries act on rigid manifolds that are not deformable into one another.
They have become especially important in describing fracton phases of matter, which have garnered significant interest as a particularly exotic form of topological order where quasiparticle excitations exhibit restricted mobility~\cite{Chamon05,Haah11,Yoshida13,vijay_fracton_2016,shirley_fracton_2018,shirley_foliated_2019,bulmash_gauging_2019,nandkishore_fractons_2019,pretko_fracton_2020}.

Higher form and subsystem symmetries demonstrate the value of understanding how symmetry can be generalized by violating the \textit{first} criterion above. Other works have studied symmetries which violate both constraints (see e.g.~\cite{BulmashBarkeshli2019,PremWilliamson2019,TJV1,tantivasadakarn_non-abelian_2021,TuChang21,HsinSlagle21,rayhaun_higher-form_2021}), though these symmetries are less well-understood. In this work, we study a symmetry which is complementary to the above, as it violates only the \textit{second} criterion, balancing physical richness and tractability.

Conventionally, a symmetry in a quantum system is realized by a set of unitary operators $U_g$ which commute with the Hamiltonian and transform as a linear representation of the symmetry group $G$, meaning that $U_gU_h=U_{gh}$. More generally, a set of symmetry operators $\{O_a\}$ can multiply according to the rules of a \textit{fusion category}~\cite{etingof2016tensor,etingof_fusion_2017} (equivalently\footnote{Fusion categories share fusion rules with fusion rings and algebras, but are equipped with additional structure, such as F-symbols which encode the degree of associativity of fusion of vector spaces defined within the category. The topic of constructing fusion categories from a fusion ring is called \textit{categorification}. Because we only use the fusion rules of the category, we are ignoring these subtleties.}, fusion ring~\cite{gepner1991fusion,gepner_rings_1994} or fusion algebra~\cite{dijkgraaf1988modular}),
\begin{equation}\label{eq:fusion_1}
	O_a O_b=\sum_{c}N_{a,b}^cO_c~,
\end{equation}
where $a$, $b$, and $c$ are simple objects in the category, and where the fusion coefficients $N^{c}_{a,b} \in \mathbb Z_{>0}$ give the number of copies of each $O_c$ appearing in the sum. We write these symmetry operators as $O$ rather than $U$ because they are not in general unitary.

Where multiplication in a group yields another element in the group ($g\cdot h =gh$), multiplication in a fusion category yields a sum of objects in the category ($a \times b=\sum_{c}N_{a,b}^c c$). Groups are special cases of fusion categories where $N_{g,h}^{k}=\delta_{k,gh}$, and when we refer to fusion category symmetries we implicitly mean symmetries corresponding to fusion categories that are not groups.

The fusion category perhaps most familiar to physicists -- though not necessarily by name -- is SU$(2)_k$~\cite{fuchs1995affine,trebst2008short,greiter2019non}, which describes the fusion of the first $k+1$ irreps of SU$(2)$~\cite{georgi2000lie}. Physically, this fusion category describes addition of angular momentum restricted to particles of spin less than or equal to $k/2$, encoding relationships like $\frac{1}{2}\otimes\frac{1}{2}=0\oplus 1$. As an explicit example, consider the multiplication table of SU$(2)_2$:
\begin{equation*}
\begin{aligned}
\begingroup
\def\arraystretch{1.3}
\begin{array}{c|c c c}
		\otimes & 0 & \frac{1}{2} & 1 \\ \hline
         0 & 0 & \frac{1}{2} & 1 \\
         \frac{1}{2} & \frac{1}{2} & 0\oplus1 & \frac{1}{2} \\
         1 & 1 & \frac{1}{2} & 0 \\
\end{array}\endgroup&\qquad\text{SU}(2)_2
\end{aligned}
\end{equation*}
Notice that the fusion rules of SU(2)$_2$ are beyond what can be described by a group. Most strikingly, they allow for the product of two objects to equal a sum of objects. This is a central feature of the generalization from groups to fusion categories, and leads to the existence of non-invertible simple objects, and therefore non-invertible symmetries.

The study of fusion categories -- and their descendants, like unitary modular tensor categories  -- has been fruitful in condensed matter physics~\cite{walker20123+,kong2014anyon,kong2014braided,wen2016theory,bernevig2017topological,cong2017defects,kong2017boundary,wen2019choreographed,barkeshli_symmetry_2019,JiWen20,chatterjee_symmetry_2023}. The foremost application is describing fusion and braiding of anyons. While abelian anyons fuse according to the rules of a group, non-abelian anyons fuse according to the rules of a fusion category~\cite{simon_topological_2021}.
Non-abelian anyons exhibit a variety of exotic physics due to their unusual fusion properties, including universality for topological quantum computation~\cite{Nayak_RMP}.

In this paper, we construct a microscopic model with fusion-category SPT order on a 1D lattice of qudits whose basis states are valued in a finite group $G$.
Our model reduces to the well-known $\mathbb{Z}_2\times\mathbb{Z}_2$ abelian SPT order: the one-dimensional cluster state~\cite{chen_symmetry-protected_2012-1,son2012topological} when $G=\mathbb{Z}_2$.
For non-abelian $G$, the second symmetry factor in the original abelian order generalizes to $\rep G$, the fusion category of representations of $G$. Such cases yield models protected by a combination of group and representation category symmetry which, to our knowledge, has not been considered before.

We show how microscopic signatures of SPT order extend to fusion category-symmetric systems through the study of the \textit{group cluster state} (or $G$ cluster state, where $G$ is a finite group), a generalization of the familiar cluster state~\cite{briegel_persistent_2001}. We contribute the following:
\begin{enumerate}
    \item A proof of the claim in Ref.~\onlinecite{brell_generalized_2015} that the $G$ cluster state is an SPT protected by a fusion category symmetry.
    \item An identification of microscopic signatures of SPT order in the $G$ cluster state, including string order, edge modes, and topological response.
    \item Further development of the group-based Pauli stabilizer formalism from Ref.~\onlinecite{albert_spin_2021}, making contact with matrix-product states (MPSs) and matrix-product operators (MPOs). By studying \textit{microscopics} of explicit models as opposed to abstractly classifying phases, we open up models with such exotic symmetries to further study and potential realization on quantum devices.
    \item An algorithm for performing measurement-based quantum computation on the $G$ cluster state for a wide class of non-abelian groups, namely those which are semidirect products of abelian groups.
\end{enumerate}

%%%%%%%%%%%%%%%%%%%%%%%%%%%%%%%%%%%%%%%%%%%%%%%%%%%
\subsection{Relation to Prior Works}

This paper appears among a growing literature studying exotic phases of matter in lattice models, as well as studying categorical symmetries~\cite{Chang19,kong_algebraic_2020,ji_categorical_2020,thorngren_fusion_2019,thorngren_fusion_2021,inamura_topological_2021,Heidenreich21Completeness,Bartsch:2022mpm,Bartsch:2022ytj,Bartsch:2023wvv,Roumpedakis23higher,inamura2023fermionization,Choi22,Choi23triality,Choi23timereversal,zhang2023anomalies,cordova2023anomalies,bhardwaj23unifying,bhardwaj23symmetryweb,bhardwaj2022universal,Bhardwaj23noninvertible,Kaidi23symTFT,bhardwaj2023generalized,bhardwaj2023gapped,bhardwaj2023categorical,choi2023self,Decoppet23,Pace2023emergent,perez2023notes}.
Non-invertible symmetry on the lattice has been discussed in several works~\cite{hu2017boundary,hu2018boundary,inamura_lattice_2022,tantivasadakarn_hierarchy_2022,SeibergShao23,Shao23,inamura2023fusion,delcamp2023higher,tantivasadakarn2023string}, though most known lattice models of \textit{SPTs} are protected by a conventional, subsystem~\cite{you_subsystem_2018,stephen_subsystem_2019-1,TantivasadakarnVijay20,DevakulShirleyWang20,StephenGarre-RubioDuaWilliamson2020,Tantivasadakarn20,Shirley20,Han23}, or higher-form symmetry~\cite{KapustinThorngren15,DelcampTiwari18,tsui_lattice_2020,Chen21}. Ref.~\onlinecite{inamura_lattice_2022} is especially relevant, as it constructs commuting projector lattice models for topological quantum field theories (TQFTs) with fusion category symmetry and shows that degenerate edge modes arise in the SPT class of these models. While this work is quite general and its results broadly applicable, it does not study a model with $G\times \rep G$ symmetry explicitly, nor does it study concrete models in great detail, so that our work is complementary. We also note that our stabilizer model contains additional structure compared to the commuting projector models constructed therein, as a stabilizer Hamiltonian can always be expressed as a sum of commuting projectors, while the contrary is not guaranteed.

Although there are few explicit lattice constructions, there is good reason to suspect that the signatures of SPT order should extend to much more general contexts. Macroscopic descriptions of SPT orders with exotic categorical symmetries have been defined within higher category theory~\cite{kong_algebraic_2020,ji_categorical_2020}. There has also been work studying fusion category symmetries in quantum field theories which have addressed the existence of fusion category-symmetric gapped phases~\cite{thorngren_fusion_2019,thorngren_fusion_2021} and gauging non-abelian symmetries to yield fusion categorical symmetries~\cite{bhardwaj_finite_2018}. The question of classifying fusion category-symmetric gapped phases has also been addressed at the categorical level~\cite{thorngren_fusion_2019,inamura_topological_2021}. In general, category theoretic methods are very powerful for studying generalized symmetries and the phases they admit. It has been shown that for a symmetry described by a tensor category $\mathcal{C}$, the exact module categories over $\mathcal{C}$ correspond to the gapped symmetric phases with symmetry $\mathcal{C}$~\cite{thorngren_fusion_2019,ostrik_module_2003}.

In this work, we use group-based Pauli operators to define microscopic lattice models, which more closely resemble spin chains. This formalism was recently used in Ref.~\onlinecite{albert_spin_2021} to study gapped edges of quantum doubles. Group-based Pauli operators preserve some useful features of the stabilizer formalism~\cite{gottesman_stabilizer_nodate} and
are natural for programming onto quantum devices. These operators also admit a graphical description as matrix product operators which makes contact with the tensor network literature on quantum phases of matter with exotic symmetries (see eg.~\cite{williamson2017symmetry,lootens_matrix_2021,molnar_matrix_2022}).

While we intentionally use group-valued degrees of freedom to generate fusion category symmetries, they arise naturally in other areas. High energy physicists have long studied gauge fields which take values in gauge symmetry groups, and lattice gauge theories work with degrees of freedom labeled by group elements living on the lattice~\cite{kogut_hamiltonian_1975,kogut_introduction_1979,harlow2021symmetries}. In condensed matter physics, these degrees of freedom are used to construct Kitaev's quantum double model, a generalization of the toric code which hosts non-abelian anyons~\cite{kitaev_fault-tolerant_2003-2}. Group-valued degrees of freedom are also used in the TQFT approach to constructing microscopic lattice models of SPTs with conventional symmetry~\cite{dijkgraaf1990topological,chen_symmetry_2013-1,GuWen14,KapustinThorngren15,Cheng18,tantivasadakarn2017dimensional,bulmash2020absolute,Tata23}.

The $G$ cluster state which we study in this work was first introduced in Ref.~\onlinecite{brell_generalized_2015}, which generalizes cluster states to group-valued qudits, as Kitaev's quantum double model does for the toric code. Ref.~\onlinecite{brell_generalized_2015} conjectures that $G$ cluster states should exhibit SPT order and be usable for measurement-based quantum computation, ideas which we establish in this paper.

We extend results from the significant body of knowledge which has emerged from the study of SPTs with conventional symmetry. We study microscopic signatures of SPT order which are well-understood in conventional 1d SPTs, including edge modes and ground state degeneracy~\cite{Pollmann10,chen_symmetry-protected_2012-1,chen_symmetry_2013-1}, string order~\cite{den_nijs_preroughening_1989,kennedy_hidden_1992,pollmann_detection_2012,Else13,Bahri14}, and topological response~\cite{hatsugai_chern_1993,senthil_integer_2013,cheng_topological_2014,zaletel_flux_2014}. We also study the $G$ cluster state as a resource state for MBQC~\cite{gross2007novel,gross_measurement-based_2007}, which is known to be intimately related to cluster states in particular~\cite{briegel_measurement-based_2009, briegel_persistent_2001, raussendorf_measurement-based_2003, raussendorf_one-way_2001} and SPT order in general~\cite{stephen_computational_2017,wei_universal_2017-2,else_symmetry-protected_2012-2}.

\section{Notations and Conventions}
In this section, we introduce the notations and conventions which we use in this paper. Especially important is Sec.~\ref{sec:Paulis} which introduces the group-based Pauli operators we use extensively.
\subsection{Group Theory}
We denote by $G$ a finite group which is in general non-abelian. For each $g\in G$, the inverse of $g$ is denoted $\bar{g}$. The identity element is denoted $e$.

Irreducible representations are denoted $\Gamma$ and defined according to the map $\Gamma:G\rightarrow \text{GL}_{d_\Gamma}(\mathbb{C})$, where $d_\Gamma$ is the dimension of the representation. $\Gamma(g)$ is the $d_\Gamma\times d_\Gamma$ matrix which is obtained when the function $\Gamma$ is evaluated at the group element $g$. Being representations, the $\Gamma$ satisfy
\begin{subequations}
\begin{align}
    \Gamma(g)\Gamma(h)&=\Gamma(gh),\\
    \Gamma(g)^\dagger&=\Gamma(\bar g),\\
    \Gamma(g)^n&=\Gamma(g^n),\\
    \Gamma(e)&=\mathbbm{1}_{d_\Gamma}.
\end{align}
\end{subequations}
We denote the trivial representation $\mathbf{1}$, so that $\mathbf{1}(g)=1$ for all $g$. The tensor product of irreps of a finite group is fully decomposable and therefore isomorphic to a direct sum of irreps
\begin{equation}\label{eq:A_fusion}
    {\Gamma_i}\otimes{\Gamma_j}\simeq\bigoplus_k N^{\Gamma_k}_{\Gamma_i,\Gamma_j}{\Gamma_k},
\end{equation}
where the factor $N^{\Gamma_k}_{\Gamma_i,\Gamma_j}\in\mathbb{Z}_{\geq0}$ is called the multiplicity and gives the number of copies of $\Gamma_k$ appearing in the direct sum.

\subsection{Matrix Product States and Operators}

Matrix product states (MPS) are tensor network states in 1d that efficiently represent area law states. A general MPS is written
\begin{equation}
    \ket{\psi_A}=\sum_{\{g_i\}}\Tr[BA^{(1)}_{g_1}\cdots A^{(N)}_{g_N}]\ket{g_1,\ldots,g_N},
\end{equation}

\noindent where $A^{(i)}_g$ is the matrix evaluated at site $i$ when $g_i=g$ and $B$ encodes the boundary conditions. Diagrammatically, we can express each piece of the MPS as a rank-three (i.e., three-legged) tensor
\begin{equation}
    \sum_{g_i}A^{(i)}_{g_i}\otimes\ket{g_i}= \ \begin{tikzpicture}
    \draw (-0.5,0)--(0.5,0);
    \draw (0,0) -- (0,0.5);
    \node[even] (t) at (0,0) {};
  \end{tikzpicture}
\end{equation}
and the entire MPS succinctly as
\begin{equation}
\ket{\psi_A} = \ \begin{tikzpicture}
		  \draw[black] (0,0) rectangle (3,-0.5);
		  \foreach \x in {0.6,2.4}{
		    \draw (\x,0) --++ (0,0.5);
		    \node[even] (t\x) at (\x,0) {};
		  }
		  \node[fill=white] at (1.5,0) {$\dots$};
		  %\node[fill=white] at (1.5,-0.5) {\text{  }};
		  %\node[edge,black] at (1.3,-0.5) {};
		  %\node[edge,black] at (1.7,-0.5) {};
		  \node[edge,black,label=below:$B$] at (1.5,-0.5) {};
		\end{tikzpicture} \ .
\end{equation}
Similarly, matrix product operators (MPO) are tensor network \textit{operators} in 1d. A general MPO $O_A$ is written
\begin{equation}
    \sum_{\{g_i,g_i'\}}\Tr[BA^{(1)}_{g_1,g_1'}\cdots A^{(N)}_{g_N,g_N'}]\ketbra{g_1,\ldots,g_N}{g_1',\ldots,g_N'},
\end{equation}
where $A^{(i)}_{g,h}$ is the matrix evaluated at site $i$ when $g_i=g$ and $g_i'=h$, and $B$ encodes the boundary conditions. Diagrammatically, we write
\begin{align*}
    \sum_{g_i}A^{(i)}_{g_i,g_i'}\otimes\ketbra{g_i}{g_i'}= \ \begin{tikzpicture}
    \draw[red] (-0.5,0)--(0.5,0);
    \draw (0,-0.5) -- (0,0.5);
    \node[mpo] (t) at (0,0) {$A$};
  \end{tikzpicture} \\
O_A = \ \begin{tikzpicture}
		  \draw[red] (0,0) rectangle (3,-0.75);
		  \foreach \x in {0.6,2.4}{
		    \draw (\x,-0.5) --++ (0,1);
		    \node[mpo] (t\x) at (\x,0) {$A$};
		  }
		  \node[fill=white] at (1.5,0) {$\dots$};
		  %\node[fill=white] at (1.5,-0.5) {\text{  }};
		  %\node[edge,black] at (1.3,-0.5) {};
		  %\node[edge,black] at (1.7,-0.5) {};
		  \node[edge,red,label=below:$B$] at (1.5,-0.75) {};
		\end{tikzpicture} \ .
\end{align*}
Note that the trace is evaluated clockwise on the red MPO virtual level.

Sometimes, when we want to emphasize that the boundary conditions of an MPS or MPO are open, we will denote them as
\begin{equation*}
    \begin{tikzpicture}[baseline=0.4]
       \draw (0,0)--(2,0);
       \foreach \x in {0.35,1.65}{
       \draw (\x,0)--(\x,0.5);
       \node[even] at (\x,0) {};
       
       }
       \node[edge] at (0,0) {};
       \node[edge] at (2,0) {};
       \node[fill=white] at (1,0) {$\dots$};
    \end{tikzpicture} \ \quad,\quad \
    \begin{tikzpicture}[baseline=0.4]
		  \draw[red] (0,0) -- (3,0);
		  \foreach \x in {0.65,2.35}{
		    \draw (\x,-.4) --++ (0,0.9);
		    \node[mpo] (t\x) at (\x,0) {$A$};
		  }
		  \node[fill=white] at (1.5,-0.5) {\text{  }};
		  \node[edge,red] at (0,0) {};
		  \node[edge,red] at (3,0) {};
		  \node[fill=white] at (1.5,0) {$\dots$};
		\end{tikzpicture} \ .
\end{equation*}

\subsection{Group-Based Pauli Operators} \label{sec:Paulis}

\begingroup
\def\arraystretch{1.7}
\begin{table*}[t]
	\centering
	\begin{tabular}{|c|c||c|c|}
        \hline
		 $\mathbb{Z}_2$&  Qubit& Group-Valued Qudit&  $G$ \\
	  	\hline\hline
	  	Elements $\{0,1\}$& $\ket{0},\ket{1}$&$\ket{g}:g\in G$&Elements $\{g\in G\}$\\\hline
	  	Irreps $\{\Gamma_+,\Gamma_-\}$&$\ket{\pm}=\frac{1}{\sqrt{2}}\left(\Gamma_{\pm}(0)\ket{0}+\Gamma_{\pm}(1)\ket{1}\right)$&$\ket{\Gamma_{\alpha\beta}} = \sqrt\frac{d_{\Gamma}}{|G|}\sum_g [\Gamma(g)]_{\alpha\beta}\ket{g}$& Irreps $\{\Gamma\in \mathrm{Rep}(G)\}$\\\hline
	  	 Bit-flip &\begin{tabular}{c}$\lx_0=\rx_0=\mathbbm{1}$\\ $\lx_1=\rx_1=X$\end{tabular} &\begin{tabular}{c}$\lx_g\ket{h} =\ket{gh}$\\$\rx_g\ket{h} = \ket{h\bar{g}}$\end{tabular}&Group Multiplication\\\hline
         --- &\begin{tabular}{c}---\end{tabular} &\begin{tabular}{c}$\cx_g\ket{h} =\ket{hg\bar h}$\end{tabular}&Conjugation\\\hline

	  	Phase $\pm 1$&\begin{tabular}{c}$\zt_{\Gamma_+}=\zt_{\Gamma_+}^\dagger=\mathbbm{1}$\\ $\zt_{\Gamma_-}=\zt_{\Gamma_-}^\dagger=Z$\end{tabular}&\begin{tabular}{c}$\Tr[\zt_{\Gamma}]\ket{g}=\Tr[\Gamma(g)]\ket{g}$\\$[\zt_{\Gamma}]_{\alpha\beta}\ket{g}=[\Gamma(g)]_{\alpha\beta}\ket{g}$\end{tabular}& Phase $[\Gamma(g)]_{\alpha\beta}$ or $\Tr\left[\Gamma(g)\right]$\\\hline
	  	Controlled $X$& $CX\ket{g}\ket{h}=\ket{g}\ket{g\oplus h}$& \begin{tabular}{c}$C\lx\ket{g}\ket{h}=\ket{g}\ket{gh}$\\$C\rx\ket{g}\ket{h}=\ket{g}\ket{h\bar{g}}$\end{tabular} & Controlled Multiplication\\
    \hline
	\end{tabular}
	\caption{Summary of the group-valued qudit formalism, reproduced with minor edits from Ref.~\onlinecite{brell_generalized_2015}. Note that the group and irrep bases outlined in the first two rows of the table are dual to one another and related by the non-abelian Fourier transform.}
    \label{tab:group_qudit}
\end{table*}
\endgroup

In this section, we discuss the notation for group-based Pauli operators which will be used throughout this work, largely following the conventions of Ref.~\onlinecite{albert_spin_2021}. This construction is summarized in \cref{tab:group_qudit}. Additional details not discussed below can be found in Refs.~\onlinecite{albert_spin_2021,brell_generalized_2015}.

We work with a local Hilbert space whose basis kets are labelled by the elements of a finite group $G$, which we typically assume is non-Abelian. The local Hilbert space is the group algebra $\mathbb{C}[G]$, and arbitrary single-site states may be written as vectors, $\ket{\psi}=\sum_{g\in G}c_g\ket{g}$, with complex coefficients $c_g$.

We make extensive use of group-based Pauli operators which extend the action of the qudit Pauli operators to group-valued degrees of freedom. The group-based $X$-type operators are labeled by group elements $g$ and act by group multiplication\footnote{Elsewhere in the literature, these operators are sometimes called $L_g$ and $R_g$. We also note that our arrow convention is consistent with~\cite{albert_spin_2021} but reverses that of~\cite{brell_generalized_2015}.}:
\begin{subequations}
\begin{align}
    \lx_g=\sum_{h\in G}\ketbra{gh}{h}\quad\Leftrightarrow\quad \lx_g|h\rangle=|gh\rangle,\\
    \rx_g=\sum_{h\in G}\ketbra{h\bar{g}}{h}\quad\Leftrightarrow\quad \rx_g|h\rangle=|h\bar{g}\rangle.
\end{align}
\end{subequations}

Diagrammatically, we have
\begin{equation*}
\lx_g= \begin{tikzpicture}
    \draw (0,-0.5) -- (0,0.5);
    \node[mpo] at (0,0) {$\lx_g$};
  \end{tikzpicture}  \ , \ \quad \rx_g= \begin{tikzpicture}
    \draw (0,-0.5) -- (0,0.5);
    \node[mpo] at (0,0) {$\rx_g$};
  \end{tikzpicture}  \ .
\end{equation*}
Observe that these operators form the left and right regular representations~\cite{ArovasGroupTheory} of $G$: $\lx_g\lx_h=\lx_{gh}$ and $\rx_g\rx_h=\rx_{gh}$. In the case $G=\mathbb{Z}_d$, these operators reduce to the familiar qudit Pauli operators $X$, $\lx = \rx^\dagger =X$. The group-based $X$ operators also satisfy the following identities:
\begin{align}
\begin{split}
    \left(\lx_g\right)^n&=\lx_{g^n}, \quad  \left(\rx_g\right)^n=\rx_{g^n}\\
    \left(\lx_g\right)^\dagger&=\lx_{\bar g}, \quad  \left(\rx_g\right)^\dagger=\rx_{\bar g}.
\end{split}
\end{align}

The two types of $X$ operators can be combined to form the \textit{conjugation} operator:
\begin{equation}
    \cx_g:=\lx_g\rx_g=\rx_g\lx_g.
\end{equation}

The group-based $\zt$-type operators are labelled by irreps $\Gamma$ and act as generalized phase gates:
\begin{align}
    \zt_\Gamma=\sum_{g\in G}\Gamma(g)\otimes\ketbra{g}{g}\quad\Leftrightarrow\quad \zt_\Gamma|g\rangle=\Gamma(g)\otimes|g\rangle,\\
    \zt_\Gamma^\dagger=\sum_{g\in G}\Gamma(\bar{g})\otimes\ketbra{g}{g}\quad\Leftrightarrow\quad\zt_\Gamma^\dagger|g\rangle=\Gamma(\bar{g})\otimes|g\rangle.
\end{align}
These are matrix product operators (MPOs), with the irrep matrices acting in the virtual space. In graphical notation, we have
\begin{equation}
\zt_{\Gamma}=
\begin{tikzpicture}
    \draw[red] (-0.5,0)--(0.5,0);
    \draw (0,-0.5) -- (0,0.5);
    \node[mpo] (t) at (0,0) {$\zt_\Gamma$};
  \end{tikzpicture}
=\sum_{g\in G}\textcolor{red}{\Gamma(g)}\otimes \ketbra{g}{g}.
\end{equation}

In order to reduce these MPOs to operators acting solely on the physical degrees of freedom, we must contract over the virtual space. We can either do this by tracing over the space -- equivalent to imposing periodic boundary conditions -- which is denoted by
\begin{widetext}
\begin{equation*}
\Tr\left[\prod^{n}_{i=1}\zt_{\Gamma}^{(i)}\right]:=\sum_{g_i\in G}{\Tr\left[\Gamma\left(g_1\cdots g_n\right)\right]} \ketbra{g_1,\ldots,g_n}{g_1,\ldots,g_n}=
\begin{tikzpicture}
		  \draw[red] (0,0) rectangle (3,-0.5);
		  \foreach \x in {0.6,2.4}{
		    \draw (\x,-.4) --++ (0,0.9);
		    \node[mpo] (t\x) at (\x,0) {$\zt_\Gamma$};
		  }
		  \node[fill=white] at (1.5,0) {$\dots$};
		\end{tikzpicture} \ ,
\end{equation*}
or by choosing a particular matrix element of $\Gamma$, denoted by
\begin{equation*}
\left[\prod_{i=1}^n\zt_{\Gamma}^{(i)}\right]_{\alpha\beta}:=\sum_{g_i\in G}{\bra{\alpha}\Gamma\left(g_1\cdots g_n\right)\ket{\beta}} \ketbra{g_1,\ldots,g_n}{g_1,\ldots,g_n}=
		\begin{tikzpicture}
		  \draw[red] (0,0) -- (3,0);
		  \foreach \x in {0.6,2.4}{
		    \draw (\x,-.5) -- (\x,0.5);
		    \node[mpo] (t\x) at (\x,0) {$\zt_\Gamma$};
		  }
		  \node[fill=white] at (1.5,0) {$\dots$};
		  \node[fill=white] at (1.5,-0.5) {\text{  }};
		  \node[edge,red] at (0,0) {};
		  \node[edge,red] at (3,0) {};
		\end{tikzpicture} \ ,
\end{equation*}
\end{widetext}
where the product is taken in ascending order of indices (clockwise along the  loop of virtual links), $\begin{tikzpicture}
		  \draw[red] (0,0) -- (0.5,0);
		  \node[edge,red] at (0,0) {};
		\end{tikzpicture}=\bra{\alpha}$, and $\begin{tikzpicture}
		  \draw[red] (0.5,0) -- (1,0);
		  \node[edge,red] at (1,0) {};
		\end{tikzpicture}=\ket{\beta}$. Here, $\ket{\alpha}$ and $\ket{\beta}$ are states in the virtual Hilbert space of dimension $d_\Gamma$, not to be confused with physical states $\ket{g}$. Notice that $\zt_\Gamma$ is an operator acting on the physical degrees of freedom, while $\Gamma(g)$ is simply a matrix.

Sums over $\zt$-type operators labelled by different $\Gamma$ can be equivalently written as projectors, and this notation is sometimes used in other works. To go from the Pauli picture to the projector picture, one uses the delta function on the group, given by
\begin{equation}\label{eq:G_delta}
    \delta^G_{g,h}=\sum_{\Gamma\in\rep{G}} \frac{d_\Gamma}{|G|} \Tr[\Gamma(\bar{g}h)],
\end{equation}
where the sum is over the irreps $\Gamma$ of $G$, and where $d_{\Gamma}$ is the dimension of the $\Gamma$ irrep. For example, a sum over irreps of single-site $\Tr[\zt_\Gamma]$ operators projects onto the identity state:
\begin{align}
\sum_{\Gamma\in\rep{G}} \frac{d_\Gamma}{|G|} \Tr[\zt_\Gamma]=\ketbra{e}{e},
\end{align} 
and the generalized two-site Ising interaction projects onto the subspace with spins aligned:
\begin{equation}
\sum_\Gamma\frac{d_\Gamma}{|G|}\Tr\left[\zt^{(1)\dagger}_\Gamma.\zt^{(2)}_\Gamma\right]=\sum_g\ketbra{g,g}{g,g},
\end{equation}
where the `.' denotes multiplication in the virtual space.

Group-based Pauli operators have more subtle commutation relations than their qudit counterparts due to non-commutativity of operators on the virtual space; the order in which the virtual space matrices are contracted matters. This complication comes from the fact that $\Gamma(g_i)$ and $\Gamma(g_j)$ do not in general commute. Indeed, for any non-abelian group, there will exist some irrep $\Gamma$ with dimension $d_\Gamma>1$ and some $g_i,g_j$ such that $\Gamma(g_i)$ and $\Gamma(g_j)$ do not commute. This is important to keep in mind when working with the operators algebraically, but becomes more obvious when working diagrammatically.

The $X$ operators commute with one another when they act from different sides or on different sites, and follow the commutation rules of the group otherwise:
\begin{equation}
\begin{split}
    \left[\lx^{(i)}_g,\rx^{(j)}_h\right]&=0\\
    \left[\lx^{(i)}_g,\lx^{(j)}_h\right]\propto\left[\rx^{(i)}_g,\rx^{(j)}_h\right]&\propto \delta_{i,j}.
\end{split}
\end{equation}
The commutation relations between $X$ and $\zt$ operators on the same site are given by
\begin{align}
    \begin{split}
        \lx_g\zt_\Gamma =  \Gamma(\bar g).\zt_\Gamma\lx_g,\\
        \rx_g\zt_\Gamma =  \zt_\Gamma.\Gamma(g)\rx_g.
    \end{split}
\end{align}
Notice that these operators fail to commute up to a matrix $\Gamma(\bar{g})$ or $\Gamma(g)$ acting on the virtual space, rather than up to a phase as is the case for qubits.

In addition to the computational basis labelled by the group elements, there exists a dual basis labeled by matrix elements of irreps. These states are defined to be
\begin{equation}
    |\Gamma_{\alpha\beta}\rangle:=\sqrt{\frac{d_\Gamma}{|G|}}\sum_{g\in G}[\Gamma(g)]_{\alpha\beta}|g\rangle.
\end{equation}

It can be shown using the Grand Orthogonality Theorem of group representations~\cite{ArovasGroupTheory} that these states form a complete basis for $\mathbb{C}[G]$. In the qubit case $G=\mathbb{Z}_2$, these states indeed reduce to the states $\ket{+}$ and $\ket{-}$, corresponding to $\Gamma$ being the trivial irrep and sign irrep, respectively\footnote{$\mathbb{Z}_2$ is abelian and therefore only has one dimensional irreps, making the use of indices $\alpha\beta$ unnecessary in this case.}.

It is also sometimes convenient to think of the set of character states
\begin{equation}
    |\Gamma\rangle:=\sqrt{\frac{d_\Gamma}{|G|}}\sum_{g\in G}\Tr[\Gamma(g)]|g\rangle,
\end{equation}
which span a subspace of $\mathbb{C}[G]$.
%%%%%%%%%%%%%%%%%%%%%%%%%%%%%%%%%%%%%%%%%%%%%%%%%%%%%%%%%%%%%%%%%%
\section{$G$ Cluster State}\label{sec:G_cluster_state}
The family of cluster states based on finite groups was introduced by Brell in Ref.~\onlinecite{brell_generalized_2015}. Brell set out to generalize cluster states to a group-valued Hilbert space, analogous to the relationship between the toric code and Kitaev's quantum double. However, the cluster states presented an additional hurdle not present in the toric code: stabilizers which mixed $X$ and $Z$ operators. To use the language of quantum error correction, the toric code is a CSS stabilizer code \eczoo[]{css}, and the $ZXZ$ cluster state is not.

The difficulty of working with mixed stabilizers arises from the fact that there is no natural isomorphism between group elements and irreducible representations of a non-abelian group. The qubit cluster state is stabilized by $ZXZ$ operators. These would naively generalize to stabilizers like $\zt^\dagger_\Gamma\lx_g\zt_\Gamma$, but it is not obvious which $g$ should be paired with which $\Gamma$ in a given term. For $\ztwo$, for example, the natural pairing between the computational and dual basis is $\ket{0}\leftrightarrow \ket{+}$ and $\ket{1}\leftrightarrow\ket{-}$. When $G$ is non-abelian, however, there is no natural pairing, in part because there are more group elements than irreps.

\begin{figure}[t]
    \includegraphics[width=0.47\textwidth]{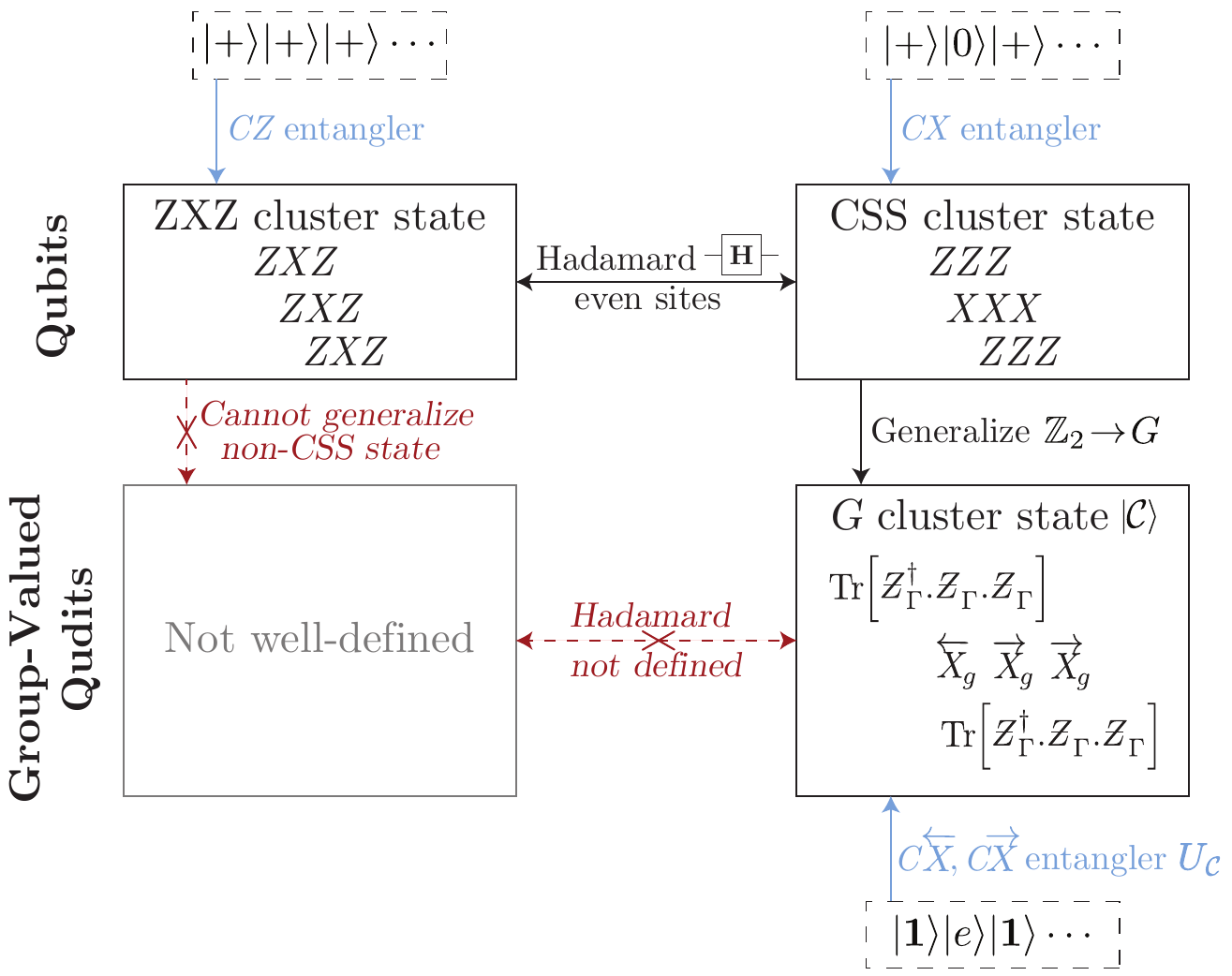}
    \caption{The relationship between the various cluster states discussed in~\cite{brell_generalized_2015}. Each state is listed along with the structure of its stabilizers. The $G$ cluster state $\ket{\Cl}$ corresponds to a particular choice of directed entangler graph.}
    \label{fig:clusters}
\end{figure}

Brell circumvented this issue by introducing the CSS cluster state (see \cref{fig:clusters}), which can be generalized from qubits to group-valued Hilbert spaces. The CSS cluster state is generated by acting on the 1d cluster state with a Hadamard gate on the even sites. The CSS cluster state has new stabilizers, which are also obtained through the action of the Hadamard gate. The stabilizers beginning on odd sites are transformed from $ZXZ$ to $ZZZ$, and the stabilizers beginning on even sites from $ZXZ$ to $XXX$. Since the stabilizers now consist of only a single type of Pauli operator, the new state is a CSS stabilizer state.

The CSS cluster state can also be prepared by acting with a circuit of $CX$ gates on the product state $\ket{+}\ket{0}\ket{+}\cdots$. This method of state preparation generalizes to the finite-group case. Noticing that, for $G=\mathbb{Z}_2$, $\Gamma_+$ is the trivial irrep and $0$ is the identity element, we immediately find that the product state used in the preparation of the $G$ cluster state should be $\ket{\mathbf{1}}\ket{e}\ket{\mathbf{1}}\cdots$, where $e\in G$ is the identity and $\mathbf{1}\in\rep G$ is the trivial irrep.

While a simple graph is sufficient to specify a typical qubit-based cluster state, the graph necessary to specify a $G$ cluster state must be bipartite and directed~\cite{brell_generalized_2015}.
There is additional information which must be included to specify a $G$ cluster state in higher dimensions, but we will forgo that discussion since we are working in one dimension.

The graph must be bipartite for the $G$ cluster state -- and even the qubit CSS cluster state -- because $CX$ and its generalizations -- $C\lx$, and $C\rx$ -- are not symmetric with respect to control and target sites, unlike $CZ$. With a bipartite graph, it is possible to assign one subset to be the control qudits and the other the targets, removing this ambiguity.

The graph must be directed because there are two different group-based $CX$ operators
\begin{equation}
\begin{split}
    C\lx^{(i,j)}\ket{g_i,g_j}:=\ket{g_i,g_ig_j},\\ C\rx^{(i,j)}\ket{g_i,g_j}:=\ket{g_i,g_j\bar{g_i}},
\end{split}
\end{equation}
and the arrow on each edge can be used to specify whether $C\lx$ or $C\rx$ should be applied. For the same reason, the Kitaev quantum double is also defined on a directed graph~\cite{kitaev_fault-tolerant_2003-2}. Though there is a unique 1d qubit CSS cluster state, there are many possible 1d $G$ cluster states because of the choice of the orientation of each edge.

The model which we study in this work is a particular group-based cluster state specified by a simple choice of graph -- odd-site controls and even-sites targets, with all edges oriented to the left -- which we call the $G$ \textit{cluster state}. It is defined on an open 1d chain with $n$ sites, each of which has local Hilbert space $\mathbb{C}[G]$.
The $G$ cluster state is given by~\cite{brell_generalized_2015}:
\begin{equation}\label{eq:true_ground}
    |\mathcal{C}\rangle=\mathcal{N}\sum_{\{g_i\}\in G}|g_1\rangle|g_1\bar{g_2}\rangle|g_2\rangle\cdots|g_{N-1}\bar{g_N}\rangle|g_N\rangle,
\end{equation}
where here and hereafter $\mathcal{N}$ is a normalization factor which ensures that states have unit norm.

In the case $G=\mathbb{Z}_2$, the above state reduces to the CSS cluster state as expected. In \cref{ap:CSS_cluster}, we overview the main results applied specifically to the CSS cluster state, which we generalize to the $G$ cluster state in the following sections.

Notice that the $G$ cluster state $\ket{\Cl}$ can be prepared from the product state $\ket{\psi_0}:=\ket{\mathbf{1}}\ket{e}\ket{\mathbf{1}}\cdots$ in finite depth by the circuit
\begin{equation}\label{eq:U_C}
    U_\mathcal{C}=\prod_{i\text{ odd}}C\lx^{(i,i+1)}C\rx^{(i-1,i)}.
\end{equation}
In Sec.~\ref{sec:repeated_U_C}, we will study the family of SPT states obtained by repeatedly applying $U_\Cl$ to $\ket{\psi_0}$.

\subsection{MPS Representation}
There exists an MPS representation for the $G$ cluster state.
By regrouping indices of the PEPS tensors given in Ref.~\onlinecite{brell_generalized_2015}, we arrive at the following MPS tensors:
\begin{equation}\label{eq:MPS_tensors}
  \begin{tikzpicture}
    % \node[irrep] at (0.2,0) {};
    \draw (-0.5,0)--(0.5,0);
    \draw (0,0) -- (0,0.5);
    \node[odd] (t) at (0,0) {};
  \end{tikzpicture}
\ =\sum_{g\in G}\ketbra{g}{g}\otimes\ket{g},\quad
  \begin{tikzpicture}
    % \node[irrep] at (0.2,0) {};
    \draw (-0.5,0)--(0.5,0);
    \draw (0,0) -- (0,0.5);
    \node[even] (t) at (0,0) {};

  \end{tikzpicture}
\ =\sum_{g\in G} \lx_g\otimes\ket{g}~,
\end{equation}
which we place on odd and even sites, respectively. The state $\ket{\Cl}$ on an open chain can then be written as
\begin{equation}\label{eq:cluster_wvfnctn}
    \ket{\Cl}=\begin{tikzpicture}
    \draw (0,0)--(5.5,0);
    \foreach \x in {0.5,2,3.5,5}{
		     \draw (\x,0) -- (\x,0.5);
		    \node[odd] at (\x,0) {};
		  }
		  \foreach \x in {1.25,4.25}{
		     \draw (\x,0) -- (\x,0.5);
		    \node[even] at (\x,0) {};
		  }
		  \node[fill=white] at (2.75,0) {$\dots$};
		  \node[edge] at (0,0) {};
		  \node[edge] at (5.5,0) {};
  \end{tikzpicture} \ .
\end{equation}

The above tensors are closely related to other objects in physics and math. When $G=\ztwo$, the tensors \cref{eq:MPS_tensors} are exactly the ``three-legged spiders'' that appear in ZX-calculus (see e.g.~\cite{vandewetering2020zxcalculus}). When $G$ is a finite group, the odd and even tensors enact comultiplication ($g\to g\otimes g$) and multiplication ($g\otimes h\to hg$), respectively (see e.g.~\cite{bauer2016symmetries}).

Here, we catalog tensor network identities for the group-based Pauli operators which we will use extensively through the rest of the paper:
\begin{subequations}
\begin{equation}\label{eq:pulling_through_main}
  \begin{tikzpicture}[baseline={(0,.4)}]
    \draw (-1,0)--(1,0);
    \draw (0,0) -- (0,1);
    \node[odd] (t) at (0,0) {};
    % \node[] at (0.2,0.2) {o};
    \node[mpo] at (0,0.5) {$\lx_g$};
    % \node[] at (0.2,0.7) {$g$};

  \end{tikzpicture}
\ = \begin{tikzpicture}[baseline={(0,.4)}]
    \draw (-1,0)--(1,0);
    \draw (0,0) -- (0,1);
    \node[odd] (t) at (0,0) {};
    % \node[] at (0.2,0.2) {o};
    \node[mpo] at (-0.5,0) {$\lx_g^\dagger$};
     \node[mpo] at (0.5,0) {$\lx_g$};
    % \node[] at (-0.3,0.2) {$g$};
  \end{tikzpicture} \
\end{equation}

\vspace{-1 em}
\begin{equation}
  \begin{tikzpicture}[baseline={(0,.4)}]
    \draw (-1,0)--(1,0);
    \draw (0,0) -- (0,1);
    \node[odd] (t) at (0,0) {};
    % \node[] at (0.2,0.2) {o};
    \node[mpo] at (0,0.5) {$\rx_g$};
    % \node[] at (0.2,0.7) {$g$};

  \end{tikzpicture}
\ = \begin{tikzpicture}[baseline={(0,.4)}]
    \draw (-1,0)--(1,0);
    \draw (0,0) -- (0,1);
    \node[odd] (t) at (0,0) {};
    % \node[] at (0.2,0.2) {o};
    \node[mpo] at (-0.5,0) {$\rx_g^\dagger$};
     \node[mpo] at (0.5,0) {$\rx_g$};
    % \node[] at (-0.3,0.2) {$g$};

  \end{tikzpicture} \
\end{equation}

\vspace{-1 em}
\begin{equation}
  \begin{tikzpicture}[baseline={(0,.4)}]
    \draw (-1,0)--(1,0);
    \draw (0,0) -- (0,1);
    \node[even] (t) at (0,0) {};
    % \node[] at (0.2,0.2) {o};
    \node[mpo] at (0,0.5) {$\lx_g$};
    % \node[] at (0.2,0.7) {$g$};

  \end{tikzpicture}
\ = \begin{tikzpicture}[baseline={(0,.4)}]
    \draw (-1,0)--(1,0);
    \draw (0,0) -- (0,1);
    \node[even] (t) at (0,0) {};
    % \node[] at (0.2,0.2) {o};
    \node[mpo] at (-0.5,0) {$\lx_g^\dagger$};
    % \node[] at (-0.3,0.2) {$g$};

  \end{tikzpicture} \
\end{equation}

\vspace{-1 em}
\begin{equation}
\begin{tikzpicture}[baseline={(0,.4)}]
    \draw (-1,0)--(1,0);
    \draw (0,0) -- (0,1);
    \node[even] (t) at (0,0) {};
    % \node[] at (0.2,0.2) {o};
    \node[mpo] at (0,0.5) {$\rx_g$};
    % \node[] at (0.2,0.7) {$g$};

  \end{tikzpicture}
\ = \begin{tikzpicture}[baseline={(0,.4)}]
    \draw (-1,0)--(1,0);
    \draw (0,0) -- (0,1);
    \node[even] (t) at (0,0) {};
    % \node[] at (0.2,0.2) {o};
    \node[mpo] at (0.5,0) {$\lx_g$};
    % \node[] at (0.7,0.2) {$\bar g$};
  \end{tikzpicture}
\end{equation}

\vspace{-1 em}
\begin{equation}
  \begin{tikzpicture}[baseline={(0,.4)}]
    \draw (-1,0)--(1,0);
    \draw (0,0) -- (0,1);
    \draw[red] (-1,0.5)--(1,0.5);
    \node[odd] (t) at (0,0) {};
    % \node[] at (0.2,0.2) {o};
    \node[mpo] at (0,0.5) {$\zt_\Gamma$};
    % \node[] at (0.2,0.7) {$g$};

  \end{tikzpicture}
\ = \begin{tikzpicture}[baseline={(0,.4)}]
    \draw (-1,0)--(1,0);
    \draw (0,0) -- (0,1);
    \draw[red] (-1,-0.2)--(1,-0.2);
    \node[odd] (t) at (0,0) {};
    % \node[] at (0.2,0.2) {o};
    \node[mpo] at (-0.5,-0.1) {$\zt_\Gamma$};
    % \node[] at (-0.3,0.2) {$g$};
  \end{tikzpicture} \
  = \begin{tikzpicture}[baseline={(0,.4)}]
    \draw (-1,0)--(1,0);
    \draw (0,0) -- (0,1);
    \draw[red] (-1,-0.2)--(1,-0.2);
    \node[odd] (t) at (0,0) {};
    % \node[] at (0.2,0.2) {o};
    \node[mpo] at (0.5,-0.1) {$\zt_\Gamma$};
    % \node[] at (-0.3,0.2) {$g$};
  \end{tikzpicture} \
\end{equation}

\vspace{-1 em}
\begin{equation}
  \begin{tikzpicture}[baseline={(0,.4)}]
    \draw (-1,0)--(1,0);
    \draw (0,0) -- (0,1);
    \draw[red] (-1,0.5)--(1,0.5);
    \node[even] (t) at (0,0) {};
    % \node[] at (0.2,0.2) {o};
    \node[mpo] at (0,0.5) {$\zt_\Gamma$};
    % \node[] at (0.2,0.7) {$g$};

  \end{tikzpicture}
\ = \begin{tikzpicture}[baseline={(0,.4)}]
    \draw (-1,0)--(1,0);
    \draw (0,0) -- (0,1);
    \draw[red] (-1,-0.2)--(1,-0.2);
    \node[even] (t) at (0,0) {};
    % \node[] at (0.2,0.2) {o};
    \node[mpo] at (-0.5,-0.1) {$\zt_\Gamma$};
     \node[mpo] at (0.5,-0.1) {$\zt_\Gamma^\dagger$};
    % \node[] at (-0.3,0.2) {$g$};
  \end{tikzpicture} \
\end{equation}
\end{subequations}
We also collect in \cref{tab:tensors} the explicit definition of each tensor.

\subsection{Stabilizer Hamiltonian and Symmetries}
We can construct a Hamiltonian with ground states $\ket{\mathcal{C}}$ as a sum of commuting stabilizers:
\begin{equation}\label{eq:ham}
\begin{split}
    H_\Cl=-\frac{1}{|G|}\sum_{i\text{ odd}}\Bigg(\sum_{\Gamma\in\rep{G}}\Tr\left[\zt^{\dagger(i)}_\Gamma.\zt^{(i+1)}_\Gamma.\zt^{(i+2)}_\Gamma\right]d_\Gamma\\+\sum_{g\in G}\rx_g^{(i+1)}\lx_g^{(i+2)}\lx_g^{(i+3)}\Bigg),
\end{split}
\end{equation}
where the sum over $\Gamma\in\rep{G}$ denotes a sum over the simple objects of $\rep{G}$, namely the irreps $\Gamma$ of $G$. We call this a (generalized) stabilizer Hamiltonian because the group-based Pauli terms all commute with one another, and the ground state space is the joint eigenspace of all operators with maximum eigenvalue. For the $X$-type terms, this maximum eigenvalue is $1/|G|$, and for the $Z$-type terms, it is $d_\Gamma^2/|G|$. We choose to normalize to these values rather than to 1 because the Hamiltonian also becomes a commuting projector Hamiltonian when written with this normalization.

The Hamiltonian \cref{eq:ham} has \textit{four} independent families of global symmetries, all of which are respected by the ground state $\ket{\mathcal{C}}$. There are two $G$ symmetries, given by tensor products of $X$-type operators, a $\rep G$ symmetry comprised of $Z$-type operators which acts as an MPO, and an $\inn G$ symmetry:
\begin{subequations}
    \begin{align}
    G_R: \overleftarrow{A}_g &:= \prod_{i\text{ odd}}\rx_g^{(i)},\\
    G_L: \overrightarrow{A}_g &:= \prod_{i\text{ odd}}\lx_g^{(i)} \cx_g^{(i+1)},\\
    \label{eq:repG} \rep G:\hat B_\Gamma &:= \Tr\left[\prod_{i\text{ even}}\zt^{(i)}_\Gamma\right],\\
    \inn G: \hat C_g&:=\prod_{i\text{ even}}\left(\sum_{g'}\rx_{g'g\bar g '}^{(i)}\delta_{g'}^{(i+1)}\lx_{g'g\bar g '}^{(i+2)}\right),
\end{align}
\end{subequations}
where $\inn G$ is the inner automorphism group of $G$. $\inn G \cong G/Z(G)$, where $Z(G)$ is the center of $G$, the subgroup of elements in $G$ which commute with every other element of $G$.

For abelian groups, the two $G$ symmetries collapse to one, and the $\inn G$ symmetry becomes trivial, yielding only two symmetries, $G\times \rep G$, that protect the SPT.
Accordingly, we have determined analytically and numerically that $G_R\times\rep G$ is the minimal subgroup which protects the SPT in the non-abelian case as well, so we focus on these symmetries for the remainder of the text.
We discuss the additional symmetries in \cref{ap:additional_sym} and reserve a full discussion of alternative avenues to protection to future work.

\begingroup
\begin{table}
\begin{tabular}{ r l | c l }
\hline\hline
  \multicolumn{2}{c}{MPS}\vline& \multicolumn{2}{c}{MPO} \\
 \hline
 \begin{tikzpicture}
    % \node[irrep] at (0.2,0) {};
    \draw (-0.5,0)--(0.5,0);
    \draw (0,0) -- (0,0.5);
    \node[odd] (t) at (0,0) {};
  \end{tikzpicture}
\ &$=\underset{g\in G}{\sum}\ketbra{g}{g}\otimes\ket{g}$ & \begin{tikzpicture}
    % \draw [white] (0,-0.6) -- (0,0.6);
    \draw (0,-0.5) -- (0,0.5);
    \node[mpo] at (0,0) {$\lx_g$};
  \end{tikzpicture}  & $=\underset{h\in G}{\sum}\mathbbm{1}\otimes\ketbra{gh}{h}$  \\
 \begin{tikzpicture}
    % \node[irrep] at (0.2,0) {};
    \draw (-0.5,0)--(0.5,0);
    \draw (0,0) -- (0,0.5);
    \node[even] (t) at (0,0) {};

  \end{tikzpicture}
\ &$=\underset{g\in G}{\sum} \lx_g\otimes\ket{g}$ & \begin{tikzpicture}
    % \draw [white] (0,-0.6) -- (0,0.6);
    \draw (0,-0.5) -- (0,0.5);
    \node[mpo] at (0,0) {$\rx_g$};
  \end{tikzpicture} & $=\underset{h\in G}{\sum}\mathbbm{1}\otimes\ketbra{h\bar{g}}{h}$   \\

 \begin{tikzpicture}
    % \node[irrep] at (0.2,0) {};
    \draw (-0.5,0)--(0.5,0);
    \draw (0,0) -- (0,0.5);
    \node[odd] (t) at (0,0) {};
    \node[oddn] (t) at (0,0) {};
  \end{tikzpicture}
\ &$=\underset{g\in G}{\sum}\ketbra{g^n}{g^n}\otimes\ket{g}$ & \begin{tikzpicture}
    % \draw [white] (0,-0.6) -- (0,0.6);
    \draw (0,-0.5) -- (0,0.5);
    \draw [red] (-0.5,0) -- (0.5,0);
    \node[mpo] at (0,0) {$\zt_\Gamma$};
  \end{tikzpicture} & $=\underset{g\in G}{\sum}\Gamma(g)\otimes \ketbra{g}{g}$   \\
 \hline\hline
\end{tabular}
\caption{Summary of matrix product state and operator tensors used throughout this work.}
\label{tab:tensors}
\end{table}

\endgroup

Notice that $\overleftarrow{A}_g$ and $\hat B_\Gamma$  commute with one another, as they are supported on different sublattices. The $\overleftarrow A$ operators form a representation of the group $G$, as
\begin{equation}
    \overleftarrow{A}_{g}\overleftarrow{A}_{h}=\overleftarrow{A}_{gh}.
\end{equation}

Diagrammatically, $\hat B_\Gamma$ can be written
\begin{equation*}
\hat B_\Gamma = \begin{tikzpicture}
		  \draw[red] (0,0) rectangle (3,-0.5);
		  \foreach \x in {0.6,2.4}{
		    \draw (\x,-.4) --++ (0,0.9);
		    \node[mpo] (t\x) at (\x,0) {$\zt_\Gamma$};
		  }
		  \node[fill=white] at (1.5,0) {$\dots$};
		\end{tikzpicture} \ .
\end{equation*}
When we act on an open chain, we will instead use the open form of the symmetry operator:
\begin{equation}
\hat B_{\Gamma_{\alpha\beta}} :=\left[\prod_{i\text{ even}}\zt^{(i)}_\Gamma\right]_{\alpha\beta}= \begin{tikzpicture}
		  \draw[red] (0,0) rectangle (3,0);
		  \foreach \x in {0.6,2.4}{
		    \draw (\x,-.4) --++ (0,0.9);
		    \node[mpo] (t\x) at (\x,0) {$\zt_\Gamma$};
		  }
		  \node[fill=white] at (1.5,0) {$\dots$};
    \node[edge,red] at (0,0) {};
    \node[edge,red] at (3,0) {};
		\end{tikzpicture} \ .
\end{equation}

The $\zt$-type operators do not generically multiply according to the rules of a group. As derived in Appendix \ref{algebraic_deriv}, the $\hat B$ operators multiply according to the rules of the fusion category $\rep G$, the category of representations of $G$:
\begin{equation}\label{eq:B_fusion}
    \hat B_{\Gamma_i}\hat B_{\Gamma_j}=\sum_k N^{\Gamma_k}_{\Gamma_i,\Gamma_j}\hat B_{\Gamma_k},
\end{equation}
where the $N^{\Gamma_k}_{\Gamma_i,\Gamma_j}$ encode the fusion rules of $\rep G$. We say that the $\hat B$ operators realize a \textit{fusion category} symmetry ~\cite{thorngren_fusion_2019,kong_algebraic_2020}. Such symmetries have also occasionally been called \textit{algebraic} symmetries in the literature.

Fusion category symmetries are also sometimes called non-invertible symmetries~\cite{heidenreich_non-invertible_2021-1,bhardwaj_universal_2022}, especially in the high energy context. This applies to our symmetry in that $\zt_\Gamma$ does not in general have an inverse. This is a consequence of the fact that for each simple object $\Gamma$ in a fusion category, there must exist a dual simple object $\bar\Gamma$ such that $\Gamma\times\bar\Gamma=\mathbf{1}+\cdots$. That is, each object must have a dual which fuses with it to yield the trivial object, but this need not be the only fusion channel.

In the case of abelian $G$, the algebraic structure in \cref{eq:B_fusion} reduces to a group structure because $\rep{G}\cong G$ when $G$ is abelian. This is to say that the character group is isomorphic to the group when $G$ is abelian. We will focus on the case where $G$ is non-abelian, so the $\zt$-type symmetry is not group-like.

\subsection{On $G\times\rep G$}
Together, the $\overleftarrow{A}$ and $\hat B$ operators realize a $G\times \rep G$ fusion category symmetry. Simple objects in this category can be labeled by the pair $(g,\Gamma)$ and satisfy the fusion rules
\begin{equation}\label{eq:cat_fusion}
    \left(g,{\Gamma_i} \right)\otimes\left(h, {\Gamma_j} \right)=\bigoplus_k N^{\Gamma_k}_{\Gamma_i,\Gamma_j}\left({gh}, {\Gamma_k}\right).
\end{equation}
Accordingly, the symmetry operators $\overleftarrow{A}_g\hat B_\Gamma$ multiply as
\begin{equation}\label{eq:op_fusion}
    \left(\overleftarrow{A}_g\hat B_{\Gamma_i} \right)\left(\overleftarrow{A}_h\hat B_{\Gamma_j} \right)=\sum_k N^{\Gamma_k}_{\Gamma_i,\Gamma_j}\overleftarrow{A}_{gh}\hat B_{\Gamma_k}.
\end{equation}
The symmetry category is a direct product of $G$ and $\rep{G}$ because the operators realizing the $G$ and $\rep G$ symmetries commute and therefore act entirely independently.

Interestingly, $G\times \rep G$ has non-commutative fusion operation $\times$, as $gh\neq hg$ in general. This is not to be confused with the statement that fusion rules are said to be non-abelian when $N^{\Gamma_k}_{\Gamma_i,\Gamma_j}$ is nonzero for multiple $\Gamma_k$, given fixed $\Gamma_i$ and $\Gamma_j$. The latter terminology comes from the fact that the fusion rules of quantum double anyons derived from non-abelian groups satisfy such a condition. However, the fusion operation of non-abelian anyons is still commutative in the sense that $a\times b=b\times a$ for two non-abelian anyons $a$ and $b$. Indeed, it is not possible to construct a physical theory of anyons based on the fusion rules given in \cref{eq:cat_fusion}. This is because commutativity of the fusion operation $\times$ is necessary for the braiding operation to be well-defined~\cite{simon_topological_2021}.

The fact that the fusion operation of $G\times\rep G$ is non-commutative also implies that it is not the fusion algebra of the irreps of any finite group: $G\times\rep G\neq \rep{G'}$ for any $G'$. This is because the fusion of irreps of any finite group is always commutative -- though, of course, not always abelian in the single-fusion-channel sense. However, every non-anomalous fusion category is isomorphic to the representation category of some semisimple Hopf algebra~\cite{etingof_fusion_2017}. This implies that there exists some semisimple Hopf algebra $H$ such that $G\times\rep G\cong \rep H$, and indeed this relation is satisfied for $H={\mathbb{C}[G]^*\otimes\mathbb{C}[G]}$. If one were to construct commuting projector Hamiltonians for phases with $G\times \rep G$ symmetry following the prescription of~\cite{inamura_lattice_2022}, this is the Hopf algebra one would use.

%%%%%%%%%%%%%%%%%%%%%%%%%%%%%%%%%%%%%%%%%%%%%%%%%%%%%%%%%%%%%%%%%%

%%%%%%%%%%%%%%%%%%%%%%%%%%%%%%%%%%%%%%%%%%%%%%%%%%%%%%%%%%%%%%%%%%%%%%%%%%%%%%%%%%%%%%%%%%%%%%%%%%%%%%%%%%%%%%%%%%%%%%%%%%%%%
\section{Distinctness from the Symmetric Product State}\label{sec:duality}
In this section, we will first discuss what it means for a state to be protected by a fusion category symmetry and understand what notion of triviality and non-triviality exists for such states. We will then construct a Kramers-Wannier duality which maps the $G\times \rep G$-symmetric phases in which we are interested to $G\times G$ spontaneous symmetry breaking phases which can be more easily understood. This duality allows us to conclude that the $G$ cluster state is non-trivial, in that it is not adiabatically connected to the symmetric $G\times \rep G$ product state via any symmetric path of Hamiltonians, and that this order is robust to weak symmetric local perturbations.

\subsection{Definition of SPT with fusion-category symmetry}

Given a fusion category $\mathcal{A}$, we define an $\mathcal{A}$-SPT state as the unique ground state of a gapped Hamiltonian which respects the symmetry $\mathcal{A}$ under periodic boundary conditions. Two $\mathcal{A}$-symmetric states belong to distinct phases if they cannot be deformed into one another by a finite time evolution of a symmetric local Hamiltonian~\cite{kong_algebraic_2020}. $\ket{\psi_0}:=\ket{\mathbf{1},e,\mathbf{1},e,\ldots}$ is a product state that respects the $G\times \rep G$ symmetry, so we can use it as a representative symmetric product state. Since the state $\ket{\mathcal{C}}$ is also a $G\times \rep G$-symmetric state, our goal is to show that it belongs to a distinct phase from the product state $\ket{\psi_0}$.

It is important to emphasize here and throughout the paper that we will be intentionally using the phrase (symmetric) \textit{product} state in lieu of the \textit{trivial} SPT state. To clarify our nomenclature, there are two possible definitions one can use to define a trivial SPT phase
\begin{enumerate}
    \item It is the identity element of the group of SPT phases with stacking as a group operation
    \item It is the phase to which the symmetric product state belongs
\end{enumerate}
In the case of invertible symmetries, these two definitions are equivalent. However, SPTs protected by non-invertible symmetries cannot be stacked, and therefore only form a set rather than a group. To see this, it is helpful to review how to stack SPT phases when the symmetry is unitary. Consider two Hilbert spaces $\mathcal H_1$ and $\mathcal H_2$ each hosting a $G$-symmetric SPT phase $\ket{\psi_1}$ and $\ket{\psi_2}$, respectively. To stack the two phases, we may consider the tensor product state $\ket{\psi_1} \otimes \ket{\psi_2}$ in $\mathcal H_1 \otimes \mathcal H_2$ and restrict the full symmetry $G\times G$ to the diagonal subgroup $G$ which acts on both copies at the same time. Then, given a choice of a local symmetry action, the trivial SPT phase then contains all wavefunctions $\ket{\psi_0}$ such that $\ket{\psi} \otimes \ket{\psi_0}$ is in the same phase as $\ket{\psi}$ as $G$-symmetric phases. In contrast, a general fusion-category symmetry $\mathcal A \times \mathcal A$ may not contain $\mathcal A$ as a diagonal subcategory. Thus there is generally no notion of stacking, and therefore the trivial SPT phase in the first sense cannot be defined.

On the other hand, having chosen a tensor product Hilbert space and a representation for the symmetry, the symmetric product state (if it exists, given that particular symmetry action) lives in one and only one of these SPT phases. In such cases, there is a ``special'' SPT phase containing the symmetric product state\footnote{If there is no symmetric product state to begin with, there is still a basis transformation that turns one of the fixed-point SPT phases into a symmetric product state, but this choice is not canonical.}. Thus, the trivial SPT phase in the second sense can be defined if the fixed symmetry action admits a symmetric product state.

 Relatedly, although such SPT phases can be unambiguously called \textit{short-range entangled} phases since they can be deformed to a product state by explicitly breaking the symmetry, it is ambiguous to define whether an SPT state is \textit{invertible}. In the first sense, invertibility is not well defined due to the lack of a group structure. In the second sense, it is possible the unitary can be used to permute between different SPT phases. Since every unitary has an inverse, these SPTs can be thought of as invertible in the second sense if the symmetric product state is chosen as a base point. However, if we do not declare the symmetric product state as a special phase, then this unitary only gives rise to a group action on the set of SPT phases (i.e. a \textit{torsor}~\cite{thorngren_fusion_2019}).

\subsection{Duality argument}\label{sec:dualityargument}
We construct a Kramers-Wannier duality~\cite{lootens2023dualities} which associates the $G$ cluster state $\ket{\mathcal{C}}$ and the $G\times\rep{G}$-symmetric product state $\ket{\psi_0}$ to two different $G\times G$ symmetry-breaking states. States with different patterns of spontaneous symmetry breaking belong to different phases, implying through the duality that $\ket{\mathcal{C}}$ and $\ket{\psi_0}$ also belong to different phases.

We begin by constructing two models which spontaneously break $G\times G$ symmetry:

\begingroup
\setlength{\tabcolsep}{0pt}
\def\arraystretch{1.7}
\begin{table*}
    \centering
    \resizebox{\textwidth}{!}{\begin{tabular}{|c||c|c|c|}
        \cline{1-3} \large Symmetry & \multicolumn{2}{c|}{\large States} & \multicolumn{1}{l}{}\\\hhline{===-}
        \multirow{8}{*}{\large{$G\times G$}} & \myred \textit{SSB1}  & \myblue \textit{SSB2}  & Phase \\\hhline{|~||---}
         & \myred
         $\left\{\ket{\mathbf{1}\gap,\gap g\gap,\gap\mathbf{1}\gap,\gap g,\gap\mathbf{1}\gap,\gap g\gap,\gap\ldots}:g\in G\right\}$ & \myblue $\left\{\sum_{\{g_i\}\in G}|g_1,g_1g,g_2,g_{2}g,g_3,g_3g,\ldots\rangle:g\in G\right\}$ & \quad\makecell{ Ground State \\ Manifold }{\quad} \\\hhline{|~||---}
         & \myred \hspace{-7.35em}$\Tr[\zt_\Gamma^\dagger.\mathbbm{1}.\zt_\Gamma]$ & \myblue \hspace{-5.3em}$\Tr[\zt_\Gamma^\dagger.\zt_\Gamma.\zt_\Gamma\overset{\leftrightarrow}{.}\zt_\Gamma^\dagger]$ & \multirow{4}{*}{Stabilizers} \\
         & \myred \hspace{-6.3em} $\lx_g$& \myblue \hspace{-1em}$\lx_g\gap\gap\gap\lx_g$&  \\
         & \myred \hspace{-2.5em} $\Tr[\zt_\Gamma^\dagger.\mathbbm{1}.\zt_\Gamma]$ & \myblue \hspace{1.2em}$\Tr[\zt_\Gamma^\dagger.\zt_\Gamma.\zt_\Gamma\overset{\leftrightarrow}{.}\zt_\Gamma^\dagger]$ &  \\
         & \myred \hspace{-.9em} $\lx_g$ & \hspace{5.8em}\myblue $\lx_g\gap\gap\gap\lx_g$ &  \\\hhline{|~||---}
        & \myred $\underset{\Gamma\neq\mathbf{1}}{\sum}\frac{d_\Gamma}{|G|-1}\Tr\left[\zt_\Gamma^{\dagger(\text{even})}.\zt_\Gamma^{(\text{even}+2k)}\right]$ & \myblue $\underset{\Gamma\neq\mathbf{1}}{\sum}\frac{d_\Gamma}{|G|-1}\Tr\left[\zt_\Gamma^{\dagger(\text{odd})}.\zt_\Gamma^{(\text{even})}.\zt_\Gamma^{\dagger(\text{even}+2k)}.\zt_\Gamma^{(\text{odd}+2k)}\right]$ & \makecell{ Order \\ Parameter } \\\hhline{|~||---}
        & \myred $\frac{1}{|G|-1}\underset{g\neq e}{\sum}\left(\overset{k}{\underset{j=0}{\prod}}\lx_g^{(\text{odd}+2j)}\right)$ & \myblue $\frac{1}{|G|-1}\underset{g\neq e}{\sum}\left(\overset{k}{\underset{j=0}{\prod}}\lx_g^{(\text{odd}+2j)}\lx_{g}^{(\text{even}+2j)}\right)$ & \makecell{ Disorder \\ Parameter } \\ \hline\hline
         \multirow{8}{*}{\quad\large{$G\times \rep G$}\quad}& \mygreen\textit{Product State} & \myyellow$G$\textit{ Cluster State (SPT)} & Phase \\\hhline{|~||---}
         & \mygreen $\ket{\mathbf{1}\gap,\gap e\gap,\gap\mathbf{1}\gap,\gap e\gap,\gap\mathbf{1}\gap,\gap e\gap,\gap\ldots}$ &  \myyellow \quad$\sum_{\{g_i\}\in G}|g_1,g_1\bar{g_2},g_2,g_{2}\bar{g_3},g_3,g_3\bar{g_4},\ldots\rangle${\quad} & \makecell{Unique \\ Ground State} \\\hhline{|~||---}
         & \hspace{-6.3em}\mygreen $\Tr[\zt_\Gamma^\dagger]$ & \hspace{-4.6em}\myyellow $\Tr[\zt_\Gamma^\dagger\gap.\gap\zt_\Gamma\gap.\gap\zt_\Gamma]$ & \multirow{4}{*}{Stabilizers} \\
         & \hspace{-2.6em}\mygreen $\lx_g$ & \hspace{0.1em}\myyellow $\rx_g\gap\gap\gap\lx_g\gap\gap\gap\lx_g$ &  \\
         & \hspace{-1.3em}\mygreen $\Tr[\zt_\Gamma^\dagger]$ & \hspace{2.7em}\myyellow $\Tr[\zt_\Gamma^\dagger\gap.\gap\zt_\Gamma\gap.\gap\zt_\Gamma]$ &  \\
         & \hspace{2.1em}\mygreen $\lx_g$ & \hspace{7.5em}\myyellow $\rx_g\gap\gap\gap\lx_g\gap\gap\gap\lx_g$ &  \\\hhline{|~||---}
         & \mygreen $\underset{\Gamma\neq\mathbf{1}}{\sum}\frac{d_\Gamma}{|G|-1}\Tr\left[\overset{k}{\underset{j=0}{\prod}}\zt_\Gamma^{\dagger(\text{even}+2j)}\right]$ & \myyellow $\underset{\Gamma\neq\mathbf{1}}{\sum}\frac{d_\Gamma}{|G|-1}\Tr\left[\zt_\Gamma^{\dagger(\text{odd})}.\left(\overset{k-1}{\underset{j=0}{\prod}}\zt_\Gamma^{(\text{even}+2j)}\right).\zt_\Gamma^{(\text{odd}+2k)}\right]$ & \makecell{ String Order \\ Parameter }\\\hhline{|~||---}
         & \mygreen $\frac{1}{|G|-1}\underset{g\neq e}{\sum}\left(\overset{k}{\underset{j=0}{\prod}}\lx_g^{(\text{odd}+2j)}\right)$ & \myyellow $\frac{1}{|G|-1}\underset{g\neq e}{\sum}\rx_g^{(\text{even})}\lx_g^{(\text{odd})}\left(\overset{k-1}{\underset{j=1}{\prod}}\cx_g^{(\text{even}+2j)}\lx_g^{(\text{odd}+2j)}\right)\lx_g^{(\text{even}+2k)}$ & \makecell{ Disorder \\ Parameter } \\\hline
    \end{tabular}}
    \caption{A summary of the four quantum phases involved in the duality argument. The double arrow in the $ZZZZ$ stabilizers of SSB2 reflects the fact that the operators corresponding to the third and fourth sites appear in reverse order in the virtual space matrix product, see \cref{eq:SSB_H}. The data associated with each phase can be derived from the data of any other phase using the quantum circuits and dualities laid out in \cref{fig:KW_b}.}
    \label{tab:KW_a}
\end{table*}
\endgroup

\begin{widetext}
\begin{equation}\label{eq:SSB_H}
\begin{aligned}
    H_\text{\textit{SSB1}} &= -\frac{1}{|G|}\sum_{i\text{ odd}}\bigg(\sum_{\Gamma\in\rep{G}}\Tr\left[\zt^{(i+1)}_\Gamma.\zt^{\dagger(i+3)}_\Gamma\right]d_\Gamma+\sum_{g\in G}\lx_g^{(i)}\bigg) = -\sum_{i\text{ odd}}\left(\delta^G_{g_{i+1},g_{i+3}}+\frac{1}{|G|}\sum_{g\in G}\lx_g^{(i)}\right),\\[10pt]
    H_\text{\textit{SSB2}} &= -\frac{1}{|G|}\sum_{i\text{ odd}}\bigg(\sum_{\Gamma\in\rep{G}}\Tr\left[\zt^{\dagger(i)}_\Gamma.\zt^{(i+1)}_\Gamma.\zt^{\dagger(i+3)}_\Gamma.\zt^{(i+2)}_\Gamma\right]d_\Gamma
    +\sum_{g\in G}\lx_g^{(i)}\lx_g^{(i+1)}\bigg).
\end{aligned}
\end{equation}
\end{widetext}
Both of these Hamiltonians have a family of symmetries on each sublattice:
\begin{align}
    {G}_R^\text{odd}: \prod_{i\text{ odd}}\rx_g^{(i)},\qquad
     {G}_R^\text{even}: \prod_{i\text{ even}}\rx_g^{(i)}.
\end{align}
These symmetries act independently, so the total symmetry group is ${G}_R^\text{odd}\times {G}_R^\text{even}$.
Symmetry is spontaneously broken in both models. The ground-state subspaces are given by
\begin{equation}
\begin{split}
   H_\text{\textit{SSB1}}&:\{\ket{\mathbf{1},g,\mathbf{1},g,\ldots}:g\in G\}  ,\\
   H_\text{\textit{SSB2}}&:\bigg\{\sum_{\{g_i\}\in G}\ket{g_1,g_1g,g_2,g_2g,\ldots}:g\in G\bigg\} .
\end{split}
\end{equation}
 In \textit{SSB1}, the unbroken symmetry is ${G}_R^\text{odd}$. In \textit{SSB2} the unbroken symmetry for the state labeled by $g$ is $G_g:=\left\{\rx_h^{(i)}\rx_{\bar g hg}^{(i+1)}:h\in G\right\}$. In each case, the symmetry is broken from $G\times G\to G$, but with different unbroken subgroups. Thus, the two families of ground states realize distinct gapped phases under the ${G}_R^\text{odd}\times {G}_R^\text{even}$ symmetry.

 Notice that \textit{SSB1} and \textit{SSB2} are related by a controlled multiplication from the odd sublattice to the even. In particular, let
 \begin{align}\label{eq:U_SSB}
    U_\text{SSB}:=\prod_{i\text{ odd}} C\lx^{(i,i+1)},
 \end{align}
 then we see that
 \begin{equation}
     \begin{split}
        &U_\text{SSB}\ket{\mathbf{1},g,\mathbf{1},g,\ldots}=\sum_{\{g_i\}\in G}\ket{g_1,g_1g,g_2,g_2g,\ldots},\\
        &U_\text{SSB}H_\textit{SSB1}\left(U_\text{SSB}\right)^{\dagger}=H_\textit{SSB2}.
     \end{split}
 \end{equation}

We now want to apply Kramers-Wannier duality to the even sites, taking the $G\times G$ symmetry to $G\times \rep G$. This duality can also be thought of as gauging the $G_R^\text{even}$ subgroup of ${G}_R^\text{odd}\times {G}_R^\text{even}$ in the sense that the original theory is pure matter and the KW dual theory is pure gauge. Gauging of finite subgroups was discussed previously in Ref.~\onlinecite{tachikawa_gauging_2020}.

It was shown in Ref.~\onlinecite{tantivasadakarn_long-range_2022} (see \cref{ap:KW} for a brief review) that the KW duality can be implemented via the cluster state MPS by flipping the physical leg of the odd-site tensors. Using the MPS for the $G$ cluster state, we find:
\begin{align*}
 \begin{tikzpicture}[baseline={(0,-.09)}]
    \draw (0,0)--(2.5,0);
    \foreach \x in {0.5,2}{
		     \draw (\x,0) -- (\x,0.5);
		    \node[even] at (\x,0) {};
		  }
		  \foreach \x in {1.25}{
		     \draw (\x,0) -- (\x,-1);
		    \node[odd] at (\x,0) {};
		    \node[mpo] (t\x) at (\x,-0.5) {$\lx_g$};
		  }
  \end{tikzpicture} \ &= \
\begin{tikzpicture}[baseline={(0,-.09)}]
    \draw (0,0)--(2.5,0);
    \foreach \x in {0.5,2}{
		     \draw (\x,0) -- (\x,1);
		    \node[even] at (\x,0) {};
		  }\
		  \node[mpo] at (0.5,0.5) {$\rx_g$};
		  \node[mpo] at (2,0.5) {$\lx_g$};
		  \foreach \x in {1.25}{
		     \draw (\x,0) -- (\x,-0.5);
		    \node[odd] at (\x,0) {};
		  }
  \end{tikzpicture}, \\
 \begin{tikzpicture}[baseline={(0,-.09)}]
    \draw (0,0)--(2.5,0);
    \draw[red] (0,-0.5) -- (2.5,-0.5);
    \foreach \x in {0.5,2}{
		     \draw (\x,0) -- (\x,-1);
		    \node[odd] at (\x,0) {};
		  }
		  \foreach \x in {1.25}{
		     \draw (\x,0) -- (\x,0.5);
		    \node[even] at (\x,0) {};
		  }
  \node[mpo] at (0.5,-0.5) {$\zt_\Gamma$};
  \node[mpo] at (2,-0.5) {$\zt^\dagger_\Gamma$};
  \end{tikzpicture} \ &= \
\begin{tikzpicture}[baseline={(0,-.09)}]
    \draw (0,0)--(2.5,0);
    \draw[red] (0,0.5) -- (2.5,0.5);
    \foreach \x in {0.5,2}{
		     \draw (\x,0) -- (\x,-0.5);
		    \node[odd] at (\x,0) {};
		  }
		  \foreach \x in {1.25}{
		     \draw (\x,0) -- (\x,1);
		    \node[even] at (\x,0) {};
		  }
  \node[mpo] at (1.25,0.5) {$\zt_\Gamma$};
  \end{tikzpicture}. \end{align*}
 From this diagramatic derivation, we read off the appropriate KW duality:
 \begin{align}\label{eq:KW}
 \begin{split}
     \zt_\Gamma^{(i+1)}.\zt_\Gamma^{(i+3)\dagger}\mapsto\zt_\Gamma^{(i+1)},\\ \lx_g^{(i+1)}\mapsto\rx_g^{(i-1)}\lx_g^{(i+1)}.
 \end{split}
 \end{align}

 \noindent This is a generalized domain wall duality.
Applying the duality to each of our SSB Hamiltonians, we find
 %\begin{widetext}
\begin{equation}
\begin{aligned}
    H_\textit{SSB1} \mapsto& -\sum_{i\text{ odd}}\sum_{\Gamma\in\rep{G}}\Tr\left[\zt^{\dagger(i+1)}_\Gamma\right]d_\Gamma+\sum_{g\in G}\lx_g^{(i)},\\[10pt]
    H_\textit{SSB2} \mapsto& -\sum_{i\text{ odd}}\sum_{\Gamma\in\rep{G}}\Tr\left[\zt^{\dagger(i)}_\Gamma.\zt^{(i+1)}_\Gamma.\zt^{(i+2)}_\Gamma\right]d_\Gamma\\&\qquad\qquad+\sum_{g\in G}\rx_g^{(i-1)}\lx_g^{(i)}\lx_g^{(i+1)}=H_\Cl,
\end{aligned}
\end{equation}
 % \end{widetext}
Notice that $H_\textit{SSB2}$ maps to $H_\Cl$ under the KW duality, while $H_\textit{SSB1}$ maps to a Hamiltonian with unique ground state $\ket{\psi_0}:=\ket{\mathbf{1},e,\mathbf{1},e,\ldots}$, our reference $G\times\rep G$-symmetric product state.

We can use this duality to prove that the $G$ cluster state is not in the same phase as the symmetric product state as follows: because $H_\text{SSB1}$ and $H_\text{SSB2}$ exhibit different patterns of spontaneous symmetry breaking, they belong to different phases. We have just shown that these Hamiltonians are mapped under the KW duality into the product state Hamiltonian and $G$ cluster state Hamiltonian, respectively. If there were a symmetric path connecting the $G$ cluster state to the product state, then pushing this path through the KW duality would give a symmetric path connecting the two SSB states, which contradicts the fact that they belong to distinct phases. This implies that there does not exist a symmetric path connecting the cluster state to the product state, so that they must be in separate phases. We can conclude that $\ket{\Cl}$ belongs to a distinct fusion category SPT phase from the symmetric product state.

\begin{figure}[tp]
    \centering
    \includegraphics[width=0.4\textwidth]{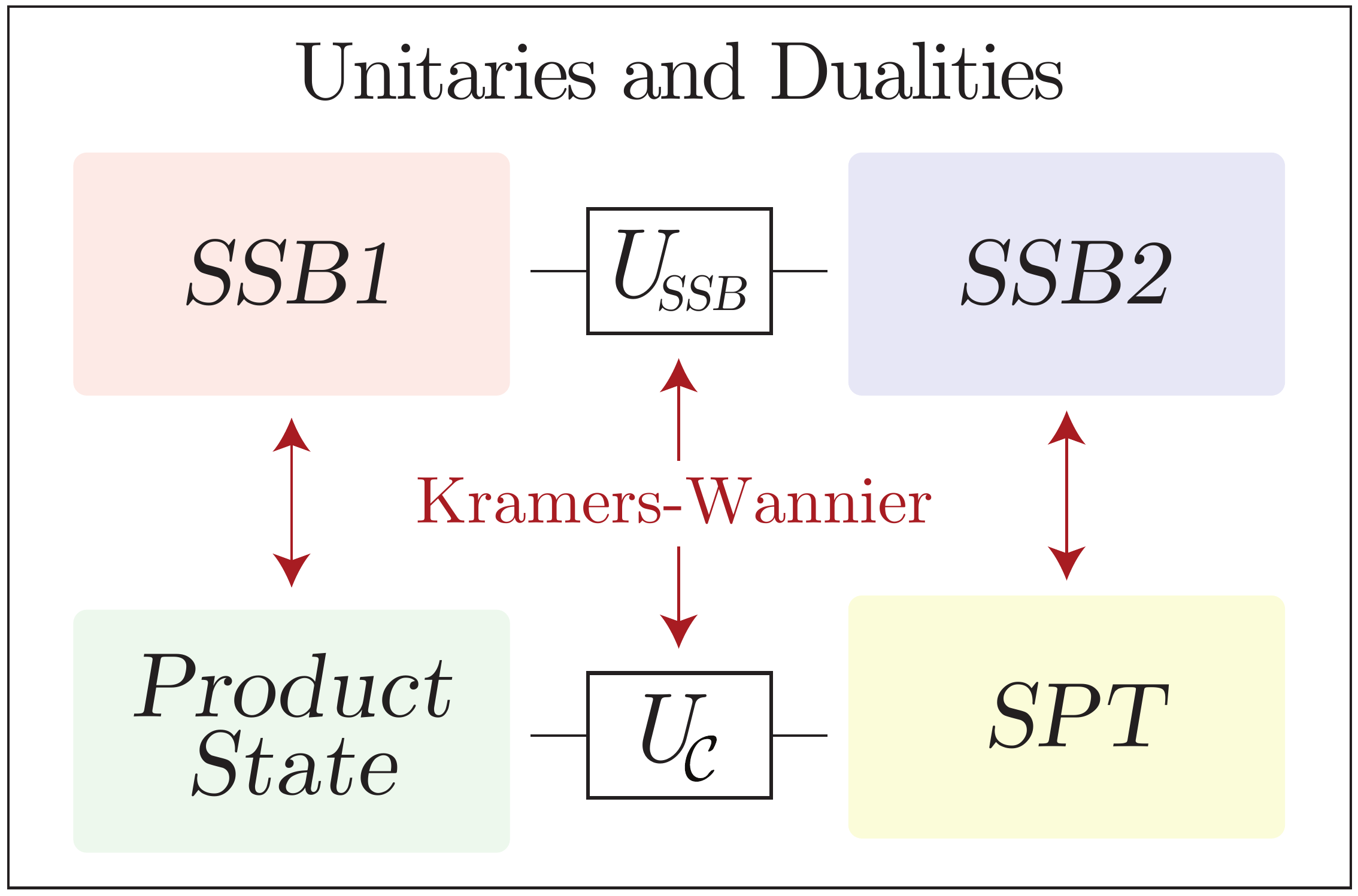}
    \caption{A depiction of the KW duality and quantum circuits relating the phases tabulated in \cref{tab:KW_a}. Note that the quantum circuits themselves -- defined in \cref{eq:U_SSB} and \cref{eq:U_C} -- are also related through the KW duality.}
    \label{fig:KW_b}
\end{figure}

We can further argue that the fusion category SPT phase to which the $G$ belongs is robust to weak symmetric local perturbations. Consider such a perturbation $\delta H_{G\times\rep G}$, where the subscript denotes the symmetry respected by the perturbation. Beginning with $H_\Cl$, the perturbed Hamiltonian becomes
\begin{equation}
    H_\Cl'=H_\Cl+\delta H_{G\times\rep G}.
\end{equation} We want to know whether $H_\Cl'$ describes the same phase as $H_\Cl$, for any choice of $\delta H_{G\times\rep G}$, which is equivalent to saying that the perturbation doesn't close the gap. Because local symmetric operators are mapped to local symmetric operators through the KW duality~\cite{lootens2023dualities}, we can map $H_\Cl'$ to
\begin{equation}
    H_\textit{SSB1}'=H_\textit{SSB1}+\delta H_{G\times G},
\end{equation}
where $H_{G\times G}$ is a local $G\times G$-symmetric perturbation. We know that the SSB order of $H_\text{SSB1}$ is robust to local symmetric perturbations, so we can conclude that the ground states of $H_\text{SSB1}$ and $H_\text{SSB1}'$ lie in the same phase. From this we know that the ground states of the Hamiltonians they map into under the KW duality -- namely $H_\Cl$ and $H_\Cl'$ -- lie in the same phase. This means that the SPT order in $\ket{\Cl}$ is robust; in particular, its ground state degeneracy -- to be discussed in the following section -- is robust to local symmetric perturbations. We verify robustness of ground-state degeneracy numerically for the simplest non-abelian case in Sec.~\ref{sec:D3_degen}.

%%%%%%%%%%%%%%%%%%%%%%%%%%%%%%%%%%%%%%%%%%%%%%%%%%%%%%%%%%%%%%%%%%%%%%%%%%%%%%%%%%%%%%%%%%%%%%%%%%%%%%%%%%%%%%%%%%%%%%%%%%%%%
\section{Signatures of SPT Order}
 In this section, we discuss how characteristic signatures of SPT order generalize to the context of fusion category symmetry. For a review of the qubit results which are generalized here, see \cref{ap:CSS_cluster}.

\subsection{Edge Modes and Ground State Degeneracy}\label{sec:edge_modes}
1d SPTs on open chains with on-site group symmetry action support edge modes that are robust to local symmetric perturbations. These edge modes are responsible for the ground state degeneracy of these phases. In the bosonic case, the edge modes transform as a projective representation of the global symmetry $G$ and can be identified with elements of the second cohomology group $\mathcal{H}^2(G,U(1))$~\cite{Pollmann10,chen_symmetry-protected_2012-1,chen_symmetry_2013-1}, which in turn provides a classification of 1d bosonic $G$-SPTs.

We can detect the presence of edge modes by acting on the state $\ket{\Cl}$ with the global symmetries $\overleftarrow{A}_g$ and $\hat{B}_\Gamma$. We find that the action of the global symmetries pulls through to the virtual level and reduces to operators acting only on the edge Hilbert spaces $\mathcal{H}_L$ and $\mathcal{H}_R$:
%\begin{widetext}
\begin{align*}
\overleftarrow A_g\ket{\Cl} =\quad& \begin{tikzpicture}[baseline={(0,.4)}]
    \draw (0,0)--(5.5,0);
    \foreach \x in {0.5,2,3.5,5}{
		     \draw (\x,0) -- (\x,1);
		    \node[mpo] (t\x) at (\x,0.5) {$\rx_g$};
		    \node[odd] at (\x,0) {};
		  }
		  \foreach \x in {1.25,4.25}{
		     \draw (\x,0) -- (\x,1);
		    \node[even] at (\x,0) {};
		  }
		  \node[fill=white] at (2.75,0) {$\dots$};
		  \node[edge] at (0,0) {};
		  \node[edge] at (5.5,0) {};
  \end{tikzpicture} \\
  =\quad&
  \begin{tikzpicture}[baseline={(0,.4)}]
    \draw (-0.5,0)--(6,0);
    \foreach \x in {0.5,2,3.5,5}{
		     \draw (\x,0) -- (\x,1);
		    \node[odd] at (\x,0) {};
		  }
		  \foreach \x in {1.25,4.25}{
		     \draw (\x,0) -- (\x,1);
		    \node[even] at (\x,0) {};
		  }
		  \node[fill=white] at (2.75,0) {$\dots$};
		  \node[edge] at (-0.5,0) {};
		  \node[edge] at (6,0) {};
		  \node[mpo] at (0,0) {$\rx_g^\dagger$};
		  \node[mpo] at (5.5,0) {$\rx_g$};
  \end{tikzpicture} ,
  \end{align*}
  \begin{align*}
\hat B_{\Gamma_{\alpha\beta}}\ket{\Cl}=\quad&\begin{tikzpicture}[baseline={(0,.4)}]
    \draw (0,0)--(5.5,0);
    \draw[red] (0,0.5)--(5.5,0.5);
    \foreach \x in {0.5,2,3.5,5}{
		     \draw (\x,0) -- (\x,0.45);
		     \draw (\x,0.55) -- (\x,1);
		    \node[odd] at (\x,0) {};
		  }
		  \foreach \x in {1.25,4.25}{
		     \draw (\x,0) -- (\x,1);
		    \node[even] at (\x,0) {};
		    \node[mpo] (t\x) at (\x,0.5) {$\zt_\Gamma$};
		  }
		  \node[fill=white] at (2.75,0) {$\dots$};
		  \node[fill=white] at (2.75,0.5) {$\dots$};
		  \node[edge] at (0,0) {};
		  \node[edge] at (5.5,0) {};
        \node[edge, red] at (0,0.5) {};
		  \node[edge, red] at (5.5,0.5) {};
  \end{tikzpicture} \\ =\quad&
  \begin{tikzpicture}[baseline={(0,.4)}]
    \draw (-0.5,0)--(6,0);
    \draw[red] (-0.5,-0.2) -- (6,-0.2);
    \foreach \x in {0.5,2,3.5,5}{
		     \draw (\x,0) -- (\x,1);
		    \node[odd] at (\x,0) {};
		  }
		  \foreach \x in {1.25,4.25}{
		     \draw (\x,0) -- (\x,1);
		    \node[even] at (\x,0) {};
		  }
		  \node[fill=white] at (2.75,0) {$\dots$};
		  \node[fill=white] at (2.75,-0.2) {$\dots$};
		  \node[edge] at (-0.5,0) {};
		  \node[edge] at (6,0) {};
		  \node[mpo] at (0,-0.1) {$\zt_\Gamma$};
		  \node[mpo] at (5.5,-0.1) {$\zt_\Gamma^\dagger$};
        \node[edge, red] at (-0.5,-0.2) {};
		  \node[edge, red] at (6,-0.2) {};
  \end{tikzpicture} \ ,
\end{align*}
%\end{widetext}
where \begin{tikzpicture}
    \draw (0,0)--(0.25,0);
		  \node[edge] at (0,0) {};
  \end{tikzpicture}
 = $\bra{\psi_L}\in\mathcal{H}_L^*$
 and
 \begin{tikzpicture}
    \draw (0,0)--(0.25,0);
		  \node[edge] at (0.25,0) {};
  \end{tikzpicture}
 = $\ket{\psi_R}\in\mathcal{H}_R$. From this diagrammatic derivation, we can read off the edge modes:
 \begin{equation}\label{eq:edge_modes}
 G: \rx_g^{(L)}\rx_g^{(R)},\quad \rep G:\left[\zt_\Gamma^{(L)}.\zt_\Gamma^{\dagger(R)}\right]_{\alpha\beta},
\end{equation}
 which act on $\mathcal{H}_L\otimes\mathcal{H}_R$. We have written $\rx_g^{(L)}$ rather than $\rx_g^{(L)\dagger}$ because $\bra{\psi_L}\rx_g^{(L)\dagger}\mapsto\rx_g^{(L)}\ket{\psi_L}$. It is also possible to derive the existence of edge modes without knowing the form of the ground state, using only the stabilizer Hamiltonian \cref{eq:ham} (see \cref{ap:edge_modes}).

Notice that the edge modes acting on each edge do not commute with one another:
\begin{equation}\label{eq:edge_comm}
\begin{split}
    \rx_g^{(L)}\left[\zt_\Gamma^{(L)}.\zt_\Gamma^{\dagger(R)}\right]_{\alpha\beta} =  \left[\zt_\Gamma^{(L)}.\Gamma(g).\zt_\Gamma^{\dagger(R)}\right]_{\alpha\beta}\rx_g^{(L)},\\ \rx_g^{(R)}\left[\zt_\Gamma^{(L)}.\zt_\Gamma^{\dagger(R)}\right]_{\alpha\beta} =  \left[\zt_\Gamma^{(L)}.\Gamma(\bar g).\zt_\Gamma^{\dagger(R)}\right]_{\alpha\beta}\rx_g^{(R)}.
\end{split}
\end{equation}
In the qubit case, the edge modes fail to commute up to a phase of $\pm 1$, reflecting the fact that the edge modes generate the Pauli matrices, a projective representation of $\mathbb{Z}_2\times\mathbb{Z}_2$.
Here, we find that the edge modes for group-valued qudits fail to commute \textit{up to a matrix} $\Gamma(g)$ or $\Gamma(\bar g)$.
%In the qubit case, this commutation up to a phase .
However, the total edge operators -- the product of the left and right operators -- still commute,
 \begin{equation}
     \rx_g^{(L)}\rx_g^{(R)}\left[\zt_\Gamma^{(L)}.\zt_\Gamma^{\dagger(R)}\right]_{\alpha\beta} =  \left[\zt_\Gamma^{(L)}.\zt_\Gamma^{\dagger(R)}\right]_{\alpha\beta}\rx_g^{(L)}\rx_g^{(R)}~,
 \end{equation}
because the extra factors of $\Gamma(g)$ and $\Gamma(\bar g)$ meet between the $\zt$ operators and multiply to the identity.

These edge modes are also manifest in the wavefunctions of the ground states. On an open chain, the ground state manifold is spanned by $|G|^2$ basis states given by
    \begin{equation}\label{eq:OBC_GS}
    |g_1\rangle\left(\mathcal{N}\sum_{g_2}\cdots\sum_{g_{N-1}}|g_1\bar{g_2}\rangle|g_2\rangle\cdots\\|g_{N-1}\bar{g_N}\rangle\right)|g_N\rangle,
    \end{equation}
for $g_1,g_N\in G$. We can denote the state in \cref{eq:OBC_GS} by $\ket{g_1^{(L)},g_N^{(R)}}$. Notice that these states, expressed as matrix product states, exhibit the same bulk structure subject to different boundary conditions.

Let us show that the symmetry action of $\overleftarrow{A}_g$ and $\hat{B}_{\Gamma_{\alpha\beta}}$ on the edge modes given in \cref{eq:edge_modes} can map between all basis states.  This can be seen by performing a basis transformation by the unitary $C\rx^{(L,R)}$, which maps
\begin{align}
    \rx_g^{(L)}\rx_g^{(R)}\mapsto\rx_g^{(L)},\quad \left[\zt_\Gamma^{(L)}.\zt_\Gamma^{\dagger(R)}\right]_{\alpha\beta}\mapsto \left[\zt_\Gamma^{\dagger(R)}\right]_{\alpha\beta}.
\end{align}
In this new basis, we may start with the basis state $\ket{e^{(L)},\mathbf{1}^{(R)}}=\sum_{g\in G}\ket{e^{(L)},g^{(R)}}$. Then by acting with each symmetry, we find 
\begin{equation}
    \overleftarrow{A}_g \hat B_{\Gamma_{\alpha\beta}}\ket{e^{(L)},\mathbf{1}^{(R)}}=\ket{g^{(L)},\Gamma_{\alpha\beta}^{(R)}}.
\end{equation}
Since $\ket{g^{(L)},\Gamma_{\alpha\beta}^{(R)}}$ forms an orthonormal basis for $\mathbb{C}[G]\otimes\mathbb{C}[G]$, we have $|G|^2$ ground states.

%%%%%%%%%%%%%%%%%%%%%%%%%%%%%%%%%%%%%%%%%%%%%%%%%%%%%%%%%%%%%%%%%%%%%%%%%%%%%%%%%%%%%%%%%%%%%%%%%%%%%%%%%%%%%%%%%%%%%%%%%%%%%
\subsection{String Order Parameters}
A large class of SPTs admit a \textit{non-local} or \textit{string order parameter} which can effectively detect SPT order~\cite{den_nijs_preroughening_1989,kennedy_hidden_1992,pollmann_detection_2012,Else13,Bahri14} (see Ref.~\cite{tasaki2018topological} for an exception).
We can derive two families of string order parameters for the $G$ cluster state by mapping the order and disorder parameters in SSB2 through the KW duality. Because these parameters distinguish SSB1 from SSB2 on one side of the duality, we expect that they will be able to distinguish $\ket{\psi_0}$ from $\ket{\Cl}$ on the other side. 

The $Z$-type string order parameter can be derived by mapping the order parameter from SSB2. It consists of a string of $\zt_\Gamma$ on even sites, with a $\zt^\dagger_\Gamma$ and $\zt_\Gamma$ on the odd sites preceding and following the string, respectively:
\begin{equation}\label{eq:string_order_param}
    \hat{\mathcal{S}}^{(i,k)}_{\zt}=\sum_{\Gamma\neq\mathbf{1}}\frac{d_\Gamma}{|G|-1}\Tr\left[\zt_\Gamma^{\dagger(i)}.\left(\prod_{j=0}^{k-1}\zt_\Gamma^{(i+1+2j)}\right).\zt_\Gamma^{(i+2k)}\right],
\end{equation}
where the superscript indexes the left endpoint $i$ and the length $2k+1$ of the string operator. Notice also that $\hat{\mathcal{S}}^{(i,k)}_\zt$ is Hermitian, so that its expectation value will always be real.

Diagrammatically, $\hat{\mathcal{S}}^{(i,k)}_\zt$ acts on the $G$ cluster state as
\begin{widetext}
\begin{equation}
\hat{\mathcal{S}}^{(i,k)}_\zt\ket{\Cl}=\sum_{\Gamma\neq\mathbf{1}}\frac{d_\Gamma}{|G|-1}\begin{tikzpicture}[baseline={(0,.4)}]
    \draw (0,0)--(5.5,0);
    \node[fill=white] at (0,0) {$\dots$};
    \node[fill=white] at (5.5,0) {$\dots$};
    \draw[red] (-0.25,0.5) rectangle (5.75,-0.25);
    \foreach \x in {5}{
	    \draw (\x,0) -- (\x,1);
		\node[odd] at (\x,0) {};
		\node[mpo] (t\x) at (\x,0.5) {$\zt_\Gamma$};
		  }
    \foreach \x in {2}{
		 \draw (\x,0) -- (\x,0.45);
		 \draw (\x,0.55) -- (\x,1);
		 \node[odd] at (\x,0) {};
		  }
		  \foreach \x in {1.25,2.75,4.25}{
		     \draw (\x,0) -- (\x,1);
		    \node[even] at (\x,0) {};
		    \node[mpo] (t\x) at (\x,0.5) {$\zt_\Gamma$};
		  }
	% \foreach \x in {5.75}{
 %        \draw (\x,0) -- (\x,0.45);
	% 	\draw (\x,0.55) -- (\x,1);
	% 	\node[even] at (\x,0) {};
	% 	  }
		  \foreach \x in {.5}{
		     \draw (\x,0) -- (\x,1);
		    \node[odd] at (\x,0) {};
		    \node[mpo] (t\x) at (\x,0.5) {$\zt^\dagger_\Gamma$};}
		  %\node[fill=white] at (3.5,0) {$\dots$};
		  %\node[fill=white] at (3.5,0.5) {$\dots$};
		  % \node[edge] at (0,0) {};
		  % \node[edge] at (5.5,0) {};

        \node[fill=white] at (3.5,0) {$\dots$};
        \node[fill=white] at (3.5,-.25) {$\dots$};
        \node[fill=white] at (3.5,.5) {$\dots$};
  \end{tikzpicture} \
  = \left(\sum_{\Gamma\neq\mathbf{1}}\frac{d_\Gamma^2}{|G|-1}\right)\ket{\Cl}
   \
  = \ket{\Cl},
\end{equation}
\end{widetext}
as $\sum_\Gamma d_\Gamma^2=|G|$ so that $\sum_{\Gamma\neq \mathbf{1} }d_\Gamma^2=|G|-1$. We can then conclude that $\bra{\Cl}\hat{\mathcal{S}}_{\zt}\ket{\Cl}=1$,
as expected for a (string) order parameter. 

We will now show that the expectation value of $\hat{\mathcal{S}}_{\zt}^{(i,k)}$ in the  $G\times\rep G$-symmetric product state $\ket{\psi_0}$ is zero, so that $\hat{\mathcal{S}}_{\zt}^{(i,k)}$ can be used to detect the SPT order in $\ket{\Cl}$. First, recall that the summation of the trace of a $Z$ operator over all irreps projects the argument to the identity element (see \cref{eq:G_delta}). We can use this to relate $\hat{\mathcal{S}}_{\zt}^{(i,k)}$ to the delta function on $G$:
\begin{equation}\label{eq:string_delta}
    \hat{\mathcal{S}}_{\zt}^{(i,k)}=\frac{|G|}{|G|-1}\delta^G_{e,\bar{g}_i\left(\prod_{j=0}^{k-1}g_{i+1+2j}\right)g_{i+2k}}-\frac{1}{|G|-1}.
\end{equation}
It directly follows that indeed $\bra{\psi_0}\hat{\mathcal{S_Z}}\ket{\psi_0}=0$, the details of which can be found in \cref{ap:string_order}.

The $X$-type string order parameter can be derived by mapping the \textit{dis}order parameter from SSB2 through the KW duality. It consists of a string of $\rx_g$ on odd sites and $\cx_g$ on even sites, with an $\lx_g$ and $\rx_g\rx_g$ on the sites preceding and following the string, respectively:
\begin{widetext}
\begin{equation}\label{eq:X_string_order_param}
    \hat{\mathcal{S}}^{(i,k)}_{X}=\frac{1}{|G|-1}\sum_{g\neq e}\rx_g^{(i)}\lx_g^{(i+1)}\left(\prod_{j=1}^{k-1}\cx_g^{(i+2j)}\lx_g^{(i+1+2j)}\right)\lx_g^{(i+2k)},
\end{equation}
\end{widetext}
where the superscript indexes the left endpoint $i$ and the length $2k+1$ of the string operator. Note that $\hat{\mathcal{S}}^{(i,k)}_{X}$ begins on an even site.

Because $\ket{\Cl}$ is a $+1$ eigenstate of each term in the summand and there are $|G|-1$ terms, we have $\bra{\Cl}\hat{\mathcal{S}}_{X}\ket{\Cl}=1$. Furthermore, it is straightforward to see that each term in the summand has expectation value zero in $\ket{\psi_0}$ because factors of $\braket{e|g}=\delta^{G}_{e,g}$ appear and the sum is over $g\neq e$. This allows us to conclude that $\bra{\psi_0}\hat{\mathcal{S}}_{X}^{(i,k)}\ket{\psi_0}=0$, so that $\hat{\mathcal{S}}_{X}^{(i,k)}$ can also be used to detect the SPT order in $\ket{\Cl}$.

\subsection{Topological Response}\label{sec:topo_response}
Topologically ordered and SPT states are known to pump quantized symmetry charge in response to the insertion of symmetry flux. This is known as \textit{charge-flux attachment}~\cite{senthil_integer_2013} or \textit{topological response}~\cite{cheng_topological_2014}. The most well known example of topological response is the quantum Hall effect, in which $U(1)$ symmetry flux in the form of magnetic flux induces an accumulation of $U(1)$ symmetry charge in the form of electric charge~\cite{hatsugai_chern_1993}. The quantized response in that case is the Hall conductivity. Topological response is an important signature which can distinguish SPT order from trivial order, the latter hosting no quantized response.

In the case of Abelian $G$, the response can be understood from the perspective of the decorated domain wall construction of the cluster state~\cite{Chen14DDW}. The wave function of the $\ztwo \times \ztwo$ cluster state can be understood as charges of the first $\ztwo$ decorated on the domain walls of the second $\ztwo$. Threading a $\ztwo$ flux creates a single domain wall from which we can detect the charge of the other $\ztwo$ (see Appendix~\ref{app:responseZ2} for further details). We will now see how this generalizes in the case of the $G$ cluster state\footnote{It was also argued in~\cite{albert_spin_2021} that charge and flux excitations in the quantum double bulk also yield twisted boundary conditions in the effective edge theory. Thus we can equivalently think of the probe in the above topological response as flux through the SPT on a ring, or as anyonic excitations in the bilayer quantum double which hosts $\ket{\Cl}$ as a gapped edge.}.

In order to define flux insertion for a 1d SPT, we place the state on a ring with periodic boundary conditions. Threading flux through a state on a ring is equivalent to twisting the boundary condition~\cite{zaletel_flux_2014}. Threading a $g$-flux through $\ket{\mathcal{C}}$ has the effect of inserting a $g$ domain wall between the first and last sites:
\begin{equation}
    |\mathcal{C}_g\rangle:=
    \mathcal{N}\sum_{{g_i}\in G}|g_1\rangle|g_1\bar{g_2}\rangle|g_2\rangle\cdots|g_{N-1}g\bar{g_1}\rangle,
\end{equation}
where $\ket{\Cl_g}$ is the $G$ cluster state on a ring with $g$ flux inserted. Diagrammatically, this corresponds to inserting an $X$-type operator on the virtual space:
\begin{equation}
    \ket{\Cl_g}=\begin{tikzpicture}
    \draw[black] (-0.5,0) rectangle (4.55,-0.5);
    \foreach \x in {0.5,2,3.5}{
		     \draw (\x,0) -- (\x,0.5);
		    \node[odd] at (\x,0) {};
		  }
		  \foreach \x in {1.25,4.25}{
		     \draw (\x,0) -- (\x,0.5);
		    \node[even] at (\x,0) {};
		  }
		  \node[fill=white] at (2.75,0) {$\dots$};
            \node[mpo] at (-0.05,0.0) {$\rx_g$};
  \end{tikzpicture} \ .
\end{equation}
We now want to show that the system responds to the insertion of $G$ flux by pumping a nontrivial $\rep G$ charge. The charge of $\ket{\Cl}$ under $\hat B_\Gamma$ is 1 because $\ket{\Cl}$ is $\rep G$-symmetric. If we act with $\hat B_\Gamma$ on $\ket{\Cl_g}$ and find an eigenvalue other than 1, then we say that $|\mathcal{C}_g\rangle$ is nontrivially charged under the $\rep G$ symmetry. Indeed, we find
\begin{equation}
    \hat B_\Gamma\ket{\Cl_g}=\Tr\left[\prod_{i\text{ even}}\zt_\Gamma^{(i)}\right]|\mathcal{C}_g\rangle=\Tr[\Gamma(g)]|\mathcal{C}_g\rangle.
\end{equation}
We see that the insertion of $g$-flux induces a response given by the character of $g$: $\Tr\left[\Gamma(g)\right]$. This also implies the response only depends on the conjugacy class of $g$, since characters are invariant on conjugacy classes. 

It is also interesting to note that $\ket{\Cl_g}$ is an eigenstate of $\hat B_\Gamma$, i.e. it transforms as a 1-dimensional irrep of the fusion category $\rep G$. In general, states which are charged under a symmetry will transform as some irrep of the symmetry. In this case, the irreps of $\rep G$ are labeled by group elements due to Tannaka-Krein duality\footnote{This is a natural generalization of \textit{Pontryagin duality}, which states that a locally compact abelian group is isomorphic to its character group. For more, see e.g. \cite{kosarew2002geometric,kirillov2012elements}.} \cite{tannaka1939dualitatssatz}. This can be understood by considering the following (reducible) representation $R$ of $\rep G$:
\begin{align}
\begin{split}
    R:&\rep G \to GL_{|G|}(\mathbb{C})\\
    &\Gamma\mapsto \Tr[\zt_\Gamma]
\end{split}
\end{align}
As can be readily verified, the matrices $R_\Gamma=\Tr[\zt_\Gamma]$ respect the fusion rules of the fusion category and therefore form a representation of the category. However, notice that
\begin{align}
\Tr[\zt_\Gamma]=\sum_{g}\Tr[\Gamma(g)]\ketbra{g}{g}
\end{align}
is diagonal for every $\Gamma\in\rep G$. This tells us that the $|G|$-dimensional representation $R$ is actually decomposable into $|G|$ one-dimensional representations. We denote the inequivalent representations as $R^{[g]}$, labeled by the conjugacy classes $[g]$ of $G$, and each representation potentially appears multiple times in the decomposition. In particular, we have
\begin{align}
\begin{gathered}
	R=\bigoplus_{[g]\in G}|[g]|R^{[g]},\\
	R^{[g]}_\Gamma=\Tr[\Gamma(g)],
\end{gathered}
\end{align}
where $|[g]|$ is a multiplicity factor equal to the number of elements in the conjugacy class $[g]$. The maps $R^{[g]}$ all satisfy the fusion rules of the fusion category and are one-dimensional, therefore they are irreps. These are precisely the irreps in which $\ket{\Cl_g}$ transforms under $\hat B_\Gamma$.

We can similarly thread a $\Gamma_{\alpha\beta}$-flux through $|\mathcal{C}\rangle$ between the first and last sites, yielding
  \begin{equation}\label{eq:Gamma_alpha_beta}
    |\mathcal{C}_{\Gamma_{\alpha\beta}}\rangle:=
    \mathcal{N}\sum_{{g_i}\in G}[\Gamma(g_1)]_{\alpha\beta}|g_1\rangle|g_1\bar{g_2}\rangle|g_2\rangle\cdots|g_{N-1}\bar{g_1}\rangle.
\end{equation}
Diagrammatically, this corresponds to inserting a $\zt$-type operator in the virtual space:
\begin{equation}
    \ket{\Cl_\Gamma}=\begin{tikzpicture}
    \draw[black] (-.8,0) rectangle (4.55,-0.5);
    \draw[red] (-0.65,-0.2) -- (0.25,-.2);
    \foreach \x in {0.5,2,3.5}{
		     \draw (\x,0) -- (\x,0.5);
		    \node[odd] at (\x,0) {};
		  }
		  \foreach \x in {1.25,4.25}{
		     \draw (\x,0) -- (\x,0.5);
		    \node[even] at (\x,0) {};
		  }
		  \node[fill=white] at (2.75,0) {$\dots$};
            \node[mpo] at (-0.2,-0.1) {$\zt_\Gamma$};
        \node[edge,red] at (-0.65,-0.2) {};
        \node[edge,red] at (0.25,-0.2) {};
  \end{tikzpicture} \ .
\end{equation}
We would like to understand how the state responds to the insertion of $\Gamma_{\alpha\beta}$ flux, so we act on it with $\overleftarrow A_g$, the representation of the $G$ symmetry. We see that $|\mathcal{C}_{\Gamma_{\alpha\beta}}\rangle$ transforms under this family of global $X$-type symmetries:
\begin{align}
\overleftarrow{A}_g\ket{\Cl_{\Gamma_{\alpha\beta}}}&=\mathcal{N}\sum_{{g_i}\in G}[\Gamma(g_1g)]_{\alpha\beta}|g_1\rangle|g_1\bar{g_2}\rangle|g_2\rangle\cdots|g_{N-1}\bar{g_1}\rangle \nonumber \\
    &=\sum_\gamma[\Gamma(g)]_{\gamma\beta}\ket{\Cl_{\Gamma_{\alpha\gamma}}}.
\end{align}
Holding $\alpha$ fixed, the states $\{\ket{\Cl_{\Gamma_{\alpha\beta}}}:\beta=1,\cdots,d_\Gamma\}$ transform in the irrep $\Gamma$ under the action of $\overleftarrow A_g$. There are $d_\Gamma$ different values $\alpha$ can take, so in total the states $\{\ket{\Cl_{\Gamma_{\alpha\beta}}}:\alpha,\beta=1,\cdots,d_\Gamma\}$ transform as the direct sum of $d_\Gamma$ copies of $\Gamma$, which we denote $d_\Gamma\Gamma$.
% \vva{This subspace is clearly transforming as $\Gamma$ irrep, no? If this is true for each $\alpha$, then there are $d_{\Gamma}$ copies of said irrep in this space. I don't think you need to diagonalize anything in the dihedral example, as this is clear from the above already.}
% Thus, $\ket{\Cl_{\Gamma_{\alpha\beta}}}$ is not necessarily an eigenstate of the $G$ symmetry. Instead, we have a subspace supporting states
% \begin{equation}
%     \left\{\overleftarrow A_g\ket{\Cl_{\Gamma_{\alpha\beta}}}:g\in G\right\}.
% \end{equation}
This is the nontrivial topological $G$ response to the insertion of $\rep G$ flux.

The product state $\ket{\psi_0}$ is unaffected by the insertion of symmetry flux. This can be understood as arising from the fact that $\ket{\psi_0}$ is a product state, with only single-site terms in its stabilizer Hamiltonian. Because there aren't any terms which span two sites, there are no terms to modify through flux insertion between the sites. Clearly, then, the product state does not have any charge response to the insertion of symmetry flux. This is yet another signature which distinguishes the state $\ket{\Cl}$ from the product state.

%%%%%%%%%%%%%%%%%%%%%%%%%%%%%%%%%%%%%%%%%%%%%%%%%%%%%%%%%%%%%%%%%%
\section{Dihedral example: $G=D_3$}\label{sec:D3_example}

We now turn to an example to illustrate the points of the previous section in a more concrete setting. We choose $G=D_3=\mathbb{Z}_3\rtimes\ztwo$, as this is the smallest non-abelian group. It can be thought of as the symmetry group of an equilateral triangle, which consists of rotations by multiples of $2\pi/3$ ($R\in\mathbb{Z}_3$) and reflections ($S\in\ztwo$). It is a semidirect product, rather than a direct product, because the rotations and reflections do not commute. The group multiplication table is given by:
\begin{equation*}
\begin{aligned}
\begin{array}{c|c c c c c c}
		  D_3 & e & R & R^2 & S & SR & SR^2 \\ \hline
        e & e & R & R^2 & S & SR & SR^2 \\
        R & R & R^2 & e & SR^2 & S & SR  \\
        R^2 & R^2 & e & R & SR & SR^2 &  S \\
        S & S & SR & SR^2 & e & R & R^2 \\
        SR & SR & SR^2 & S & R^2 & e & R \\
        SR^2 & SR^2 & S & SR & R & R^2 & e \\
\end{array}&\quad\begin{tikzpicture}
    \node[triangle]{};
    \node[] at (0,-0.675) {$S$};
    \node[] at (0,.95) {$R$};
    \draw[<->,red] (-.7,-0.5)--(.7,-.5);
    \draw[dashed] (0,-0.35)--(0,0.65);
    \draw[<->,blue] (60:0.8) arc(60:120:0.8);
\end{tikzpicture}
\end{aligned}
\end{equation*}

There are three irreducible representations of $D_3$: the trivial irrep $\mathbf{1}$, the sign irrep $\Gamma_{s}$, and the 2d irrep $\Gamma_{2d}$. They are given by:
\begin{align}\label{eq:D3_irreps}
    \begin{array}{c|ccc}
          & \mathbf{1} & \Gamma_\text{s} & \Gamma_\text{2d} \\\hline
        e & 1 & 1 & \begin{pmatrix} 1 & 0 \\ 0 & 1\\ \end{pmatrix} \\
        R & 1 & 1 & \begin{pmatrix} -\frac{1}{2} & -\frac{\sqrt{3}}{2} \\ \frac{\sqrt{3}}{2} & -\frac{1}{2}\\ \end{pmatrix} \\
        R^2 & 1 & 1 &  \begin{pmatrix} -\frac{1}{2} & \frac{\sqrt{3}}{2} \\ -\frac{\sqrt{3}}{2} & -\frac{1}{2}\\ \end{pmatrix} \\
        S & 1 & -1 &  \begin{pmatrix} \frac{1}{2} & -\frac{\sqrt{3}}{2} \\ -\frac{\sqrt{3}}{2} & -\frac{1}{2}\\ \end{pmatrix} \\
        SR & 1 & -1 & \begin{pmatrix} -1 & 0 \\ 0 & 1\\ \end{pmatrix}\\
        SR^2 & 1 & -1 & \begin{pmatrix} \frac{1}{2} & \frac{\sqrt{3}}{2} \\ \frac{\sqrt{3}}{2} & -\frac{1}{2}\\ \end{pmatrix} \\
    \end{array}
\end{align}

We can also consider the tensor product of irreps, which decomposes into direct sums of irreps according to the rules of $\rep{D_3}$:
\begin{equation*}
\begin{aligned}\rep{D_3}:
\begin{array}{r|c c c c c c}
		  \otimes & \mathbf{1} & \Gamma_{s} & \Gamma_{2d} \\ \hline
        \mathbf{1} & \mathbf{1} & \Gamma_{s} & \Gamma_{2d} \\
        \Gamma_{s} & \Gamma_{s} & \mathbf{1} & \Gamma_{2d} \\
        \Gamma_{2d} & \Gamma_{2d} & \Gamma_{2d} & \mathbf{1}\oplus\Gamma_{s}\oplus\Gamma_{2d}
\end{array}
\end{aligned}
\end{equation*}
Notice that $\mathbf{1}$ and $\Gamma_{s}$ have inverses in the conventional sense, while $\Gamma_{2d}$ does not. This means that $\zt_{\Gamma_{2d}}$ does not have an inverse, so that the $\rep{D_3}$ symmetry is non-invertible. Namely,
\begin{equation}
    \Tr[\zt_{\Gamma_{2d}}]\Tr[\zt_{\Gamma_{2d}}]=\Tr[\zt_{\mathbf{1}}]+\Tr[\zt_{\Gamma_{s}}]+\Tr[\zt_{\Gamma_{2d}}]\neq\mathbbm{1}.
\end{equation}
More explicitly, consider the $\Tr[\zt_{\Gamma_{2d}}]$ operator acting on a single site:
\begin{equation}
    \Tr[\zt_{\Gamma_{2d}}]=\left(
\begin{array}{cccccc}
 2 & 0 & 0 & 0 & 0 & 0 \\
 0 & -1 & 0 & 0 & 0 & 0 \\
 0 & 0 & -1 & 0 & 0 & 0 \\
 0 & 0 & 0 & 0 & 0 & 0 \\
 0 & 0 & 0 & 0 & 0 & 0 \\
 0 & 0 & 0 & 0 & 0 & 0 \\
\end{array}
\right).
\end{equation}
Clearly this matrix has determinant zero and is therefore not invertible.

\subsection{Ground State Degeneracy}\label{sec:D3_degen}

\begin{figure*}[btp]
    \centering
    \includegraphics[]{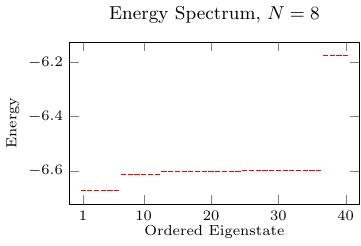}
    %%%%% FIGURE 1 %%%%%
   %\begin{tikzpicture}
   %\begin{axis}[
   %height = 0.75\columnwidth/1.5,
   %width = 0.75\columnwidth,
   %title = {Energy Spectrum, $N=8$},
   %xmin=-1, xmax=42,
   %xlabel = {Ordered Eigenstate},
   %ylabel = {Energy},
   %x label style={at={(axis description cs:0.5,0.1)},anchor=north},
   %y label style={at={(axis description cs:0.1,.5)},anchor=south},
   %xtick  = {1,10,20,30,40},
   %% ytick  = {-8.4,-8.7},
   %title style = {font=\small}]
   % \addplot[mark=-,mark size=1.25,draw=red] table [x=point, y=energy, col sep=comma, only marks] {data/ftt_8_2.csv};
   %\end{axis}
   %\end{tikzpicture}
    %%%%% FIGURE 2 %%%%%
    \includegraphics[]{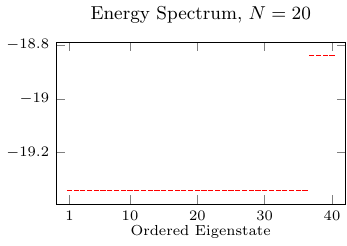}
    %\begin{tikzpicture}
    %\begin{axis}[
    %height = 0.75\columnwidth/1.5,
    %width = 0.75\columnwidth,
    %title = {Energy Spectrum, $N=20$},
    %xmin=-1, xmax=42,
    %xlabel = {Ordered Eigenstate},
    %% ylabel = {Energy},
    %x label style={at={(axis description cs:0.5,0.1)},anchor=north},
    %xtick  = {1,10,20,30,40},
    %title style = {font=\small}]
    % \addplot[mark=-,mark size=1.25,draw=red] table [x=point, y=energy, col sep=comma, only marks] {data/ftt_20_2.csv};
    %\end{axis}
    %\end{tikzpicture}
    %%%%% FIGURE 3 %%%%%
    %%% SUBPLOT %%%
    \includegraphics[]{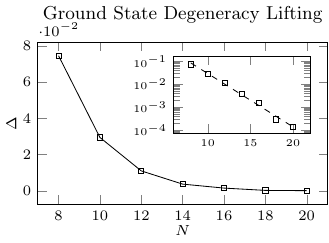}
    %\begin{tikzpicture}
    %\begin{axis}[xshift=2.3cm,yshift=1.2cm,
    %height = 0.5\columnwidth/1.5,
    %width = 0.5*.9*\columnwidth,
    %title = {},
    %ymode=log,
    %xmin=6,
    %xmax=22,
    %xlabel = {},
    %ylabel = {},
    %xtick  = {},
    %tick label style = {font=\fontsize{2}{2}\selectfont}
    %]
    % \addplot[mark=square, mark size=1.25,draw=black] table [x=point, y=delta, only marks] {data/splitting.csv};
    % \addplot [dashed] table[x=point,
    %y={create col/linear regression={y=delta}}]
    %{data/splitting.csv};
    %\end{axis}
    %%%% PLOT %%%
    %\begin{axis}[
    %height = 0.75\columnwidth/1.5,
    %width = 0.75*\columnwidth,
    %title = {Ground State Degeneracy Lifting},
    %xmin=7, xmax=21,
    %xlabel = {$N$},
    %ylabel = {$\Delta$},
    %x label style={at={(axis description cs:0.5,0.1)},anchor=north},
    %y label style={at={(axis description cs:0.2,.5)},anchor=south},
    %xtick  = {8,10,12,14,16,18,20},
    %title style = {font=\small}]
    % \addplot[mark=square, mark size=1.25,draw=black] table [x=point, y=delta] {data/splitting.csv};
    %\end{axis}
    %\end{tikzpicture}
    %%%%%%%%%%%%%%%%%%%

    \caption{\textbf{Perturbed $H_\Cl'$.} We have numerically calculated the spectrum of $H_\Cl'$ on open chains of even sizes between $N=8$ and $N=20$, with the spectrum for the smallest and largest chains plotted. We have set $\alpha=4\cdot 10^{-2}$. The inset of the rightmost plot shows the same data on a log-linear scale with line of best fit in order to more obviously illustrate the exponential decay.}
    \label{fig:d3_perturbed}
\end{figure*}

With this concrete model in hand, we explored several of the features discussed above numerically by constructing the Hamiltonian \cref{eq:ham} and finding ground states using DMRG in ITensor~\cite{ITensor}. We found $|D_3|^2=36$ ground states, consistent with the presence of a group-valued qudit at each edge.

We then tested the stability of the phase under random symmetric single-site perturbations. We formed the odd sublattice perturbations by first selecting a random Hermitian matrix $\mathcal{O}_\text{odd}$ then symmetrizing it under $D_3$:
\begin{equation}
    \mathcal{O}_\text{odd}^\text{sym}=\sum_{g\in D_3}\rx_g\mathcal{O}_\text{odd}\rx_g^\dagger.
\end{equation}
Symmetrizing an operator with respect to an MPO symmetry is more subtle, but one can derive using category theoretic methods that the most general single-site operator preserving $\rep{G}$ symmetry is a random real diagonal matrix~\cite{Bridgeman_2023}. We selected such a matrix $\mathcal{O}_\text{even}^\text{sym}$ as our even sublattice perturbation and considered the Hamiltonian
\begin{equation}\label{eq:noisy_H_itensors}
    H_\Cl'=H_\Cl-\alpha\sum_{i\text{ odd}}\mathcal{O}_\text{odd}^{\text{sym}(i)}-\alpha\sum_{i\text{ even}}\mathcal{O}_\text{even}^{\text{sym}(i)},
\end{equation}
where $\alpha$ is the strength of the perturbation. For a symmetric perturbation with fixed small $\alpha$, we expect that the ground state degeneracy is split by $\Delta\sim e^{-N}$ for finite system size $N$. In the thermodynamic limit $\Delta\to0$ and the degeneracy is restored. This is illustrated in \cref{fig:d3_perturbed}.

\subsection{Entanglement Spectrum Degeneracy}
We have also studied numerically the entanglement spectrum. Entanglement spectrum degeneracy is another signature of SPT order which arises from the projective action of the symmetry at the edge~\cite{levin_detecting_2006,pollmann_entanglement_2010}. For a bipartitioned state $\ket{\psi_{AB}}\in\mathcal{H}_A\otimes\mathcal{H}_B$, the entanglement spectrum is defined as the eigenspectrum of the reduced density matrix $\rho_A=\Tr_B[\ketbra{\psi_{AB}}{\psi_{AB}}]$.

We can identify the subsystems $A$ and $B$ with the left and right halves of an infinite chain. In order to calculate the entanglement spectrum of the chain, we would trace out one half of the chain, introducing a boundary of the remaining half-infinite chain where the two halves met. Because SPT states support unpinned degrees of freedom at their boundaries, we expect degeneracy in the energy spectrum of the remaining half-infinite chain, and therefore in the entanglement spectrum. 

\begin{figure}[tp]
    \centering
    \includegraphics[]{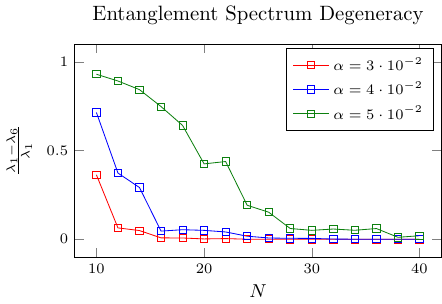}
    %\begin{tikzpicture}
    %\begin{axis}[
    %height = 0.9\columnwidth/1.5,
    %width = 0.9\columnwidth,
    %title=Entanglement Spectrum Degeneracy,
    %xmin=8, xmax=42,
    %ymin=-0.1, ymax=1.1,
    %xlabel = {$N$},
    %ylabel = {$\frac{\lambda_1-\lambda_6}{\lambda_1}$},
    %xtick  = {10,20,30,40},
    %label style = {font=\small},
    %title style = {font=\normalsize},
    %x label style={at={(axis description cs:0.5,0.05)},anchor=north},
    %y label style={at={(axis description cs:0.1,.5)},anchor=south},]
    % \addplot[mark=square,mark size=1.75,draw=red] table [x=point, y=alpha_3, col sep=comma] {data/ent_spec.csv};
    % \addplot[mark=square,mark size=1.75,draw=blue] table [x=point, y=alpha_4, col sep=comma] {data/ent_spec.csv};
    % \addplot[mark=square,mark size=1.75,draw=dartmouthgreen] table [x=point, y=alpha_5, col sep=comma] {data/ent_spec.csv};
    % \legend{$\alpha=3\cdot 10^{-2}$,$\alpha=4\cdot 10^{-2}$,$\alpha=5\cdot 10^{-2}$};
    %\end{axis}
    %\end{tikzpicture}
    \caption{Numerical results for the normalized splitting between the largest eigenvalue $\lambda_1$ and sixth largest eigenvalue $\lambda_6$ in the entanglement spectrum of the ground state of \cref{eq:noisy_H_itensors} on open chains of various lengths $N$, and as a function of the strength of symmetric noise $\alpha$.}
    \label{fig:d3_ent_spec}
\end{figure}

Numerically, we of course only have access to chains of finite length. However, we can think of the tensors deep in the bulk of a long-but-finite chain as approximating the tensors of an infinite chain, with the approximation converging as a function of system size. Accordingly, we can use a tensor deep in the bulk to calculate the entanglement spectrum degeneracy of the finite chain, the results of which are plotted in \cref{fig:d3_ent_spec}. In particular, we plot the normalized splitting between the largest eigenvalue $\lambda_1$ and sixth largest eigenvalue $\lambda_6$ in the entanglement spectrum of the ground state of the perturbed $H_\Cl'$ on open chains of various lengths. We find that this splitting tends to zero in the thermodynamic limit, indicative of a sixfold entanglement spectrum degeneracy. This is consistent with the physical picture of an unpinned $D_3$-qudit hosted at the edge.

\subsection{Topological Response}
Recall from \cref{sec:topo_response} that the $G$ cluster state exhibits a topological response to the insertion of symmetry charge when placed on a ring. In this section, we discuss in more detail the case $G=D_3$.
% In this case, the $G$ flux-inserted state $\ket{\Cl_g}$ acquires a $\rep G$ charge $\Gamma_{ij}(g)$ and the $\rep G$ flux-inserted state $\ket{\Cl_{\Gamma_{ij}}}$ transforms in a reducible representation of the $G$ symmetry.
In particular, we have
\begin{equation}\label{eq:D3_response}
    \begin{split}
    \hat B_\Gamma\ket{\Cl_g}=[\Gamma(g)]_{\alpha\beta}\ket{\Cl_g},\text{ for all } g\in D_3,\text{ }\Gamma\in\rep {D_3},\\
    \overleftarrow A_g\ket{\Cl_\Gamma}=\Gamma(g)\ket{\Cl_\Gamma},\text{ for all } g\in D_3,\text{ }\Gamma\in\{\mathbf{1},\Gamma_s\},
    \end{split}
\end{equation}
where the $\Gamma(g)$ and $[\Gamma(g)]_{\alpha\beta}$ are given in \cref{eq:D3_irreps}. The only case not covered in \cref{eq:D3_response} is the response to the insertion of $[\Gamma_{2d}]_{\alpha\beta}$ flux. In this case, the charge of the state $\ket{\Cl_{[\Gamma_{2d}]_{\alpha\beta}}}$ under $\overleftarrow A_g$ is non-abelian in character, i.e. we have a multiplet of states 
\begin{equation}
    \{\ket{\Cl_{[\Gamma_{2d}]_{\alpha\beta}}}:\alpha,\beta=1,2\} 
\end{equation}
transforming in the four-dimensional reducible representation $\Gamma_{2d}\oplus\Gamma_{2d}$. 

We can see explicitly that this is the case. If we define the matrices $M_g$ via
\begin{equation}
    [M_g]_{(\alpha,\beta),(\alpha',\beta')}:=\bra{\Cl_{[\Gamma_{2d}]_{\alpha'\beta'}}}\overleftarrow A_g\ket{\Cl_{[\Gamma_{2d}]_{\alpha\beta}}},
\end{equation}
then we can use \cref{eq:Gamma_alpha_beta,eq:D3_irreps} to compute
\begin{align*}
    \begin{gathered}
        M_e=\left(
\begin{array}{cccc}
 1 & 0 & 0 & 0 \\
 0 & 1 & 0 & 0 \\
 0 & 0 & 1 & 0 \\
 0 & 0 & 0 & 1 \\
\end{array}
\right)=\Gamma_{2d}(e)\oplus\Gamma_{2d}(e),
\\
M_R=\left(
\begin{array}{cccc}
 \frac{-1}{2} & \frac{-\sqrt{3}}{2}  & 0 & 0 \\
 \frac{\sqrt{3}}{2}  & \frac{-1}{2} & 0 & 0 \\
 0 & 0 & \frac{-1}{2} & \frac{-\sqrt{3}}{2}  \\
 0 & 0 & \frac{\sqrt{3}}{2}  & \frac{-1}{2} \\
\end{array}
\right)=\Gamma_{2d}(R)\oplus\Gamma_{2d}(R),\\
M_{R^2}=\left(
\begin{array}{cccc}
 \frac{-1}{2} & \frac{\sqrt{3}}{2}  & 0 & 0 \\
 \frac{-\sqrt{3}}{2}  & \frac{-1}{2} & 0 & 0 \\
 0 & 0 & \frac{-1}{2} & \frac{\sqrt{3}}{2}  \\
 0 & 0 & \frac{-\sqrt{3}}{2}  & \frac{-1}{2} \\
\end{array}
\right)=\Gamma_{2d}(R^2)\oplus\Gamma_{2d}(R^2),\\
M_S=\left(
\begin{array}{cccc}
 -1 & 0 & 0 & 0 \\
 0 & 1 & 0 & 0 \\
 0 & 0 & -1 & 0 \\
 0 & 0 & 0 & 1 \\
\end{array}
\right)=\Gamma_{2d}(S)\oplus\Gamma_{2d}(S),\\
    \end{gathered}
\end{align*}
\begin{align*}
    \begin{gathered}
M_{SR}=\left(
\begin{array}{cccc}
 \frac{1}{2} & \frac{-\sqrt{3}}{2}  & 0 & 0 \\
 \frac{-\sqrt{3}}{2}  & \frac{-1}{2} & 0 & 0 \\
 0 & 0 & \frac{1}{2} & \frac{-\sqrt{3}}{2}  \\
 0 & 0 & \frac{-\sqrt{3}}{2}  & \frac{-1}{2} \\
\end{array}
\right)=\Gamma_{2d}(SR)\oplus\Gamma_{2d}(SR),\\
M_{SR^2}=\left(
\begin{array}{cccc}
 \frac{1}{2} & \frac{\sqrt{3}}{2}  & 0 & 0 \\
 \frac{\sqrt{3}}{2}  & \frac{-1}{2} & 0 & 0 \\
 0 & 0 & \frac{1}{2} & \frac{\sqrt{3}}{2}  \\
 0 & 0 & \frac{\sqrt{3}}{2}  & \frac{-1}{2} \\
\end{array}
\right)=\Gamma_{2d}(SR^2)\oplus\Gamma_{2d}(SR^2).
    \end{gathered}
\end{align*}
As expected, we find that the states $\{\ket{\Cl_{[\Gamma_{2d}]_{\alpha\beta}}}:\alpha,\beta=1,2\}$ transform in the reducible representation $\Gamma_\text{2d}\oplus\Gamma_\text{2d}$.
This is a special case of the fact that in general, a $G$ cluster state with $\Gamma_{\alpha\beta}$ flux inserted transforms as $d_\Gamma\Gamma$, namely, $d_\Gamma$ copies of the direct sum of $\Gamma$.

%%%%%%%%%%%%%%%%%%%%%%%%%%%%%%%%%%%%%%%%%%%%%%%%%%%%%%%%%%%%%%%%%%
\section{Repeated Action of $U_\Cl$}\label{sec:repeated_U_C}
By repeatedly applying the state preparation unitary to $\ket{\psi_0}$, we can define a family of SPT states $\ket{\Cl^n}:=(U_\mathcal{C})^n\ket{\psi_0}$ with the same symmetry and different projective actions. We find
\begin{equation}
    \ket{\Cl^n}=\mathcal{N}\sum_{\{g_i\}}\ket{g_1}\ket{(g_1)^n(\bar{g_2})^n}\ket{g_2}\cdots\ket{(g_{N-1})^n(\bar{g_N})^n}\ket{g_N}.
\end{equation}

The MPS tensors for this state are given by
\begin{equation}
  \begin{tikzpicture}
    % \node[irrep] at (0.2,0) {};
    \draw (-0.5,0)--(0.5,0);
    \draw (0,0) -- (0,0.5);
    \node[odd] (t) at (0,0) {};
    \node[oddn] (t) at (0,0) {};
  \end{tikzpicture}
\ =\sum_{g}\ketbra{g^n}{g^n}\otimes\ket{g},\quad
  \begin{tikzpicture}
    % \node[irrep] at (0.2,0) {};
    \draw (-0.5,0)--(0.5,0);
    \draw (0,0) -- (0,0.5);
    \node[even] (t) at (0,0) {};

  \end{tikzpicture}
\ =\sum_g \lx_g\otimes\ket{g}~,
\end{equation}
on odd and even sites respectively. The even sites are the same as those of $\ket{\Cl}$ and are denoted the same. The odd sites differ, so we introduce the above numbered tensor. This tensor satisfies

\begin{gather*}
  \begin{tikzpicture}[baseline={(0,.4)}]
    \draw (-1,0)--(1,0);
    \draw (0,0) -- (0,1);
    \node[odd] (t) at (0,0) {};
    \node[oddn] (t) at (0,0) {};
    % \node[] at (0.2,0.2) {o};
    \node[mpo] at (0,0.5) {$\lx_g$};
    % \node[] at (0.2,0.7) {$g$};
  \end{tikzpicture}
\ = \begin{tikzpicture}[baseline={(0,.4)}]
    \draw (-1.2,0)--(1.2,0);
    \draw (0,0) -- (0,1);
    \node[odd] (t) at (0,0) {};
    \node[oddn] (t) at (0,0) {};
    % \node[] at (0.2,0.2) {o};
    \node[mpo] at (-0.6,0) {$\lx_{g^n}^\dagger$};
     \node[mpo] at (0.6,0) {$\lx_{g^n}$};
    % \node[] at (-0.3,0.2) {$g$};
  \end{tikzpicture}~, \\
  \begin{tikzpicture}[baseline={(0,.4)}]
    \draw (-1,0)--(1,0);
    \draw (0,0) -- (0,1);
    \node[odd] (t) at (0,0) {};
    \node[oddn] (t) at (0,0) {};
    % \node[] at (0.2,0.2) {o};
    \node[mpo] at (0,0.5) {$\rx_g$};
    % \node[] at (0.2,0.7) {$g$};
  \end{tikzpicture} \ = \begin{tikzpicture}[baseline={(0,.4)}]
    \draw (-1.2,0)--(1.2,0);
    \draw (0,0) -- (0,1);
    \node[odd] (t) at (0,0) {};
    \node[oddn] (t) at (0,0) {};
    % \node[] at (0.2,0.2) {o};
    \node[mpo] at (-0.6,0) {$\rx_{g^n}^\dagger$};
     \node[mpo] at (0.6,0) {$\rx_{g^n}$};
    % \node[] at (-0.3,0.2) {$g$};
  \end{tikzpicture}~,
  \end{gather*}
  \begin{gather*}
  \begin{tikzpicture}[baseline={(0,.4)}]
    \draw (-1,0)--(1,0);
    \draw (0,0) -- (0,1);
    \draw[red] (-1,0.5)--(1,0.5);
    \node[odd] (t) at (0,0) {};
    \node[oddn] (t) at (0,0) {};
    % \node[] at (0.2,0.2) {o};
    \node[mpo] at (0,0.5) {$\zt_{\Gamma^{n}}$};
    % \node[] at (0.2,0.7) {$g$};
  \end{tikzpicture}
\ = \begin{tikzpicture}[baseline={(0,.4)}]
    \draw (-1,0)--(1,0);
    \draw (0,0) -- (0,1);
    \draw[red] (-1,-0.2)--(1,-0.2);
    \node[odd] (t) at (0,0) {};
    \node[oddn] (t) at (0,0) {};
    % \node[] at (0.2,0.2) {o};
    \node[mpo] at (-0.5,-0.1) {$\zt_\Gamma$};
    % \node[] at (-0.3,0.2) {$g$};
  \end{tikzpicture} \
  = \begin{tikzpicture}[baseline={(0,.4)}]
    \draw (-1,0)--(1,0);
    \draw (0,0) -- (0,1);
    \draw[red] (-1,-0.2)--(1,-0.2);
    \node[odd] (t) at (0,0) {};
    \node[oddn] (t) at (0,0) {};
    % \node[] at (0.2,0.2) {o};
    \node[mpo] at (0.5,-0.1) {$\zt_\Gamma$};
    % \node[] at (-0.3,0.2) {$g$};
  \end{tikzpicture}~. \
\end{gather*}

This state is still $G_R\times \rep{G}$-symmetric, with symmetry operators $\overleftarrow A_g$ and $\hat B_\Gamma$. We can repeat the analysis of Sec.~\ref{sec:edge_modes} to determine the edge modes of this new state:

\begin{align*}
\overleftarrow A_g\ket{\Cl^n} =& \begin{tikzpicture}[baseline={(0,.4)}]
    \draw (0,0)--(5.5,0);
    \foreach \x in {0.5,2,3.5,5}{
		     \draw (\x,0) -- (\x,1);
		    \node[mpo] (t\x) at (\x,0.5) {$\rx_g$};
		    \node[odd] at (\x,0) {};
		    \node[oddn] at (\x,0) {};
		  }
		  \foreach \x in {1.25,4.25}{
		     \draw (\x,0) -- (\x,1);
		    \node[even] at (\x,0) {};
		  }
		  \node[fill=white] at (2.75,0) {$\dots$};
		  \node[edge] at (0,0) {};
		  \node[edge] at (5.5,0) {};
  \end{tikzpicture}\\
  =& \begin{tikzpicture}[baseline={(0,.4)}]
    \draw (-.8,0)--(6.3,0);
    \foreach \x in {0.5,2,3.5,5}{
		     \draw (\x,0) -- (\x,1);
		    \node[odd] at (\x,0) {};
		    \node[oddn] at (\x,0) {};
		  }
		  \foreach \x in {1.25,4.25}{
		     \draw (\x,0) -- (\x,1);
		    \node[even] at (\x,0) {};
		  }
		  \node[fill=white] at (2.75,0) {$\dots$};
		  \node[edge] at (-.8,0) {};
		  \node[edge] at (6.3,0) {};
		  \node[mpo] at (-0.15,0) {$\rx_{g^n}^\dagger$};
		  \node[mpo] at (5.65,0) {$\rx_{g^n}$};
  \end{tikzpicture} \
\end{align*}

\begin{align*}
\hat B_{\Gamma_{\alpha\beta}}\ket{\Cl^n}=\ &\begin{tikzpicture}[baseline={(0,.4)}]
    \draw (0,0)--(5.5,0);
    \draw[red] (0,0.5) rectangle (5.5,0.5);
    \foreach \x in {0.5,2,3.5,5}{
		     \draw (\x,0) -- (\x,0.45);
		     \draw (\x,0.55) -- (\x,1);
		    \node[odd] at (\x,0) {};
		    \node[oddn] at (\x,0) {};
		  }
		  \foreach \x in {1.25,4.25}{
		     \draw (\x,0) -- (\x,1);
		    \node[even] at (\x,0) {};
		    \node[mpo] (t\x) at (\x,0.5) {$\zt_\Gamma$};
		  }
		  \node[fill=white] at (2.75,0) {$\dots$};
		  \node[fill=white] at (2.75,0.5) {$\dots$};
		  \node[edge] at (0,0) {};
		  \node[edge] at (5.5,0) {};
        \node[edge,red] at (0,0.5) {};
		  \node[edge,red] at (5.5,0.5) {};
  \end{tikzpicture} \\ = \ &
  \begin{tikzpicture}[baseline={(0,.4)}]
    \draw (-0.5,0)--(6,0);
    \draw[red] (-0.5,-0.2)--(6,-0.2);
    \foreach \x in {0.5,2,3.5,5}{
		     \draw (\x,0) -- (\x,1);
		    \node[odd] at (\x,0) {};
		    \node[oddn] at (\x,0) {};
		  }
		  \foreach \x in {1.25,4.25}{
		     \draw (\x,0) -- (\x,1);
		    \node[even] at (\x,0) {};
		  }
		  \node[fill=white] at (2.75,0) {$\dots$};
		  \node[fill=white] at (2.75,-0.2) {$\dots$};
		  \node[edge] at (-0.5,0) {};
		  \node[edge] at (6,0) {};
		  \node[mpo] at (0,-0.1) {$\zt_\Gamma$};
		  \node[mpo] at (5.5,-0.1) {$\zt_\Gamma^\dagger$};
        \node[edge,red] at (-0.5,-0.2) {};
		  \node[edge,red] at (6,-0.2) {};
  \end{tikzpicture} \
\end{align*}
From this diagrammatic derivation, we can read off the edge modes
 \begin{equation}
 G: \rx_{g^n}^{(L)}\rx_{g^n}^{(R)},\quad
 \rep G:\left[\zt_\Gamma^{(L)}.\zt_\Gamma^{\dagger(R)}\right]_{\alpha\beta},
\end{equation}
which satisfy the commutation relations
\begin{equation}\label{eq:n_edge_comm}
\begin{split}
    \rx_{g^n}^{(L)}\left[\zt_\Gamma^{(L)}.\zt_\Gamma^{\dagger(R)}\right]_{\alpha\beta} =  \left[\zt_\Gamma^{(L)}.\Gamma(g^n).\zt_\Gamma^{\dagger(R)}\right]_{\alpha\beta}\rx_{g^n}^{(L)},\\ \rx_{g^n}^{(R)}\left[\zt_\Gamma^{(L)}.\zt_\Gamma^{\dagger(R)}\right]_{\alpha\beta} =  \left[\zt_\Gamma^{(L)}.\Gamma(\bar{g}^n).\zt_\Gamma^{\dagger(R)}\right]_{\alpha\beta}\rx_{g^n}^{(R)}.
\end{split}
\end{equation}

Let $n_*$ be the exponent of the group, defined to be the least common multiple of the order of each element in the group. This implies that $g^{n_*}=e$ for all $g\in G$, and $n_*$ is the smallest integer for which this is true. When $n=n_*$, we have
\begin{equation}
    \ket{\Cl^{n_*}}=\ket{\mathbf{1},e,\mathbf{1},\ldots,\mathbf{1},e,\mathbf{1}}=\ket{\psi_0}.
\end{equation}
This shows that applying $U_\Cl$ to $\ket{\psi_0}$ $n_*$ times returns $\ket{\psi_0}$. This can also be seen in the fact that the edge mode algebra \cref{eq:n_edge_comm} trivializes -- that is, the edge operators all commute -- when $n=n_*$. We can therefore create a family of $n_*$ SPT states with symmetry $G_R\times \rep G$  by repeatedly applying $U_\Cl$. In the case of Abelian $G$, these states give rise to distinct SPT phases, and moreover realizes all SPT states that require protection by both $G$ and $\rep{G}$. When $G$ is non-abelian, analysis of the fiber functors (i.e. short-range entangled phases~\cite{thorngren_fusion_2019}) of $G\times \rep G$ leads to the conclusion that there are additional $G\times \rep G$ SPTs beyond the states $\ket{\Cl^n}$. For example, the SPTs could be distinct when protected individually by only $G$ or $\rep G$. Furthermore, the states $\ket{\Cl^n}$ do not necessarily belong to $n^*$ unique phases for general $G$. We comment on the $G=D_3$ case in \cref{ap:D3_SymTFT} and leave a more detailed study of these SPTs to future work.

%%%%%%%%%%%%%%%%%%%%%%%%%%%%%%%%%%%%%%%%%%%%%%%%%%%%%%%%%%%%%%%%%%%%%%%%%%%%%%%%%%%%%%%%%%%%%%%%%%%%%%%%%%%%%%%%%%%%%%%
\section{Utility for Measurement-Based Quantum Computation}
Measurement-based quantum computation (MBQC) is an alternative paradigm for quantum computing which enacts unitary gates through measurement. Unlike traditional methods of quantum computation, MBQC begins with a highly entangled resource state, rather than generating entanglement during the computation. The choice of measurement basis enables the quantum programmer to implement particular unitary gates (often up to probabilistic byproducts) and is in general conditioned on outcomes of earlier measurements. Recently, it has also been shown that MBQC can be formulated in terms of a gauge theory~\cite{wong_gauge_2022-1}.

Cluster states were the first states used for MBQC, and they remain the paradigmatic resource state~\cite{briegel_measurement-based_2009, briegel_persistent_2001, raussendorf_measurement-based_2003, raussendorf_one-way_2001}. Later works have shown that the computational power of cluster states is intimately related to their SPT order, such that entire SPT phases are equivalent resource states for MBQC~\cite{else_symmetry-protected_2012-2,Raussendorf17,stephen_computational_2017,stephen_subsystem_2019-1,Raussendorf19}. This observation has given rise to the notion of \textit{computational phases of matter}, significantly expanding the conception of what can be an MBQC resource state. Because of the correspondence between SPT order and MBQC universality, we expect that the $G$ cluster state should be useful as an MBQC resource state.

In this section, we will present an algorithm for MBQC on $G$ cluster states based on abelian groups and non-abelian groups which are semidirect products of abelian groups, as well as lay out the idea to perform MBQC when $G$ is a \textit{solvable group}: a group formed via extensions of abelian groups. The abelian case is relatively straightforward, and we will show that the more complex cases can be understood as several parallel instances of the abelian case.

\subsection{MBQC with MPS}

The language of matrix product states naturally provides an elegant implementation of MBQC in 1d~\cite{gross2007novel,gross_measurement-based_2007}. This is called the correlation or virtual space picture of MBQC. In this picture, the logical qudit is the state in the virtual space encoding the boundary condition on the right end of the chain, and the computation takes place in the virtual space.

Consider an MPS
\begin{equation}
    \ket{\psi} = \sum_{\{g_i\}}\bra{v_L}A^{(1)}_{g_1}\cdots A^{(n)}_{g_n}\ket{v_R}|g_1,\ldots,g_n\rangle,
\end{equation}
where $\ket{v_R}$ is the right boundary condition and logical qudit. If we measure the physical qudit at site $n$ and obtain the outcome $\ket{s}$, we find that the new state is
\begin{equation}
    \ket{\psi} \xrightarrow{\ket{s}} \sum_{\{g_i\}}\bra{v_L}A^{(1)}_{g_1}\cdots A^{(n-1)}_{g_{n-1}}A^{(n)}_s\ket{v_R}|g_1,\ldots,g_{n-1}\rangle\ket{s},
\end{equation}
where $A^{n}_s\:=\sum_{g_n} \braket{s|g_n} A_{g_n}^{(n)}$. We can think of the $n-1$ qudits which we have not yet measured as an MPS
\begin{equation}
    \ket{\psi'} = \sum_{\{g_i\}}\bra{v_L}A^{(1)}_{g_1}\cdots A^{(n-1)}_{g_{n-1}}\ket{v_R'}|g_1,\ldots,g_{n-1}\rangle,
\end{equation}
where $\ket{v_R'}:=A^{(n)}_s\ket{v_R}$.

We see that this measurement effectively acts on the virtual state $\ket{v_R}$ with the operator $A^{(n)}_s$, yielding the state $\ket{v_R'}:=A^{(n)}_s\ket{v_R}$. The evolution from $\ket{v_R}$ to $\ket{v_R'}$ is one step of computation within the MBQC paradigm, akin to one layer of a quantum circuit.
An MPS is said to be \textit{universal} if this evolution can be used to apply any $SU(\chi)$ gate, where $\chi$ is the dimension of the virtual space.

\subsection{$G$ cluster state as an MBQC Resource State}

Recall the explicit form of the MPS tensors for the $G$ cluster state:
\begin{equation}
  \begin{tikzpicture}
    % \node[irrep] at (0.2,0) {};
    \draw (-0.5,0)--(0.5,0);
    \draw (0,0) -- (0,0.5);
    \node[odd] (t) at (0,0) {};
  \end{tikzpicture}
\ =\sum_{g\in G}\ketbra{g}{g}\otimes\ket{g},\quad
  \begin{tikzpicture}
    % \node[irrep] at (0.2,0) {};
    \draw (-0.5,0)--(0.5,0);
    \draw (0,0) -- (0,0.5);
    \node[even] (t) at (0,0) {};

  \end{tikzpicture}
\ =\sum_{g\in G} \lx_g\otimes\ket{g}.
\end{equation}
If we measure this state in the irrep matrix element basis on odd sites and group basis on even sites, we find the action on the virtual Hilbert space to be
\begin{equation}
\begin{split}
    \text{Odd: }&\ket{v_R} \overset{\ket{\Gamma_{\alpha\beta}}}{\longrightarrow} \left[\zt_{\Gamma}\right]_{\alpha\beta} \ket{v_R},\\ \text{Even: }& \ket{v_R} \overset{\ket{g}}{\longrightarrow} \lx_g \ket{v_R},
\end{split}
\end{equation}
where $\ket{\Gamma_{\alpha\beta}}$ and $\ket{g}$ are the measurement outcomes. Diagrammatically, we can write this as
\begin{equation}
    \begin{tikzpicture}[baseline={(0,.4)}]
    \draw (.25,0)--(1.75,0);
    \foreach \x in {1}{
		     \draw (\x,0) -- (\x,.75);
       \draw (\x-.15,.75) -- (\x+.15,.75);
	       \draw (\x-.15,.8) -- (\x+.15,.8);}
		    \node[even] at (1,0) {};
    \node at (1,.95) {$g$};
    \node[edge] at (1.75,0) {};
  \end{tikzpicture}\ = \begin{tikzpicture}[baseline={(0,.4)}]
    \draw (.5,0)--(1.5,0);
	\node[mpo] at (1,0) {$\lx_g$};
 \node[edge] at (1.5,0) {};
  \end{tikzpicture} \ ,\quad \ \begin{tikzpicture}[baseline={(0,.4)}]
    \draw (.25,0)--(1.75,0);
    \foreach \x in {1}{
		     \draw (\x,0) -- (\x,.75);
       \draw (\x-.15,.75) -- (\x+.15,.75);
	       \draw (\x-.15,.8) -- (\x+.15,.8);}
		    \node[odd] at (1,0) {};
    \node at (1,.95) {$\Gamma_{\alpha\beta}$};
    \node[edge] at (1.75,0) {};
  \end{tikzpicture}\ = \begin{tikzpicture}[baseline={(0,.4)}]
    \draw (.5,0)--(1.5,0);
    \draw[red] (.5,-.3)--(1.5,-.3);
	\node[mpo] at (1,-.15) {$\zt_\Gamma$};
    \node[edge] at (1.5,0) {};
    \node[edge,red] at (1.5,-.3) {};
    \node[edge,red] at (0.5,-.3) {};
  \end{tikzpicture}\ ,
\end{equation}

This poses a complication: the $Z$-type Pauli errors are not in general unitary, and the $X$-type Pauli errors do not in general commute. This renders the straightforward generalization of the abelian protocol intractable. However, we find that the $G$ cluster state is still universal for MBQC for a wide class of non-abelian groups, though some additional modifications from the abelian case must be made.

\subsubsection{Warm-Up: G Abelian}\label{sec:abelian_case}
We first show that the $G$ cluster state is universal when the group is abelian. Here, we generalize the construction in Ref.~\onlinecite{clark_valence_2006} for performing MBQC with the qudit cluster state. We translate the protocol to the correlation space picture of MBQC and generalized from qudits $(G=\mathbb{Z}_d)$ to arbitrary abelian $G$.

\textbf{Implementing Gates Virtually.} The universal single group-valued qudit gate set which we want to implement virtually is
    \begin{equation}
        R_G=\left\{R_g^{(\pm)}(\theta),R_\Gamma^{(\pm)}(\theta):g\in G,\Gamma\in\rep G\right\},
    \end{equation}
where $\theta\in U(1)$ is a rotation angle. The rotation operators are given by
\begin{equation}\label{eq:left_rot}
\begin{split}
        \overrightarrow R_g^{(\pm)}(\theta)=\left\{\begin{tabular}{rl}
             $e^{i\theta(\lx_g+\lx_{\bar g})}$, & $g\neq \bar g,+$  \\
             $e^{-\theta (\lx_g-\lx_{\bar g})}$, & $g\neq \bar g,-$  \\
             $e^{i\theta\lx_g}$, & $g=\bar g$
        \end{tabular}\right. ,\\
        R_\Gamma^{(\pm)}(\theta)=\left\{\begin{tabular}{rl}
             $e^{i\theta(\zt_\Gamma+\zt_\Gamma^\dagger)}$, & $\Gamma\text{ complex},+$  \\
             $e^{-\theta (\zt_\Gamma-\zt_\Gamma^\dagger)}$, & $\Gamma\text{ complex},-$  \\
             $e^{i\theta\zt_\Gamma}$, & $\Gamma\text{ real}$
        \end{tabular}\right. .
\end{split}
\end{equation}
We prove in \cref{ap:universality} that this gate set is indeed universal for a single group-valued qudit.

We will now explain how to implement any gate in $R_G$ on the virtual level by measuring the $G$ cluster state in appropriately rotated bases. Essential to this protocol is the fact that we can implement these gates on the virtual level by applying an appropriate unitary to the physical legs. In particular, we find that

\begin{equation}\label{eq:phys_to_virt}
\begin{split}
\begin{tikzpicture}[baseline={(0,.4)}]
    \draw (0.25,0)--(1.75,0);
    \foreach \x in {1}{
		     \draw (\x,0) -- (\x,1);}
		    \node[even] at (1,0) {};
		  \node[mpo] at (1,.5) {\normalsize $\overleftarrow R_g^{(\pm)}$};
		  % \node[edge] at (1.75,0) {};
  \end{tikzpicture} \ = \ \begin{tikzpicture}[baseline={(0,.4)}]
    \draw (.25,0)--(2.75,0);
    \foreach \x in {1}{
		     \draw (\x,0) -- (\x,1);}
		    \node[even] at (1,0) {};
		  \node[mpo] at (2,0) {\normalsize$\overrightarrow R_g^{(\pm)}$};
		  % \node[edge] at (2.75,0) {};
  \end{tikzpicture},\\
  \begin{tikzpicture}[baseline={(0,.4)}]
    \draw (0.25,0)--(1.75,0);
    \foreach \x in {1}{
		     \draw (\x,0) -- (\x,1);}
		    \node[odd] at (1,0) {};
		  \node[mpo] at (1,.5) {\normalsize $R_\Gamma^{(\pm)}$};
		  % \node[edge] at (1.75,0) {};
  \end{tikzpicture} \ = \ \begin{tikzpicture}[baseline={(0,.4)}]
    \draw (0.25,0)--(2.75,0);
    \foreach \x in {1}{
		     \draw (\x,0) -- (\x,1);}
		    \node[odd] at (1,0) {};
		  \node[mpo] at (2,0) {\normalsize$R_\Gamma^{(\pm)}$};
		  % \node[edge] at (2.75,0) {};
  \end{tikzpicture},
\end{split}
\end{equation}
 where $\overleftarrow R_g^{(\pm)}$ is defined as in \cref{eq:left_rot} but with right multiplication instead of left. These relations follow straightforwardly from \cref{eq:pulling_through_main}.

To implement the gate $\overrightarrow R_g^{(\pm)}(\theta)$ on the virtual level, we can measure an even site in the basis $\left\{\overleftarrow R_g^{(\pm)\dagger}(\theta)\ket{h}:h\in G\right\}$:
\begin{align}
\begin{tikzpicture}[baseline={(0,.4)}]
    \draw (0.25,0)--(1.75,0);
    \foreach \x in {1}{
		     \draw (\x,0) -- (\x,1);
          \draw (\x-.15,1) -- (\x+.15,1);
	       \draw (\x-.15,1.05) -- (\x+.15,1.05);}
		    \node[even] at (1,0) {};
		  \node[mpo] at (1,.5) {\normalsize $\overleftarrow R_g^{(\pm)}$};
		  % \node[edge] at (1.75,0) {};
    \node at (1,1.2) {$h$};
  \end{tikzpicture} \ = \ \begin{tikzpicture}[baseline={(0,.4)}]
    \draw (.25,0)--(2.75,0);
    \foreach \x in {1}{
		     \draw (\x,0) -- (\x,.75);
       \draw (\x-.15,.75) -- (\x+.15,.75);
	       \draw (\x-.15,.8) -- (\x+.15,.8);}
		    \node[even] at (1,0) {};
		  \node[mpo] at (2,0) {\normalsize$\overrightarrow R_g^{(\pm)}$};
		  % \node[edge] at (2.75,0) {};
    \node at (1,.95) {$h$};
  \end{tikzpicture}
\end{align}

Similarly, to implement the gate $R_\Gamma^{(\pm)}(\theta)$ on the virtual level, we can measure an odd site in the basis\footnote{Because the group is abelian, the set of irreducible representations form a group isomorphic to the original group and can thus be labeled by the group elements. This is the reasoning for the irrep notation $\Gamma_x$: $\Gamma_x$ is the irrep that is associated with $x\in G$ through the Fourier transform.}
$\left\{R_\Gamma^{(\pm)\dagger}(\theta)\ket{\Gamma_x}:x\in G\right\}$:
\begin{align}
\begin{tikzpicture}[baseline={(0,.4)}]
    \draw (0.25,0)--(1.75,0);
    \foreach \x in {1}{
		     \draw (\x,0) -- (\x,1);
          \draw (\x-.15,1) -- (\x+.15,1);
	       \draw (\x-.15,1.05) -- (\x+.15,1.05);}
		    \node[odd] at (1,0) {};
		  \node[mpo] at (1,.5) {\normalsize $R_\Gamma^{(\pm)}$};
		  % \node[edge] at (1.75,0) {};
    \node at (1,1.2) {$\Gamma_x$};
  \end{tikzpicture} \ = \ \begin{tikzpicture}[baseline={(0,.4)}]
    \draw (.25,0)--(2.75,0);
    \foreach \x in {1}{
		     \draw (\x,0) -- (\x,.75);
       \draw (\x-.15,.75) -- (\x+.15,.75);
	       \draw (\x-.15,.8) -- (\x+.15,.8);}
		    \node[odd] at (1,0) {};
		  \node[mpo] at (2,0) {\normalsize$R_\Gamma^{(\pm)}$};
		  % \node[edge] at (2.75,0) {};
    \node at (1,.95) {$\Gamma_x$};
  \end{tikzpicture}
\end{align}
Recalling that measuring the even and odd MPS tensors has the following action:
\begin{align}
    \begin{tikzpicture}[baseline={(0,.4)}]
    \draw (.25,0)--(1.75,0);
    \foreach \x in {1}{
		     \draw (\x,0) -- (\x,.75);
       \draw (\x-.15,.75) -- (\x+.15,.75);
	       \draw (\x-.15,.8) -- (\x+.15,.8);}
		    \node[even] at (1,0) {};
    \node at (1,.95) {$g$};
  \end{tikzpicture}\ = \begin{tikzpicture}[baseline={(0,.4)}]
    \draw (.5,0)--(1.5,0);
	\node[mpo] at (1,0) {$\lx_g$};
  \end{tikzpicture} \ ,\quad \ \begin{tikzpicture}[baseline={(0,.4)}]
    \draw (.25,0)--(1.75,0);
    \foreach \x in {1}{
		     \draw (\x,0) -- (\x,.75);
       \draw (\x-.15,.75) -- (\x+.15,.75);
	       \draw (\x-.15,.8) -- (\x+.15,.8);}
		    \node[odd] at (1,0) {};
    \node at (1,.95) {$\Gamma$};
  \end{tikzpicture}\ = \begin{tikzpicture}[baseline={(0,.4)}]
    \draw (.5,0)--(1.5,0);
	\node[mpo] at (1,0) {$\zt_\Gamma$};
  \end{tikzpicture}\ ,
\end{align}
we see that the total effect of the rotated measurement on the boundary state is
\begin{align}\label{eq:total_meas}
\begin{split}
    \ket{v_R} \xrightarrow{\overleftarrow R_g^{(\pm)\dagger}\otimes R_\Gamma^{(\pm)\dagger}\ket{h,\Gamma_x}} \lx_h \overrightarrow R_g^{(\pm)} \zt_{\Gamma_x} R_\Gamma^{(\pm)} \ket{v_R}.
\end{split}
\end{align}
We are therefore able to deterministically implement the entire gate set $R_G$ up to group-based Pauli byproducts which are dependent on the measurement outcomes.

\textbf{Adaptive Measurements.} Notice that Pauli by-products lie between the logical gates in \cref{eq:total_meas}. We would prefer for all logical gates to be to the right, acting on the logical qudit, with Pauli by-products to the left. We can ensure this form by adapting later measurement bases to compensate for earlier Pauli errors.

Consider an example with three sites which illustrates the general strategy. The first measurement is performed on an odd site in the usual basis:
\begin{widetext}
    \begin{equation}
        \begin{tikzpicture}[baseline={(0,.4)}]
    \draw (3,0)--(5.5,0);
    \foreach \x in {3.5,5}{

		     \draw (\x,0) -- (\x,1);
		    \node[odd] at (\x,0) {};
		  }
		  \foreach \x in {4.25}{
		     \draw (\x,0) -- (\x,1);
		    \node[even] at (\x,0) {};
		  }
		  \node[edge] (ref) at (5.5,0) {};
    \node[fill=white] at (2.75,0) {$\dots$};
    \coordinate (base) at (0,5);
  \end{tikzpicture} \ \mapsto \
  \begin{tikzpicture}[baseline={(0,.4)}]
    \draw (3,0)--(5.5,0);
    \foreach \x in {3.5,5}{

		     \draw (\x,0) -- (\x,1);
		    \node[odd] at (\x,0) {};
		  }
		  \foreach \x in {4.25}{
		     \draw (\x,0) -- (\x,1);
		    \node[even] at (\x,0) {};
		  }
        \node[mpo] at (5,0.5) {\normalsize$R_\Gamma^{(\pm)}$};
        \draw (5-.15,1) -- (5+.15,1);
	    \draw (5-.15,1.05) -- (5+.15,1.05);
        \node at (5,1.2) {$\Gamma_x$};
		  \node[edge] at (5.5,0) {};
    \node[fill=white] at (2.75,0) {$\dots$};
  \end{tikzpicture}\ = \
  \begin{tikzpicture}[baseline={(0,.4)}]
    \draw (3,0)--(6.5,0);
    \foreach \x in {3.5,5}{
		     \draw (\x,0) -- (\x,1);
		    \node[odd] at (\x,0) {};
		  }
		  \foreach \x in {4.25}{
		     \draw (\x,0) -- (\x,1);
		    \node[even] at (\x,0) {};
		  }
        \node[mpo] at (5.75,0) {\normalsize$R_\Gamma^{(\pm)}$};
        \draw (5-.15,1) -- (5+.15,1);
	    \draw (5-.15,1.05) -- (5+.15,1.05);
        \node at (5,1.2) {$\Gamma_x$};
		  \node[edge] at (6.5,0) {};
    \node[fill=white] at (2.75,0) {$\dots$};
  \end{tikzpicture}\ = \
  \begin{tikzpicture}[baseline={(0,.4)}]
    \draw (3,0)--(6.5,0);
    \foreach \x in {3.5}{
		     \draw (\x,0) -- (\x,1);
		    \node[odd] at (\x,0) {};
		  }
		  \foreach \x in {4.25}{
		     \draw (\x,0) -- (\x,1);
		    \node[even] at (\x,0) {};
		  }
        \node[mpo] at (5.9,0) {\normalsize$R_\Gamma^{(\pm)}$};
        \node[mpo] at (4.9,0) {\normalsize$\zt_{\Gamma_x}$};
		  \node[edge] at (6.5,0) {};
    \node[fill=white] at (2.75,0) {$\dots$};
  \end{tikzpicture}\ .
    \end{equation}
In order to compensate for the $\zt_{\Gamma_x}$ error on the virtual level, we measure in the basis $\left\{\left(\overleftarrow R_g^{(\pm)}(\theta)Z_{\Gamma_x}\right)^\dagger\ket{h}:h\in G\right\}$, which will propagate the error past the next logical operation:
\begin{equation*}
    \mapsto \
    \begin{tikzpicture}[baseline={(0,.4)}]
    \draw (3,0)--(6.5,0);
    \foreach \x in {3.5}{
		     \draw (\x,0) -- (\x,1);
		    \node[odd] at (\x,0) {};
		  }
		  \foreach \x in {4.25}{
		     \draw (\x,0) -- (\x,2);
		    \node[even] at (\x,0) {};
		  }
        \node[mpo] at (4.25,1.5) {\normalsize$\overleftarrow R_g^{(\pm)}$};
        \node[mpo] at (4.25,.75) {\normalsize$\zt_{\Gamma_x}$};
        \draw (4.25-.15,2) -- (4.25+.15,2);
	    \draw (4.25-.15,2.05) -- (4.25+.15,2.05);
        \node at (4.25,2.2) {$h$};

        \node[mpo] at (5.9,0) {\normalsize$R_\Gamma^{(\pm)}$};
        \node[mpo] at (4.9,0) {\normalsize$\zt_{\Gamma_x}$};
		  \node[edge] at (6.5,0) {};
    \node[fill=white] at (2.75,0) {$\dots$};
    \end{tikzpicture}\ = \
    \begin{tikzpicture}[baseline={(0,.4)}]
    \draw (3,0)--(6.5,0);
    \foreach \x in {3.5}{
		     \draw (\x,0) -- (\x,1);
		    \node[odd] at (\x,0) {};
		  }
		  \foreach \x in {5}{
		     \draw (\x,0) -- (\x,1.25);
		    \node[even] at (\x,0) {};
		  }
        \node[mpo] at (5,.75) {\normalsize$\overleftarrow R_g^{(\pm)}$};
        \draw (5-.15,1.25) -- (5+.15,1.25);
	    \draw (5-.15,1.3) -- (5+.15,1.3);
        \node at (5,1.45) {$h$};

        \node[mpo] at (5.75,0) {\normalsize$R_\Gamma^{(\pm)}$};
        \node[mpo] at (4.25,0) {\normalsize$\zt_{\Gamma_x}$};
		  \node[edge] at (6.5,0) {};
    \node[fill=white] at (2.75,0) {$\dots$};
    \end{tikzpicture} \ = \
    \begin{tikzpicture}[baseline={(0,.4)}]
    \draw (3,0)--(8,0);
    \foreach \x in {3.5}{
		     \draw (\x,0) -- (\x,1);
		    \node[odd] at (\x,0) {};
		  }
        \node[mpo] at (6.25,0) {\normalsize$\overrightarrow R_g^{(\pm)}$};
        \node[mpo] at (5.25,0) {\normalsize$\lx_h$};

        \node[mpo] at (7.25,0) {\normalsize$R_\Gamma^{(\pm)}$};
        \node[mpo] at (4.25,0) {\normalsize$\zt_{\Gamma_x}$};
		  \node[edge] at (8,0) {};
    \node[fill=white] at (2.75,0) {$\dots$};
    \end{tikzpicture}
\end{equation*}
We can use the properties of the tensors to move the $\zt_{\Gamma_x}$ to the left of the next odd tensor, after which we can make a measurement in the basis $\left\{\left(R_{\Gamma'}^{(\pm)}(\theta)\lx_{\bar h}\right)^\dagger\ket{h}:h\in G\right\}$ which moves the $\lx_h$ error to the left:
\begin{equation*}
    = \ \begin{tikzpicture}[baseline={(0,.4)}]
    \draw (3,0)--(7.5,0);
    \foreach \x in {4.25}{
		     \draw (\x,0) -- (\x,1);
		    \node[odd] at (\x,0) {};
		  }
        \node[mpo] at (5.75,0) {\normalsize$\overrightarrow R_g^{(\pm)}$};
        \node[mpo] at (4.85,0) {\normalsize$\lx_h$};
        \node[mpo] at (6.75,0) {\normalsize$R_\Gamma^{(\pm)}$};
        \node[mpo] at (3.6,0) {\normalsize$\zt_{\Gamma_x}$};
		  \node[edge] at (7.5,0) {};
    \node[fill=white] at (2.75,0) {$\dots$};
    \end{tikzpicture} \ \mapsto \
    \begin{tikzpicture}[baseline={(0,.4)}]
    \draw (3,0)--(7.5,0);
    \foreach \x in {4.25}{
		     \draw (\x,0) -- (\x,2);
		    \node[odd] at (\x,0) {};
		  }
        \node[mpo] at (5.75,0) {\normalsize$\overrightarrow R_g^{(\pm)}$};
        \node[mpo] at (4.85,0) {\normalsize$\lx_h$};
        \node[mpo] at (6.75,0) {\normalsize$R_\Gamma^{(\pm)}$};
        \node[mpo] at (3.6,0) {\normalsize$\zt_{\Gamma_x}$};
		  \node[edge] at (7.5,0) {};

        \node[mpo] at (4.25,1.5) {\normalsize$R_{\Gamma'}^{(\pm)}$};
        \node[mpo] at (4.25,.75) {\normalsize$\lx_{\bar h}$};
        \draw (4.25-.15,2) -- (4.25+.15,2);
	    \draw (4.25-.15,2.05) -- (4.25+.15,2.05);
        \node at (4.25,2.2) {$\Gamma_y$};

    \node[fill=white] at (2.75,0) {$\dots$};
    \end{tikzpicture} \ = \
    \begin{tikzpicture}[baseline={(0,.4)}]
    \draw (3,0)--(7.5,0);
    \foreach \x in {5}{
		     \draw (\x,0) -- (\x,1.25);
		    \node[odd] at (\x,0) {};
		  }
        \node[mpo] at (5.75,0) {\normalsize$\overrightarrow R_g^{(\pm)}$};
        \node[mpo] at (4.4,0) {\normalsize$\lx_h$};
        \node[mpo] at (6.75,0) {\normalsize$R_\Gamma^{(\pm)}$};
        \node[mpo] at (3.6,0) {\normalsize$\zt_{\Gamma_x}$};
		  \node[edge] at (7.5,0) {};

        \node[mpo] at (5,.75) {\normalsize$R_{\Gamma'}^{(\pm)}$};
        \draw (5-.15,1.25) -- (5+.15,1.25);
	    \draw (5-.15,1.3) -- (5+.15,1.3);
        \node at (5,1.45) {$\Gamma_y$};
    \node[fill=white] at (2.75,0) {$\dots$};
    \end{tikzpicture}
\end{equation*}
\begin{equation*}
    = \ \begin{tikzpicture}[baseline={(0,.4)}]
    \draw (2,0)--(8.5,0);
    \foreach \x in {2.5}{
		     \draw (\x,0) -- (\x,1);
		    \node[even] at (\x,0) {};
		  }

        \node[mpo] at (4.85,0) {\normalsize$\zt_{\Gamma_y}$};
        \node[mpo] at (6.75,0) {\normalsize$\overrightarrow R_g^{(\pm)}$};
        \node[mpo] at (4.05,0) {\normalsize$\lx_h$};
        \node[mpo] at (7.75,0) {\normalsize$R_\Gamma^{(\pm)}$};
        \node[mpo] at (3.25,0) {\normalsize$\zt_{\Gamma_x}$};
		  \node[edge] at (8.5,0) {};
        \node[mpo] at (5.75,0) {\normalsize$R_{\Gamma'}^{(\pm)}$};
    \node[fill=white] at (1.75,0) {$\dots$};
    \end{tikzpicture} \ = \Gamma_x(h)\
    \begin{tikzpicture}[baseline={(0,.4)}]
    \draw (2.5,0)--(8.5,0);
    \foreach \x in {3.25}{
		     \draw (\x,0) -- (\x,1);
		    \node[even] at (\x,0) {};
		  }

        \node[mpo] at (4.8,0) {\normalsize$\zt_{\Gamma_{xy}}$};
        \node[mpo] at (6.75,0) {\normalsize$\overrightarrow R_g^{(\pm)}$};
        \node[mpo] at (3.95,0) {\normalsize$\lx_h$};
        \node[mpo] at (7.75,0) {\normalsize$R_\Gamma^{(\pm)}$};
		  \node[edge] at (8.5,0) {};
        \node[mpo] at (5.75,0) {\normalsize$R_{\Gamma'}^{(\pm)}$};
    \node[fill=white] at (2.5,0) {$\dots$};
    \end{tikzpicture} \ ,
\end{equation*}
\end{widetext}
where we have used the group-based Pauli commutation relation $\zt_{\Gamma_x}\lx_g=\Gamma_x(g)\lx_g\zt_{\Gamma_x}$ and multiplication of abelian irreps $\zt_{\Gamma_x}\zt_{\Gamma_y}=\zt_{\Gamma_{xy}}$. Because we have taken $G$ to be abelian, the factor $\Gamma_x(h)$ is a global phase which can be discarded.

We have demonstrated the essential aspects of the strategy, which can be employed indefinitely to perform any number of logical operations. As desired, we see that conditioning measurement bases on the outcomes of previous measurements allows us to propagate all Pauli errors to the left of the logical operators.

\textbf{Outcome of Computation.} Once we have applied the entire sequence of gates, we will have mapped the edge state from $\ket{v_R}$ to $\ket{v_{R,\text{target}}}$, up to a global phase which we can disregard and a Pauli byproduct $\lx_{g'}\zt_{\Gamma_{h'}}$. $g'$ and $h'$ will be a function of the measurement outcomes and entirely determined by the relations \cref{eq:pulling_through_main}. To correct this Pauli error and complete the computation, we can use classical post-processing, or if available experimentally, act on the physical legs of the next measurement site with $(\zt_{\Gamma_{h'}}^{(\bullet)\dagger}.\zt_{\Gamma_{h'}}^{(\circ)})(\mathbbm{1}\otimes\rx_{\bar{g}'})$, which pulls through to $\zt_{\Gamma_{h'}}^\dagger\lx_{\bar{g}'}$:
\begin{equation}
    \begin{tikzpicture}
    \draw (0,0)--(2,0);
    \foreach \x in {0.5,1.5}{
		     \draw (\x,0) -- (\x,1.75);
		     % \draw (\x-.15,1) -- (\x+.15,1);
	% \draw (\x-.15,1.05) -- (\x+.15,1.05);
 }
	\node[odd] at (0.5,0) {};
	\node[even] at (1.5,0) {};
    \node[mpo] at (1.5,0.5) {$\rx_{\bar{g}'}$};
    \node[mpo] at (1.5,1.25) {$\zt_{\Gamma_{h'}}$};
    \node[mpo] at (0.5,1.25) {$\zt^\dagger_{\Gamma_{h'}}$};
	\node[edge] at (2,0) {};
  \end{tikzpicture} \ = \ \begin{tikzpicture}
    \draw (0,0)--(3.5,0);
    \foreach \x in {0.5,1.5}{
		     \draw (\x,0) -- (\x,.75);
       }
    \node[odd] at (0.5,0) {};
    \node[even] at (1.5,0) {};
    \node[mpo] at (2.25,0) {$\zt^\dagger_{\Gamma_{h'}}$};
    \node[mpo] at (3,0) {$\lx_{\bar{g}'}$};
    \node[edge] at (3.5,0) {};
  \end{tikzpicture} \ .
\end{equation}

In summary, we have shown that the abelian $G$ cluster state is a suitable resource state for universal single group-valued qudit MBQC. In particular, we can implement each gate in the universal gate set $R_G$ (up to group-based Pauli byproducts) through measurements in rotated bases. We have shown that we can use knowledge of measurement results and classical feedback to propagate byproducts to the left and apply a series of gates sequentially. In the next section, we will show that only small modifications are necessary to adapt this protocol to utilize certain non-abelian $G$ cluster states for MBQC.

\subsubsection{G a Semidirect Product of Abelian Groups}
We approach non-abelian groups by considering those which can be constructed as semidirect products of abelian groups. Recently, a systematic construction of topological order preparation through finite-depth local unitaries and measurements was put forth in Ref.~\onlinecite{tantivasadakarn_hierarchy_2022}. En route to their results, the authors present a general formalism for implementing KW duality on a solvable group by applying the duality to a sequence of abelian subgroups. In this construction, unitaries encoding the particular extension are intertwined between abelian KW MPOs to form the non-abelian KW MPO. Groups constructed as semidirect products of abelian groups are a special case of solvable groups, so we may apply these results to the present task.

Recalling that the KW MPO is simply the cluster state MPS with odd physical legs flipped~\cite{tantivasadakarn_long-range_2022}, this construction in turn provides a construction of the non-abelian $G$ cluster state in terms of abelian cluster states and unitaries (see \cref{eq:QN_decomp} for a special case). We will use this decomposition to build up MBQC with non-abelian $G$ cluster states from the abelian computation protocol laid out in the preceding section.

A particularly simple class of non-abelian groups are those which can be constructed as a \textit{split extension} of two abelian groups. Mathematically, we say that some group $G$ is an \textit{extension} of a quotient group $Q$ by a normal subgroup $N$ if there exists a short exact sequence
\begin{equation}
    1\rightarrow N\rightarrow G\rightarrow Q\rightarrow 1.
\end{equation}
We say that the extension is \textit{split} if $G$ is a semidirect product of $Q$ and $N$. In this case, the decomposition of a non-abelian $G$-qudit into abelian $N$- and $Q$-qudits is relatively simple, as presented in \cref{eq:QN_decomp}.

As Hilbert spaces, there exists an isomorphism $\mathbb{C}[G]\cong \mathbb{C}[\text{\textcolor{red}{$Q$}}]\otimes\mathbb{C}[\text{\textcolor{blue}{$N$}}]$ 
which enables the decomposition
\begin{equation}
    \ket{g}\rightarrow \ket{q_g,n_g}.
\end{equation}
However, this does not yet capture the fact that $G$ is a \textit{semi}direct product of $Q$ and $N$. This is encoded in the action of $Q$ on $N$ given by a map $\sigma:Q\rightarrow \text{Aut}(N)$ (i.e., $\sigma^q:N\rightarrow N$ is an automorphism). We may define a unitary operator which implements this group action as a controlled operator $\Sigma = \sum_{q,n}\ketbra{q,\sigma^q[n]}{q,n}$. One can show that using this unitary, the MPS tensors can be decomposed as~\cite{tantivasadakarn_hierarchy_2022}
\begin{equation}\label{eq:QN_decomp}
  \begin{tikzpicture}
    \draw (0,0)--(2.125,0);
    \foreach \x in {0.75,1.5}{
		     \draw (\x,0) -- (\x,.75);}
		    \node[even] at (0.75,0) {};
		    \node[odd] at (1.5,0) {};
		  % \node[edge] at (2.125,0) {};
		  % \node[fill=white] at (-.2,0) {$\dots$};
  \end{tikzpicture} \
  =
  \ \begin{tikzpicture}
    \draw[color=blue] (2.75,0)--(4.625,0);
    \draw[color=red] (2.75,.5)--(4.625,.5);
    %%%%%%%%%%%%%%%%%%%%%%%%%%%%%%%%%%%%%%%%%%%%%
    \draw[color=blue] (4.625-2*.375,0)--(4.625-2*.375,1.65);
    \draw[color=red] (4.625-3*.375,0.5)--(4.625-3*.375,1.65);

    \draw[fill=white] (3.3,.8) rectangle (4.05,1.45);
    \node at (7.35/2,1.125) {\normalsize$\Sigma$};

    \draw[color=blue] (4.625-1*.375,0)--(4.625-1*.375,1.65);
    \draw[color=red] (4.625-4*.375,0.5)--(4.625-4*.375,1.65);

    %%%%%%%%%%%%%%%%%%%%%%%%%%%%%%%%%%%%%%%%%%%%%
    \foreach \x in {4.625-2*.375}{   \node[even,color=blue,fill=white] at (\x,0) {};}
	\foreach \x in {5-2*.375}{
	    \node[odd,color=blue] at (\x,0) {};
		  }
		  % \node[edge,color=blue] at (4.625,0) {};
		  % \node[fill=white] at (2.5,0) {$\dots$};
    %%%%%%%%%%%%%%%%%%%%%%%%%%%%%%%%%%%%%%%%%%%%%
	\foreach \x in {3.5-.375}{ \node[even,color=red,fill=white] at (\x,.5) {};
		  }
		  \foreach \x in {4.25-2*.375}{
		    \node[odd,color=red] at (\x,.5) {};
		  }
		  % \node[edge,color=red] at (4.625,.5) {};
		  % \node[fill=white] at (2.5,.5) {$\dots$};
	\end{tikzpicture} \ ,
\end{equation}
where the red and blue MPSs are group cluster states corresponding to the abelian groups $Q$ and $N$m respectively:
\begin{align}
\begin{tikzpicture}
    \draw[red] (-0.5,0)--(0.5,0);
    \draw[red] (0,0) -- (0,0.5);
    \node[odd,red] (t) at (0,0) {};
  \end{tikzpicture}
 &=\sum_{q\in Q}\ketbra{q}{q}\otimes\ket{q},&
  \begin{tikzpicture}
    \draw[red] (-0.5,0)--(0.5,0);
    \draw[red] (0,0) -- (0,0.5);
    \node[even,red,fill=white] (t) at (0,0) {};
  \end{tikzpicture}
 &=\sum_{q\in Q} \lx_q\otimes\ket{q}, \nonumber\\
\begin{tikzpicture}
    \draw[blue] (-0.5,0)--(0.5,0);
    \draw[blue] (0,0) -- (0,0.5);
    \node[odd,blue] (t) at (0,0) {};
  \end{tikzpicture}
& =\sum_{n\in N}\ketbra{n}{n}\otimes\ket{n},&
  \begin{tikzpicture}
    \draw[blue] (-0.5,0)--(0.5,0);
    \draw[blue] (0,0) -- (0,0.5);
    \node[even,blue,fill=white] (t) at (0,0) {};
  \end{tikzpicture}
& =\sum_{n\in N} \lx_n\otimes\ket{n}.
\end{align}
In order to compensate for the additional $\Sigma$ operator, we will only have to slightly modify the approach we took in the abelian case above. 

Due to the decomposition $\mathbb C[G] = \mathbb C[Q] \otimes \mathbb C[N]$ we may perform a measurement in each factor separately.

We first measure the blue odd site in the basis:
\begin{equation}\label{eq:odd_blue}
    \left\{\left(R_{\Gamma_1}^{\pm}\right)^\dagger\ket{\Gamma_n}:\Gamma_n\in \rep N\right\}.
\end{equation}
The effect of this measurement is
\newcommand\legheight{3.35}
\begin{widetext}
\begin{equation*}\label{eq:split_meas}
  \begin{tikzpicture}[baseline={(0,.4)}]
    \draw[color=blue] (2.5,0)--(4.625,0);
    \draw[color=red] (2.5,.6)--(4.625,.6);
    %%%%%%%%%%%%%
    \draw[color=blue] (4.625-2*.375,0)--(4.625-2*.375,1.65);
    \draw[color=red] (4.625-3*.375,0.6)--(4.625-3*.375,1.65);

    \draw[fill=white] (3.3,.8) rectangle (4.05,1.45);
    \node at (7.35/2,1.125) {\normalsize$\Sigma$};

    \draw[color=blue] (4.625-1*.375,0)--(4.625-1*.375,1.65);
    \draw[color=red] (4.625-4*.375,0.6)--(4.625-4*.375,1.65);

    %%%%%%%%%%%%%%
    \foreach \x in {4.625-2*.375}{   \node[even,color=blue,fill=white] at (\x,0) {};}
	\foreach \x in {5-2*.375}{
	    \node[odd,color=blue] at (\x,0) {};
		  }
		  \node[edge,color=blue] at (4.625,0) {};
		  \node[fill=white] at (2.5,0) {$\dots$};
    %%%%%%%%%%%%%%%
	\foreach \x in {3.5-.375}{ \node[even,color=red,fill=white] at (\x,.6) {};
		  }
		  \foreach \x in {4.25-2*.375}{
		    \node[odd,color=red] at (\x,.6) {};
		  }
		  \node[edge,color=red] at (4.625,.6) {};
		  \node[fill=white] at (2.5,.6) {$\dots$};
	\end{tikzpicture} \ \mapsto \
 %%%%%%%%%%%%%%%%%%%%%%%%%%%%%%%%%%%%%%%%%%%%%%%%%%%%%%%%%%%%%%%%%%%
 \begin{tikzpicture}[baseline={(0,.4)}]
    \draw[color=blue] (2.5,0)--(4.625,0);
    \draw[color=red] (2.5,.6)--(4.625,.6);
    %%%%%%%%%%%%%%
    \draw[color=blue] (4.625-2*.375,0)--(4.625-2*.375,2.25);
    \draw[color=red] (4.625-3*.375,0.6)--(4.625-3*.375,2.25);

    \draw[fill=white] (3.3,.8) rectangle (4.05,1.45);
    \node at (7.35/2,1.125) {\normalsize$\Sigma$};

    \draw[color=blue] (4.625-1*.375,0)--(4.625-1*.375,2.25);
    \draw[color=red] (4.625-4*.375,0.6)--(4.625-4*.375,2.25);

    %%%%%%%%%%
    \foreach \x in {4.625-2*.375}{   \node[even,color=blue,fill=white] at (\x,0) {};}
	\foreach \x in {5-2*.375}{
	    \node[odd,color=blue] at (\x,0) {};
		  }
		  \node[edge,color=blue] at (4.625,0) {};
		  \node[fill=white] at (2.5,0) {$\dots$};
    %%%%%%%%%%
	\foreach \x in {3.5-.375}{ \node[even,color=red,fill=white] at (\x,.6) {};
		  }
		  \foreach \x in {4.25-2*.375}{
		    \node[odd,color=red] at (\x,.6) {};
		  }
		  \node[edge,color=red] at (4.625,.6) {};
		  \node[fill=white] at (2.5,.6) {$\dots$};
    %%%%%%%%%%
    \node[mpo] at (5-2*.375,1.85) {$R_{\Gamma_1}^{\pm}$};
    \draw[color=blue] (5-2*.375-.15,2.25)--(5-2*.375+.15,2.25);
    \draw[color=blue] (5-2*.375-.15,2.3)--(5-2*.375+.15,2.3);
    \node at (5-2*.375,2.45) {$\Gamma_n$};
	\end{tikzpicture}
	\ = \
 %%%%%%%%%%%%%%%%%%%%%%%%%%%%%%%%%%%%%%%%%%%%%%%%%%%%%%%%%%%%%%%%%%%
 \begin{tikzpicture}[baseline={(0,.4)}]
    \draw[color=blue] (2.5,0)--(5.25,0);
    \draw[color=red] (2.5,.6)--(5.25,.6);
    %%%%%%%%%%%%%%
    \draw[color=blue] (4.625-2*.375,0)--(4.625-2*.375,1.65);
    \draw[color=red] (4.625-3*.375,0.6)--(4.625-3*.375,1.65);

    \draw[fill=white] (3.3,.8) rectangle (4.05,1.45);
    \node at (7.35/2,1.125) {\normalsize$\Sigma$};

    \draw[color=blue] (4.625-1*.375,0)--(4.625-1*.375,1.65);
    \draw[color=red] (4.625-4*.375,0.6)--(4.625-4*.375,1.65);

    %%%%%%%%%%
    \foreach \x in {4.625-2*.375}{   \node[even,color=blue,fill=white] at (\x,0) {};}
	\foreach \x in {5-2*.375}{
	    \node[odd,color=blue] at (\x,0) {};
		  }
		  \node[edge,color=blue] at (5.25,0) {};
		  \node[fill=white] at (2.5,0) {$\dots$};
    %%%%%%%%%%
	\foreach \x in {3.5-.375}{ \node[even,color=red,fill=white] at (\x,.6) {};
		  }
		  \foreach \x in {4.25-2*.375}{
		    \node[odd,color=red] at (\x,.6) {};
		  }
		  \node[edge,color=red] at (5.25,.6) {};
		  \node[fill=white] at (2.5,.6) {$\dots$};
    %%%%%%%%%%
    \node[mpo] at (4.75,0) {$R_{\Gamma_1}^{\pm}$};
    \draw[color=blue] (5-2*.375-.15,1.65)--(5-2*.375+.15,1.65);
    \draw[color=blue] (5-2*.375-.15,1.7)--(5-2*.375+.15,1.7);
    \node at (5-2*.375,1.85) {$\Gamma_n$};
	\end{tikzpicture}
	\ = \
 %%%%%%%%%%%%%%%%%%%%%%%%%%%%%%%%%%%%%%%%%%%%%%%%%%%%%%%%%%%%%%%%%%%
 \begin{tikzpicture}[baseline={(0,.4)}]
    \draw[color=blue] (2.5,0)--(5.7,0);
    \draw[color=red] (2.5,.6)--(5.7,.6);
    %%%%%%%%%%%%%%
    \draw[color=blue] (4.625-2*.375,0)--(4.625-2*.375,1.65);
    \draw[color=red] (4.625-3*.375,0.6)--(4.625-3*.375,1.65);

    \draw[fill=white] (3.3,.8) rectangle (4.05,1.45);
    \node at (7.35/2,1.125) {\normalsize$\Sigma$};
    \draw[color=red] (4.625-4*.375,0.6)--(4.625-4*.375,1.65);

    %%%%%%%%%%
    \foreach \x in {4.625-2*.375}{   \node[even,color=blue,fill=white] at (\x,0) {};}
		  \node[edge,color=blue] at (5.7,0) {};
		  \node[fill=white] at (2.5,0) {$\dots$};
    %%%%%%%%%%
	\foreach \x in {3.5-.375}{ \node[even,color=red,fill=white] at (\x,.6) {};
		  }
		  \foreach \x in {4.25-2*.375}{
		    \node[odd,color=red] at (\x,.6) {};
		  }
		  \node[edge,color=red] at (5.7,.6) {};
		  \node[fill=white] at (2.5,.6) {$\dots$};
    %%%%%%%%%%

    \node[mpo] at (5.2,0) {$R_{\Gamma_1}^{\pm}$};
    \node[mpo] at (4.45,0) {$\zt_{\Gamma_n}^\dagger$};
	\end{tikzpicture} \ . \
\end{equation*}
We next measure the two central qudits so that we can explicitly counteract the gate $\Sigma$. We measure in the basis
\begin{equation}
    \left\{\left[\left(R_{\Gamma_2}^{\pm}\otimes \overleftarrow{R}_{n_1}^{\pm}\right)\left(\mathbbm{1}\otimes \zt_{\Gamma_n}^\dagger\right)\Sigma^\dagger\right]^\dagger\ket{\Gamma_q,n}:\Gamma_q\in \rep Q,n\in N\right\}.
\end{equation}
The effect of this measurement is
\begin{equation*}\label{eq:split_meas}
   \ \mapsto \
 %%%%%%%%%%%%%%%%%%%%%%%%%%%%%%%%%%%%%%%%%%%%%%%%%%%%%%%%%%%%%%%%%%%
 \begin{tikzpicture}[baseline={(0,.4)}]
    \draw[color=blue] (2,0)--(5.7,0);
    \draw[color=red] (2,.6)--(5.7,.6);
    %%%%%%%%%%%%%%
    \draw[color=blue] (4.625-2*.375,0)--(4.625-2*.375,3.55);
    \draw[color=red] (4.625-5*.375,0.6)--(4.625-5*.375,3.55);
    \draw[color=red] (4.625-4*.375,0.6)--(4.625-4*.375,3.55);
    
    \draw[fill=white] (2.975,.8) rectangle (4.05,1.45);
    \node at (2.975/2+4.05/2,1.125) {\normalsize$\Sigma$};
    \draw[fill=white] (2.975,1.55) rectangle (4.05,2.2);
    \node at (2.975/2+4.05/2,1.875) {\normalsize$\Sigma^\dagger$};
    \node[mpo] at (5-3*.375,2.55) {$\zt_{\Gamma_n}^\dagger$};

    \node[mpo] at (5-3*.375,3.15) {$\overleftarrow R_{n_1}^{\pm}$};
    \node[mpo] at (5-5*.375,3.15) {$R_{\Gamma_2}^{\pm}$};

    \draw[color=red] (4.25-3*.375-.15,3.55)--(4.25-3*.375+.15,3.55);
    \draw[color=red] (4.25-3*.375-.15,3.6)--(4.25-3*.375+.15,3.6);
    \node at (4.25-3*.375,3.75) {$\Gamma_q$};

    \draw[color=blue] (4.25-1*.375-.15,3.55)--(4.25-1*.375+.15,3.55);
    \draw[color=blue] (4.25-1*.375-.15,3.6)--(4.25-1*.375+.15,3.6);
    \node at (4.25-1*.375,3.75) {$n$};
    %%%%%%%%%%
    \foreach \x in {4.625-2*.375}{   \node[even,color=blue,fill=white] at (\x,0) {};}
		  \node[edge,color=blue] at (5.7,0) {};
		  \node[fill=white] at (2,0) {$\dots$};
    %%%%%%%%%%
	\foreach \x in {3.5-2*.375}{ \node[even,color=red,fill=white] at (\x,.6) {};
		  }
		  \foreach \x in {4.25-3*.375}{
		    \node[odd,color=red] at (\x,.6) {};
		  }
		  \node[edge,color=red] at (5.7,.6) {};
		  \node[fill=white] at (2,.6) {$\dots$};
    %%%%%%%%%%

    \node[mpo] at (5.2,0) {$R_{\Gamma_1}^{\pm}$};
    \node[mpo] at (4.45,0) {$\zt_{\Gamma_n}^\dagger$};
	\end{tikzpicture}
	\ = \
 %%%%%%%%%%%%%%%%%%%%%%%%%%%%%%%%%%%%%%%%%%%%%%%%%%%%%%%%%%%%%%%%%%%
 \begin{tikzpicture}[baseline={(0,.4)}]
    \draw[color=blue] (2,0)--(5.7,0);
    \draw[color=red] (2,.6)--(5.7,.6);
    %%%%%%%%%%%%%%
    \draw[color=blue] (4.625-2*.375,0)--(4.625-2*.375,2.05);
    \draw[color=red] (4.625-5*.375,0.6)--(4.625-5*.375,2.05);
    \draw[color=red] (4.625-4*.375,0.6)--(4.625-4*.375,2.05);
    
    \node[mpo] at (5-3*.375,1.05) {$\zt_{\Gamma_n}^\dagger$};

    \node[mpo] at (5-3*.375,1.65) {$\overleftarrow R_{n_1}^{\pm}$};
    \node[mpo] at (5-5*.375,1.65) {$R_{\Gamma_2}^{\pm}$};

    \draw[color=red] (4.25-3*.375-.15,2.05)--(4.25-3*.375+.15,2.05);
    \draw[color=red] (4.25-3*.375-.15,2.1)--(4.25-3*.375+.15,2.1);
    \node at (4.25-3*.375,2.25) {$\Gamma_q$};

    \draw[color=blue] (4.25-1*.375-.15,2.05)--(4.25-1*.375+.15,2.05);
    \draw[color=blue] (4.25-1*.375-.15,2.1)--(4.25-1*.375+.15,2.1);
    \node at (4.25-1*.375,2.25) {$n$};
    %%%%%%%%%%
    \foreach \x in {4.625-2*.375}{   \node[even,color=blue,fill=white] at (\x,0) {};}
		  \node[edge,color=blue] at (5.7,0) {};
		  \node[fill=white] at (2,0) {$\dots$};
    %%%%%%%%%%
	\foreach \x in {3.5-2*.375}{ \node[even,color=red,fill=white] at (\x,.6) {};
		  }
		  \foreach \x in {4.25-3*.375}{
		    \node[odd,color=red] at (\x,.6) {};
		  }
		  \node[edge,color=red] at (5.7,.6) {};
		  \node[fill=white] at (2,.6) {$\dots$};
    %%%%%%%%%%

    \node[mpo] at (5.2,0) {$R_{\Gamma_1}^{\pm}$};
    \node[mpo] at (4.45,0) {$\zt_{\Gamma_n}^\dagger$};
	\end{tikzpicture}
	\ = \
 %%%%%%%%%%%%%%%%%%%%%%%%%%%%%%%%%%%%%%%%%%%%%%%%%%%%%%%%%%%%%%%%%%%
 \begin{tikzpicture}[baseline={(0,.4)}]
    \draw[color=blue] (2,0)--(4.95,0);
    \draw[color=red] (2,.6)--(4.95,.6);
    %%%%%%%%%%%%%%
    \draw[color=blue] (4.625-2*.375,0)--(4.625-2*.375,1.45);
    \draw[color=red] (4.625-5*.375,0.6)--(4.625-5*.375,1.45);
    \draw[color=red] (4.625-4*.375,0.6)--(4.625-4*.375,1.45);
    
    % \node[mpo] at (5-3*.375,2.55) {$\zt_{\Gamma_n}^\dagger$};

    \node[mpo] at (5-3*.375,1.05) {$\overleftarrow R_{n_1}^{\pm}$};
    \node[mpo] at (5-5*.375,1.05) {$R_{\Gamma_2}^{\pm}$};

    \draw[color=red] (4.25-3*.375-.15,1.45)--(4.25-3*.375+.15,1.45);
    \draw[color=red] (4.25-3*.375-.15,1.5)--(4.25-3*.375+.15,1.5);
    \node at (4.25-3*.375,1.65) {$\Gamma_q$};

    \draw[color=blue] (4.25-1*.375-.15,1.45)--(4.25-1*.375+.15,1.45);
    \draw[color=blue] (4.25-1*.375-.15,1.5)--(4.25-1*.375+.15,1.5);
    \node at (4.25-1*.375,1.65) {$n$};
    %%%%%%%%%%
    \foreach \x in {4.625-2*.375}{   \node[even,color=blue,fill=white] at (\x,0) {};}
		  \node[edge,color=blue] at (4.95,0) {};
		  \node[fill=white] at (2,0) {$\dots$};
    %%%%%%%%%%
	\foreach \x in {3.5-2*.375}{ \node[even,color=red,fill=white] at (\x,.6) {};
		  }
		  \foreach \x in {4.25-3*.375}{
		    \node[odd,color=red] at (\x,.6) {};
		  }
		  \node[edge,color=red] at (4.95,.6) {};
		  \node[fill=white] at (2,.6) {$\dots$};
    %%%%%%%%%%

    \node[mpo] at (4.45,0) {$R_{\Gamma_1}^{\pm}$};
    \node[mpo] at (3.3,0) {$\zt_{\Gamma_n}^\dagger$};
	\end{tikzpicture} \ = \
 %%%%%%%%%%%%%%%%%%%%%%%%%%%%%%%%%%%%%%%%%%%%%%%%%%%%%%%%%%%%%%%%%%%
  \begin{tikzpicture}[baseline={(0,.4)}]
    \draw[color=blue] (1.5,0)--(4.9,0);
    \draw[color=red] (1.5,.6)--(4.9,.6);
    %%%%%%%%%%%%%%
    \node[edge,color=blue] at (4.9,0) {};
	\node[fill=white] at (1.5,0) {$\dots$};
    %%%%%%%%%%
	\node[edge,color=red] at (4.9,.6) {};
	\node[fill=white] at (1.5,.6) {$\dots$};
    %%%%%%%%%%
    \node[mpo] at (2.25,0) {$\zt_{\Gamma_n}^\dagger$};
    \node[mpo] at (2.95,0) {$\lx_n$};
    \node[mpo] at (3.65,0) {$\overrightarrow R_{n_1}^{\pm}$};
    \node[mpo] at (4.4,0) {$R_{\Gamma_1}^{\pm}$};
    
    \node[mpo] at (3.65,0.6) {$\zt_{\Gamma_q}^\dagger$};
    \node[mpo] at (4.4,.6) {$R_{\Gamma_2}^{\pm}$};

    \draw[color=red] (2.625,0.6)--(2.625,1.45);
    \node[even,color=red,fill=white] at (2.625,.6) {};
    \end{tikzpicture}\ .
\end{equation*}
Finally, we measure the red even site in the basis:
\begin{equation}
    \left\{\left(\overleftarrow{R}_{q_1}^{\pm}\zt_{\Gamma_q}^\dagger\right)^\dagger\ket{q}:q\in Q\right\}.
\end{equation}
The effect of this measurement is
\begin{equation*}\label{eq:split_meas_2}
\mapsto
    %%%%%%%%%%%%%%%%%%%%%%%%%%%%%%%%%%%%%%%%%%%%%%%%%%%%%%%%%%%%%%%%%%%
    \begin{tikzpicture}[baseline={(0,.4)}]
    \draw[color=blue] (1.5,0)--(4.9,0);
    \draw[color=red] (1.5,.6)--(4.9,.6);
    %%%%%%%%%%%%%%
    \node[edge,color=blue] at (4.9,0) {};
	\node[fill=white] at (1.5,0) {$\dots$};
    %%%%%%%%%%
	\node[edge,color=red] at (4.9,.6) {};
	\node[fill=white] at (1.5,.6) {$\dots$};
    %%%%%%%%%%
    \node[mpo] at (2.25,0) {$\zt_{\Gamma_n}^\dagger$};
    \node[mpo] at (2.95,0) {$\lx_n$};
    \node[mpo] at (3.65,0) {$\overrightarrow R_{n_1}^{\pm}$};
    \node[mpo] at (4.4,0) {$R_{\Gamma_1}^{\pm}$};
    
    \node[mpo] at (3.65,0.6) {$\zt_{\Gamma_q}^\dagger$};
    \node[mpo] at (4.4,.6) {$R_{\Gamma_2}^{\pm}$};

    \draw[color=red] (2.625,0.6)--(2.625,2.05);
    \node[even,color=red,fill=white] at (2.625,.6) {};

    \node[mpo] at (2.625,1.05) {$\zt_{\Gamma_q}^\dagger$};
    \node[mpo] at (2.625,1.65) {$\overleftarrow R_{q_1}^{\pm}$};

    \draw[color=red] (2.625-.15,2.05)--(2.625+.15,2.05);
    \draw[color=red] (2.625-.15,2.1)--(2.625+.15,2.1);
    \node at (2.625,2.25) {$q$};
    \end{tikzpicture} \ = \
    %%%%%%%%%%%%%%%%%%%%%%%%%%%%%%%%%%%%%%%%%%%%%%%%%%%%%%%%%%%%%%%%%%%
    \begin{tikzpicture}[baseline={(0,.4)}]
    \draw[color=blue] (1.5,0)--(4.9,0);
    \draw[color=red] (1.5,.6)--(4.9,.6);
    %%%%%%%%%%%%%%
    \node[edge,color=blue] at (4.9,0) {};
	\node[fill=white] at (1.5,0) {$\dots$};
    %%%%%%%%%%
	\node[edge,color=red] at (4.9,.6) {};
	\node[fill=white] at (1.5,.6) {$\dots$};
    %%%%%%%%%%
    \node[mpo] at (2.25,0) {$\zt_{\Gamma_n}^\dagger$};
    \node[mpo] at (2.95,0) {$\lx_n$};
    \node[mpo] at (3.65,0) {$\overrightarrow R_{n_1}^{\pm}$};
    \node[mpo] at (4.4,0) {$R_{\Gamma_1}^{\pm}$};
    
    \node[mpo] at (2.25,0.6) {$\zt_{\Gamma_q}^\dagger$};
    \node[mpo] at (4.4,.6) {$R_{\Gamma_2}^{\pm}$};

    \draw[color=red] (2.95,0.6)--(2.95,1.45);
    \node[even,color=red,fill=white] at (2.95,.6) {};

    \node[mpo] at (2.95,1.05) {$\overleftarrow R_{q_1}^{\pm}$};

    \draw[color=red] (2.95-.15,1.45)--(2.95+.15,1.45);
    \draw[color=red] (2.95-.15,1.5)--(2.95+.15,1.5);
    \node at (2.95,1.65) {$q$};
    \end{tikzpicture} \ = \
    %%%%%%%%%%%%%%%%%%%%%%%%%%%%%%%%%%%%%%%%%%%%%%%%%%%%%%%%%%%%%%%%%%%
    \begin{tikzpicture}[baseline={(0,.4)}]
    \draw[color=blue] (1.5,0)--(4.9,0);
    \draw[color=red] (1.5,.6)--(4.9,.6);
    %%%%%%%%%%%%%%

    %%%%%%%%%%
    \node[edge,color=blue] at (4.9,0) {};
	\node[fill=white] at (1.5,0) {$\dots$};
    %%%%%%%%%%
	\node[edge,color=red] at (4.9,.6) {};
	\node[fill=white] at (1.5,.6) {$\dots$};
    %%%%%%%%%%
    \node[mpo] at (2.25,0) {$\zt_{\Gamma_n}$};
    \node[mpo] at (2.95,0) {$\lx_n$};
    \node[mpo] at (3.65,0) {$\overrightarrow R_{n_1}^{\pm}$};
    \node[mpo] at (4.4,0) {$R_{\Gamma_1}^{\pm}$};

    \node[mpo] at (2.25,0.6) {$\zt_{\Gamma_q}$};
    \node[mpo] at (2.95,0.6) {$\lx_q$};
    \node[mpo] at (3.65,0.6) {$\overrightarrow R_{q_1}^{\pm}$};
    \node[mpo] at (4.4,.6) {$R_{\Gamma_2}^{\pm}$};
    \end{tikzpicture} \ .
\end{equation*}
\end{widetext}
We now see that we can perform universal single-qudit quantum computation in the $Q$ and $N$ virtual spaces separately. Thus far, the only difference from the abelian case of Sec.~\ref{sec:abelian_case} is the additional rotation by $\Sigma^\dagger$ in the measurement basis on the center two sites, which was necessary to compensate for the $\Sigma$ gate.

However, in order to achieve single-qudit universality on the $G$ qudit, we must achieve two-qudit universality on the pair of $Q$ and $N$ qudits. Performing universal single-qudit rotations on the $Q$ and $N$ virtual spaces is not sufficient; we must also introduce a two-qudit imprimitive gate -- a gate which generates entanglement -- acting between the $Q$ and $N$ virtual spaces.

The imprimitive gate which we define is a controlled-multiplication acting between the groups:
\begin{align}
    \begin{split}
    \overrightarrow\Lambda := \sum_{q}\ketbra{q}{q}\otimes\lx_{\lambda(q)},\\
    \overleftarrow\Lambda := \sum_{q}\ketbra{q}{q}\otimes\rx_{\lambda(q)},
    \end{split}
\end{align}
where $\lambda:Q\rightarrow N$ is a map from $Q$ to $N$ which generates entanglement. Notice that $\overleftarrow\Lambda$ acting on the two central qudits can be pulled from the physical legs to the virtual:
\begin{align}\label{eq:lambda_pulling_through}
  \begin{tikzpicture}[baseline={(0,.4)}]
    \draw[color=blue] (2.75,0)--(4.625,0);
    \draw[color=red] (2.75,.5)--(4.625,.5);
    %%%%%%%%%%
    \draw[color=red] (4.625-3*.375,0.5)--(4.625-3*.375,1.65);
    \draw[color=blue] (4.625-2*.375,0)--(4.625-2*.375,1.65);
    \node[mpowide] at (7.35/2,1.125) {\normalsize$\overleftarrow\Lambda$};
    %%%%%%%%%%
    \foreach \x in {4.625-2*.375}{   \node[odd,color=blue,fill=white] at (\x,0) {};}
    %%%%%%%%%%
		  \foreach \x in {4.25-2*.375}{
		    \node[odd,color=red] at (\x,.5) {};
		  }
	\end{tikzpicture} \ = \
 %%%%%%%%%%%%%%%%%%%%%%%%%%%%%%%%%%%%%%%%%%%%%%%%%%%%%%%%%%%%%%%%%%%
 \begin{tikzpicture}[baseline={(0,0.4)}]
    \draw[color=blue] (2.75,0)--(5.125,0);
    \draw[color=red] (2.75,.5)--(5.125,.5);
    %%%%%%%%%%
    \draw[color=red] (4.625-3*.375,0.5)--(4.625-3*.375,1.65);
    \draw[color=blue] (4.625-2*.375,0)--(4.625-2*.375,1.65);
    % \node[mpotall] at (2.85,0.25) {\normalsize$\Sigma^\dagger$};
    \node[mpotall] at (4.525,0.25) {\normalsize$\overrightarrow\Lambda$};
    %%%%%%%%%%
    \foreach \x in {4.625-2*.375}{   \node[odd,color=blue,fill=white] at (\x,0) {};}
    %%%%%%%%%%
		  \foreach \x in {4.25-2*.375}{
		    \node[odd,color=red] at (\x,.5) {};
		  }
	\end{tikzpicture}
 .
\end{align}

We can repeat the preceding analysis and show that we can enact $\overrightarrow\Lambda$ on the virtual space via an appropriate series of measurements. We first measure the odd blue site in the basis 
$\left\{\ket{\Gamma_n}:\Gamma_n\in \rep N\right\}$,
resulting in the following action:

\begin{align*}
    \begin{tikzpicture}[baseline={(0,.4)}]
    \draw[color=blue] (2.5,0)--(4.625,0);
    \draw[color=red] (2.5,.6)--(4.625,.6);
    %%%%%%%%%%%%%
    \draw[color=blue] (4.625-2*.375,0)--(4.625-2*.375,1.65);
    \draw[color=red] (4.625-3*.375,0.6)--(4.625-3*.375,1.65);
    \draw[fill=white] (3.3,.8) rectangle (4.05,1.45);
    \node at (7.35/2,1.125) {\normalsize$\Sigma$};
    \draw[color=blue] (4.625-1*.375,0)--(4.625-1*.375,1.65);
    \draw[color=red] (4.625-4*.375,0.6)--(4.625-4*.375,1.65);
    %%%%%%%%%%%%%%
    \foreach \x in {4.625-2*.375}{   \node[even,color=blue,fill=white] at (\x,0) {};}
	\foreach \x in {5-2*.375}{
	    \node[odd,color=blue] at (\x,0) {};
		  }
		  \node[edge,color=blue] at (4.625,0) {};
		  \node[fill=white] at (2.5,0) {$\dots$};
    %%%%%%%%%%%%%%%
	\foreach \x in {3.5-.375}{ \node[even,color=red,fill=white] at (\x,.6) {};
		  }
		  \foreach \x in {4.25-2*.375}{
		    \node[odd,color=red] at (\x,.6) {};
		  }
		  \node[edge,color=red] at (4.625,.6) {};
		  \node[fill=white] at (2.5,.6) {$\dots$};
	\end{tikzpicture} \ \mapsto \
 %%%%%%%%%%%%%%%%%%%%%%%%%%%%%%%%%%%%%%%%%%%%%%%%%%%%%%%%%%%%%%%%%%%
 \begin{tikzpicture}[baseline={(0,.4)}]
    \draw[color=blue] (2.5,0)--(4.975,0);
    \draw[color=red] (2.5,.6)--(4.975,.6);
    %%%%%%%%%%%%%%
    \draw[color=blue] (4.625-2*.375,0)--(4.625-2*.375,1.65);
    \draw[color=red] (4.625-3*.375,0.6)--(4.625-3*.375,1.65);
    \draw[fill=white] (3.3,.8) rectangle (4.05,1.45);
    \node at (7.35/2,1.125) {\normalsize$\Sigma$};
    \draw[color=red] (4.625-4*.375,0.6)--(4.625-4*.375,1.65);
    %%%%%%%%%%
    \foreach \x in {4.625-2*.375}{   \node[even,color=blue,fill=white] at (\x,0) {};}
		  \node[edge,color=blue] at (4.975,0) {};
		  \node[fill=white] at (2.5,0) {$\dots$};
    %%%%%%%%%%
	\foreach \x in {3.5-.375}{ \node[even,color=red,fill=white] at (\x,.6) {};
		  }
		  \foreach \x in {4.25-2*.375}{
		    \node[odd,color=red] at (\x,.6) {};
		  }
		  \node[edge,color=red] at (4.975,.6) {};
		  \node[fill=white] at (2.5,.6) {$\dots$};
    %%%%%%%%%%
    \node[mpo] at (4.45,0) {$\zt_{\Gamma_n}^\dagger$};
	\end{tikzpicture} \ .
\end{align*}
\noindent Next, we measure the two center sites in the basis 
\begin{align}\label{eq:Lambda_basis}
    \left\{\left[\overleftarrow{\Lambda}\left(\mathbbm{1}\otimes \zt_{\Gamma_n}^\dagger\right)\Sigma^\dagger\right]^\dagger\ket{\Gamma_q,n}:\Gamma_q\in \rep Q,n\in N\right\},
\end{align}

\begin{widetext}
resulting in
    \begin{equation*}
    \ \mapsto \ 
    \begin{tikzpicture}[baseline={(0,.4)}]
    \draw[color=blue] (2,0)--(4.975,0);
    \draw[color=red] (2,.6)--(4.975,.6);
    %%%%%%%%%%%%%%
    \draw[color=blue] (4.625-2*.375,0)--(4.625-2*.375,3.7);
    \draw[color=red] (4.625-5*.375,0.6)--(4.625-5*.375,3.7);
    \draw[color=red] (4.625-4*.375,0.6)--(4.625-4*.375,3.7);
    \draw[fill=white] (2.975,.8) rectangle (4.05,1.45);
    \node at (2.975/2+4.05/2,1.125) {\normalsize$\Sigma$};
    \draw[fill=white] (2.975,1.55) rectangle (4.05,2.2);
    \node at (2.975/2+4.05/2,1.875) {\normalsize$\Sigma^\dagger$};
    \node[mpo] at (5-3*.375,2.55) {$\zt_{\Gamma_n}^\dagger$};
    
    \draw[fill=white] (2.975,2.9) rectangle (4.05,3.55);
    \node at (2.975/2+4.05/2,3.55/2+2.9/2) {\normalsize$\overleftarrow{\Lambda}$};

    \draw[color=red] (4.25-3*.375-.15,3.7)--(4.25-3*.375+.15,3.7);
    \draw[color=red] (4.25-3*.375-.15,3.75)--(4.25-3*.375+.15,3.75);
    \node at (4.25-3*.375,3.9) {$\Gamma_q$};

    \draw[color=blue] (4.25-1*.375-.15,3.7)--(4.25-1*.375+.15,3.7);
    \draw[color=blue] (4.25-1*.375-.15,3.75)--(4.25-1*.375+.15,3.75);
    \node at (4.25-1*.375,3.9) {$n$};
    %%%%%%%%%%
    \foreach \x in {4.625-2*.375}{   \node[even,color=blue,fill=white] at (\x,0) {};}
		  \node[edge,color=blue] at (4.975,0) {};
		  \node[fill=white] at (2,0) {$\dots$};
    %%%%%%%%%%
	\foreach \x in {3.5-2*.375}{ \node[even,color=red,fill=white] at (\x,.6) {};
		  }
		  \foreach \x in {4.25-3*.375}{
		    \node[odd,color=red] at (\x,.6) {};
		  }
		  \node[edge,color=red] at (4.975,.6) {};
		  \node[fill=white] at (2,.6) {$\dots$};
    %%%%%%%%%%

    \node[mpo] at (4.45,0) {$\zt_{\Gamma_n}^\dagger$};
	\end{tikzpicture} \ = \
%%%%%%%%%%%%%%%%%%%%%%%%%%%%%%%%%%%%%%%%%%%%%%%%%%%%%%%%%%%%%%%%%%%%%%%%%%%%%%%%%%%%
 % \begin{tikzpicture}[baseline={(0,.4)}]
 %    \draw[color=blue] (2,0)--(4.975,0);
 %    \draw[color=red] (2,.6)--(4.975,.6);
 %    %%%%%%%%%%%%%%
 %    \draw[color=blue] (4.625-2*.375,0)--(4.625-2*.375,2.2);
 %    \draw[color=red] (4.625-5*.375,0.6)--(4.625-5*.375,2.2);
 %    \draw[color=red] (4.625-4*.375,0.6)--(4.625-4*.375,2.2);
 %    \node[mpo] at (5-3*.375,1.05) {$\zt_{\Gamma_n}^\dagger$};
 %    \draw[fill=white] (2.975,1.4) rectangle (4.05,2.05);
 %    \node at (2.975/2+4.05/2,3.45/2) {\normalsize$\overleftarrow{\Lambda}$};
 %    \draw[color=red] (4.25-3*.375-.15,2.2)--(4.25-3*.375+.15,2.2);
 %    \draw[color=red] (4.25-3*.375-.15,2.25)--(4.25-3*.375+.15,2.25);
 %    \node at (4.25-3*.375,2.4) {$\Gamma_q$};
 %    \draw[color=blue] (4.25-1*.375-.15,2.2)--(4.25-1*.375+.15,2.2);
 %    \draw[color=blue] (4.25-1*.375-.15,2.25)--(4.25-1*.375+.15,2.25);
 %    \node at (4.25-1*.375,2.4) {$n$};
 %    %%%%%%%%%%
 %    \foreach \x in {4.625-2*.375}{   \node[even,color=blue,fill=white] at (\x,0) {};}
	% 	  \node[edge,color=blue] at (4.975,0) {};
	% 	  \node[fill=white] at (2,0) {$\dots$};
 %    %%%%%%%%%%
	% \foreach \x in {3.5-2*.375}{ \node[even,color=red,fill=white] at (\x,.6) {};
	% 	  }
	% 	  \foreach \x in {4.25-3*.375}{
	% 	    \node[odd,color=red] at (\x,.6) {};
	% 	  }
	% 	  \node[edge,color=red] at (4.975,.6) {};
	% 	  \node[fill=white] at (2,.6) {$\dots$};
 %    %%%%%%%%%%
 %    \node[mpo] at (4.45,0) {$\zt_{\Gamma_n}^\dagger$};
	% \end{tikzpicture} \ = \
 %%%%%%%%%%%%%%%%%%%%%%%%%%%%%%%%%%%%%%%%%%%%%%%%%%%%%%%%%%%%%%%%%%%%%%%%%%%%%%%%%%%%
 \begin{tikzpicture}[baseline={(0,.4)}]
    \draw[color=blue] (2,0)--(4.2,0);
    \draw[color=red] (2,.6)--(4.2,.6);
    %%%%%%%%%%%%%%
    \draw[color=blue] (4.625-2*.375,0)--(4.625-2*.375,1.6);
    \draw[color=red] (4.625-5*.375,0.6)--(4.625-5*.375,1.6);
    \draw[color=red] (4.625-4*.375,0.6)--(4.625-4*.375,1.6);
    
    \draw[fill=white] (2.975,.8) rectangle (4.05,1.45);
    \node at (2.975/2+4.05/2,1.125) {\normalsize$\overleftarrow{\Lambda}$};

    \draw[color=red] (4.25-3*.375-.15,1.6)--(4.25-3*.375+.15,1.6);
    \draw[color=red] (4.25-3*.375-.15,1.65)--(4.25-3*.375+.15,1.65);
    \node at (4.25-3*.375,1.8) {$\Gamma_q$};

    \draw[color=blue] (4.25-1*.375-.15,1.6)--(4.25-1*.375+.15,1.6);
    \draw[color=blue] (4.25-1*.375-.15,1.65)--(4.25-1*.375+.15,1.65);
    \node at (4.25-1*.375,1.8) {$n$};
    %%%%%%%%%%
    \foreach \x in {4.625-2*.375}{   \node[even,color=blue,fill=white] at (\x,0) {};}
		  \node[edge,color=blue] at (4.2,0) {};
		  \node[fill=white] at (2,0) {$\dots$};
    %%%%%%%%%%
	\foreach \x in {3.5-2*.375}{ \node[even,color=red,fill=white] at (\x,.6) {};
		  }
		  \foreach \x in {4.25-3*.375}{
		    \node[odd,color=red] at (\x,.6) {};
		  }
		  \node[edge,color=red] at (4.2,.6) {};
		  \node[fill=white] at (2,.6) {$\dots$};
    %%%%%%%%%%
    \node[mpo] at (3.3,0) {$\zt_{\Gamma_n}^\dagger$};
	\end{tikzpicture}   \ = \
 %%%%%%%%%%%%%%%%%%%%%%%%%%%%%%%%%%%%%%%%%%%%%%%%%%%%%%%%%%%%%%%%%%%%%%%%%%%%%%%%%%%%
 \begin{tikzpicture}[baseline={(0,.4)}]
    \draw[color=blue] (2,0)--(5.05,0);
    \draw[color=red] (2,.6)--(5.05,.6);
    %%%%%%%%%%%%%%
    \draw[color=blue] (4.625-2*.375,0)--(4.625-2*.375,1.6);
    \draw[color=red] (4.625-5*.375,0.6)--(4.625-5*.375,1.6);
    \draw[color=red] (4.625-4*.375,0.6)--(4.625-4*.375,1.6);

    \draw[fill=white] (4.15,.3-1.075/2) rectangle (4.8,.3+1.075/2);
    
    % \draw[fill=white] (2.975,.8) rectangle (4.05,1.45);
    \node at (4.15/2+4.8/2,0.3) {\normalsize$\overrightarrow{\Lambda}$};

    \draw[color=red] (4.25-3*.375-.15,1.6)--(4.25-3*.375+.15,1.6);
    \draw[color=red] (4.25-3*.375-.15,1.65)--(4.25-3*.375+.15,1.65);
    \node at (4.25-3*.375,1.8) {$\Gamma_q$};

    \draw[color=blue] (4.25-1*.375-.15,1.6)--(4.25-1*.375+.15,1.6);
    \draw[color=blue] (4.25-1*.375-.15,1.65)--(4.25-1*.375+.15,1.65);
    \node at (4.25-1*.375,1.8) {$n$};
    %%%%%%%%%%
    \foreach \x in {4.625-2*.375}{   \node[even,color=blue,fill=white] at (\x,0) {};}
		  \node[edge,color=blue] at (5.05,0) {};
		  \node[fill=white] at (2,0) {$\dots$};
    %%%%%%%%%%
	\foreach \x in {3.5-2*.375}{ \node[even,color=red,fill=white] at (\x,.6) {};
		  }
		  \foreach \x in {4.25-3*.375}{
		    \node[odd,color=red] at (\x,.6) {};
		  }
		  \node[edge,color=red] at (5.05,.6) {};
		  \node[fill=white] at (2,.6) {$\dots$};
    %%%%%%%%%%
    \node[mpo] at (3.3,0) {$\zt_{\Gamma_n}^\dagger$};
	\end{tikzpicture} \ = \
 %%%%%%%%%%%%%%%%%%%%%%%%%%%%%%%%%%%%%%%%%%%%%%%%%%%%%%%%%%%%%%%%%%%%%%%%%%%%%%%%%%%%
 \begin{tikzpicture}[baseline={(0,.4)}]
    \draw[color=blue] (2,0)--(5.05,0);
    \draw[color=red] (2,.6)--(5.05,.6);
    %%%%%%%%%%%%%%
    \draw[color=red] (3.025,0.6)--(3.025,1.6);

    \draw[fill=white] (4.15,.3-1.075/2) rectangle (4.8,.3+1.075/2);
    
    % \draw[fill=white] (2.975,.8) rectangle (4.05,1.45);
    \node at (4.15/2+4.8/2,0.3) {\normalsize$\overrightarrow{\Lambda}$};

    %%%%%%%%%%
	\foreach \x in {3.025}{ \node[even,color=red,fill=white] at (\x,.6) {};
		  }
		  \node[edge,color=red] at (5.05,.6) {};
        \node[edge,color=blue] at (5.05,0) {};
		  \node[fill=white] at (2.,.6) {$\dots$};
        \node[fill=white] at (2,0) {$\dots$};
    %%%%%%%%%%
    \node[mpo] at (3.025,0) {$\zt_{\Gamma_n}^\dagger$};
     \node[mpo] at (3.75,0) {$\lx_n$};

     \node[mpo] at (3.725,0.6) {$\zt_{\Gamma_q}^\dagger$};
	\end{tikzpicture} \ . \
\end{equation*}

\noindent Finally, we measure the remaining red even site in the basis:
\begin{equation}
    \left\{\left(\zt_{\Gamma_q}^\dagger\right)^\dagger\ket{q}:q\in Q\right\},
\end{equation}
resulting in
\begin{align*}
    \ \mapsto \ \begin{tikzpicture}[baseline={(0,.4)}]
    \draw[color=blue] (2,0)--(5.05,0);
    \draw[color=red] (2,.6)--(5.05,.6);
    %%%%%%%%%%%%%%
    \draw[color=red] (3.025,0.6)--(3.025,1.6);
    \draw[fill=white] (4.15,.3-1.075/2) rectangle (4.8,.3+1.075/2);
    % \draw[fill=white] (2.975,.8) rectangle (4.05,1.45);
    \node at (4.15/2+4.8/2,0.3) {\normalsize$\overrightarrow{\Lambda}$};
    %%%%%%%%%%
	\foreach \x in {3.025}{ \node[even,color=red,fill=white] at (\x,.6) {};
		  }
		  \node[edge,color=red] at (5.05,.6) {};
        \node[edge,color=blue] at (5.05,0) {};
		  \node[fill=white] at (2.,.6) {$\dots$};
        \node[fill=white] at (2,0) {$\dots$};
    %%%%%%%%%%
    \node[mpo] at (3.025,0) {$\zt_{\Gamma_n}^\dagger$};
    \node[mpo] at (3.025,0) {$\zt_{\Gamma_n}^\dagger$};
     \node[mpo] at (3.75,0) {$\lx_n$};
     \node[mpo] at (3.725,0.6) {$\zt_{\Gamma_q}^\dagger$};
     \draw[color=red] (3.025-.15,1.6)--(3.025+.15,1.6);
    \draw[color=red] (3.025-.15,1.65)--(3.025+.15,1.65);
    \node at (3.025,1.8) {$q$};
    \node[mpo] at (3.025,1.2) {$\zt_{\Gamma_q}^\dagger$};
	\end{tikzpicture} \ = \
%%%%%%%%%%%%%%%%%%%%%%%%%%%%%%%%%%%%%%%%%%%%%%%%%%%%%%%%%%%%%%%%%%%%%%%%%%%%%%%%%%%%
 \begin{tikzpicture}[baseline={(0,.4)}]
    \draw[color=blue] (2,0)--(5.05,0);
    \draw[color=red] (2,.6)--(5.05,.6);
    %%%%%%%%%%%%%%
    \draw[fill=white] (4.15,.3-1.075/2) rectangle (4.8,.3+1.075/2);
    % \draw[fill=white] (2.975,.8) rectangle (4.05,1.45);
    \node at (4.15/2+4.8/2,0.3) {\normalsize$\overrightarrow{\Lambda}$};
    %%%%%%%%%%
	\foreach \x in {3.025}{ \node[even,color=red,fill=white] at (\x,.6) {};
		  }
		  \node[edge,color=red] at (5.05,.6) {};
        \node[edge,color=blue] at (5.05,0) {};
		  \node[fill=white] at (2.,.6) {$\dots$};
        \node[fill=white] at (2,0) {$\dots$};
    %%%%%%%%%%
    \node[mpo] at (3.025,0) {$\zt_{\Gamma_n}^\dagger$};
    \node[mpo] at (3.025,0) {$\zt_{\Gamma_n}^\dagger$};
     \node[mpo] at (3.75,0) {$\lx_n$};
     \node[mpo] at (3.75,0.6) {$\lx_q$};
    \node[mpo] at (3.025,0.6) {$\zt_{\Gamma_q}^\dagger$};
	\end{tikzpicture} \ . \
\end{align*}
\end{widetext}
We now see that we can implement $\overrightarrow{\Lambda}$ virtually by measuring the center two sites in the basis \cref{eq:Lambda_basis}. We prove in \cref{ap:imprimitive} that this gate is imprimitive so long as $\lambda(q)$ depends non-trivially on $q$.

\begin{figure}[htp]
    \centering
      \includegraphics[width=0.45\textwidth]{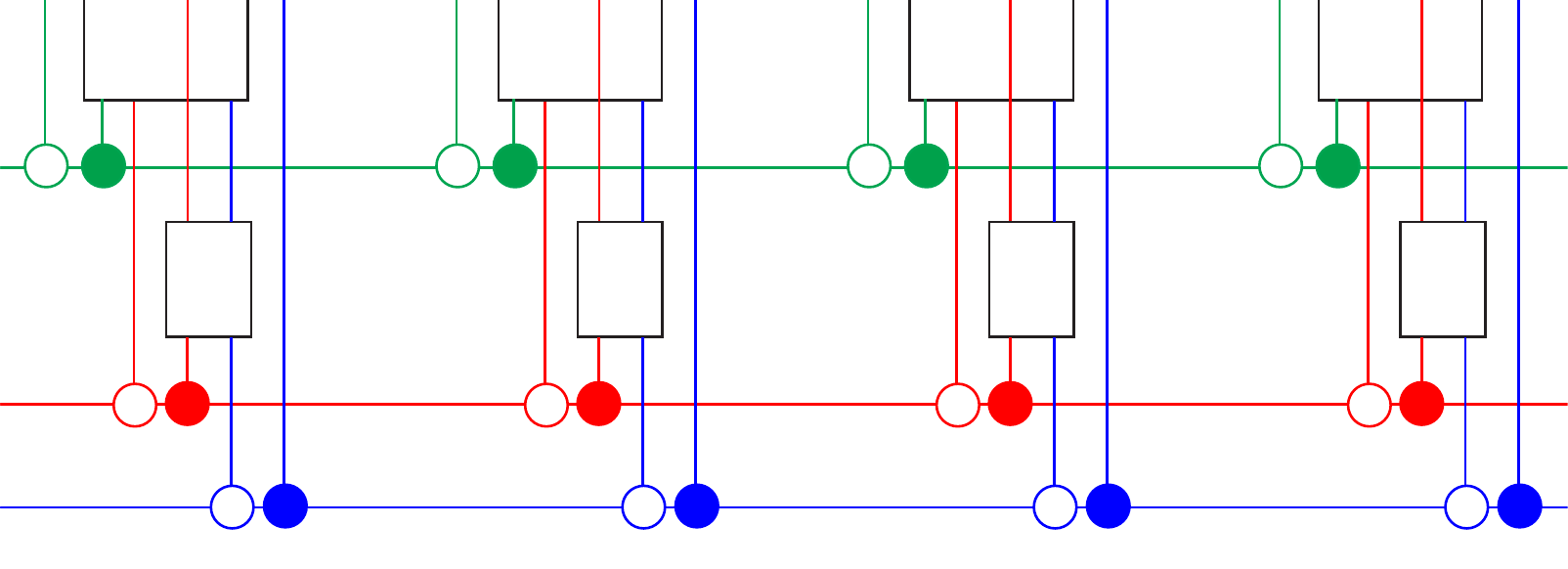}
    \caption{MPS decomposition for the $G$ cluster state based on a group constructed from split extensions of abelian groups. Different colored tensors are cluster states corresponding to different abelian groups and odd sites are control qudits for even site(s).}
    \label{fig:split_MPS}
\end{figure}

In summary, we have shown that MBQC in the case where $G$ is a split extension of abelian groups can be understood as two instances of the abelian MBQC protocol running in parallel. We saw that we can compensate for the entangling gate $\Sigma$ in our choice of measurement bases and that we can generate entanglement between the two abelian MBQC rails via the gate $\overrightarrow\Lambda$. It is also possible to treat the case where $G$ is a successive split extension involving several abelian groups similarly, which is depicted schematically in \cref{fig:split_MPS}.

\subsubsection{G Solvable}
\begin{figure}[htp]
    \centering
      \includegraphics[width=0.45\textwidth]{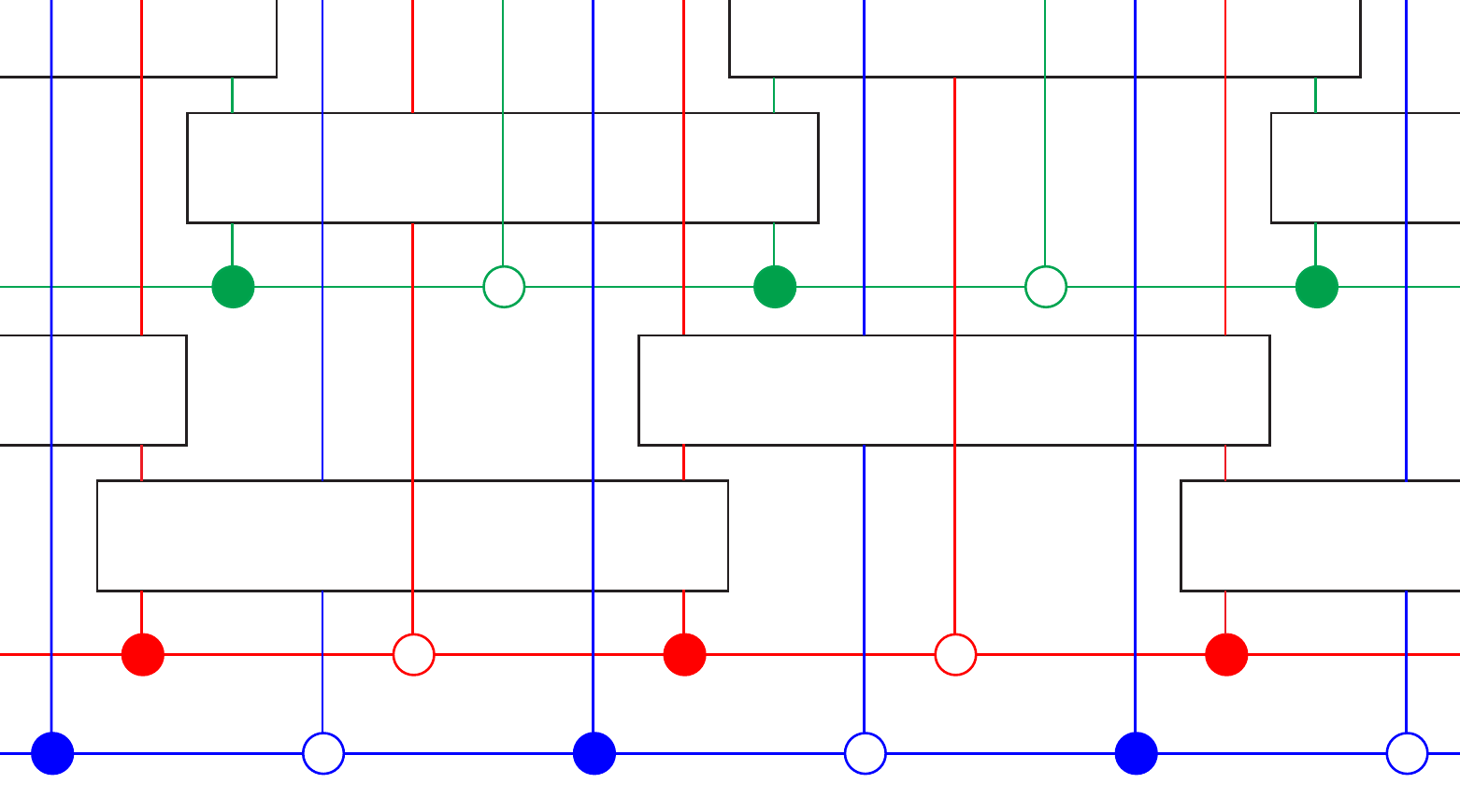}
    \caption{MPS decomposition for the $G$ cluster state based on some solvable group. Once again, different colored tensors are cluster states corresponding to different abelian groups and odd sites are control qudits for even sites.}
    \label{fig:solvable_MPS}

\end{figure}

If $G$ is solvable, it can be constructed from abelian groups using extensions. The example discussed in the previous section -- namely, a group $G$ formed via a split extension of the abelian groups $Q$ and $N$ -- is a simple example of a solvable group. For any solvable group, the cluster state MPS tensors can be decomposed into a set of abelian MPS tensors (corresponding to the set of abelian groups) and unitaries (capturing the details of the extensions of these abelian groups), generalizing \cref{eq:QN_decomp}. In general, this decomposition will be of the form \cref{fig:solvable_MPS}. 

It is not obvious how to generalize the strategy discussed in the previous section to this class of groups because the unitaries in the decomposition have overlapping support. However, because cluster states derived from solvable groups still admit a decomposition into abelian groups and unitaries, we suspect that it should be possible to devise a protocol which performs universal MBQC using such a state as resource. We leave the details of this construction to future work.

\subsubsection{G not Solvable}
In the case where $G$ is not solvable, we cannot decompose $G$ into a set of abelian groups. We are therefore met with the same problems which arise when trying to design an MBQC algorithm for non-abelian $G$ directly: the $Z$-type Pauli errors are not in general unitary, and the $X$-type Pauli errors do not in general commute. We leave the challenge of designing an MBQC algorithm for these cases to future work.

\section{Discussion and Outlook}\label{sec:discussion}

We have studied the $G$ cluster state as an SPT state protected by a fusion category symmetry. Our contribution to the literature is a study of the microscopic signatures of fusion category SPT order in a relatively simple lattice model. This allows us to explore fusion category SPT order in greater detail and with more direct applicability to quantum information applications than in prior works. This study would be useful in developing a general microscopic theory of fusion category SPTs as successful as that of group SPTs. It also presents several technical contributions, including graphical methods for manipulating the group-based Pauli operators which could be useful in the study of other exotic lattice models.

\textbf{Group-Based Pauli Spin Chains.} Pauli spin chains are extremely versatile toy models which, despite their simplicity, exhibit many interesting phenomena. As we have pointed out, relatively few of these phenomena have been generalized to group-valued Hilbert spaces. It is beneficial to do so not only to study fusion category symmetry as we have done, but also to design systems with on-demand finite group symmetries, to study gauge theories, or to design more complex quantum computational schemes. Group-based Pauli spin chains also retain the properties that have made Pauli spin chains so ubiquitous: they are simple to write down and highly versatile.

The techniques we have used to study the $G$ cluster state are highly applicable in the theoretical analysis of other group-based Pauli spin chains.  We have here developed both diagrammatic and algebraic methods for working with group-based Pauli spin chains, and have already seen how to generalize symmetries, stabilizers, and edge modes.

With spin degrees of freedom and operators comprised of Pauli strings, spin chains are also natural models to study on quantum hardware. This is particularly relevant for researchers interested in quantum simulation and quantum computing. Group-based spin chains comprise a class of exotic models which could be realized in a quantum simulator. A proposal already exists for encoding qudits valued in the dihedral group $D_3$ into quantum hardware~\cite{brennen_simulations_2009-1}. Such proposals also allow for any quantum computational protocols designed using group-valued qudits to be implemented directly.

\textbf{Algebraic Classification of $\mathcal{A}$-SPTs.} While classifications of fusion category SPTs in terms of categorical data exist~\cite{thorngren_fusion_2019,inamura_topological_2021}, it is not yet known whether fusion categories may be classified by an algebraic object (such as a cohomology group), as is the case for many conventional SPTs. The classification of 1d bosonic SPTs protected by group symmetries in terms of group cohomology~\cite{chen_symmetry-protected_2012-1,chen_symmetry_2013-1} has been extremely successful. It has been essential in formulating the connections between SPTs and measurement-based quantum computing~\cite{else_symmetry-protected_2012-2,stephen_computational_2017} and the generation of long-range entanglement~\cite{tantivasadakarn_long-range_2022}, as well as in placing SPT states within the modern categorical understanding of topological order~\cite{barkeshli_symmetry_2019}. It is therefore a very interesting open question to determine whether SPTs protected by categorical symmetries can be classified in terms of some algebraic object. Our discussion of edge modes in Sec.~\ref{sec:edge_modes} and of constructing other $G\times \rep{G}$-SPTs in Sec.~\ref{sec:repeated_U_C} should shed light on this search, just as similar inquiries in group-symmetric SPTs point toward the group cohomology classification.

We suspect that it should be possible to derive a classification of fusion category SPTs in terms of some algebraic object using the methods developed in Ref.~\onlinecite{garre-rubio_classifying_2022}. This paper considers MPSs symmetric under the action of MPOs, which can be cast into the language of bimodule categories. By choosing appropriate categories to describe finite group symmetry as an MPO, this formalism recovers the classification in terms of group cohomology as the solution to a coupled pentagon equation. Repeating this procedure with a different choice of categories, chosen to describe fusion category symmetry as an MPO, will return a different constraint. We conjecture that this constraint will be a generalized cocycle condition which should be satisfied by our matrix ``phases'' $\Gamma(g)$. With this condition in hand, it may then be possible to recognize it as defining some algebraic object, just as the 2-cocycle condition (and 2-coboundaries) define the second cohomology group. It would also be of significant practical value to understand how such an object is manifest microscopically, if one exists.

\textbf{$G$ cluster state as Gapped Boundary of 2d TO.} We expect that the $G$ cluster state $\ket{\mathcal{C}}$ can be derived as a gapped boundary of the quantum double $\mathcal D(G\times \rep{G})$. It is known that the qubit cluster state can be realized as the gapped boundary of the bilayer toric code where $e_1m_2$ and $m_1e_2$ anyons are condensed~\cite{yoshida_topological_2016,lichtman_bulk_2021}. Similarly, the CSS cluster state corresponds to the boundary where $e_1e_2$ and $m_1m_2$ are condensed. More generally for group $G$ case, we know that $H_{SSB2}$ in \cref{eq:SSB_H} corresponds to the gapped boundary of $\mathcal D(G\times G)$ where $a_1\bar a_2$ condenses for all $a \in \mathcal D(G)$ (which corresponds to the trivial interface between $\mathcal D(G)$ and itself after unfolding~\cite{kitaev_models_2012}. Performing an isomorphism relating anyons in $\mathcal D(G)$ to $\mathcal D(\rep{G})$ on the second copy gives a construction of the $G$ cluster state as a gapped boundary of $\mathcal D(G\times \rep{G})$. We provide a microscopic construction relating our model to this viewpoint in Appendix~\ref{app:SymTFT}.

\textbf{$\ket{\Cl}$ as an Error Correcting Code.} We have repeatedly referred to $\ket{\Cl}$ as a generalized stabilizer state, and it would be interesting to study it as a generalized stabilizer error correcting code. One immediate complication is that the group-based Pauli operators $\lx_g$, $\rx_g$, and $\zt_\Gamma$ no longer form a group when $G$ is non-abelian. It is unclear what algebraic object these operators form under multiplication, and this question is also directly related to the classification of the edge modes, as the edge modes form the same object.

Despite these complications, the stabilizers still serve to uniquely specify the $G$ cluster state and could therefore protect information in a well-defined codespace. We leave to future work the task of rigorously defining group-based Pauli stabilizer codes and exploring where they may offer an advantage over traditional stabilizer codes.

\textbf{$\ket{\Cl}$ as a Higgs phase of a gauge theory.}
It has been recently appreciated that the $\mathbb Z_d$ cluster state is naturally thought of as the Higgs phase of a $\mathbb Z_d$ gauge theory~\cite{verresen_higgs_2022}, where the even/odd sites corresponds to $\mathbb Z_d$ matter/gauge fields, respectively. The two stabilizers of the cluster state can then be thought of as imposing a Gauss law of the gauge theory, while the other realizes the Higgs condensation by proliferating matter fields connected by Wilson loops. It would be interesting to establish this viewpoint for non-Abelian groups as well, both in terms of a $G$-gauge theory, and perhaps more interestingly in terms of a non-invertible gauge symmetry $\rep{G}$. 

\textbf{Group-based cluster states in higher dimensions as SPT phases}.
We expect the analysis in higher dimensions to be straightforward. In particular, in $d$ spatial dimensions, one can construct a generalized cluster state by putting $Z$-type tensors on vertices and $X$-type tensors on directed edges of a cellulation of the manifold. We expect that this cluster state is an SPT state protected by $G \times \rep{G}^{(d-1)}$ symmetry, where $\rep{G}^{(d-1)}$ is a $(d-1)$-form non-invertible symmetry.

\textbf{$\ket{\Cl}$ for Hopf Algebras as SPT phases.} In general, generalized cluster states can be defined for any Hopf algebra~\cite{brell_generalized_2015}. Unlike typical anyon chains~\cite{Feiguin07}, such models live in a local tensor product Hilbert space. It would be interesting to analyze the SPT properties of these states in full generality.

\begin{acknowledgments} We thank Dave Aasen, Maissam Barkeshli,  Wenjie Ji, Laurens Lootens, Sahand Seifnashri, David T. Stephen, Ruben Verresen, and Brayden Ware for helpful discussions. CF thanks the Joint Quantum Institute at the University of Maryland for support through a JQI fellowship. CF was supported in part by the NSF STAQ program and the Institute for Robust Quantum Simulation (RQS). NT is supported by the Walter Burke Institute for Theoretical Physics at Caltech. VVA acknowledges support from NSF QLCI grant OMA-2120757. Contributions to this work by NIST, an agency of the US government, are not subject to US copyright.
Any mention of commercial products does not indicate endorsement by NIST.
VVA thanks Olga Albert and Ryhor Kandratsenia for providing daycare support throughout this work.

\end{acknowledgments}

\bibliography{SPT_Master}

%apsrev4-2.bst 2019-01-14 (MD) hand-edited version of apsrev4-1.bst
%Control: key (0)
%Control: author (8) initials jnrlst
%Control: editor formatted (1) identically to author
%Control: production of article title (0) allowed
%Control: page (0) single
%Control: year (1) truncated
%Control: production of eprint (0) enabled
\begin{thebibliography}{180}%
\makeatletter
\providecommand \@ifxundefined [1]{%
 \@ifx{#1\undefined}
}%
\providecommand \@ifnum [1]{%
 \ifnum #1\expandafter \@firstoftwo
 \else \expandafter \@secondoftwo
 \fi
}%
\providecommand \@ifx [1]{%
 \ifx #1\expandafter \@firstoftwo
 \else \expandafter \@secondoftwo
 \fi
}%
\providecommand \natexlab [1]{#1}%
\providecommand \enquote  [1]{``#1''}%
\providecommand \bibnamefont  [1]{#1}%
\providecommand \bibfnamefont [1]{#1}%
\providecommand \citenamefont [1]{#1}%
\providecommand \href@noop [0]{\@secondoftwo}%
\providecommand \href [0]{\begingroup \@sanitize@url \@href}%
\providecommand \@href[1]{\@@startlink{#1}\@@href}%
\providecommand \@@href[1]{\endgroup#1\@@endlink}%
\providecommand \@sanitize@url [0]{\catcode `\\12\catcode `\$12\catcode
  `\&12\catcode `\#12\catcode `\^12\catcode `\_12\catcode `\%12\relax}%
\providecommand \@@startlink[1]{}%
\providecommand \@@endlink[0]{}%
\providecommand \url  [0]{\begingroup\@sanitize@url \@url }%
\providecommand \@url [1]{\endgroup\@href {#1}{\urlprefix }}%
\providecommand \urlprefix  [0]{URL }%
\providecommand \Eprint [0]{\href }%
\providecommand \doibase [0]{https://doi.org/}%
\providecommand \selectlanguage [0]{\@gobble}%
\providecommand \bibinfo  [0]{\@secondoftwo}%
\providecommand \bibfield  [0]{\@secondoftwo}%
\providecommand \translation [1]{[#1]}%
\providecommand \BibitemOpen [0]{}%
\providecommand \bibitemStop [0]{}%
\providecommand \bibitemNoStop [0]{.\EOS\space}%
\providecommand \EOS [0]{\spacefactor3000\relax}%
\providecommand \BibitemShut  [1]{\csname bibitem#1\endcsname}%
\let\auto@bib@innerbib\@empty
%</preamble>
\bibitem [{\citenamefont {McGreevy}(2023)}]{mcgreevy2023generalized}%
  \BibitemOpen
  \bibfield  {author} {\bibinfo {author} {\bibfnamefont {J.}~\bibnamefont
  {McGreevy}},\ }\bibfield  {title} {\bibinfo {title} {Generalized symmetries
  in condensed matter},\ }\href
  {https://doi.org/10.1146/annurev-conmatphys-040721-021029} {\bibfield
  {journal} {\bibinfo  {journal} {Annual Review of Condensed Matter Physics}\
  }\textbf {\bibinfo {volume} {14}},\ \bibinfo {pages} {57} (\bibinfo {year}
  {2023})}\BibitemShut {NoStop}%
\bibitem [{\citenamefont {Cordova}\ \emph {et~al.}(2022)\citenamefont
  {Cordova}, \citenamefont {Dumitrescu}, \citenamefont {Intriligator},\ and\
  \citenamefont {Shao}}]{Cordova22}%
  \BibitemOpen
  \bibfield  {author} {\bibinfo {author} {\bibfnamefont {C.}~\bibnamefont
  {Cordova}}, \bibinfo {author} {\bibfnamefont {T.~T.}\ \bibnamefont
  {Dumitrescu}}, \bibinfo {author} {\bibfnamefont {K.}~\bibnamefont
  {Intriligator}},\ and\ \bibinfo {author} {\bibfnamefont {S.-H.}\ \bibnamefont
  {Shao}},\ }\href@noop {} {\bibinfo {title} {Snowmass white paper: Generalized
  symmetries in quantum field theory and beyond}} (\bibinfo {year} {2022}),\
  \Eprint {https://arxiv.org/abs/2205.09545} {arXiv:2205.09545 [hep-th]}
  \BibitemShut {NoStop}%
\bibitem [{\citenamefont {Gaiotto}\ \emph {et~al.}(2015)\citenamefont
  {Gaiotto}, \citenamefont {Kapustin}, \citenamefont {Seiberg},\ and\
  \citenamefont {Willett}}]{gaiotto_generalized_2015}%
  \BibitemOpen
  \bibfield  {author} {\bibinfo {author} {\bibfnamefont {D.}~\bibnamefont
  {Gaiotto}}, \bibinfo {author} {\bibfnamefont {A.}~\bibnamefont {Kapustin}},
  \bibinfo {author} {\bibfnamefont {N.}~\bibnamefont {Seiberg}},\ and\ \bibinfo
  {author} {\bibfnamefont {B.}~\bibnamefont {Willett}},\ }\bibfield  {title}
  {\bibinfo {title} {Generalized global symmetries},\ }\href
  {https://doi.org/10.1007/JHEP02(2015)172} {\bibfield  {journal} {\bibinfo
  {journal} {Journal of High Energy Physics}\ }\textbf {\bibinfo {volume}
  {2015}},\ \bibinfo {pages} {172} (\bibinfo {year} {2015})}\BibitemShut
  {NoStop}%
\bibitem [{\citenamefont {Kapustin}\ and\ \citenamefont
  {Thorngren}(2017{\natexlab{a}})}]{KapustinThorngren2017}%
  \BibitemOpen
  \bibfield  {author} {\bibinfo {author} {\bibfnamefont {A.}~\bibnamefont
  {Kapustin}}\ and\ \bibinfo {author} {\bibfnamefont {R.}~\bibnamefont
  {Thorngren}},\ }\bibfield  {title} {\bibinfo {title} {Higher symmetry and
  gapped phases of gauge theories},\ }in\ \href
  {https://doi.org/10.1007/978-3-319-59939-7_5} {\emph {\bibinfo {booktitle}
  {Algebra, Geometry, and Physics in the 21st Century}}}\ (\bibinfo
  {publisher} {Springer},\ \bibinfo {year} {2017})\ pp.\ \bibinfo {pages}
  {177--202}\BibitemShut {NoStop}%
\bibitem [{\citenamefont {Yoshida}(2016)}]{yoshida_topological_2016}%
  \BibitemOpen
  \bibfield  {author} {\bibinfo {author} {\bibfnamefont {B.}~\bibnamefont
  {Yoshida}},\ }\bibfield  {title} {\bibinfo {title} {Topological phases with
  generalized global symmetries},\ }\href
  {https://doi.org/10.1103/PhysRevB.93.155131} {\bibfield  {journal} {\bibinfo
  {journal} {Physical Review B}\ }\textbf {\bibinfo {volume} {93}},\ \bibinfo
  {pages} {155131} (\bibinfo {year} {2016})}\BibitemShut {NoStop}%
\bibitem [{\citenamefont {Tsui}\ and\ \citenamefont
  {Wen}(2020)}]{tsui_lattice_2020}%
  \BibitemOpen
  \bibfield  {author} {\bibinfo {author} {\bibfnamefont {L.}~\bibnamefont
  {Tsui}}\ and\ \bibinfo {author} {\bibfnamefont {X.-G.}\ \bibnamefont {Wen}},\
  }\bibfield  {title} {\bibinfo {title} {Lattice models that realize
  $\mathbb{Z}_{n}$-1 symmetry-protected topological states for even $n$},\
  }\href {https://doi.org/10.1103/PhysRevB.101.035101} {\bibfield  {journal}
  {\bibinfo  {journal} {Phys. Rev. B}\ }\textbf {\bibinfo {volume} {101}},\
  \bibinfo {pages} {035101} (\bibinfo {year} {2020})}\BibitemShut {NoStop}%
\bibitem [{\citenamefont {Newman}\ and\ \citenamefont
  {Moore}(1999)}]{NewmanMoore1999}%
  \BibitemOpen
  \bibfield  {author} {\bibinfo {author} {\bibfnamefont {M.~E.~J.}\
  \bibnamefont {Newman}}\ and\ \bibinfo {author} {\bibfnamefont
  {C.}~\bibnamefont {Moore}},\ }\bibfield  {title} {\bibinfo {title} {Glassy
  dynamics and aging in an exactly solvable spin model},\ }\href
  {https://doi.org/10.1103/PhysRevE.60.5068} {\bibfield  {journal} {\bibinfo
  {journal} {Phys. Rev. E}\ }\textbf {\bibinfo {volume} {60}},\ \bibinfo
  {pages} {5068} (\bibinfo {year} {1999})}\BibitemShut {NoStop}%
\bibitem [{\citenamefont {Xu}\ and\ \citenamefont {Moore}(2004)}]{XuMoore2004}%
  \BibitemOpen
  \bibfield  {author} {\bibinfo {author} {\bibfnamefont {C.}~\bibnamefont
  {Xu}}\ and\ \bibinfo {author} {\bibfnamefont {J.~E.}\ \bibnamefont {Moore}},\
  }\bibfield  {title} {\bibinfo {title} {Strong-weak coupling self-duality in
  the two-dimensional quantum phase transition of $p+ip$ superconducting
  arrays},\ }\href {https://doi.org/10.1103/PhysRevLett.93.047003} {\bibfield
  {journal} {\bibinfo  {journal} {Phys. Rev. Lett.}\ }\textbf {\bibinfo
  {volume} {93}},\ \bibinfo {pages} {047003} (\bibinfo {year}
  {2004})}\BibitemShut {NoStop}%
\bibitem [{\citenamefont {Vijay}\ \emph {et~al.}(2016)\citenamefont {Vijay},
  \citenamefont {Haah},\ and\ \citenamefont {Fu}}]{vijay_fracton_2016}%
  \BibitemOpen
  \bibfield  {author} {\bibinfo {author} {\bibfnamefont {S.}~\bibnamefont
  {Vijay}}, \bibinfo {author} {\bibfnamefont {J.}~\bibnamefont {Haah}},\ and\
  \bibinfo {author} {\bibfnamefont {L.}~\bibnamefont {Fu}},\ }\bibfield
  {title} {\bibinfo {title} {Fracton topological order, generalized lattice
  gauge theory, and duality},\ }\href
  {https://doi.org/10.1103/PhysRevB.94.235157} {\bibfield  {journal} {\bibinfo
  {journal} {Physical Review B}\ }\textbf {\bibinfo {volume} {94}},\ \bibinfo
  {pages} {235157} (\bibinfo {year} {2016})}\BibitemShut {NoStop}%
\bibitem [{\citenamefont {You}\ \emph {et~al.}(2018{\natexlab{a}})\citenamefont
  {You}, \citenamefont {Devakul}, \citenamefont {Burnell},\ and\ \citenamefont
  {Sondhi}}]{Youetal2018}%
  \BibitemOpen
  \bibfield  {author} {\bibinfo {author} {\bibfnamefont {Y.}~\bibnamefont
  {You}}, \bibinfo {author} {\bibfnamefont {T.}~\bibnamefont {Devakul}},
  \bibinfo {author} {\bibfnamefont {F.~J.}\ \bibnamefont {Burnell}},\ and\
  \bibinfo {author} {\bibfnamefont {S.~L.}\ \bibnamefont {Sondhi}},\ }\bibfield
   {title} {\bibinfo {title} {Subsystem symmetry protected topological order},\
  }\href {https://doi.org/10.1103/PhysRevB.98.035112} {\bibfield  {journal}
  {\bibinfo  {journal} {Phys. Rev. B}\ }\textbf {\bibinfo {volume} {98}},\
  \bibinfo {pages} {035112} (\bibinfo {year} {2018}{\natexlab{a}})}\BibitemShut
  {NoStop}%
\bibitem [{\citenamefont {Devakul}\ \emph {et~al.}(2019)\citenamefont
  {Devakul}, \citenamefont {You}, \citenamefont {Burnell},\ and\ \citenamefont
  {Sondhi}}]{Devakuletal2018}%
  \BibitemOpen
  \bibfield  {author} {\bibinfo {author} {\bibfnamefont {T.}~\bibnamefont
  {Devakul}}, \bibinfo {author} {\bibfnamefont {Y.}~\bibnamefont {You}},
  \bibinfo {author} {\bibfnamefont {F.~J.}\ \bibnamefont {Burnell}},\ and\
  \bibinfo {author} {\bibfnamefont {S.~L.}\ \bibnamefont {Sondhi}},\ }\bibfield
   {title} {\bibinfo {title} {{Fractal Symmetric Phases of Matter}},\ }\href
  {https://doi.org/10.21468/SciPostPhys.6.1.007} {\bibfield  {journal}
  {\bibinfo  {journal} {SciPost Phys.}\ }\textbf {\bibinfo {volume} {6}},\
  \bibinfo {pages} {7} (\bibinfo {year} {2019})}\BibitemShut {NoStop}%
\bibitem [{\citenamefont {Yoshida}(2013)}]{Yoshida13}%
  \BibitemOpen
  \bibfield  {author} {\bibinfo {author} {\bibfnamefont {B.}~\bibnamefont
  {Yoshida}},\ }\bibfield  {title} {\bibinfo {title} {Exotic topological order
  in fractal spin liquids},\ }\href
  {https://doi.org/10.1103/PhysRevB.88.125122} {\bibfield  {journal} {\bibinfo
  {journal} {Phys. Rev. B}\ }\textbf {\bibinfo {volume} {88}},\ \bibinfo
  {pages} {125122} (\bibinfo {year} {2013})}\BibitemShut {NoStop}%
\bibitem [{\citenamefont {Williamson}(2016)}]{Williamson2016}%
  \BibitemOpen
  \bibfield  {author} {\bibinfo {author} {\bibfnamefont {D.~J.}\ \bibnamefont
  {Williamson}},\ }\bibfield  {title} {\bibinfo {title} {Fractal symmetries:
  Ungauging the cubic code},\ }\href
  {https://doi.org/10.1103/PhysRevB.94.155128} {\bibfield  {journal} {\bibinfo
  {journal} {Phys. Rev. B}\ }\textbf {\bibinfo {volume} {94}},\ \bibinfo
  {pages} {155128} (\bibinfo {year} {2016})}\BibitemShut {NoStop}%
\bibitem [{\citenamefont {Yoshida}(2015)}]{Yoshida15}%
  \BibitemOpen
  \bibfield  {author} {\bibinfo {author} {\bibfnamefont {B.}~\bibnamefont
  {Yoshida}},\ }\bibfield  {title} {\bibinfo {title} {Topological color code
  and symmetry-protected topological phases},\ }\href
  {https://doi.org/10.1103/PhysRevB.91.245131} {\bibfield  {journal} {\bibinfo
  {journal} {Phys. Rev. B}\ }\textbf {\bibinfo {volume} {91}},\ \bibinfo
  {pages} {245131} (\bibinfo {year} {2015})}\BibitemShut {NoStop}%
\bibitem [{\citenamefont {Yoshida}(2017)}]{yoshida_gapped_2017}%
  \BibitemOpen
  \bibfield  {author} {\bibinfo {author} {\bibfnamefont {B.}~\bibnamefont
  {Yoshida}},\ }\bibfield  {title} {\bibinfo {title} {Gapped boundaries, group
  cohomology and fault-tolerant logical gates},\ }\href
  {https://doi.org/10.1016/j.aop.2016.12.014} {\bibfield  {journal} {\bibinfo
  {journal} {Annals of Physics}\ }\textbf {\bibinfo {volume} {377}},\ \bibinfo
  {pages} {387} (\bibinfo {year} {2017})},\ \Eprint
  {https://arxiv.org/abs/1509.03626} {arXiv:1509.03626 [cond-mat,
  physics:quant-ph]} \BibitemShut {NoStop}%
\bibitem [{\citenamefont {Barkeshli}\ \emph
  {et~al.}(2023{\natexlab{a}})\citenamefont {Barkeshli}, \citenamefont {Chen},
  \citenamefont {Huang}, \citenamefont {Kobayashi}, \citenamefont
  {Tantivasadakarn},\ and\ \citenamefont {Zhu}}]{barkeshli_codimension-2_2022}%
  \BibitemOpen
  \bibfield  {author} {\bibinfo {author} {\bibfnamefont {M.}~\bibnamefont
  {Barkeshli}}, \bibinfo {author} {\bibfnamefont {Y.-A.}\ \bibnamefont {Chen}},
  \bibinfo {author} {\bibfnamefont {S.-J.}\ \bibnamefont {Huang}}, \bibinfo
  {author} {\bibfnamefont {R.}~\bibnamefont {Kobayashi}}, \bibinfo {author}
  {\bibfnamefont {N.}~\bibnamefont {Tantivasadakarn}},\ and\ \bibinfo {author}
  {\bibfnamefont {G.}~\bibnamefont {Zhu}},\ }\bibfield  {title} {\bibinfo
  {title} {{Codimension-2 defects and higher symmetries in (3+1)D topological
  phases}},\ }\href {https://doi.org/10.21468/SciPostPhys.14.4.065} {\bibfield
  {journal} {\bibinfo  {journal} {SciPost Phys.}\ }\textbf {\bibinfo {volume}
  {14}},\ \bibinfo {pages} {065} (\bibinfo {year}
  {2023}{\natexlab{a}})}\BibitemShut {NoStop}%
\bibitem [{\citenamefont {Barkeshli}\ \emph
  {et~al.}(2023{\natexlab{b}})\citenamefont {Barkeshli}, \citenamefont {Chen},
  \citenamefont {Hsin},\ and\ \citenamefont
  {Kobayashi}}]{barkeshli2023highergroup}%
  \BibitemOpen
  \bibfield  {author} {\bibinfo {author} {\bibfnamefont {M.}~\bibnamefont
  {Barkeshli}}, \bibinfo {author} {\bibfnamefont {Y.-A.}\ \bibnamefont {Chen}},
  \bibinfo {author} {\bibfnamefont {P.-S.}\ \bibnamefont {Hsin}},\ and\
  \bibinfo {author} {\bibfnamefont {R.}~\bibnamefont {Kobayashi}},\ }\href@noop
  {} {\bibinfo {title} {Higher-group symmetry in finite gauge theory and
  stabilizer codes}} (\bibinfo {year} {2023}{\natexlab{b}}),\ \Eprint
  {https://arxiv.org/abs/2211.11764} {arXiv:2211.11764 [cond-mat.str-el]}
  \BibitemShut {NoStop}%
\bibitem [{\citenamefont {Zhu}\ \emph {et~al.}(2022)\citenamefont {Zhu},
  \citenamefont {Jochym-O'Connor},\ and\ \citenamefont {Dua}}]{Zhu22}%
  \BibitemOpen
  \bibfield  {author} {\bibinfo {author} {\bibfnamefont {G.}~\bibnamefont
  {Zhu}}, \bibinfo {author} {\bibfnamefont {T.}~\bibnamefont
  {Jochym-O'Connor}},\ and\ \bibinfo {author} {\bibfnamefont {A.}~\bibnamefont
  {Dua}},\ }\bibfield  {title} {\bibinfo {title} {Topological order, quantum
  codes, and quantum computation on fractal geometries},\ }\href
  {https://doi.org/10.1103/PRXQuantum.3.030338} {\bibfield  {journal} {\bibinfo
   {journal} {PRX Quantum}\ }\textbf {\bibinfo {volume} {3}},\ \bibinfo {pages}
  {030338} (\bibinfo {year} {2022})}\BibitemShut {NoStop}%
\bibitem [{\citenamefont {Zhu}\ \emph {et~al.}(2023)\citenamefont {Zhu},
  \citenamefont {Sikander}, \citenamefont {Portnoy}, \citenamefont {Cross},\
  and\ \citenamefont {Brown}}]{Zhu23}%
  \BibitemOpen
  \bibfield  {author} {\bibinfo {author} {\bibfnamefont {G.}~\bibnamefont
  {Zhu}}, \bibinfo {author} {\bibfnamefont {S.}~\bibnamefont {Sikander}},
  \bibinfo {author} {\bibfnamefont {E.}~\bibnamefont {Portnoy}}, \bibinfo
  {author} {\bibfnamefont {A.~W.}\ \bibnamefont {Cross}},\ and\ \bibinfo
  {author} {\bibfnamefont {B.~J.}\ \bibnamefont {Brown}},\ }\bibfield  {title}
  {\bibinfo {title} {Non-clifford and parallelizable fault-tolerant logical
  gates on constant and almost-constant rate homological quantum ldpc codes via
  higher symmetries},\ }\href@noop {} {\bibfield  {journal} {\bibinfo
  {journal} {arXiv preprint arXiv:2310.16982}\ } (\bibinfo {year}
  {2023})}\BibitemShut {NoStop}%
\bibitem [{\citenamefont {Barkeshli}\ \emph
  {et~al.}(2023{\natexlab{c}})\citenamefont {Barkeshli}, \citenamefont {Hsin},\
  and\ \citenamefont {Kobayashi}}]{barkeshli2023higher}%
  \BibitemOpen
  \bibfield  {author} {\bibinfo {author} {\bibfnamefont {M.}~\bibnamefont
  {Barkeshli}}, \bibinfo {author} {\bibfnamefont {P.-S.}\ \bibnamefont
  {Hsin}},\ and\ \bibinfo {author} {\bibfnamefont {R.}~\bibnamefont
  {Kobayashi}},\ }\bibfield  {title} {\bibinfo {title} {Higher-group symmetry
  of (3+1) d fermionic $\mathbb{Z}_2$ gauge theory: logical ccz, cs, and t
  gates from higher symmetry},\ }\href@noop {} {\bibfield  {journal} {\bibinfo
  {journal} {arXiv preprint arXiv:2311.05674}\ } (\bibinfo {year}
  {2023}{\natexlab{c}})}\BibitemShut {NoStop}%
\bibitem [{\citenamefont {Chamon}(2005)}]{Chamon05}%
  \BibitemOpen
  \bibfield  {author} {\bibinfo {author} {\bibfnamefont {C.}~\bibnamefont
  {Chamon}},\ }\bibfield  {title} {\bibinfo {title} {Quantum glassiness in
  strongly correlated clean systems: An example of topological
  overprotection},\ }\href {https://doi.org/10.1103/PhysRevLett.94.040402}
  {\bibfield  {journal} {\bibinfo  {journal} {Phys. Rev. Lett.}\ }\textbf
  {\bibinfo {volume} {94}},\ \bibinfo {pages} {040402} (\bibinfo {year}
  {2005})}\BibitemShut {NoStop}%
\bibitem [{\citenamefont {Haah}(2011)}]{Haah11}%
  \BibitemOpen
  \bibfield  {author} {\bibinfo {author} {\bibfnamefont {J.}~\bibnamefont
  {Haah}},\ }\bibfield  {title} {\bibinfo {title} {Local stabilizer codes in
  three dimensions without string logical operators},\ }\href
  {https://doi.org/10.1103/PhysRevA.83.042330} {\bibfield  {journal} {\bibinfo
  {journal} {Phys. Rev. A}\ }\textbf {\bibinfo {volume} {83}},\ \bibinfo
  {pages} {042330} (\bibinfo {year} {2011})}\BibitemShut {NoStop}%
\bibitem [{\citenamefont {Shirley}\ \emph {et~al.}(2018)\citenamefont
  {Shirley}, \citenamefont {Slagle}, \citenamefont {Wang},\ and\ \citenamefont
  {Chen}}]{shirley_fracton_2018}%
  \BibitemOpen
  \bibfield  {author} {\bibinfo {author} {\bibfnamefont {W.}~\bibnamefont
  {Shirley}}, \bibinfo {author} {\bibfnamefont {K.}~\bibnamefont {Slagle}},
  \bibinfo {author} {\bibfnamefont {Z.}~\bibnamefont {Wang}},\ and\ \bibinfo
  {author} {\bibfnamefont {X.}~\bibnamefont {Chen}},\ }\bibfield  {title}
  {\bibinfo {title} {Fracton {{Models}} on {{General Three-Dimensional
  Manifolds}}},\ }\href {https://doi.org/10.1103/PhysRevX.8.031051} {\bibfield
  {journal} {\bibinfo  {journal} {Physical Review X}\ }\textbf {\bibinfo
  {volume} {8}},\ \bibinfo {pages} {031051} (\bibinfo {year}
  {2018})}\BibitemShut {NoStop}%
\bibitem [{\citenamefont {Shirley}\ \emph {et~al.}(2019)\citenamefont
  {Shirley}, \citenamefont {Slagle},\ and\ \citenamefont
  {Chen}}]{shirley_foliated_2019}%
  \BibitemOpen
  \bibfield  {author} {\bibinfo {author} {\bibfnamefont {W.}~\bibnamefont
  {Shirley}}, \bibinfo {author} {\bibfnamefont {K.}~\bibnamefont {Slagle}},\
  and\ \bibinfo {author} {\bibfnamefont {X.}~\bibnamefont {Chen}},\ }\bibfield
  {title} {\bibinfo {title} {Foliated fracton order from gauging subsystem
  symmetries},\ }\href {https://doi.org/10.21468/SciPostPhys.6.4.041}
  {\bibfield  {journal} {\bibinfo  {journal} {SciPost Physics}\ }\textbf
  {\bibinfo {volume} {6}},\ \bibinfo {pages} {041} (\bibinfo {year}
  {2019})}\BibitemShut {NoStop}%
\bibitem [{\citenamefont {Bulmash}\ and\ \citenamefont
  {Barkeshli}(2019{\natexlab{a}})}]{bulmash_gauging_2019}%
  \BibitemOpen
  \bibfield  {author} {\bibinfo {author} {\bibfnamefont {D.}~\bibnamefont
  {Bulmash}}\ and\ \bibinfo {author} {\bibfnamefont {M.}~\bibnamefont
  {Barkeshli}},\ }\bibfield  {title} {\bibinfo {title} {Gauging fractons:
  {{Immobile}} non-{{Abelian}} quasiparticles, fractals, and position-dependent
  degeneracies},\ }\href {https://doi.org/10.1103/PhysRevB.100.155146}
  {\bibfield  {journal} {\bibinfo  {journal} {Physical Review B}\ }\textbf
  {\bibinfo {volume} {100}},\ \bibinfo {pages} {155146} (\bibinfo {year}
  {2019}{\natexlab{a}})}\BibitemShut {NoStop}%
\bibitem [{\citenamefont {Nandkishore}\ and\ \citenamefont
  {Hermele}(2019)}]{nandkishore_fractons_2019}%
  \BibitemOpen
  \bibfield  {author} {\bibinfo {author} {\bibfnamefont {R.~M.}\ \bibnamefont
  {Nandkishore}}\ and\ \bibinfo {author} {\bibfnamefont {M.}~\bibnamefont
  {Hermele}},\ }\bibfield  {title} {\bibinfo {title} {Fractons},\ }\href
  {https://doi.org/10.1146/annurev-conmatphys-031218-013604} {\bibfield
  {journal} {\bibinfo  {journal} {Annual Review of Condensed Matter Physics}\
  }\textbf {\bibinfo {volume} {10}},\ \bibinfo {pages} {295} (\bibinfo {year}
  {2019})}\BibitemShut {NoStop}%
\bibitem [{\citenamefont {Pretko}\ \emph {et~al.}(2020)\citenamefont {Pretko},
  \citenamefont {Chen},\ and\ \citenamefont {You}}]{pretko_fracton_2020}%
  \BibitemOpen
  \bibfield  {author} {\bibinfo {author} {\bibfnamefont {M.}~\bibnamefont
  {Pretko}}, \bibinfo {author} {\bibfnamefont {X.}~\bibnamefont {Chen}},\ and\
  \bibinfo {author} {\bibfnamefont {Y.}~\bibnamefont {You}},\ }\bibfield
  {title} {\bibinfo {title} {Fracton phases of matter},\ }\href
  {https://doi.org/10.1142/S0217751X20300033} {\bibfield  {journal} {\bibinfo
  {journal} {International Journal of Modern Physics A}\ }\textbf {\bibinfo
  {volume} {35}},\ \bibinfo {pages} {2030003} (\bibinfo {year}
  {2020})}\BibitemShut {NoStop}%
\bibitem [{\citenamefont {Bulmash}\ and\ \citenamefont
  {Barkeshli}(2019{\natexlab{b}})}]{BulmashBarkeshli2019}%
  \BibitemOpen
  \bibfield  {author} {\bibinfo {author} {\bibfnamefont {D.}~\bibnamefont
  {Bulmash}}\ and\ \bibinfo {author} {\bibfnamefont {M.}~\bibnamefont
  {Barkeshli}},\ }\bibfield  {title} {\bibinfo {title} {Gauging fractons:
  Immobile non-abelian quasiparticles, fractals, and position-dependent
  degeneracies},\ }\href {https://doi.org/10.1103/PhysRevB.100.155146}
  {\bibfield  {journal} {\bibinfo  {journal} {Phys. Rev. B}\ }\textbf {\bibinfo
  {volume} {100}},\ \bibinfo {pages} {155146} (\bibinfo {year}
  {2019}{\natexlab{b}})}\BibitemShut {NoStop}%
\bibitem [{\citenamefont {Prem}\ and\ \citenamefont
  {Williamson}(2019)}]{PremWilliamson2019}%
  \BibitemOpen
  \bibfield  {author} {\bibinfo {author} {\bibfnamefont {A.}~\bibnamefont
  {Prem}}\ and\ \bibinfo {author} {\bibfnamefont {D.~J.}\ \bibnamefont
  {Williamson}},\ }\bibfield  {title} {\bibinfo {title} {{Gauging permutation
  symmetries as a route to non-Abelian fractons}},\ }\href
  {https://doi.org/10.21468/SciPostPhys.7.5.068} {\bibfield  {journal}
  {\bibinfo  {journal} {SciPost Phys.}\ }\textbf {\bibinfo {volume} {7}},\
  \bibinfo {pages} {68} (\bibinfo {year} {2019})}\BibitemShut {NoStop}%
\bibitem [{\citenamefont {Tantivasadakarn}\ \emph
  {et~al.}(2021{\natexlab{a}})\citenamefont {Tantivasadakarn}, \citenamefont
  {Ji},\ and\ \citenamefont {Vijay}}]{TJV1}%
  \BibitemOpen
  \bibfield  {author} {\bibinfo {author} {\bibfnamefont {N.}~\bibnamefont
  {Tantivasadakarn}}, \bibinfo {author} {\bibfnamefont {W.}~\bibnamefont
  {Ji}},\ and\ \bibinfo {author} {\bibfnamefont {S.}~\bibnamefont {Vijay}},\
  }\bibfield  {title} {\bibinfo {title} {Hybrid fracton phases: Parent orders
  for liquid and nonliquid quantum phases},\ }\href
  {https://doi.org/10.1103/PhysRevB.103.245136} {\bibfield  {journal} {\bibinfo
   {journal} {Phys. Rev. B}\ }\textbf {\bibinfo {volume} {103}},\ \bibinfo
  {pages} {245136} (\bibinfo {year} {2021}{\natexlab{a}})}\BibitemShut
  {NoStop}%
\bibitem [{\citenamefont {Tantivasadakarn}\ \emph
  {et~al.}(2021{\natexlab{b}})\citenamefont {Tantivasadakarn}, \citenamefont
  {Ji},\ and\ \citenamefont {Vijay}}]{tantivasadakarn_non-abelian_2021}%
  \BibitemOpen
  \bibfield  {author} {\bibinfo {author} {\bibfnamefont {N.}~\bibnamefont
  {Tantivasadakarn}}, \bibinfo {author} {\bibfnamefont {W.}~\bibnamefont
  {Ji}},\ and\ \bibinfo {author} {\bibfnamefont {S.}~\bibnamefont {Vijay}},\
  }\bibfield  {title} {\bibinfo {title} {Non-{{Abelian}} hybrid fracton
  orders},\ }\href {https://doi.org/10.1103/PhysRevB.104.115117} {\bibfield
  {journal} {\bibinfo  {journal} {Physical Review B}\ }\textbf {\bibinfo
  {volume} {104}},\ \bibinfo {pages} {115117} (\bibinfo {year}
  {2021}{\natexlab{b}})}\BibitemShut {NoStop}%
\bibitem [{\citenamefont {Tu}\ and\ \citenamefont {Chang}(2021)}]{TuChang21}%
  \BibitemOpen
  \bibfield  {author} {\bibinfo {author} {\bibfnamefont {Y.-T.}\ \bibnamefont
  {Tu}}\ and\ \bibinfo {author} {\bibfnamefont {P.-Y.}\ \bibnamefont {Chang}},\
  }\bibfield  {title} {\bibinfo {title} {Non-abelian fracton order from gauging
  a mixture of subsystem and global symmetries},\ }\href
  {https://doi.org/10.1103/PhysRevResearch.3.043084} {\bibfield  {journal}
  {\bibinfo  {journal} {Phys. Rev. Res.}\ }\textbf {\bibinfo {volume} {3}},\
  \bibinfo {pages} {043084} (\bibinfo {year} {2021})}\BibitemShut {NoStop}%
\bibitem [{\citenamefont {Hsin}\ and\ \citenamefont
  {Slagle}(2021)}]{HsinSlagle21}%
  \BibitemOpen
  \bibfield  {author} {\bibinfo {author} {\bibfnamefont {P.-S.}\ \bibnamefont
  {Hsin}}\ and\ \bibinfo {author} {\bibfnamefont {K.}~\bibnamefont {Slagle}},\
  }\bibfield  {title} {\bibinfo {title} {{Comments on foliated gauge theories
  and dualities in 3+1d}},\ }\href
  {https://doi.org/10.21468/SciPostPhys.11.2.032} {\bibfield  {journal}
  {\bibinfo  {journal} {SciPost Phys.}\ }\textbf {\bibinfo {volume} {11}},\
  \bibinfo {pages} {032} (\bibinfo {year} {2021})}\BibitemShut {NoStop}%
\bibitem [{\citenamefont {Rayhaun}\ and\ \citenamefont
  {Williamson}(2023)}]{rayhaun_higher-form_2021}%
  \BibitemOpen
  \bibfield  {author} {\bibinfo {author} {\bibfnamefont {B.~C.}\ \bibnamefont
  {Rayhaun}}\ and\ \bibinfo {author} {\bibfnamefont {D.~J.}\ \bibnamefont
  {Williamson}},\ }\bibfield  {title} {\bibinfo {title} {{Higher-form subsystem
  symmetry breaking: Subdimensional criticality and fracton phase
  transitions}},\ }\href {https://doi.org/10.21468/SciPostPhys.15.1.017}
  {\bibfield  {journal} {\bibinfo  {journal} {SciPost Phys.}\ }\textbf
  {\bibinfo {volume} {15}},\ \bibinfo {pages} {017} (\bibinfo {year}
  {2023})}\BibitemShut {NoStop}%
\bibitem [{\citenamefont {Etingof}\ \emph {et~al.}(2016)\citenamefont
  {Etingof}, \citenamefont {Gelaki}, \citenamefont {Nikshych},\ and\
  \citenamefont {Ostrik}}]{etingof2016tensor}%
  \BibitemOpen
  \bibfield  {author} {\bibinfo {author} {\bibfnamefont {P.}~\bibnamefont
  {Etingof}}, \bibinfo {author} {\bibfnamefont {S.}~\bibnamefont {Gelaki}},
  \bibinfo {author} {\bibfnamefont {D.}~\bibnamefont {Nikshych}},\ and\
  \bibinfo {author} {\bibfnamefont {V.}~\bibnamefont {Ostrik}},\ }\href@noop {}
  {\emph {\bibinfo {title} {Tensor categories}}},\ Vol.\ \bibinfo {volume}
  {205}\ (\bibinfo  {publisher} {American Mathematical Soc.},\ \bibinfo {year}
  {2016})\BibitemShut {NoStop}%
\bibitem [{\citenamefont {Etingof}\ \emph {et~al.}(2005)\citenamefont
  {Etingof}, \citenamefont {Nikshych},\ and\ \citenamefont
  {Ostrik}}]{etingof_fusion_2017}%
  \BibitemOpen
  \bibfield  {author} {\bibinfo {author} {\bibfnamefont {P.}~\bibnamefont
  {Etingof}}, \bibinfo {author} {\bibfnamefont {D.}~\bibnamefont {Nikshych}},\
  and\ \bibinfo {author} {\bibfnamefont {V.}~\bibnamefont {Ostrik}},\
  }\bibfield  {title} {\bibinfo {title} {On fusion categories},\ }\href
  {http://www.jstor.org/stable/20159926} {\bibfield  {journal} {\bibinfo
  {journal} {Annals of Mathematics}\ }\textbf {\bibinfo {volume} {162}},\
  \bibinfo {pages} {581} (\bibinfo {year} {2005})}\BibitemShut {NoStop}%
\bibitem [{\citenamefont {Gepner}(1991)}]{gepner1991fusion}%
  \BibitemOpen
  \bibfield  {author} {\bibinfo {author} {\bibfnamefont {D.}~\bibnamefont
  {Gepner}},\ }\bibfield  {title} {\bibinfo {title} {Fusion rings and
  geometry},\ }\href {https://doi.org/10.1007/BF02101511} {\bibfield  {journal}
  {\bibinfo  {journal} {Communications in Mathematical Physics}\ }\textbf
  {\bibinfo {volume} {141}},\ \bibinfo {pages} {381} (\bibinfo {year}
  {1991})}\BibitemShut {NoStop}%
\bibitem [{\citenamefont {Gepner}\ and\ \citenamefont
  {Kapustin}(1995)}]{gepner_rings_1994}%
  \BibitemOpen
  \bibfield  {author} {\bibinfo {author} {\bibfnamefont {D.}~\bibnamefont
  {Gepner}}\ and\ \bibinfo {author} {\bibfnamefont {A.}~\bibnamefont
  {Kapustin}},\ }\bibfield  {title} {\bibinfo {title} {{On the classification
  of fusion rings}},\ }\href {https://doi.org/10.1016/0370-2693(95)00172-H}
  {\bibfield  {journal} {\bibinfo  {journal} {Phys. Lett. B}\ }\textbf
  {\bibinfo {volume} {349}},\ \bibinfo {pages} {71} (\bibinfo {year}
  {1995})}\BibitemShut {NoStop}%
\bibitem [{\citenamefont {Dijkgraaf}\ and\ \citenamefont
  {Verlinde}(1988)}]{dijkgraaf1988modular}%
  \BibitemOpen
  \bibfield  {author} {\bibinfo {author} {\bibfnamefont {R.}~\bibnamefont
  {Dijkgraaf}}\ and\ \bibinfo {author} {\bibfnamefont {E.}~\bibnamefont
  {Verlinde}},\ }\bibfield  {title} {\bibinfo {title} {Modular invariance and
  the fusion algebra},\ }\href {https://doi.org/10.1016/0920-5632(88)90371-4}
  {\bibfield  {journal} {\bibinfo  {journal} {Nuclear Physics B-Proceedings
  Supplements}\ }\textbf {\bibinfo {volume} {5}},\ \bibinfo {pages} {87}
  (\bibinfo {year} {1988})}\BibitemShut {NoStop}%
\bibitem [{\citenamefont {Fuchs}(1995)}]{fuchs1995affine}%
  \BibitemOpen
  \bibfield  {author} {\bibinfo {author} {\bibfnamefont {J.}~\bibnamefont
  {Fuchs}},\ }\href@noop {} {\emph {\bibinfo {title} {Affine Lie algebras and
  quantum groups: An Introduction, with applications in conformal field
  theory}}}\ (\bibinfo  {publisher} {Cambridge university press},\ \bibinfo
  {year} {1995})\BibitemShut {NoStop}%
\bibitem [{\citenamefont {Trebst}\ \emph {et~al.}(2008)\citenamefont {Trebst},
  \citenamefont {Troyer}, \citenamefont {Wang},\ and\ \citenamefont
  {Ludwig}}]{trebst2008short}%
  \BibitemOpen
  \bibfield  {author} {\bibinfo {author} {\bibfnamefont {S.}~\bibnamefont
  {Trebst}}, \bibinfo {author} {\bibfnamefont {M.}~\bibnamefont {Troyer}},
  \bibinfo {author} {\bibfnamefont {Z.}~\bibnamefont {Wang}},\ and\ \bibinfo
  {author} {\bibfnamefont {A.~W.}\ \bibnamefont {Ludwig}},\ }\bibfield  {title}
  {\bibinfo {title} {A short introduction to fibonacci anyon models},\ }\href
  {https://doi.org/10.1143/PTPS.176.384} {\bibfield  {journal} {\bibinfo
  {journal} {Progress of Theoretical Physics Supplement}\ }\textbf {\bibinfo
  {volume} {176}},\ \bibinfo {pages} {384} (\bibinfo {year}
  {2008})}\BibitemShut {NoStop}%
\bibitem [{\citenamefont {Greiter}\ \emph {et~al.}(2019)\citenamefont
  {Greiter}, \citenamefont {Haldane},\ and\ \citenamefont
  {Thomale}}]{greiter2019non}%
  \BibitemOpen
  \bibfield  {author} {\bibinfo {author} {\bibfnamefont {M.}~\bibnamefont
  {Greiter}}, \bibinfo {author} {\bibfnamefont {F.~D.~M.}\ \bibnamefont
  {Haldane}},\ and\ \bibinfo {author} {\bibfnamefont {R.}~\bibnamefont
  {Thomale}},\ }\bibfield  {title} {\bibinfo {title} {Non-abelian statistics in
  one dimension: Topological momentum spacings and su(2) level-$k$ fusion
  rules},\ }\href {https://doi.org/10.1103/PhysRevB.100.115107} {\bibfield
  {journal} {\bibinfo  {journal} {Phys. Rev. B}\ }\textbf {\bibinfo {volume}
  {100}},\ \bibinfo {pages} {115107} (\bibinfo {year} {2019})}\BibitemShut
  {NoStop}%
\bibitem [{\citenamefont {Georgi}(2000)}]{georgi2000lie}%
  \BibitemOpen
  \bibfield  {author} {\bibinfo {author} {\bibfnamefont {H.}~\bibnamefont
  {Georgi}},\ }\href@noop {} {\emph {\bibinfo {title} {Lie algebras in particle
  physics: from isospin to unified theories}}}\ (\bibinfo  {publisher} {Taylor
  \& Francis},\ \bibinfo {year} {2000})\BibitemShut {NoStop}%
\bibitem [{\citenamefont {Walker}\ and\ \citenamefont
  {Wang}(2012)}]{walker20123+}%
  \BibitemOpen
  \bibfield  {author} {\bibinfo {author} {\bibfnamefont {K.}~\bibnamefont
  {Walker}}\ and\ \bibinfo {author} {\bibfnamefont {Z.}~\bibnamefont {Wang}},\
  }\bibfield  {title} {\bibinfo {title} {(3+1)-tqfts and topological
  insulators},\ }\href {https://doi.org/10.1007/s11467-011-0194-z} {\bibfield
  {journal} {\bibinfo  {journal} {Frontiers of Physics}\ }\textbf {\bibinfo
  {volume} {7}},\ \bibinfo {pages} {150} (\bibinfo {year} {2012})}\BibitemShut
  {NoStop}%
\bibitem [{\citenamefont {Kong}(2014)}]{kong2014anyon}%
  \BibitemOpen
  \bibfield  {author} {\bibinfo {author} {\bibfnamefont {L.}~\bibnamefont
  {Kong}},\ }\bibfield  {title} {\bibinfo {title} {Anyon condensation and
  tensor categories},\ }\href {https://doi.org/10.1016/j.nuclphysb.2014.07.003}
  {\bibfield  {journal} {\bibinfo  {journal} {Nuclear Physics B}\ }\textbf
  {\bibinfo {volume} {886}},\ \bibinfo {pages} {436} (\bibinfo {year}
  {2014})}\BibitemShut {NoStop}%
\bibitem [{\citenamefont {Kong}\ and\ \citenamefont
  {Wen}(2014)}]{kong2014braided}%
  \BibitemOpen
  \bibfield  {author} {\bibinfo {author} {\bibfnamefont {L.}~\bibnamefont
  {Kong}}\ and\ \bibinfo {author} {\bibfnamefont {X.-G.}\ \bibnamefont {Wen}},\
  }\bibfield  {title} {\bibinfo {title} {Braided fusion categories,
  gravitational anomalies, and the mathematical framework for topological
  orders in any dimensions},\ }\href@noop {} {\bibfield  {journal} {\bibinfo
  {journal} {arXiv preprint arXiv:1405.5858}\ } (\bibinfo {year}
  {2014})}\BibitemShut {NoStop}%
\bibitem [{\citenamefont {Wen}(2016)}]{wen2016theory}%
  \BibitemOpen
  \bibfield  {author} {\bibinfo {author} {\bibfnamefont {X.-G.}\ \bibnamefont
  {Wen}},\ }\bibfield  {title} {\bibinfo {title} {A theory of 2+1d bosonic
  topological orders},\ }\href {https://doi.org/10.1093/nsr/nwv077} {\bibfield
  {journal} {\bibinfo  {journal} {National Science Review}\ }\textbf {\bibinfo
  {volume} {3}},\ \bibinfo {pages} {68} (\bibinfo {year} {2016})}\BibitemShut
  {NoStop}%
\bibitem [{\citenamefont {Bernevig}\ and\ \citenamefont
  {Neupert}(2017)}]{bernevig2017topological}%
  \BibitemOpen
  \bibfield  {author} {\bibinfo {author} {\bibfnamefont {A.}~\bibnamefont
  {Bernevig}}\ and\ \bibinfo {author} {\bibfnamefont {T.}~\bibnamefont
  {Neupert}},\ }\bibfield  {title} {\bibinfo {title} {{Topological
  superconductors and category theory}},\ }in\ \href
  {https://doi.org/10.1093/acprof:oso/9780198785781.003.0002} {\emph {\bibinfo
  {booktitle} {{Topological Aspects of Condensed Matter Physics: Lecture Notes
  of the Les Houches Summer School: Volume 103, August 2014}}}}\ (\bibinfo
  {publisher} {Oxford University Press},\ \bibinfo {year} {2017})\BibitemShut
  {NoStop}%
\bibitem [{\citenamefont {Cong}\ \emph {et~al.}(2017)\citenamefont {Cong},
  \citenamefont {Cheng},\ and\ \citenamefont {Wang}}]{cong2017defects}%
  \BibitemOpen
  \bibfield  {author} {\bibinfo {author} {\bibfnamefont {I.}~\bibnamefont
  {Cong}}, \bibinfo {author} {\bibfnamefont {M.}~\bibnamefont {Cheng}},\ and\
  \bibinfo {author} {\bibfnamefont {Z.}~\bibnamefont {Wang}},\ }\bibfield
  {title} {\bibinfo {title} {Defects between gapped boundaries in
  two-dimensional topological phases of matter},\ }\href
  {https://doi.org/10.1103/PhysRevB.96.195129} {\bibfield  {journal} {\bibinfo
  {journal} {Phys. Rev. B}\ }\textbf {\bibinfo {volume} {96}},\ \bibinfo
  {pages} {195129} (\bibinfo {year} {2017})}\BibitemShut {NoStop}%
\bibitem [{\citenamefont {Kong}\ \emph {et~al.}(2017)\citenamefont {Kong},
  \citenamefont {Wen},\ and\ \citenamefont {Zheng}}]{kong2017boundary}%
  \BibitemOpen
  \bibfield  {author} {\bibinfo {author} {\bibfnamefont {L.}~\bibnamefont
  {Kong}}, \bibinfo {author} {\bibfnamefont {X.-G.}\ \bibnamefont {Wen}},\ and\
  \bibinfo {author} {\bibfnamefont {H.}~\bibnamefont {Zheng}},\ }\bibfield
  {title} {\bibinfo {title} {Boundary-bulk relation in topological orders},\
  }\href {https://doi.org/10.1016/j.nuclphysb.2017.06.023} {\bibfield
  {journal} {\bibinfo  {journal} {Nuclear Physics B}\ }\textbf {\bibinfo
  {volume} {922}},\ \bibinfo {pages} {62} (\bibinfo {year} {2017})}\BibitemShut
  {NoStop}%
\bibitem [{\citenamefont {Wen}(2019)}]{wen2019choreographed}%
  \BibitemOpen
  \bibfield  {author} {\bibinfo {author} {\bibfnamefont {X.-G.}\ \bibnamefont
  {Wen}},\ }\bibfield  {title} {\bibinfo {title} {Choreographed entanglement
  dances: Topological states of quantum matter},\ }\href
  {https://doi.org/10.1126/science.aal3099} {\bibfield  {journal} {\bibinfo
  {journal} {Science}\ }\textbf {\bibinfo {volume} {363}},\ \bibinfo {pages}
  {eaal3099} (\bibinfo {year} {2019})}\BibitemShut {NoStop}%
\bibitem [{\citenamefont {Barkeshli}\ \emph {et~al.}(2019)\citenamefont
  {Barkeshli}, \citenamefont {Bonderson}, \citenamefont {Cheng},\ and\
  \citenamefont {Wang}}]{barkeshli_symmetry_2019}%
  \BibitemOpen
  \bibfield  {author} {\bibinfo {author} {\bibfnamefont {M.}~\bibnamefont
  {Barkeshli}}, \bibinfo {author} {\bibfnamefont {P.}~\bibnamefont
  {Bonderson}}, \bibinfo {author} {\bibfnamefont {M.}~\bibnamefont {Cheng}},\
  and\ \bibinfo {author} {\bibfnamefont {Z.}~\bibnamefont {Wang}},\ }\bibfield
  {title} {\bibinfo {title} {Symmetry fractionalization, defects, and gauging
  of topological phases},\ }\href {https://doi.org/10.1103/PhysRevB.100.115147}
  {\bibfield  {journal} {\bibinfo  {journal} {Physical Review B}\ }\textbf
  {\bibinfo {volume} {100}},\ \bibinfo {pages} {115147} (\bibinfo {year}
  {2019})}\BibitemShut {NoStop}%
\bibitem [{\citenamefont {Ji}\ and\ \citenamefont
  {Wen}(2020{\natexlab{a}})}]{JiWen20}%
  \BibitemOpen
  \bibfield  {author} {\bibinfo {author} {\bibfnamefont {W.}~\bibnamefont
  {Ji}}\ and\ \bibinfo {author} {\bibfnamefont {X.-G.}\ \bibnamefont {Wen}},\
  }\bibfield  {title} {\bibinfo {title} {Categorical symmetry and noninvertible
  anomaly in symmetry-breaking and topological phase transitions},\ }\href
  {https://doi.org/10.1103/PhysRevResearch.2.033417} {\bibfield  {journal}
  {\bibinfo  {journal} {Phys. Rev. Res.}\ }\textbf {\bibinfo {volume} {2}},\
  \bibinfo {pages} {033417} (\bibinfo {year} {2020}{\natexlab{a}})}\BibitemShut
  {NoStop}%
\bibitem [{\citenamefont {Chatterjee}\ and\ \citenamefont
  {Wen}(2023)}]{chatterjee_symmetry_2023}%
  \BibitemOpen
  \bibfield  {author} {\bibinfo {author} {\bibfnamefont {A.}~\bibnamefont
  {Chatterjee}}\ and\ \bibinfo {author} {\bibfnamefont {X.-G.}\ \bibnamefont
  {Wen}},\ }\bibfield  {title} {\bibinfo {title} {Symmetry as a shadow of
  topological order and a derivation of topological holographic principle},\
  }\href {https://doi.org/10.1103/PhysRevB.107.155136} {\bibfield  {journal}
  {\bibinfo  {journal} {Phys. Rev. B}\ }\textbf {\bibinfo {volume} {107}},\
  \bibinfo {pages} {155136} (\bibinfo {year} {2023})}\BibitemShut {NoStop}%
\bibitem [{\citenamefont {Simon}(2023)}]{simon_topological_2021}%
  \BibitemOpen
  \bibfield  {author} {\bibinfo {author} {\bibfnamefont {S.~H.}\ \bibnamefont
  {Simon}},\ }\href@noop {} {\emph {\bibinfo {title} {Topological Quantum}}}\
  (\bibinfo  {publisher} {Oxford University Press},\ \bibinfo {year}
  {2023})\BibitemShut {NoStop}%
\bibitem [{\citenamefont {Nayak}\ \emph {et~al.}(2008)\citenamefont {Nayak},
  \citenamefont {Simon}, \citenamefont {Stern}, \citenamefont {Freedman},\ and\
  \citenamefont {Das~Sarma}}]{Nayak_RMP}%
  \BibitemOpen
  \bibfield  {author} {\bibinfo {author} {\bibfnamefont {C.}~\bibnamefont
  {Nayak}}, \bibinfo {author} {\bibfnamefont {S.~H.}\ \bibnamefont {Simon}},
  \bibinfo {author} {\bibfnamefont {A.}~\bibnamefont {Stern}}, \bibinfo
  {author} {\bibfnamefont {M.}~\bibnamefont {Freedman}},\ and\ \bibinfo
  {author} {\bibfnamefont {S.}~\bibnamefont {Das~Sarma}},\ }\bibfield  {title}
  {\bibinfo {title} {Non-abelian anyons and topological quantum computation},\
  }\href {https://doi.org/10.1103/RevModPhys.80.1083} {\bibfield  {journal}
  {\bibinfo  {journal} {Rev. Mod. Phys.}\ }\textbf {\bibinfo {volume} {80}},\
  \bibinfo {pages} {1083} (\bibinfo {year} {2008})}\BibitemShut {NoStop}%
\bibitem [{\citenamefont {Chen}\ \emph {et~al.}(2012)\citenamefont {Chen},
  \citenamefont {Gu}, \citenamefont {Liu},\ and\ \citenamefont
  {Wen}}]{chen_symmetry-protected_2012-1}%
  \BibitemOpen
  \bibfield  {author} {\bibinfo {author} {\bibfnamefont {X.}~\bibnamefont
  {Chen}}, \bibinfo {author} {\bibfnamefont {Z.-C.}\ \bibnamefont {Gu}},
  \bibinfo {author} {\bibfnamefont {Z.-X.}\ \bibnamefont {Liu}},\ and\ \bibinfo
  {author} {\bibfnamefont {X.-G.}\ \bibnamefont {Wen}},\ }\bibfield  {title}
  {\bibinfo {title} {Symmetry-{{Protected Topological Orders}} in {{Interacting
  Bosonic Systems}}},\ }\href {https://doi.org/10.1126/science.1227224}
  {\bibfield  {journal} {\bibinfo  {journal} {Science}\ }\textbf {\bibinfo
  {volume} {338}},\ \bibinfo {pages} {1604} (\bibinfo {year}
  {2012})}\BibitemShut {NoStop}%
\bibitem [{\citenamefont {Son}\ \emph {et~al.}(2012)\citenamefont {Son},
  \citenamefont {Amico},\ and\ \citenamefont {Vedral}}]{son2012topological}%
  \BibitemOpen
  \bibfield  {author} {\bibinfo {author} {\bibfnamefont {W.}~\bibnamefont
  {Son}}, \bibinfo {author} {\bibfnamefont {L.}~\bibnamefont {Amico}},\ and\
  \bibinfo {author} {\bibfnamefont {V.}~\bibnamefont {Vedral}},\ }\bibfield
  {title} {\bibinfo {title} {Topological order in 1d cluster state protected by
  symmetry},\ }\href {https://doi.org/10.1007/s11128-011-0346-7} {\bibfield
  {journal} {\bibinfo  {journal} {Quantum Information Processing}\ }\textbf
  {\bibinfo {volume} {11}},\ \bibinfo {pages} {1961} (\bibinfo {year}
  {2012})}\BibitemShut {NoStop}%
\bibitem [{\citenamefont {Briegel}\ and\ \citenamefont
  {Raussendorf}(2001)}]{briegel_persistent_2001}%
  \BibitemOpen
  \bibfield  {author} {\bibinfo {author} {\bibfnamefont {H.~J.}\ \bibnamefont
  {Briegel}}\ and\ \bibinfo {author} {\bibfnamefont {R.}~\bibnamefont
  {Raussendorf}},\ }\bibfield  {title} {\bibinfo {title} {Persistent
  {{Entanglement}} in {{Arrays}} of {{Interacting Particles}}},\ }\href
  {https://doi.org/10.1103/PhysRevLett.86.910} {\bibfield  {journal} {\bibinfo
  {journal} {Physical Review Letters}\ }\textbf {\bibinfo {volume} {86}},\
  \bibinfo {pages} {910} (\bibinfo {year} {2001})}\BibitemShut {NoStop}%
\bibitem [{\citenamefont {Brell}(2015)}]{brell_generalized_2015}%
  \BibitemOpen
  \bibfield  {author} {\bibinfo {author} {\bibfnamefont {C.~G.}\ \bibnamefont
  {Brell}},\ }\bibfield  {title} {\bibinfo {title} {Generalized {{Cluster
  States Based}} on {{Finite Groups}}},\ }\href
  {https://doi.org/10.1088/1367-2630/17/2/023029} {\bibfield  {journal}
  {\bibinfo  {journal} {New Journal of Physics}\ }\textbf {\bibinfo {volume}
  {17}},\ \bibinfo {pages} {023029} (\bibinfo {year} {2015})},\ \Eprint
  {https://arxiv.org/abs/1408.6237} {arXiv:1408.6237} \BibitemShut {NoStop}%
\bibitem [{\citenamefont {Albert}\ \emph {et~al.}(2021)\citenamefont {Albert},
  \citenamefont {Aasen}, \citenamefont {Xu}, \citenamefont {Ji}, \citenamefont
  {Alicea},\ and\ \citenamefont {Preskill}}]{albert_spin_2021}%
  \BibitemOpen
  \bibfield  {author} {\bibinfo {author} {\bibfnamefont {V.~V.}\ \bibnamefont
  {Albert}}, \bibinfo {author} {\bibfnamefont {D.}~\bibnamefont {Aasen}},
  \bibinfo {author} {\bibfnamefont {W.}~\bibnamefont {Xu}}, \bibinfo {author}
  {\bibfnamefont {W.}~\bibnamefont {Ji}}, \bibinfo {author} {\bibfnamefont
  {J.}~\bibnamefont {Alicea}},\ and\ \bibinfo {author} {\bibfnamefont
  {J.}~\bibnamefont {Preskill}},\ }\bibfield  {title} {\bibinfo {title} {Spin
  chains, defects, and quantum wires for the quantum-double edge},\ }\href
  {https://arxiv.org/abs/2111.12096} {\bibfield  {journal} {\bibinfo  {journal}
  {arXiv preprint arXiv:2111.12096}\ } (\bibinfo {year} {2021})}\BibitemShut
  {NoStop}%
\bibitem [{\citenamefont {Chang}\ \emph {et~al.}(2019)\citenamefont {Chang},
  \citenamefont {Lin}, \citenamefont {Shao}, \citenamefont {Wang},\ and\
  \citenamefont {Yin}}]{Chang19}%
  \BibitemOpen
  \bibfield  {author} {\bibinfo {author} {\bibfnamefont {C.-M.}\ \bibnamefont
  {Chang}}, \bibinfo {author} {\bibfnamefont {Y.-H.}\ \bibnamefont {Lin}},
  \bibinfo {author} {\bibfnamefont {S.-H.}\ \bibnamefont {Shao}}, \bibinfo
  {author} {\bibfnamefont {Y.}~\bibnamefont {Wang}},\ and\ \bibinfo {author}
  {\bibfnamefont {X.}~\bibnamefont {Yin}},\ }\bibfield  {title} {\bibinfo
  {title} {Topological defect lines and renormalization group flows in two
  dimensions},\ }\href {https://doi.org/10.1007/JHEP01(2019)026} {\bibfield
  {journal} {\bibinfo  {journal} {Journal of High Energy Physics}\ }\textbf
  {\bibinfo {volume} {2019}},\ \bibinfo {pages} {26} (\bibinfo {year}
  {2019})}\BibitemShut {NoStop}%
\bibitem [{\citenamefont {Kong}\ \emph {et~al.}(2020)\citenamefont {Kong},
  \citenamefont {Lan}, \citenamefont {Wen}, \citenamefont {Zhang},\ and\
  \citenamefont {Zheng}}]{kong_algebraic_2020}%
  \BibitemOpen
  \bibfield  {author} {\bibinfo {author} {\bibfnamefont {L.}~\bibnamefont
  {Kong}}, \bibinfo {author} {\bibfnamefont {T.}~\bibnamefont {Lan}}, \bibinfo
  {author} {\bibfnamefont {X.-G.}\ \bibnamefont {Wen}}, \bibinfo {author}
  {\bibfnamefont {Z.-H.}\ \bibnamefont {Zhang}},\ and\ \bibinfo {author}
  {\bibfnamefont {H.}~\bibnamefont {Zheng}},\ }\bibfield  {title} {\bibinfo
  {title} {Algebraic higher symmetry and categorical symmetry -- a holographic
  and entanglement view of symmetry},\ }\href
  {https://doi.org/10.1103/PhysRevResearch.2.043086} {\bibfield  {journal}
  {\bibinfo  {journal} {Physical Review Research}\ }\textbf {\bibinfo {volume}
  {2}},\ \bibinfo {pages} {043086} (\bibinfo {year} {2020})},\ \Eprint
  {https://arxiv.org/abs/2005.14178} {arXiv:2005.14178} \BibitemShut {NoStop}%
\bibitem [{\citenamefont {Ji}\ and\ \citenamefont
  {Wen}(2020{\natexlab{b}})}]{ji_categorical_2020}%
  \BibitemOpen
  \bibfield  {author} {\bibinfo {author} {\bibfnamefont {W.}~\bibnamefont
  {Ji}}\ and\ \bibinfo {author} {\bibfnamefont {X.-G.}\ \bibnamefont {Wen}},\
  }\bibfield  {title} {\bibinfo {title} {Categorical symmetry and
  non-invertible anomaly in symmetry-breaking and topological phase
  transitions},\ }\href {https://doi.org/10.1103/PhysRevResearch.2.033417}
  {\bibfield  {journal} {\bibinfo  {journal} {Physical Review Research}\
  }\textbf {\bibinfo {volume} {2}},\ \bibinfo {pages} {033417} (\bibinfo {year}
  {2020}{\natexlab{b}})},\ \Eprint {https://arxiv.org/abs/1912.13492}
  {arXiv:1912.13492} \BibitemShut {NoStop}%
\bibitem [{\citenamefont {Thorngren}\ and\ \citenamefont
  {Wang}(2024)}]{thorngren_fusion_2019}%
  \BibitemOpen
  \bibfield  {author} {\bibinfo {author} {\bibfnamefont {R.}~\bibnamefont
  {Thorngren}}\ and\ \bibinfo {author} {\bibfnamefont {Y.}~\bibnamefont
  {Wang}},\ }\bibfield  {title} {\bibinfo {title} {Fusion category symmetry.
  part i. anomaly in-flow and gapped phases},\ }\href
  {https://doi.org/10.1007/JHEP04(2024)132} {\bibfield  {journal} {\bibinfo
  {journal} {Journal of High Energy Physics}\ }\textbf {\bibinfo {volume}
  {2024}},\ \bibinfo {pages} {132} (\bibinfo {year} {2024})}\BibitemShut
  {NoStop}%
\bibitem [{\citenamefont {Thorngren}\ and\ \citenamefont
  {Wang}(2021)}]{thorngren_fusion_2021}%
  \BibitemOpen
  \bibfield  {author} {\bibinfo {author} {\bibfnamefont {R.}~\bibnamefont
  {Thorngren}}\ and\ \bibinfo {author} {\bibfnamefont {Y.}~\bibnamefont
  {Wang}},\ }\href {https://doi.org/10.48550/arXiv.2106.12577} {\bibinfo
  {title} {Fusion {{Category Symmetry II}}: {{Categoriosities}} at c = 1 and
  {{Beyond}}}} (\bibinfo {year} {2021}),\ \Eprint
  {https://arxiv.org/abs/2106.12577} {arXiv:2106.12577 [cond-mat,
  physics:hep-th]} \BibitemShut {NoStop}%
\bibitem [{\citenamefont {Inamura}(2021)}]{inamura_topological_2021}%
  \BibitemOpen
  \bibfield  {author} {\bibinfo {author} {\bibfnamefont {K.}~\bibnamefont
  {Inamura}},\ }\bibfield  {title} {\bibinfo {title} {Topological field
  theories and symmetry protected topological phases with fusion category
  symmetries},\ }\href {https://doi.org/10.1007/JHEP05(2021)204} {\bibfield
  {journal} {\bibinfo  {journal} {Journal of High Energy Physics}\ }\textbf
  {\bibinfo {volume} {2021}},\ \bibinfo {pages} {204} (\bibinfo {year}
  {2021})}\BibitemShut {NoStop}%
\bibitem [{\citenamefont {Heidenreich}\ \emph
  {et~al.}(2021{\natexlab{a}})\citenamefont {Heidenreich}, \citenamefont
  {McNamara}, \citenamefont {Montero}, \citenamefont {Reece}, \citenamefont
  {Rudelius},\ and\ \citenamefont {Valenzuela}}]{Heidenreich21Completeness}%
  \BibitemOpen
  \bibfield  {author} {\bibinfo {author} {\bibfnamefont {B.}~\bibnamefont
  {Heidenreich}}, \bibinfo {author} {\bibfnamefont {J.}~\bibnamefont
  {McNamara}}, \bibinfo {author} {\bibfnamefont {M.}~\bibnamefont {Montero}},
  \bibinfo {author} {\bibfnamefont {M.}~\bibnamefont {Reece}}, \bibinfo
  {author} {\bibfnamefont {T.}~\bibnamefont {Rudelius}},\ and\ \bibinfo
  {author} {\bibfnamefont {I.}~\bibnamefont {Valenzuela}},\ }\bibfield  {title}
  {\bibinfo {title} {Non-invertible global symmetries and completeness of the
  spectrum},\ }\href {https://doi.org/10.1007/JHEP09(2021)203} {\bibfield
  {journal} {\bibinfo  {journal} {Journal of High Energy Physics}\ }\textbf
  {\bibinfo {volume} {2021}},\ \bibinfo {pages} {203} (\bibinfo {year}
  {2021}{\natexlab{a}})}\BibitemShut {NoStop}%
\bibitem [{\citenamefont {Bartsch}\ \emph
  {et~al.}(2022{\natexlab{a}})\citenamefont {Bartsch}, \citenamefont
  {Bullimore}, \citenamefont {Ferrari},\ and\ \citenamefont
  {Pearson}}]{Bartsch:2022mpm}%
  \BibitemOpen
  \bibfield  {author} {\bibinfo {author} {\bibfnamefont {T.}~\bibnamefont
  {Bartsch}}, \bibinfo {author} {\bibfnamefont {M.}~\bibnamefont {Bullimore}},
  \bibinfo {author} {\bibfnamefont {A.~E.~V.}\ \bibnamefont {Ferrari}},\ and\
  \bibinfo {author} {\bibfnamefont {J.}~\bibnamefont {Pearson}},\ }\bibfield
  {title} {\bibinfo {title} {{Non-invertible Symmetries and Higher
  Representation Theory I}},\ }\href@noop {} {\  (\bibinfo {year}
  {2022}{\natexlab{a}})},\ \Eprint {https://arxiv.org/abs/2208.05993}
  {arXiv:2208.05993 [hep-th]} \BibitemShut {NoStop}%
\bibitem [{\citenamefont {Bartsch}\ \emph
  {et~al.}(2022{\natexlab{b}})\citenamefont {Bartsch}, \citenamefont
  {Bullimore}, \citenamefont {Ferrari},\ and\ \citenamefont
  {Pearson}}]{Bartsch:2022ytj}%
  \BibitemOpen
  \bibfield  {author} {\bibinfo {author} {\bibfnamefont {T.}~\bibnamefont
  {Bartsch}}, \bibinfo {author} {\bibfnamefont {M.}~\bibnamefont {Bullimore}},
  \bibinfo {author} {\bibfnamefont {A.~E.~V.}\ \bibnamefont {Ferrari}},\ and\
  \bibinfo {author} {\bibfnamefont {J.}~\bibnamefont {Pearson}},\ }\bibfield
  {title} {\bibinfo {title} {{Non-invertible Symmetries and Higher
  Representation Theory II}},\ }\href@noop {} {\  (\bibinfo {year}
  {2022}{\natexlab{b}})},\ \Eprint {https://arxiv.org/abs/2212.07393}
  {arXiv:2212.07393 [hep-th]} \BibitemShut {NoStop}%
\bibitem [{\citenamefont {Bartsch}\ \emph {et~al.}(2023)\citenamefont
  {Bartsch}, \citenamefont {Bullimore},\ and\ \citenamefont
  {Grigoletto}}]{Bartsch:2023wvv}%
  \BibitemOpen
  \bibfield  {author} {\bibinfo {author} {\bibfnamefont {T.}~\bibnamefont
  {Bartsch}}, \bibinfo {author} {\bibfnamefont {M.}~\bibnamefont {Bullimore}},\
  and\ \bibinfo {author} {\bibfnamefont {A.}~\bibnamefont {Grigoletto}},\
  }\bibfield  {title} {\bibinfo {title} {{Representation theory for categorical
  symmetries}},\ }\href@noop {} {\  (\bibinfo {year} {2023})},\ \Eprint
  {https://arxiv.org/abs/2305.17165} {arXiv:2305.17165 [hep-th]} \BibitemShut
  {NoStop}%
\bibitem [{\citenamefont {Roumpedakis}\ \emph {et~al.}(2023)\citenamefont
  {Roumpedakis}, \citenamefont {Seifnashri},\ and\ \citenamefont
  {Shao}}]{Roumpedakis23higher}%
  \BibitemOpen
  \bibfield  {author} {\bibinfo {author} {\bibfnamefont {K.}~\bibnamefont
  {Roumpedakis}}, \bibinfo {author} {\bibfnamefont {S.}~\bibnamefont
  {Seifnashri}},\ and\ \bibinfo {author} {\bibfnamefont {S.-H.}\ \bibnamefont
  {Shao}},\ }\bibfield  {title} {\bibinfo {title} {Higher gauging and
  non-invertible condensation defects},\ }\href
  {https://doi.org/10.1007/s00220-023-04706-9} {\bibfield  {journal} {\bibinfo
  {journal} {Communications in Mathematical Physics}\ }\textbf {\bibinfo
  {volume} {401}},\ \bibinfo {pages} {3043} (\bibinfo {year}
  {2023})}\BibitemShut {NoStop}%
\bibitem [{\citenamefont {Inamura}(2023)}]{inamura2023fermionization}%
  \BibitemOpen
  \bibfield  {author} {\bibinfo {author} {\bibfnamefont {K.}~\bibnamefont
  {Inamura}},\ }\bibfield  {title} {\bibinfo {title} {Fermionization of fusion
  category symmetries in 1+1 dimensions},\ }\href@noop {} {\bibfield  {journal}
  {\bibinfo  {journal} {Journal of High Energy Physics}\ }\textbf {\bibinfo
  {volume} {2023}},\ \bibinfo {pages} {1} (\bibinfo {year} {2023})}\BibitemShut
  {NoStop}%
\bibitem [{\citenamefont {Choi}\ \emph {et~al.}(2022)\citenamefont {Choi},
  \citenamefont {C\'ordova}, \citenamefont {Hsin}, \citenamefont {Lam},\ and\
  \citenamefont {Shao}}]{Choi22}%
  \BibitemOpen
  \bibfield  {author} {\bibinfo {author} {\bibfnamefont {Y.}~\bibnamefont
  {Choi}}, \bibinfo {author} {\bibfnamefont {C.}~\bibnamefont {C\'ordova}},
  \bibinfo {author} {\bibfnamefont {P.-S.}\ \bibnamefont {Hsin}}, \bibinfo
  {author} {\bibfnamefont {H.~T.}\ \bibnamefont {Lam}},\ and\ \bibinfo {author}
  {\bibfnamefont {S.-H.}\ \bibnamefont {Shao}},\ }\bibfield  {title} {\bibinfo
  {title} {Noninvertible duality defects in $3+1$ dimensions},\ }\href
  {https://doi.org/10.1103/PhysRevD.105.125016} {\bibfield  {journal} {\bibinfo
   {journal} {Phys. Rev. D}\ }\textbf {\bibinfo {volume} {105}},\ \bibinfo
  {pages} {125016} (\bibinfo {year} {2022})}\BibitemShut {NoStop}%
\bibitem [{\citenamefont {Choi}\ \emph
  {et~al.}(2023{\natexlab{a}})\citenamefont {Choi}, \citenamefont
  {C{\'o}rdova}, \citenamefont {Hsin}, \citenamefont {Lam},\ and\ \citenamefont
  {Shao}}]{Choi23triality}%
  \BibitemOpen
  \bibfield  {author} {\bibinfo {author} {\bibfnamefont {Y.}~\bibnamefont
  {Choi}}, \bibinfo {author} {\bibfnamefont {C.}~\bibnamefont {C{\'o}rdova}},
  \bibinfo {author} {\bibfnamefont {P.-S.}\ \bibnamefont {Hsin}}, \bibinfo
  {author} {\bibfnamefont {H.~T.}\ \bibnamefont {Lam}},\ and\ \bibinfo {author}
  {\bibfnamefont {S.-H.}\ \bibnamefont {Shao}},\ }\bibfield  {title} {\bibinfo
  {title} {Non-invertible condensation, duality, and triality defects in 3+1
  dimensions},\ }\href {https://doi.org/10.1007/s00220-023-04727-4} {\bibfield
  {journal} {\bibinfo  {journal} {Communications in Mathematical Physics}\
  }\textbf {\bibinfo {volume} {402}},\ \bibinfo {pages} {489} (\bibinfo {year}
  {2023}{\natexlab{a}})}\BibitemShut {NoStop}%
\bibitem [{\citenamefont {Choi}\ \emph
  {et~al.}(2023{\natexlab{b}})\citenamefont {Choi}, \citenamefont {Lam},\ and\
  \citenamefont {Shao}}]{Choi23timereversal}%
  \BibitemOpen
  \bibfield  {author} {\bibinfo {author} {\bibfnamefont {Y.}~\bibnamefont
  {Choi}}, \bibinfo {author} {\bibfnamefont {H.~T.}\ \bibnamefont {Lam}},\ and\
  \bibinfo {author} {\bibfnamefont {S.-H.}\ \bibnamefont {Shao}},\ }\bibfield
  {title} {\bibinfo {title} {Noninvertible time-reversal symmetry},\ }\href
  {https://doi.org/10.1103/PhysRevLett.130.131602} {\bibfield  {journal}
  {\bibinfo  {journal} {Phys. Rev. Lett.}\ }\textbf {\bibinfo {volume} {130}},\
  \bibinfo {pages} {131602} (\bibinfo {year} {2023}{\natexlab{b}})}\BibitemShut
  {NoStop}%
\bibitem [{\citenamefont {Zhang}\ and\ \citenamefont
  {C{\'o}rdova}(2023)}]{zhang2023anomalies}%
  \BibitemOpen
  \bibfield  {author} {\bibinfo {author} {\bibfnamefont {C.}~\bibnamefont
  {Zhang}}\ and\ \bibinfo {author} {\bibfnamefont {C.}~\bibnamefont
  {C{\'o}rdova}},\ }\bibfield  {title} {\bibinfo {title} {Anomalies of $(1+1)d
  $ categorical symmetries},\ }\href {https://arxiv.org/abs/2304.01262}
  {\bibfield  {journal} {\bibinfo  {journal} {arXiv preprint arXiv:2304.01262}\
  } (\bibinfo {year} {2023})}\BibitemShut {NoStop}%
\bibitem [{\citenamefont {Cordova}\ \emph {et~al.}(2023)\citenamefont
  {Cordova}, \citenamefont {Hsin},\ and\ \citenamefont
  {Zhang}}]{cordova2023anomalies}%
  \BibitemOpen
  \bibfield  {author} {\bibinfo {author} {\bibfnamefont {C.}~\bibnamefont
  {Cordova}}, \bibinfo {author} {\bibfnamefont {P.-S.}\ \bibnamefont {Hsin}},\
  and\ \bibinfo {author} {\bibfnamefont {C.}~\bibnamefont {Zhang}},\ }\bibfield
   {title} {\bibinfo {title} {Anomalies of non-invertible symmetries in
  (3+1)d},\ }\href {https://arxiv.org/abs/2308.11706} {\bibfield  {journal}
  {\bibinfo  {journal} {arXiv preprint arXiv:2308.11706}\ } (\bibinfo {year}
  {2023})}\BibitemShut {NoStop}%
\bibitem [{\citenamefont {Bhardwaj}\ \emph
  {et~al.}(2023{\natexlab{a}})\citenamefont {Bhardwaj}, \citenamefont
  {Sch{\"a}fer-Nameki},\ and\ \citenamefont {Tiwari}}]{bhardwaj23unifying}%
  \BibitemOpen
  \bibfield  {author} {\bibinfo {author} {\bibfnamefont {L.}~\bibnamefont
  {Bhardwaj}}, \bibinfo {author} {\bibfnamefont {S.}~\bibnamefont
  {Sch{\"a}fer-Nameki}},\ and\ \bibinfo {author} {\bibfnamefont
  {A.}~\bibnamefont {Tiwari}},\ }\bibfield  {title} {\bibinfo {title}
  {{Unifying constructions of non-invertible symmetries}},\ }\href
  {https://doi.org/10.21468/SciPostPhys.15.3.122} {\bibfield  {journal}
  {\bibinfo  {journal} {SciPost Phys.}\ }\textbf {\bibinfo {volume} {15}},\
  \bibinfo {pages} {122} (\bibinfo {year} {2023}{\natexlab{a}})}\BibitemShut
  {NoStop}%
\bibitem [{\citenamefont {Bhardwaj}\ \emph
  {et~al.}(2023{\natexlab{b}})\citenamefont {Bhardwaj}, \citenamefont
  {Bottini}, \citenamefont {Sch{\"a}fer-Nameki},\ and\ \citenamefont
  {Tiwari}}]{bhardwaj23symmetryweb}%
  \BibitemOpen
  \bibfield  {author} {\bibinfo {author} {\bibfnamefont {L.}~\bibnamefont
  {Bhardwaj}}, \bibinfo {author} {\bibfnamefont {L.~E.}\ \bibnamefont
  {Bottini}}, \bibinfo {author} {\bibfnamefont {S.}~\bibnamefont
  {Sch{\"a}fer-Nameki}},\ and\ \bibinfo {author} {\bibfnamefont
  {A.}~\bibnamefont {Tiwari}},\ }\bibfield  {title} {\bibinfo {title}
  {{Non-invertible symmetry webs}},\ }\href
  {https://doi.org/10.21468/SciPostPhys.15.4.160} {\bibfield  {journal}
  {\bibinfo  {journal} {SciPost Phys.}\ }\textbf {\bibinfo {volume} {15}},\
  \bibinfo {pages} {160} (\bibinfo {year} {2023}{\natexlab{b}})}\BibitemShut
  {NoStop}%
\bibitem [{\citenamefont {Bhardwaj}\ \emph
  {et~al.}(2022{\natexlab{a}})\citenamefont {Bhardwaj}, \citenamefont
  {Sch{\"a}fer-Nameki},\ and\ \citenamefont {Wu}}]{bhardwaj2022universal}%
  \BibitemOpen
  \bibfield  {author} {\bibinfo {author} {\bibfnamefont {L.}~\bibnamefont
  {Bhardwaj}}, \bibinfo {author} {\bibfnamefont {S.}~\bibnamefont
  {Sch{\"a}fer-Nameki}},\ and\ \bibinfo {author} {\bibfnamefont
  {J.}~\bibnamefont {Wu}},\ }\bibfield  {title} {\bibinfo {title} {Universal
  non-invertible symmetries},\ }\href {https://doi.org/10.1002/prop.202200143}
  {\bibfield  {journal} {\bibinfo  {journal} {Fortschritte der Physik}\
  }\textbf {\bibinfo {volume} {70}},\ \bibinfo {pages} {2200143} (\bibinfo
  {year} {2022}{\natexlab{a}})}\BibitemShut {NoStop}%
\bibitem [{\citenamefont {Bhardwaj}\ \emph
  {et~al.}(2023{\natexlab{c}})\citenamefont {Bhardwaj}, \citenamefont
  {Bottini}, \citenamefont {Sch{\"a}fer-Nameki},\ and\ \citenamefont
  {Tiwari}}]{Bhardwaj23noninvertible}%
  \BibitemOpen
  \bibfield  {author} {\bibinfo {author} {\bibfnamefont {L.}~\bibnamefont
  {Bhardwaj}}, \bibinfo {author} {\bibfnamefont {L.~E.}\ \bibnamefont
  {Bottini}}, \bibinfo {author} {\bibfnamefont {S.}~\bibnamefont
  {Sch{\"a}fer-Nameki}},\ and\ \bibinfo {author} {\bibfnamefont
  {A.}~\bibnamefont {Tiwari}},\ }\bibfield  {title} {\bibinfo {title}
  {{Non-invertible higher-categorical symmetries}},\ }\href
  {https://doi.org/10.21468/SciPostPhys.14.1.007} {\bibfield  {journal}
  {\bibinfo  {journal} {SciPost Phys.}\ }\textbf {\bibinfo {volume} {14}},\
  \bibinfo {pages} {007} (\bibinfo {year} {2023}{\natexlab{c}})}\BibitemShut
  {NoStop}%
\bibitem [{\citenamefont {Kaidi}\ \emph {et~al.}(2023)\citenamefont {Kaidi},
  \citenamefont {Ohmori},\ and\ \citenamefont {Zheng}}]{Kaidi23symTFT}%
  \BibitemOpen
  \bibfield  {author} {\bibinfo {author} {\bibfnamefont {J.}~\bibnamefont
  {Kaidi}}, \bibinfo {author} {\bibfnamefont {K.}~\bibnamefont {Ohmori}},\ and\
  \bibinfo {author} {\bibfnamefont {Y.}~\bibnamefont {Zheng}},\ }\bibfield
  {title} {\bibinfo {title} {Symmetry tfts for non-invertible defects},\ }\href
  {https://doi.org/10.1007/s00220-023-04859-7} {\bibfield  {journal} {\bibinfo
  {journal} {Communications in Mathematical Physics}\ }\textbf {\bibinfo
  {volume} {404}},\ \bibinfo {pages} {1021} (\bibinfo {year}
  {2023})}\BibitemShut {NoStop}%
\bibitem [{\citenamefont {Bhardwaj}\ and\ \citenamefont
  {Schafer-Nameki}(2023{\natexlab{a}})}]{bhardwaj2023generalized}%
  \BibitemOpen
  \bibfield  {author} {\bibinfo {author} {\bibfnamefont {L.}~\bibnamefont
  {Bhardwaj}}\ and\ \bibinfo {author} {\bibfnamefont {S.}~\bibnamefont
  {Schafer-Nameki}},\ }\bibfield  {title} {\bibinfo {title} {Generalized
  charges, part ii: Non-invertible symmetries and the symmetry tft},\ }\href
  {https://arxiv.org/abs/2305.17159} {\bibfield  {journal} {\bibinfo  {journal}
  {arXiv preprint arXiv:2305.17159}\ } (\bibinfo {year}
  {2023}{\natexlab{a}})}\BibitemShut {NoStop}%
\bibitem [{\citenamefont {Bhardwaj}\ \emph
  {et~al.}(2023{\natexlab{d}})\citenamefont {Bhardwaj}, \citenamefont
  {Bottini}, \citenamefont {Pajer},\ and\ \citenamefont
  {Schafer-Nameki}}]{bhardwaj2023gapped}%
  \BibitemOpen
  \bibfield  {author} {\bibinfo {author} {\bibfnamefont {L.}~\bibnamefont
  {Bhardwaj}}, \bibinfo {author} {\bibfnamefont {L.~E.}\ \bibnamefont
  {Bottini}}, \bibinfo {author} {\bibfnamefont {D.}~\bibnamefont {Pajer}},\
  and\ \bibinfo {author} {\bibfnamefont {S.}~\bibnamefont {Schafer-Nameki}},\
  }\bibfield  {title} {\bibinfo {title} {Gapped phases with non-invertible
  symmetries:(1+1)d},\ }\href {https://arxiv.org/abs/2310.03784} {\bibfield
  {journal} {\bibinfo  {journal} {arXiv preprint arXiv:2310.03784}\ } (\bibinfo
  {year} {2023}{\natexlab{d}})}\BibitemShut {NoStop}%
\bibitem [{\citenamefont {Bhardwaj}\ \emph
  {et~al.}(2023{\natexlab{e}})\citenamefont {Bhardwaj}, \citenamefont
  {Bottini}, \citenamefont {Pajer},\ and\ \citenamefont
  {Schafer-Nameki}}]{bhardwaj2023categorical}%
  \BibitemOpen
  \bibfield  {author} {\bibinfo {author} {\bibfnamefont {L.}~\bibnamefont
  {Bhardwaj}}, \bibinfo {author} {\bibfnamefont {L.~E.}\ \bibnamefont
  {Bottini}}, \bibinfo {author} {\bibfnamefont {D.}~\bibnamefont {Pajer}},\
  and\ \bibinfo {author} {\bibfnamefont {S.}~\bibnamefont {Schafer-Nameki}},\
  }\bibfield  {title} {\bibinfo {title} {Categorical landau paradigm for gapped
  phases},\ }\href {https://arxiv.org/abs/2310.03786} {\bibfield  {journal}
  {\bibinfo  {journal} {arXiv preprint arXiv:2310.03786}\ } (\bibinfo {year}
  {2023}{\natexlab{e}})}\BibitemShut {NoStop}%
\bibitem [{\citenamefont {Choi}\ \emph
  {et~al.}(2023{\natexlab{c}})\citenamefont {Choi}, \citenamefont {Lu},\ and\
  \citenamefont {Sun}}]{choi2023self}%
  \BibitemOpen
  \bibfield  {author} {\bibinfo {author} {\bibfnamefont {Y.}~\bibnamefont
  {Choi}}, \bibinfo {author} {\bibfnamefont {D.-C.}\ \bibnamefont {Lu}},\ and\
  \bibinfo {author} {\bibfnamefont {Z.}~\bibnamefont {Sun}},\ }\bibfield
  {title} {\bibinfo {title} {Self-duality under gauging a non-invertible
  symmetry},\ }\href {https://arxiv.org/abs/2310.19867} {\bibfield  {journal}
  {\bibinfo  {journal} {arXiv preprint arXiv:2310.19867}\ } (\bibinfo {year}
  {2023}{\natexlab{c}})}\BibitemShut {NoStop}%
\bibitem [{\citenamefont {D{\'e}coppet}\ and\ \citenamefont
  {Yu}(2023)}]{Decoppet23}%
  \BibitemOpen
  \bibfield  {author} {\bibinfo {author} {\bibfnamefont {T.~D.}\ \bibnamefont
  {D{\'e}coppet}}\ and\ \bibinfo {author} {\bibfnamefont {M.}~\bibnamefont
  {Yu}},\ }\bibfield  {title} {\bibinfo {title} {Gauging noninvertible defects:
  a 2-categorical perspective},\ }\href
  {https://doi.org/10.1007/s11005-023-01655-1} {\bibfield  {journal} {\bibinfo
  {journal} {Letters in Mathematical Physics}\ }\textbf {\bibinfo {volume}
  {113}},\ \bibinfo {pages} {36} (\bibinfo {year} {2023})}\BibitemShut
  {NoStop}%
\bibitem [{\citenamefont {Pace}(2023)}]{Pace2023emergent}%
  \BibitemOpen
  \bibfield  {author} {\bibinfo {author} {\bibfnamefont {S.~D.}\ \bibnamefont
  {Pace}},\ }\bibfield  {title} {\bibinfo {title} {Emergent generalized
  symmetries in ordered phases},\ }\href {https://arxiv.org/abs/2308.05730}
  {\bibfield  {journal} {\bibinfo  {journal} {arXiv preprint arXiv:2308.05730}\
  } (\bibinfo {year} {2023})}\BibitemShut {NoStop}%
\bibitem [{\citenamefont {Perez-Lona}\ \emph {et~al.}(2023)\citenamefont
  {Perez-Lona}, \citenamefont {Robbins}, \citenamefont {Sharpe}, \citenamefont
  {Vandermeulen},\ and\ \citenamefont {Yu}}]{perez2023notes}%
  \BibitemOpen
  \bibfield  {author} {\bibinfo {author} {\bibfnamefont {A.}~\bibnamefont
  {Perez-Lona}}, \bibinfo {author} {\bibfnamefont {D.}~\bibnamefont {Robbins}},
  \bibinfo {author} {\bibfnamefont {E.}~\bibnamefont {Sharpe}}, \bibinfo
  {author} {\bibfnamefont {T.}~\bibnamefont {Vandermeulen}},\ and\ \bibinfo
  {author} {\bibfnamefont {X.}~\bibnamefont {Yu}},\ }\bibfield  {title}
  {\bibinfo {title} {Notes on gauging noninvertible symmetries, part 1:
  Multiplicity-free cases},\ }\href {https://arxiv.org/abs/2311.16230}
  {\bibfield  {journal} {\bibinfo  {journal} {arXiv preprint arXiv:2311.16230}\
  } (\bibinfo {year} {2023})}\BibitemShut {NoStop}%
\bibitem [{\citenamefont {Hu}\ \emph {et~al.}(2017)\citenamefont {Hu},
  \citenamefont {Wan},\ and\ \citenamefont {Wu}}]{hu2017boundary}%
  \BibitemOpen
  \bibfield  {author} {\bibinfo {author} {\bibfnamefont {Y.}~\bibnamefont
  {Hu}}, \bibinfo {author} {\bibfnamefont {Y.}~\bibnamefont {Wan}},\ and\
  \bibinfo {author} {\bibfnamefont {Y.-S.}\ \bibnamefont {Wu}},\ }\bibfield
  {title} {\bibinfo {title} {Boundary hamiltonian theory for gapped topological
  orders},\ }\href {https://doi.org/10.1088/0256-307X/34/7/077103} {\bibfield
  {journal} {\bibinfo  {journal} {Chinese Physics Letters}\ }\textbf {\bibinfo
  {volume} {34}},\ \bibinfo {pages} {077103} (\bibinfo {year}
  {2017})}\BibitemShut {NoStop}%
\bibitem [{\citenamefont {Hu}\ \emph {et~al.}(2018)\citenamefont {Hu},
  \citenamefont {Luo}, \citenamefont {Pankovich}, \citenamefont {Wan},\ and\
  \citenamefont {Wu}}]{hu2018boundary}%
  \BibitemOpen
  \bibfield  {author} {\bibinfo {author} {\bibfnamefont {Y.}~\bibnamefont
  {Hu}}, \bibinfo {author} {\bibfnamefont {Z.-X.}\ \bibnamefont {Luo}},
  \bibinfo {author} {\bibfnamefont {R.}~\bibnamefont {Pankovich}}, \bibinfo
  {author} {\bibfnamefont {Y.}~\bibnamefont {Wan}},\ and\ \bibinfo {author}
  {\bibfnamefont {Y.-S.}\ \bibnamefont {Wu}},\ }\bibfield  {title} {\bibinfo
  {title} {Boundary hamiltonian theory for gapped topological phases on an open
  surface},\ }\href {https://doi.org/10.1007/JHEP01(2018)134} {\bibfield
  {journal} {\bibinfo  {journal} {Journal of High Energy Physics}\ }\textbf
  {\bibinfo {volume} {2018}},\ \bibinfo {pages} {1} (\bibinfo {year}
  {2018})}\BibitemShut {NoStop}%
\bibitem [{\citenamefont {Inamura}(2022)}]{inamura_lattice_2022}%
  \BibitemOpen
  \bibfield  {author} {\bibinfo {author} {\bibfnamefont {K.}~\bibnamefont
  {Inamura}},\ }\bibfield  {title} {\bibinfo {title} {On lattice models of
  gapped phases with fusion category symmetries},\ }\href
  {https://doi.org/10.1007/JHEP03(2022)036} {\bibfield  {journal} {\bibinfo
  {journal} {Journal of High Energy Physics}\ }\textbf {\bibinfo {volume}
  {2022}},\ \bibinfo {pages} {36} (\bibinfo {year} {2022})},\ \Eprint
  {https://arxiv.org/abs/2110.12882} {arXiv:2110.12882} \BibitemShut {NoStop}%
\bibitem [{\citenamefont {Tantivasadakarn}\ \emph {et~al.}(2023)\citenamefont
  {Tantivasadakarn}, \citenamefont {Vishwanath},\ and\ \citenamefont
  {Verresen}}]{tantivasadakarn_hierarchy_2022}%
  \BibitemOpen
  \bibfield  {author} {\bibinfo {author} {\bibfnamefont {N.}~\bibnamefont
  {Tantivasadakarn}}, \bibinfo {author} {\bibfnamefont {A.}~\bibnamefont
  {Vishwanath}},\ and\ \bibinfo {author} {\bibfnamefont {R.}~\bibnamefont
  {Verresen}},\ }\bibfield  {title} {\bibinfo {title} {Hierarchy of topological
  order from finite-depth unitaries, measurement, and feedforward},\ }\href
  {https://doi.org/10.1103/PRXQuantum.4.020339} {\bibfield  {journal} {\bibinfo
   {journal} {PRX Quantum}\ }\textbf {\bibinfo {volume} {4}},\ \bibinfo {pages}
  {020339} (\bibinfo {year} {2023})}\BibitemShut {NoStop}%
\bibitem [{\citenamefont {Seiberg}\ and\ \citenamefont
  {Shao}(2023)}]{SeibergShao23}%
  \BibitemOpen
  \bibfield  {author} {\bibinfo {author} {\bibfnamefont {N.}~\bibnamefont
  {Seiberg}}\ and\ \bibinfo {author} {\bibfnamefont {S.-H.}\ \bibnamefont
  {Shao}},\ }\bibfield  {title} {\bibinfo {title} {Majorana chain and ising
  model--(non-invertible) translations, anomalies, and emanant symmetries},\
  }\href {https://arxiv.org/abs/2307.02534} {\bibfield  {journal} {\bibinfo
  {journal} {arXiv preprint arXiv:2307.02534}\ } (\bibinfo {year}
  {2023})}\BibitemShut {NoStop}%
\bibitem [{\citenamefont {Shao}(2023)}]{Shao23}%
  \BibitemOpen
  \bibfield  {author} {\bibinfo {author} {\bibfnamefont {S.-H.}\ \bibnamefont
  {Shao}},\ }\bibfield  {title} {\bibinfo {title} {What's done cannot be
  undone: Tasi lectures on non-invertible symmetry},\ }\href
  {https://arxiv.org/abs/2308.00747} {\bibfield  {journal} {\bibinfo  {journal}
  {arXiv preprint arXiv:2308.00747}\ } (\bibinfo {year} {2023})}\BibitemShut
  {NoStop}%
\bibitem [{\citenamefont {Inamura}\ and\ \citenamefont
  {Ohmori}(2023)}]{inamura2023fusion}%
  \BibitemOpen
  \bibfield  {author} {\bibinfo {author} {\bibfnamefont {K.}~\bibnamefont
  {Inamura}}\ and\ \bibinfo {author} {\bibfnamefont {K.}~\bibnamefont
  {Ohmori}},\ }\bibfield  {title} {\bibinfo {title} {Fusion surface models:
  2+1d lattice models from fusion 2-categories},\ }\href
  {https://arxiv.org/abs/2305.05774} {\bibfield  {journal} {\bibinfo  {journal}
  {arXiv preprint arXiv:2305.05774}\ } (\bibinfo {year} {2023})}\BibitemShut
  {NoStop}%
\bibitem [{\citenamefont {Delcamp}\ and\ \citenamefont
  {Tiwari}(2023)}]{delcamp2023higher}%
  \BibitemOpen
  \bibfield  {author} {\bibinfo {author} {\bibfnamefont {C.}~\bibnamefont
  {Delcamp}}\ and\ \bibinfo {author} {\bibfnamefont {A.}~\bibnamefont
  {Tiwari}},\ }\bibfield  {title} {\bibinfo {title} {Higher categorical
  symmetries and gauging in two-dimensional spin systems},\ }\href
  {https://arxiv.org/abs/2301.01259} {\bibfield  {journal} {\bibinfo  {journal}
  {arXiv preprint arXiv:2301.01259}\ }\textbf {\bibinfo {volume} {4}} (\bibinfo
  {year} {2023})}\BibitemShut {NoStop}%
\bibitem [{\citenamefont {Tantivasadakarn}\ and\ \citenamefont
  {Chen}(2023)}]{tantivasadakarn2023string}%
  \BibitemOpen
  \bibfield  {author} {\bibinfo {author} {\bibfnamefont {N.}~\bibnamefont
  {Tantivasadakarn}}\ and\ \bibinfo {author} {\bibfnamefont {X.}~\bibnamefont
  {Chen}},\ }\bibfield  {title} {\bibinfo {title} {String operators for
  cheshire strings in topological phases},\ }\href
  {https://arxiv.org/abs/2307.03180} {\bibfield  {journal} {\bibinfo  {journal}
  {arXiv preprint arXiv:2307.03180}\ } (\bibinfo {year} {2023})}\BibitemShut
  {NoStop}%
\bibitem [{\citenamefont {You}\ \emph {et~al.}(2018{\natexlab{b}})\citenamefont
  {You}, \citenamefont {Devakul}, \citenamefont {Burnell},\ and\ \citenamefont
  {Sondhi}}]{you_subsystem_2018}%
  \BibitemOpen
  \bibfield  {author} {\bibinfo {author} {\bibfnamefont {Y.}~\bibnamefont
  {You}}, \bibinfo {author} {\bibfnamefont {T.}~\bibnamefont {Devakul}},
  \bibinfo {author} {\bibfnamefont {F.~J.}\ \bibnamefont {Burnell}},\ and\
  \bibinfo {author} {\bibfnamefont {S.~L.}\ \bibnamefont {Sondhi}},\ }\bibfield
   {title} {\bibinfo {title} {Subsystem symmetry protected topological order},\
  }\href {https://doi.org/10.1103/PhysRevB.98.035112} {\bibfield  {journal}
  {\bibinfo  {journal} {Physical Review B}\ }\textbf {\bibinfo {volume} {98}},\
  \bibinfo {pages} {035112} (\bibinfo {year} {2018}{\natexlab{b}})}\BibitemShut
  {NoStop}%
\bibitem [{\citenamefont {Stephen}\ \emph {et~al.}(2019)\citenamefont
  {Stephen}, \citenamefont {Nautrup}, \citenamefont {{Bermejo-Vega}},
  \citenamefont {Eisert},\ and\ \citenamefont
  {Raussendorf}}]{stephen_subsystem_2019-1}%
  \BibitemOpen
  \bibfield  {author} {\bibinfo {author} {\bibfnamefont {D.~T.}\ \bibnamefont
  {Stephen}}, \bibinfo {author} {\bibfnamefont {H.~P.}\ \bibnamefont
  {Nautrup}}, \bibinfo {author} {\bibfnamefont {J.}~\bibnamefont
  {{Bermejo-Vega}}}, \bibinfo {author} {\bibfnamefont {J.}~\bibnamefont
  {Eisert}},\ and\ \bibinfo {author} {\bibfnamefont {R.}~\bibnamefont
  {Raussendorf}},\ }\bibfield  {title} {\bibinfo {title} {Subsystem symmetries,
  quantum cellular automata, and computational phases of quantum matter},\
  }\href {https://doi.org/10.22331/q-2019-05-20-142} {\bibfield  {journal}
  {\bibinfo  {journal} {Quantum}\ }\textbf {\bibinfo {volume} {3}},\ \bibinfo
  {pages} {142} (\bibinfo {year} {2019})}\BibitemShut {NoStop}%
\bibitem [{\citenamefont {Tantivasadakarn}\ and\ \citenamefont
  {Vijay}(2020)}]{TantivasadakarnVijay20}%
  \BibitemOpen
  \bibfield  {author} {\bibinfo {author} {\bibfnamefont {N.}~\bibnamefont
  {Tantivasadakarn}}\ and\ \bibinfo {author} {\bibfnamefont {S.}~\bibnamefont
  {Vijay}},\ }\bibfield  {title} {\bibinfo {title} {Searching for fracton
  orders via symmetry defect condensation},\ }\href
  {https://doi.org/10.1103/PhysRevB.101.165143} {\bibfield  {journal} {\bibinfo
   {journal} {Phys. Rev. B}\ }\textbf {\bibinfo {volume} {101}},\ \bibinfo
  {pages} {165143} (\bibinfo {year} {2020})}\BibitemShut {NoStop}%
\bibitem [{\citenamefont {Devakul}\ \emph {et~al.}(2020)\citenamefont
  {Devakul}, \citenamefont {Shirley},\ and\ \citenamefont
  {Wang}}]{DevakulShirleyWang20}%
  \BibitemOpen
  \bibfield  {author} {\bibinfo {author} {\bibfnamefont {T.}~\bibnamefont
  {Devakul}}, \bibinfo {author} {\bibfnamefont {W.}~\bibnamefont {Shirley}},\
  and\ \bibinfo {author} {\bibfnamefont {J.}~\bibnamefont {Wang}},\ }\bibfield
  {title} {\bibinfo {title} {Strong planar subsystem symmetry-protected
  topological phases and their dual fracton orders},\ }\href
  {https://doi.org/10.1103/PhysRevResearch.2.012059} {\bibfield  {journal}
  {\bibinfo  {journal} {Phys. Rev. Research}\ }\textbf {\bibinfo {volume}
  {2}},\ \bibinfo {pages} {012059} (\bibinfo {year} {2020})}\BibitemShut
  {NoStop}%
\bibitem [{\citenamefont {Stephen}\ \emph {et~al.}(2020)\citenamefont
  {Stephen}, \citenamefont {Garre-Rubio}, \citenamefont {Dua},\ and\
  \citenamefont {Williamson}}]{StephenGarre-RubioDuaWilliamson2020}%
  \BibitemOpen
  \bibfield  {author} {\bibinfo {author} {\bibfnamefont {D.~T.}\ \bibnamefont
  {Stephen}}, \bibinfo {author} {\bibfnamefont {J.}~\bibnamefont
  {Garre-Rubio}}, \bibinfo {author} {\bibfnamefont {A.}~\bibnamefont {Dua}},\
  and\ \bibinfo {author} {\bibfnamefont {D.~J.}\ \bibnamefont {Williamson}},\
  }\bibfield  {title} {\bibinfo {title} {Subsystem symmetry enriched
  topological order in three dimensions},\ }\href
  {https://doi.org/10.1103/PhysRevResearch.2.033331} {\bibfield  {journal}
  {\bibinfo  {journal} {Phys. Rev. Research}\ }\textbf {\bibinfo {volume}
  {2}},\ \bibinfo {pages} {033331} (\bibinfo {year} {2020})}\BibitemShut
  {NoStop}%
\bibitem [{\citenamefont {Tantivasadakarn}(2020)}]{Tantivasadakarn20}%
  \BibitemOpen
  \bibfield  {author} {\bibinfo {author} {\bibfnamefont {N.}~\bibnamefont
  {Tantivasadakarn}},\ }\bibfield  {title} {\bibinfo {title} {Jordan-wigner
  dualities for translation-invariant hamiltonians in any dimension: Emergent
  fermions in fracton topological order},\ }\href
  {https://doi.org/10.1103/PhysRevResearch.2.023353} {\bibfield  {journal}
  {\bibinfo  {journal} {Phys. Rev. Research}\ }\textbf {\bibinfo {volume}
  {2}},\ \bibinfo {pages} {023353} (\bibinfo {year} {2020})}\BibitemShut
  {NoStop}%
\bibitem [{\citenamefont {Shirley}(2020)}]{Shirley20}%
  \BibitemOpen
  \bibfield  {author} {\bibinfo {author} {\bibfnamefont {W.}~\bibnamefont
  {Shirley}},\ }\bibfield  {title} {\bibinfo {title} {Fractonic order and
  emergent fermionic gauge theory},\ }\href {https://arxiv.org/abs/2002.12026}
  {\bibfield  {journal} {\bibinfo  {journal} {arXiv preprint arXiv:2002.12026}\
  } (\bibinfo {year} {2020})}\BibitemShut {NoStop}%
\bibitem [{\citenamefont {Han}\ \emph {et~al.}(2023)\citenamefont {Han},
  \citenamefont {Lake}, \citenamefont {Lam}, \citenamefont {Verresen},\ and\
  \citenamefont {You}}]{Han23}%
  \BibitemOpen
  \bibfield  {author} {\bibinfo {author} {\bibfnamefont {J.~H.}\ \bibnamefont
  {Han}}, \bibinfo {author} {\bibfnamefont {E.}~\bibnamefont {Lake}}, \bibinfo
  {author} {\bibfnamefont {H.~T.}\ \bibnamefont {Lam}}, \bibinfo {author}
  {\bibfnamefont {R.}~\bibnamefont {Verresen}},\ and\ \bibinfo {author}
  {\bibfnamefont {Y.}~\bibnamefont {You}},\ }\bibfield  {title} {\bibinfo
  {title} {Topological quantum chains protected by dipolar and other modulated
  symmetries},\ }\href {https://arxiv.org/abs/2309.10036} {\bibfield  {journal}
  {\bibinfo  {journal} {arXiv preprint arXiv:2309.10036}\ } (\bibinfo {year}
  {2023})}\BibitemShut {NoStop}%
\bibitem [{\citenamefont {Kapustin}\ and\ \citenamefont
  {Thorngren}(2017{\natexlab{b}})}]{KapustinThorngren15}%
  \BibitemOpen
  \bibfield  {author} {\bibinfo {author} {\bibfnamefont {A.}~\bibnamefont
  {Kapustin}}\ and\ \bibinfo {author} {\bibfnamefont {R.}~\bibnamefont
  {Thorngren}},\ }\bibinfo {title} {Higher symmetry and gapped phases of gauge
  theories},\ in\ \href {https://doi.org/10.1007/978-3-319-59939-7_5} {\emph
  {\bibinfo {booktitle} {Algebra, Geometry, and Physics in the 21st Century:
  Kontsevich Festschrift}}},\ \bibinfo {editor} {edited by\ \bibinfo {editor}
  {\bibfnamefont {D.}~\bibnamefont {Auroux}}, \bibinfo {editor} {\bibfnamefont
  {L.}~\bibnamefont {Katzarkov}}, \bibinfo {editor} {\bibfnamefont
  {T.}~\bibnamefont {Pantev}}, \bibinfo {editor} {\bibfnamefont
  {Y.}~\bibnamefont {Soibelman}},\ and\ \bibinfo {editor} {\bibfnamefont
  {Y.}~\bibnamefont {Tschinkel}}}\ (\bibinfo  {publisher} {Springer
  International Publishing},\ \bibinfo {address} {Cham},\ \bibinfo {year}
  {2017})\ pp.\ \bibinfo {pages} {177--202}\BibitemShut {NoStop}%
\bibitem [{\citenamefont {Delcamp}\ and\ \citenamefont
  {Tiwari}(2018)}]{DelcampTiwari18}%
  \BibitemOpen
  \bibfield  {author} {\bibinfo {author} {\bibfnamefont {C.}~\bibnamefont
  {Delcamp}}\ and\ \bibinfo {author} {\bibfnamefont {A.}~\bibnamefont
  {Tiwari}},\ }\bibfield  {title} {\bibinfo {title} {From gauge to higher gauge
  models of topological phases},\ }\href
  {https://doi.org/10.1007/JHEP10(2018)049} {\bibfield  {journal} {\bibinfo
  {journal} {Journal of High Energy Physics}\ }\textbf {\bibinfo {volume}
  {2018}},\ \bibinfo {pages} {49} (\bibinfo {year} {2018})}\BibitemShut
  {NoStop}%
\bibitem [{\citenamefont {Chen}\ \emph {et~al.}(2021)\citenamefont {Chen},
  \citenamefont {Ellison},\ and\ \citenamefont {Tantivasadakarn}}]{Chen21}%
  \BibitemOpen
  \bibfield  {author} {\bibinfo {author} {\bibfnamefont {Y.-A.}\ \bibnamefont
  {Chen}}, \bibinfo {author} {\bibfnamefont {T.~D.}\ \bibnamefont {Ellison}},\
  and\ \bibinfo {author} {\bibfnamefont {N.}~\bibnamefont {Tantivasadakarn}},\
  }\bibfield  {title} {\bibinfo {title} {Disentangling supercohomology
  symmetry-protected topological phases in three spatial dimensions},\ }\href
  {https://doi.org/10.1103/PhysRevResearch.3.013056} {\bibfield  {journal}
  {\bibinfo  {journal} {Phys. Rev. Res.}\ }\textbf {\bibinfo {volume} {3}},\
  \bibinfo {pages} {013056} (\bibinfo {year} {2021})}\BibitemShut {NoStop}%
\bibitem [{\citenamefont {Bhardwaj}\ and\ \citenamefont
  {Tachikawa}(2018)}]{bhardwaj_finite_2018}%
  \BibitemOpen
  \bibfield  {author} {\bibinfo {author} {\bibfnamefont {L.}~\bibnamefont
  {Bhardwaj}}\ and\ \bibinfo {author} {\bibfnamefont {Y.}~\bibnamefont
  {Tachikawa}},\ }\bibfield  {title} {\bibinfo {title} {On finite symmetries
  and their gauging in two dimensions},\ }\href
  {https://doi.org/10.1007/JHEP03(2018)189} {\bibfield  {journal} {\bibinfo
  {journal} {Journal of High Energy Physics}\ }\textbf {\bibinfo {volume}
  {2018}},\ \bibinfo {pages} {189} (\bibinfo {year} {2018})}\BibitemShut
  {NoStop}%
\bibitem [{\citenamefont {Ostrik}(2003{\natexlab{a}})}]{ostrik_module_2003}%
  \BibitemOpen
  \bibfield  {author} {\bibinfo {author} {\bibfnamefont {V.}~\bibnamefont
  {Ostrik}},\ }\bibfield  {title} {\bibinfo {title} {Module categories, weak
  {{Hopf}} algebras and modular invariants},\ }\href
  {https://doi.org/10.1007/s00031-003-0515-6} {\bibfield  {journal} {\bibinfo
  {journal} {Transformation Groups}\ }\textbf {\bibinfo {volume} {8}},\
  \bibinfo {pages} {177} (\bibinfo {year} {2003}{\natexlab{a}})}\BibitemShut
  {NoStop}%
\bibitem [{\citenamefont {Gottesman}()}]{gottesman_stabilizer_nodate}%
  \BibitemOpen
  \bibfield  {author} {\bibinfo {author} {\bibfnamefont {D.}~\bibnamefont
  {Gottesman}},\ }\emph {\bibinfo {title} {Stabilizer Codes and Quantum Error
  Correction}},\ \href {https://doi.org/10.7907/rzr7-dt72} {Ph.D. thesis},\
  \bibinfo  {school} {California Institute of Technology}, \bibinfo {address}
  {{United States \textendash{} California}}\BibitemShut {NoStop}%
\bibitem [{\citenamefont {Williamson}\ \emph {et~al.}(2017)\citenamefont
  {Williamson}, \citenamefont {Bultinck},\ and\ \citenamefont
  {Verstraete}}]{williamson2017symmetry}%
  \BibitemOpen
  \bibfield  {author} {\bibinfo {author} {\bibfnamefont {D.~J.}\ \bibnamefont
  {Williamson}}, \bibinfo {author} {\bibfnamefont {N.}~\bibnamefont
  {Bultinck}},\ and\ \bibinfo {author} {\bibfnamefont {F.}~\bibnamefont
  {Verstraete}},\ }\bibfield  {title} {\bibinfo {title} {Symmetry-enriched
  topological order in tensor networks: Defects, gauging and anyon
  condensation},\ }\href {https://arxiv.org/abs/1711.07982} {\bibfield
  {journal} {\bibinfo  {journal} {arXiv preprint arXiv:1711.07982}\ } (\bibinfo
  {year} {2017})}\BibitemShut {NoStop}%
\bibitem [{\citenamefont {Lootens}\ \emph {et~al.}(2021)\citenamefont
  {Lootens}, \citenamefont {Fuchs}, \citenamefont {Haegeman}, \citenamefont
  {Schweigert},\ and\ \citenamefont {Verstraete}}]{lootens_matrix_2021}%
  \BibitemOpen
  \bibfield  {author} {\bibinfo {author} {\bibfnamefont {L.}~\bibnamefont
  {Lootens}}, \bibinfo {author} {\bibfnamefont {J.}~\bibnamefont {Fuchs}},
  \bibinfo {author} {\bibfnamefont {J.}~\bibnamefont {Haegeman}}, \bibinfo
  {author} {\bibfnamefont {C.}~\bibnamefont {Schweigert}},\ and\ \bibinfo
  {author} {\bibfnamefont {F.}~\bibnamefont {Verstraete}},\ }\bibfield  {title}
  {\bibinfo {title} {Matrix product operator symmetries and intertwiners in
  string-nets with domain walls},\ }\href
  {https://doi.org/10.21468/SciPostPhys.10.3.053} {\bibfield  {journal}
  {\bibinfo  {journal} {SciPost Physics}\ }\textbf {\bibinfo {volume} {10}},\
  \bibinfo {pages} {053} (\bibinfo {year} {2021})}\BibitemShut {NoStop}%
\bibitem [{\citenamefont {Molnar}\ \emph {et~al.}(2022)\citenamefont {Molnar},
  \citenamefont {de~Alarcon}, \citenamefont {Garre-Rubio}, \citenamefont
  {Schuch}, \citenamefont {Cirac},\ and\ \citenamefont
  {Perez-Garcia}}]{molnar_matrix_2022}%
  \BibitemOpen
  \bibfield  {author} {\bibinfo {author} {\bibfnamefont {A.}~\bibnamefont
  {Molnar}}, \bibinfo {author} {\bibfnamefont {A.~R.}\ \bibnamefont
  {de~Alarcon}}, \bibinfo {author} {\bibfnamefont {J.}~\bibnamefont
  {Garre-Rubio}}, \bibinfo {author} {\bibfnamefont {N.}~\bibnamefont {Schuch}},
  \bibinfo {author} {\bibfnamefont {J.~I.}\ \bibnamefont {Cirac}},\ and\
  \bibinfo {author} {\bibfnamefont {D.}~\bibnamefont {Perez-Garcia}},\
  }\bibfield  {title} {\bibinfo {title} {Matrix product operator algebras i:
  representations of weak hopf algebras and projected entangled pair states},\
  }\href {https://arxiv.org/abs/2204.05940} {\bibfield  {journal} {\bibinfo
  {journal} {arXiv preprint arXiv:2204.05940}\ } (\bibinfo {year}
  {2022})}\BibitemShut {NoStop}%
\bibitem [{\citenamefont {Kogut}\ and\ \citenamefont
  {Susskind}(1975)}]{kogut_hamiltonian_1975}%
  \BibitemOpen
  \bibfield  {author} {\bibinfo {author} {\bibfnamefont {J.}~\bibnamefont
  {Kogut}}\ and\ \bibinfo {author} {\bibfnamefont {L.}~\bibnamefont
  {Susskind}},\ }\bibfield  {title} {\bibinfo {title} {Hamiltonian formulation
  of {{Wilson}}'s lattice gauge theories},\ }\href
  {https://doi.org/10.1103/PhysRevD.11.395} {\bibfield  {journal} {\bibinfo
  {journal} {Physical Review D}\ }\textbf {\bibinfo {volume} {11}},\ \bibinfo
  {pages} {395} (\bibinfo {year} {1975})}\BibitemShut {NoStop}%
\bibitem [{\citenamefont {Kogut}(1979)}]{kogut_introduction_1979}%
  \BibitemOpen
  \bibfield  {author} {\bibinfo {author} {\bibfnamefont {J.~B.}\ \bibnamefont
  {Kogut}},\ }\bibfield  {title} {\bibinfo {title} {An introduction to lattice
  gauge theory and spin systems},\ }\href
  {https://doi.org/10.1103/RevModPhys.51.659} {\bibfield  {journal} {\bibinfo
  {journal} {Reviews of Modern Physics}\ }\textbf {\bibinfo {volume} {51}},\
  \bibinfo {pages} {659} (\bibinfo {year} {1979})}\BibitemShut {NoStop}%
\bibitem [{\citenamefont {Harlow}\ and\ \citenamefont
  {Ooguri}(2021)}]{harlow2021symmetries}%
  \BibitemOpen
  \bibfield  {author} {\bibinfo {author} {\bibfnamefont {D.}~\bibnamefont
  {Harlow}}\ and\ \bibinfo {author} {\bibfnamefont {H.}~\bibnamefont
  {Ooguri}},\ }\bibfield  {title} {\bibinfo {title} {Symmetries in quantum
  field theory and quantum gravity},\ }\href
  {https://doi.org/10.1007/s00220-021-04040-y} {\bibfield  {journal} {\bibinfo
  {journal} {Communications in Mathematical Physics}\ }\textbf {\bibinfo
  {volume} {383}},\ \bibinfo {pages} {1669} (\bibinfo {year}
  {2021})}\BibitemShut {NoStop}%
\bibitem [{\citenamefont {Kitaev}(2003)}]{kitaev_fault-tolerant_2003-2}%
  \BibitemOpen
  \bibfield  {author} {\bibinfo {author} {\bibfnamefont {A.~Y.}\ \bibnamefont
  {Kitaev}},\ }\bibfield  {title} {\bibinfo {title} {Fault-tolerant quantum
  computation by anyons},\ }\href
  {https://doi.org/10.1016/S0003-4916(02)00018-0} {\bibfield  {journal}
  {\bibinfo  {journal} {Annals of Physics}\ }\textbf {\bibinfo {volume}
  {303}},\ \bibinfo {pages} {2} (\bibinfo {year} {2003})}\BibitemShut {NoStop}%
\bibitem [{\citenamefont {Dijkgraaf}\ and\ \citenamefont
  {Witten}(1990)}]{dijkgraaf1990topological}%
  \BibitemOpen
  \bibfield  {author} {\bibinfo {author} {\bibfnamefont {R.}~\bibnamefont
  {Dijkgraaf}}\ and\ \bibinfo {author} {\bibfnamefont {E.}~\bibnamefont
  {Witten}},\ }\bibfield  {title} {\bibinfo {title} {Topological gauge theories
  and group cohomology},\ }\href {https://doi.org/10.1007/BF02096988}
  {\bibfield  {journal} {\bibinfo  {journal} {Communications in Mathematical
  Physics}\ }\textbf {\bibinfo {volume} {129}},\ \bibinfo {pages} {393}
  (\bibinfo {year} {1990})}\BibitemShut {NoStop}%
\bibitem [{\citenamefont {Chen}\ \emph {et~al.}(2013)\citenamefont {Chen},
  \citenamefont {Gu}, \citenamefont {Liu},\ and\ \citenamefont
  {Wen}}]{chen_symmetry_2013-1}%
  \BibitemOpen
  \bibfield  {author} {\bibinfo {author} {\bibfnamefont {X.}~\bibnamefont
  {Chen}}, \bibinfo {author} {\bibfnamefont {Z.-C.}\ \bibnamefont {Gu}},
  \bibinfo {author} {\bibfnamefont {Z.-X.}\ \bibnamefont {Liu}},\ and\ \bibinfo
  {author} {\bibfnamefont {X.-G.}\ \bibnamefont {Wen}},\ }\bibfield  {title}
  {\bibinfo {title} {Symmetry protected topological orders and the group
  cohomology of their symmetry group},\ }\href
  {https://doi.org/10.1103/PhysRevB.87.155114} {\bibfield  {journal} {\bibinfo
  {journal} {Physical Review B}\ }\textbf {\bibinfo {volume} {87}},\ \bibinfo
  {pages} {155114} (\bibinfo {year} {2013})}\BibitemShut {NoStop}%
\bibitem [{\citenamefont {Gu}\ and\ \citenamefont {Wen}(2014)}]{GuWen14}%
  \BibitemOpen
  \bibfield  {author} {\bibinfo {author} {\bibfnamefont {Z.-C.}\ \bibnamefont
  {Gu}}\ and\ \bibinfo {author} {\bibfnamefont {X.-G.}\ \bibnamefont {Wen}},\
  }\bibfield  {title} {\bibinfo {title} {Symmetry-protected topological orders
  for interacting fermions: Fermionic topological nonlinear
  $\ensuremath{\sigma}$ models and a special group supercohomology theory},\
  }\href {https://doi.org/10.1103/PhysRevB.90.115141} {\bibfield  {journal}
  {\bibinfo  {journal} {Phys. Rev. B}\ }\textbf {\bibinfo {volume} {90}},\
  \bibinfo {pages} {115141} (\bibinfo {year} {2014})}\BibitemShut {NoStop}%
\bibitem [{\citenamefont {Cheng}\ \emph {et~al.}(2018)\citenamefont {Cheng},
  \citenamefont {Tantivasadakarn},\ and\ \citenamefont {Wang}}]{Cheng18}%
  \BibitemOpen
  \bibfield  {author} {\bibinfo {author} {\bibfnamefont {M.}~\bibnamefont
  {Cheng}}, \bibinfo {author} {\bibfnamefont {N.}~\bibnamefont
  {Tantivasadakarn}},\ and\ \bibinfo {author} {\bibfnamefont {C.}~\bibnamefont
  {Wang}},\ }\bibfield  {title} {\bibinfo {title} {Loop braiding statistics and
  interacting fermionic symmetry-protected topological phases in three
  dimensions},\ }\href {https://doi.org/10.1103/PhysRevX.8.011054} {\bibfield
  {journal} {\bibinfo  {journal} {Phys. Rev. X}\ }\textbf {\bibinfo {volume}
  {8}},\ \bibinfo {pages} {011054} (\bibinfo {year} {2018})}\BibitemShut
  {NoStop}%
\bibitem [{\citenamefont
  {Tantivasadakarn}(2017)}]{tantivasadakarn2017dimensional}%
  \BibitemOpen
  \bibfield  {author} {\bibinfo {author} {\bibfnamefont {N.}~\bibnamefont
  {Tantivasadakarn}},\ }\bibfield  {title} {\bibinfo {title} {Dimensional
  reduction and topological invariants of symmetry-protected topological
  phases},\ }\href {https://doi.org/10.1103/PhysRevB.96.195101} {\bibfield
  {journal} {\bibinfo  {journal} {Phys. Rev. B}\ }\textbf {\bibinfo {volume}
  {96}},\ \bibinfo {pages} {195101} (\bibinfo {year} {2017})}\BibitemShut
  {NoStop}%
\bibitem [{\citenamefont {Bulmash}\ and\ \citenamefont
  {Barkeshli}(2020)}]{bulmash2020absolute}%
  \BibitemOpen
  \bibfield  {author} {\bibinfo {author} {\bibfnamefont {D.}~\bibnamefont
  {Bulmash}}\ and\ \bibinfo {author} {\bibfnamefont {M.}~\bibnamefont
  {Barkeshli}},\ }\bibfield  {title} {\bibinfo {title} {Absolute anomalies in
  (2+1)d symmetry-enriched topological states and exact (3+1)d constructions},\
  }\href {https://doi.org/10.1103/PhysRevResearch.2.043033} {\bibfield
  {journal} {\bibinfo  {journal} {Physical Review Research}\ }\textbf {\bibinfo
  {volume} {2}},\ \bibinfo {pages} {043033} (\bibinfo {year}
  {2020})}\BibitemShut {NoStop}%
\bibitem [{\citenamefont {Tata}\ \emph {et~al.}(2023)\citenamefont {Tata},
  \citenamefont {Kobayashi}, \citenamefont {Bulmash},\ and\ \citenamefont
  {Barkeshli}}]{Tata23}%
  \BibitemOpen
  \bibfield  {author} {\bibinfo {author} {\bibfnamefont {S.}~\bibnamefont
  {Tata}}, \bibinfo {author} {\bibfnamefont {R.}~\bibnamefont {Kobayashi}},
  \bibinfo {author} {\bibfnamefont {D.}~\bibnamefont {Bulmash}},\ and\ \bibinfo
  {author} {\bibfnamefont {M.}~\bibnamefont {Barkeshli}},\ }\bibfield  {title}
  {\bibinfo {title} {Anomalies in (2+1)d fermionic topological phases and
  (3+1)d path integral state sums for fermionic spts},\ }\href
  {https://doi.org/10.1007/s00220-022-04484-w} {\bibfield  {journal} {\bibinfo
  {journal} {Communications in Mathematical Physics}\ }\textbf {\bibinfo
  {volume} {397}},\ \bibinfo {pages} {199} (\bibinfo {year}
  {2023})}\BibitemShut {NoStop}%
\bibitem [{\citenamefont {Pollmann}\ \emph
  {et~al.}(2010{\natexlab{a}})\citenamefont {Pollmann}, \citenamefont {Turner},
  \citenamefont {Berg},\ and\ \citenamefont {Oshikawa}}]{Pollmann10}%
  \BibitemOpen
  \bibfield  {author} {\bibinfo {author} {\bibfnamefont {F.}~\bibnamefont
  {Pollmann}}, \bibinfo {author} {\bibfnamefont {A.~M.}\ \bibnamefont
  {Turner}}, \bibinfo {author} {\bibfnamefont {E.}~\bibnamefont {Berg}},\ and\
  \bibinfo {author} {\bibfnamefont {M.}~\bibnamefont {Oshikawa}},\ }\bibfield
  {title} {\bibinfo {title} {Entanglement spectrum of a topological phase in
  one dimension},\ }\href {https://doi.org/10.1103/PhysRevB.81.064439}
  {\bibfield  {journal} {\bibinfo  {journal} {Phys. Rev. B}\ }\textbf {\bibinfo
  {volume} {81}},\ \bibinfo {pages} {064439} (\bibinfo {year}
  {2010}{\natexlab{a}})}\BibitemShut {NoStop}%
\bibitem [{\citenamefont {den Nijs}\ and\ \citenamefont
  {Rommelse}(1989)}]{den_nijs_preroughening_1989}%
  \BibitemOpen
  \bibfield  {author} {\bibinfo {author} {\bibfnamefont {M.}~\bibnamefont {den
  Nijs}}\ and\ \bibinfo {author} {\bibfnamefont {K.}~\bibnamefont {Rommelse}},\
  }\bibfield  {title} {\bibinfo {title} {Preroughening transitions in crystal
  surfaces and valence-bond phases in quantum spin chains},\ }\href
  {https://doi.org/10.1103/PhysRevB.40.4709} {\bibfield  {journal} {\bibinfo
  {journal} {Phys. Rev. B}\ }\textbf {\bibinfo {volume} {40}},\ \bibinfo
  {pages} {4709} (\bibinfo {year} {1989})}\BibitemShut {NoStop}%
\bibitem [{\citenamefont {Kennedy}\ and\ \citenamefont
  {Tasaki}(1992)}]{kennedy_hidden_1992}%
  \BibitemOpen
  \bibfield  {author} {\bibinfo {author} {\bibfnamefont {T.}~\bibnamefont
  {Kennedy}}\ and\ \bibinfo {author} {\bibfnamefont {H.}~\bibnamefont
  {Tasaki}},\ }\bibfield  {title} {\bibinfo {title} {Hidden
  ${\mathrm{z}}_{2}$\ifmmode\times\else\texttimes\fi{}${\mathrm{z}}_{2}$
  symmetry breaking in haldane-gap antiferromagnets},\ }\href
  {https://doi.org/10.1103/PhysRevB.45.304} {\bibfield  {journal} {\bibinfo
  {journal} {Phys. Rev. B}\ }\textbf {\bibinfo {volume} {45}},\ \bibinfo
  {pages} {304} (\bibinfo {year} {1992})}\BibitemShut {NoStop}%
\bibitem [{\citenamefont {Pollmann}\ and\ \citenamefont
  {Turner}(2012)}]{pollmann_detection_2012}%
  \BibitemOpen
  \bibfield  {author} {\bibinfo {author} {\bibfnamefont {F.}~\bibnamefont
  {Pollmann}}\ and\ \bibinfo {author} {\bibfnamefont {A.~M.}\ \bibnamefont
  {Turner}},\ }\bibfield  {title} {\bibinfo {title} {Detection of
  symmetry-protected topological phases in one dimension},\ }\href
  {https://doi.org/10.1103/PhysRevB.86.125441} {\bibfield  {journal} {\bibinfo
  {journal} {Phys. Rev. B}\ }\textbf {\bibinfo {volume} {86}},\ \bibinfo
  {pages} {125441} (\bibinfo {year} {2012})}\BibitemShut {NoStop}%
\bibitem [{\citenamefont {Else}\ \emph {et~al.}(2013)\citenamefont {Else},
  \citenamefont {Bartlett},\ and\ \citenamefont {Doherty}}]{Else13}%
  \BibitemOpen
  \bibfield  {author} {\bibinfo {author} {\bibfnamefont {D.~V.}\ \bibnamefont
  {Else}}, \bibinfo {author} {\bibfnamefont {S.~D.}\ \bibnamefont {Bartlett}},\
  and\ \bibinfo {author} {\bibfnamefont {A.~C.}\ \bibnamefont {Doherty}},\
  }\bibfield  {title} {\bibinfo {title} {Hidden symmetry-breaking picture of
  symmetry-protected topological order},\ }\href
  {https://doi.org/10.1103/PhysRevB.88.085114} {\bibfield  {journal} {\bibinfo
  {journal} {Phys. Rev. B}\ }\textbf {\bibinfo {volume} {88}},\ \bibinfo
  {pages} {085114} (\bibinfo {year} {2013})}\BibitemShut {NoStop}%
\bibitem [{\citenamefont {Bahri}\ and\ \citenamefont
  {Vishwanath}(2014)}]{Bahri14}%
  \BibitemOpen
  \bibfield  {author} {\bibinfo {author} {\bibfnamefont {Y.}~\bibnamefont
  {Bahri}}\ and\ \bibinfo {author} {\bibfnamefont {A.}~\bibnamefont
  {Vishwanath}},\ }\bibfield  {title} {\bibinfo {title} {Detecting majorana
  fermions in quasi-one-dimensional topological phases using nonlocal order
  parameters},\ }\href {https://doi.org/10.1103/PhysRevB.89.155135} {\bibfield
  {journal} {\bibinfo  {journal} {Phys. Rev. B}\ }\textbf {\bibinfo {volume}
  {89}},\ \bibinfo {pages} {155135} (\bibinfo {year} {2014})}\BibitemShut
  {NoStop}%
\bibitem [{\citenamefont {Hatsugai}(1993)}]{hatsugai_chern_1993}%
  \BibitemOpen
  \bibfield  {author} {\bibinfo {author} {\bibfnamefont {Y.}~\bibnamefont
  {Hatsugai}},\ }\bibfield  {title} {\bibinfo {title} {Chern number and edge
  states in the integer quantum {{Hall}} effect},\ }\href
  {https://doi.org/10.1103/PhysRevLett.71.3697} {\bibfield  {journal} {\bibinfo
   {journal} {Physical Review Letters}\ }\textbf {\bibinfo {volume} {71}},\
  \bibinfo {pages} {3697} (\bibinfo {year} {1993})}\BibitemShut {NoStop}%
\bibitem [{\citenamefont {Senthil}\ and\ \citenamefont
  {Levin}(2013)}]{senthil_integer_2013}%
  \BibitemOpen
  \bibfield  {author} {\bibinfo {author} {\bibfnamefont {T.}~\bibnamefont
  {Senthil}}\ and\ \bibinfo {author} {\bibfnamefont {M.}~\bibnamefont
  {Levin}},\ }\bibfield  {title} {\bibinfo {title} {Integer {{Quantum Hall
  Effect}} for {{Bosons}}},\ }\href
  {https://doi.org/10.1103/PhysRevLett.110.046801} {\bibfield  {journal}
  {\bibinfo  {journal} {Physical Review Letters}\ }\textbf {\bibinfo {volume}
  {110}},\ \bibinfo {pages} {046801} (\bibinfo {year} {2013})}\BibitemShut
  {NoStop}%
\bibitem [{\citenamefont {Cheng}\ and\ \citenamefont
  {Gu}(2014)}]{cheng_topological_2014}%
  \BibitemOpen
  \bibfield  {author} {\bibinfo {author} {\bibfnamefont {M.}~\bibnamefont
  {Cheng}}\ and\ \bibinfo {author} {\bibfnamefont {Z.-C.}\ \bibnamefont {Gu}},\
  }\bibfield  {title} {\bibinfo {title} {Topological {{Response Theory}} of
  {{Abelian Symmetry-Protected Topological Phases}} in {{Two Dimensions}}},\
  }\href {https://doi.org/10.1103/PhysRevLett.112.141602} {\bibfield  {journal}
  {\bibinfo  {journal} {Physical Review Letters}\ }\textbf {\bibinfo {volume}
  {112}},\ \bibinfo {pages} {141602} (\bibinfo {year} {2014})}\BibitemShut
  {NoStop}%
\bibitem [{\citenamefont {Zaletel}\ \emph {et~al.}(2014)\citenamefont
  {Zaletel}, \citenamefont {Mong},\ and\ \citenamefont
  {Pollmann}}]{zaletel_flux_2014}%
  \BibitemOpen
  \bibfield  {author} {\bibinfo {author} {\bibfnamefont {M.~P.}\ \bibnamefont
  {Zaletel}}, \bibinfo {author} {\bibfnamefont {R.~S.~K.}\ \bibnamefont
  {Mong}},\ and\ \bibinfo {author} {\bibfnamefont {F.}~\bibnamefont
  {Pollmann}},\ }\bibfield  {title} {\bibinfo {title} {Flux insertion,
  entanglement, and quantized responses},\ }\href
  {https://doi.org/10.1088/1742-5468/2014/10/P10007} {\bibfield  {journal}
  {\bibinfo  {journal} {Journal of Statistical Mechanics: Theory and
  Experiment}\ }\textbf {\bibinfo {volume} {2014}},\ \bibinfo {pages} {P10007}
  (\bibinfo {year} {2014})}\BibitemShut {NoStop}%
\bibitem [{\citenamefont {Gross}\ and\ \citenamefont
  {Eisert}(2007)}]{gross2007novel}%
  \BibitemOpen
  \bibfield  {author} {\bibinfo {author} {\bibfnamefont {D.}~\bibnamefont
  {Gross}}\ and\ \bibinfo {author} {\bibfnamefont {J.}~\bibnamefont {Eisert}},\
  }\bibfield  {title} {\bibinfo {title} {Novel schemes for measurement-based
  quantum computation},\ }\href {https://doi.org/10.1103/PhysRevLett.98.220503}
  {\bibfield  {journal} {\bibinfo  {journal} {Phys. Rev. Lett.}\ }\textbf
  {\bibinfo {volume} {98}},\ \bibinfo {pages} {220503} (\bibinfo {year}
  {2007})}\BibitemShut {NoStop}%
\bibitem [{\citenamefont {Gross}\ \emph {et~al.}(2007)\citenamefont {Gross},
  \citenamefont {Eisert}, \citenamefont {Schuch},\ and\ \citenamefont
  {{Perez-Garcia}}}]{gross_measurement-based_2007}%
  \BibitemOpen
  \bibfield  {author} {\bibinfo {author} {\bibfnamefont {D.}~\bibnamefont
  {Gross}}, \bibinfo {author} {\bibfnamefont {J.}~\bibnamefont {Eisert}},
  \bibinfo {author} {\bibfnamefont {N.}~\bibnamefont {Schuch}},\ and\ \bibinfo
  {author} {\bibfnamefont {D.}~\bibnamefont {{Perez-Garcia}}},\ }\bibfield
  {title} {\bibinfo {title} {Measurement-based quantum computation beyond the
  one-way model},\ }\href {https://doi.org/10.1103/PhysRevA.76.052315}
  {\bibfield  {journal} {\bibinfo  {journal} {Physical Review A}\ }\textbf
  {\bibinfo {volume} {76}},\ \bibinfo {pages} {052315} (\bibinfo {year}
  {2007})}\BibitemShut {NoStop}%
\bibitem [{\citenamefont {Briegel}\ \emph {et~al.}(2009)\citenamefont
  {Briegel}, \citenamefont {Browne}, \citenamefont {D{\"u}r}, \citenamefont
  {Raussendorf},\ and\ \citenamefont {{Van den
  Nest}}}]{briegel_measurement-based_2009}%
  \BibitemOpen
  \bibfield  {author} {\bibinfo {author} {\bibfnamefont {H.~J.}\ \bibnamefont
  {Briegel}}, \bibinfo {author} {\bibfnamefont {D.~E.}\ \bibnamefont {Browne}},
  \bibinfo {author} {\bibfnamefont {W.}~\bibnamefont {D{\"u}r}}, \bibinfo
  {author} {\bibfnamefont {R.}~\bibnamefont {Raussendorf}},\ and\ \bibinfo
  {author} {\bibfnamefont {M.}~\bibnamefont {{Van den Nest}}},\ }\bibfield
  {title} {\bibinfo {title} {Measurement-based quantum computation},\ }\href
  {https://doi.org/10.1038/nphys1157} {\bibfield  {journal} {\bibinfo
  {journal} {Nature Physics}\ }\textbf {\bibinfo {volume} {5}},\ \bibinfo
  {pages} {19} (\bibinfo {year} {2009})}\BibitemShut {NoStop}%
\bibitem [{\citenamefont {Raussendorf}\ \emph {et~al.}(2003)\citenamefont
  {Raussendorf}, \citenamefont {Browne},\ and\ \citenamefont
  {Briegel}}]{raussendorf_measurement-based_2003}%
  \BibitemOpen
  \bibfield  {author} {\bibinfo {author} {\bibfnamefont {R.}~\bibnamefont
  {Raussendorf}}, \bibinfo {author} {\bibfnamefont {D.~E.}\ \bibnamefont
  {Browne}},\ and\ \bibinfo {author} {\bibfnamefont {H.~J.}\ \bibnamefont
  {Briegel}},\ }\bibfield  {title} {\bibinfo {title} {Measurement-based quantum
  computation on cluster states},\ }\href
  {https://doi.org/10.1103/PhysRevA.68.022312} {\bibfield  {journal} {\bibinfo
  {journal} {Physical Review A}\ }\textbf {\bibinfo {volume} {68}},\ \bibinfo
  {pages} {022312} (\bibinfo {year} {2003})}\BibitemShut {NoStop}%
\bibitem [{\citenamefont {Raussendorf}\ and\ \citenamefont
  {Briegel}(2001)}]{raussendorf_one-way_2001}%
  \BibitemOpen
  \bibfield  {author} {\bibinfo {author} {\bibfnamefont {R.}~\bibnamefont
  {Raussendorf}}\ and\ \bibinfo {author} {\bibfnamefont {H.~J.}\ \bibnamefont
  {Briegel}},\ }\bibfield  {title} {\bibinfo {title} {A {{One-Way Quantum
  Computer}}},\ }\href {https://doi.org/10.1103/PhysRevLett.86.5188} {\bibfield
   {journal} {\bibinfo  {journal} {Physical Review Letters}\ }\textbf {\bibinfo
  {volume} {86}},\ \bibinfo {pages} {5188} (\bibinfo {year}
  {2001})}\BibitemShut {NoStop}%
\bibitem [{\citenamefont {Stephen}\ \emph {et~al.}(2017)\citenamefont
  {Stephen}, \citenamefont {Wang}, \citenamefont {Prakash}, \citenamefont
  {Wei},\ and\ \citenamefont {Raussendorf}}]{stephen_computational_2017}%
  \BibitemOpen
  \bibfield  {author} {\bibinfo {author} {\bibfnamefont {D.~T.}\ \bibnamefont
  {Stephen}}, \bibinfo {author} {\bibfnamefont {D.-S.}\ \bibnamefont {Wang}},
  \bibinfo {author} {\bibfnamefont {A.}~\bibnamefont {Prakash}}, \bibinfo
  {author} {\bibfnamefont {T.-C.}\ \bibnamefont {Wei}},\ and\ \bibinfo {author}
  {\bibfnamefont {R.}~\bibnamefont {Raussendorf}},\ }\bibfield  {title}
  {\bibinfo {title} {Computational {{Power}} of {{Symmetry-Protected
  Topological Phases}}},\ }\href
  {https://doi.org/10.1103/PhysRevLett.119.010504} {\bibfield  {journal}
  {\bibinfo  {journal} {Physical Review Letters}\ }\textbf {\bibinfo {volume}
  {119}},\ \bibinfo {pages} {010504} (\bibinfo {year} {2017})}\BibitemShut
  {NoStop}%
\bibitem [{\citenamefont {Wei}\ and\ \citenamefont
  {Huang}(2017)}]{wei_universal_2017-2}%
  \BibitemOpen
  \bibfield  {author} {\bibinfo {author} {\bibfnamefont {T.-C.}\ \bibnamefont
  {Wei}}\ and\ \bibinfo {author} {\bibfnamefont {C.-Y.}\ \bibnamefont
  {Huang}},\ }\bibfield  {title} {\bibinfo {title} {Universal measurement-based
  quantum computation in two-dimensional {{SPT}} phases},\ }\href
  {https://doi.org/10.1103/PhysRevA.96.032317} {\bibfield  {journal} {\bibinfo
  {journal} {Physical Review A}\ }\textbf {\bibinfo {volume} {96}},\ \bibinfo
  {pages} {032317} (\bibinfo {year} {2017})},\ \Eprint
  {https://arxiv.org/abs/1705.06833} {arXiv:1705.06833} \BibitemShut {NoStop}%
\bibitem [{\citenamefont {Else}\ \emph {et~al.}(2012)\citenamefont {Else},
  \citenamefont {Schwarz}, \citenamefont {Bartlett},\ and\ \citenamefont
  {Doherty}}]{else_symmetry-protected_2012-2}%
  \BibitemOpen
  \bibfield  {author} {\bibinfo {author} {\bibfnamefont {D.~V.}\ \bibnamefont
  {Else}}, \bibinfo {author} {\bibfnamefont {I.}~\bibnamefont {Schwarz}},
  \bibinfo {author} {\bibfnamefont {S.~D.}\ \bibnamefont {Bartlett}},\ and\
  \bibinfo {author} {\bibfnamefont {A.~C.}\ \bibnamefont {Doherty}},\
  }\bibfield  {title} {\bibinfo {title} {Symmetry-protected phases for
  measurement-based quantum computation},\ }\href
  {https://doi.org/10.1103/PhysRevLett.108.240505} {\bibfield  {journal}
  {\bibinfo  {journal} {Physical Review Letters}\ }\textbf {\bibinfo {volume}
  {108}},\ \bibinfo {pages} {240505} (\bibinfo {year} {2012})},\ \Eprint
  {https://arxiv.org/abs/1201.4877} {arXiv:1201.4877} \BibitemShut {NoStop}%
\bibitem [{\citenamefont {Arovas}(2023)}]{ArovasGroupTheory}%
  \BibitemOpen
  \bibfield  {author} {\bibinfo {author} {\bibfnamefont {D.}~\bibnamefont
  {Arovas}},\ }\href
  {https://courses.physics.ucsd.edu/2016/Spring/physics220/LECTURES/GROUP_THEORY.pdf}
  {\bibinfo {title} {Lecture notes on group theory in physics}} (\bibinfo
  {year} {2023})\BibitemShut {NoStop}%
\bibitem [{\citenamefont {van~de
  Wetering}(2020)}]{vandewetering2020zxcalculus}%
  \BibitemOpen
  \bibfield  {author} {\bibinfo {author} {\bibfnamefont {J.}~\bibnamefont
  {van~de Wetering}},\ }\href@noop {} {\bibinfo {title} {Zx-calculus for the
  working quantum computer scientist}} (\bibinfo {year} {2020}),\ \Eprint
  {https://arxiv.org/abs/2012.13966} {arXiv:2012.13966 [quant-ph]} \BibitemShut
  {NoStop}%
\bibitem [{\citenamefont {Bauer}(2016)}]{bauer2016symmetries}%
  \BibitemOpen
  \bibfield  {author} {\bibinfo {author} {\bibfnamefont {A.}~\bibnamefont
  {Bauer}},\ }\emph {\bibinfo {title} {Symmetries and excitations in the
  quantum double models-a tensor network approach}},\ \href
  {https://schuch.univie.ac.at/fileadmin/user_upload/a_schuch/theses/master_thesis_andreas_bauer.pdf}
  {Master's thesis},\ \bibinfo  {school} {Ludwig Maximilians Universit{\"a}t
  M{\"u}nchen} (\bibinfo {year} {2016})\BibitemShut {NoStop}%
\bibitem [{\citenamefont {Heidenreich}\ \emph
  {et~al.}(2021{\natexlab{b}})\citenamefont {Heidenreich}, \citenamefont
  {McNamara}, \citenamefont {Montero}, \citenamefont {Reece}, \citenamefont
  {Rudelius},\ and\ \citenamefont
  {Valenzuela}}]{heidenreich_non-invertible_2021-1}%
  \BibitemOpen
  \bibfield  {author} {\bibinfo {author} {\bibfnamefont {B.}~\bibnamefont
  {Heidenreich}}, \bibinfo {author} {\bibfnamefont {J.}~\bibnamefont
  {McNamara}}, \bibinfo {author} {\bibfnamefont {M.}~\bibnamefont {Montero}},
  \bibinfo {author} {\bibfnamefont {M.}~\bibnamefont {Reece}}, \bibinfo
  {author} {\bibfnamefont {T.}~\bibnamefont {Rudelius}},\ and\ \bibinfo
  {author} {\bibfnamefont {I.}~\bibnamefont {Valenzuela}},\ }\bibfield  {title}
  {\bibinfo {title} {Non-invertible global symmetries and completeness of the
  spectrum},\ }\href {https://doi.org/10.1007/JHEP09(2021)203} {\bibfield
  {journal} {\bibinfo  {journal} {Journal of High Energy Physics}\ }\textbf
  {\bibinfo {volume} {2021}},\ \bibinfo {pages} {203} (\bibinfo {year}
  {2021}{\natexlab{b}})}\BibitemShut {NoStop}%
\bibitem [{\citenamefont {Bhardwaj}\ \emph
  {et~al.}(2022{\natexlab{b}})\citenamefont {Bhardwaj}, \citenamefont
  {Schafer-Nameki},\ and\ \citenamefont {Wu}}]{bhardwaj_universal_2022}%
  \BibitemOpen
  \bibfield  {author} {\bibinfo {author} {\bibfnamefont {L.}~\bibnamefont
  {Bhardwaj}}, \bibinfo {author} {\bibfnamefont {S.}~\bibnamefont
  {Schafer-Nameki}},\ and\ \bibinfo {author} {\bibfnamefont {J.}~\bibnamefont
  {Wu}},\ }\bibfield  {title} {\bibinfo {title} {Universal non-invertible
  symmetries},\ }\href {https://doi.org/10.1002/prop.202200143} {\bibfield
  {journal} {\bibinfo  {journal} {Fortschritte der Physik}\ }\textbf {\bibinfo
  {volume} {70}},\ \bibinfo {pages} {2200143} (\bibinfo {year}
  {2022}{\natexlab{b}})}\BibitemShut {NoStop}%
\bibitem [{\citenamefont {Lootens}\ \emph {et~al.}(2023)\citenamefont
  {Lootens}, \citenamefont {Delcamp}, \citenamefont {Ortiz},\ and\
  \citenamefont {Verstraete}}]{lootens2023dualities}%
  \BibitemOpen
  \bibfield  {author} {\bibinfo {author} {\bibfnamefont {L.}~\bibnamefont
  {Lootens}}, \bibinfo {author} {\bibfnamefont {C.}~\bibnamefont {Delcamp}},
  \bibinfo {author} {\bibfnamefont {G.}~\bibnamefont {Ortiz}},\ and\ \bibinfo
  {author} {\bibfnamefont {F.}~\bibnamefont {Verstraete}},\ }\bibfield  {title}
  {\bibinfo {title} {Dualities in one-dimensional quantum lattice models:
  Symmetric hamiltonians and matrix product operator intertwiners},\ }\href
  {https://doi.org/10.1103/PRXQuantum.4.020357} {\bibfield  {journal} {\bibinfo
   {journal} {PRX Quantum}\ }\textbf {\bibinfo {volume} {4}},\ \bibinfo {pages}
  {020357} (\bibinfo {year} {2023})}\BibitemShut {NoStop}%
\bibitem [{\citenamefont {Tachikawa}(2020)}]{tachikawa_gauging_2020}%
  \BibitemOpen
  \bibfield  {author} {\bibinfo {author} {\bibfnamefont {Y.}~\bibnamefont
  {Tachikawa}},\ }\bibfield  {title} {\bibinfo {title} {On gauging finite
  subgroups},\ }\href {https://doi.org/10.21468/SciPostPhys.8.1.015} {\bibfield
   {journal} {\bibinfo  {journal} {SciPost Physics}\ }\textbf {\bibinfo
  {volume} {8}},\ \bibinfo {pages} {015} (\bibinfo {year} {2020})}\BibitemShut
  {NoStop}%
\bibitem [{\citenamefont {Tantivasadakarn}\ \emph {et~al.}(2022)\citenamefont
  {Tantivasadakarn}, \citenamefont {Thorngren}, \citenamefont {Vishwanath},\
  and\ \citenamefont {Verresen}}]{tantivasadakarn_long-range_2022}%
  \BibitemOpen
  \bibfield  {author} {\bibinfo {author} {\bibfnamefont {N.}~\bibnamefont
  {Tantivasadakarn}}, \bibinfo {author} {\bibfnamefont {R.}~\bibnamefont
  {Thorngren}}, \bibinfo {author} {\bibfnamefont {A.}~\bibnamefont
  {Vishwanath}},\ and\ \bibinfo {author} {\bibfnamefont {R.}~\bibnamefont
  {Verresen}},\ }\href {https://doi.org/10.48550/arXiv.2112.01519} {\bibinfo
  {title} {Long-range entanglement from measuring symmetry-protected
  topological phases}} (\bibinfo {year} {2022}),\ \Eprint
  {https://arxiv.org/abs/2112.01519} {arXiv:2112.01519 [cond-mat,
  physics:quant-ph]} \BibitemShut {NoStop}%
\bibitem [{\citenamefont {Tasaki}(2018)}]{tasaki2018topological}%
  \BibitemOpen
  \bibfield  {author} {\bibinfo {author} {\bibfnamefont {H.}~\bibnamefont
  {Tasaki}},\ }\bibfield  {title} {\bibinfo {title} {Topological phase
  transition and z 2 index for s= 1 quantum spin chains},\ }\href@noop {}
  {\bibfield  {journal} {\bibinfo  {journal} {Physical Review Letters}\
  }\textbf {\bibinfo {volume} {121}},\ \bibinfo {pages} {140604} (\bibinfo
  {year} {2018})}\BibitemShut {NoStop}%
\bibitem [{\citenamefont {Chen}\ \emph {et~al.}(2014)\citenamefont {Chen},
  \citenamefont {Lu},\ and\ \citenamefont {Vishwanath}}]{Chen14DDW}%
  \BibitemOpen
  \bibfield  {author} {\bibinfo {author} {\bibfnamefont {X.}~\bibnamefont
  {Chen}}, \bibinfo {author} {\bibfnamefont {Y.-M.}\ \bibnamefont {Lu}},\ and\
  \bibinfo {author} {\bibfnamefont {A.}~\bibnamefont {Vishwanath}},\ }\bibfield
   {title} {\bibinfo {title} {Symmetry-protected topological phases from
  decorated domain walls},\ }\href {https://doi.org/10.1038/ncomms4507}
  {\bibfield  {journal} {\bibinfo  {journal} {Nature communications}\ }\textbf
  {\bibinfo {volume} {5}},\ \bibinfo {pages} {3507} (\bibinfo {year}
  {2014})}\BibitemShut {NoStop}%
\bibitem [{\citenamefont {Kosarew}(2002)}]{kosarew2002geometric}%
  \BibitemOpen
  \bibfield  {author} {\bibinfo {author} {\bibfnamefont {S.}~\bibnamefont
  {Kosarew}},\ }\bibfield  {title} {\bibinfo {title} {Geometric and categorical
  nonabelian duality in complex geometry},\ }\href@noop {} {\bibfield
  {journal} {\bibinfo  {journal} {Annali della Scuola Normale Superiore di
  Pisa-Classe di Scienze}\ }\textbf {\bibinfo {volume} {1}},\ \bibinfo {pages}
  {769} (\bibinfo {year} {2002})}\BibitemShut {NoStop}%
\bibitem [{\citenamefont {Kirillov}(2012)}]{kirillov2012elements}%
  \BibitemOpen
  \bibfield  {author} {\bibinfo {author} {\bibfnamefont {A.~A.}\ \bibnamefont
  {Kirillov}},\ }\href@noop {} {\emph {\bibinfo {title} {Elements of the Theory
  of Representations}}},\ Vol.\ \bibinfo {volume} {220}\ (\bibinfo  {publisher}
  {Springer Science \& Business Media},\ \bibinfo {year} {2012})\BibitemShut
  {NoStop}%
\bibitem [{\citenamefont {Tannaka}(1939)}]{tannaka1939dualitatssatz}%
  \BibitemOpen
  \bibfield  {author} {\bibinfo {author} {\bibfnamefont {T.}~\bibnamefont
  {Tannaka}},\ }\bibfield  {title} {\bibinfo {title} {{\"U}ber den
  dualit{\"a}tssatz der nichtkommutativen topologischen gruppen},\ }\href@noop
  {} {\bibfield  {journal} {\bibinfo  {journal} {Tohoku Mathematical Journal,
  First Series}\ }\textbf {\bibinfo {volume} {45}},\ \bibinfo {pages} {1}
  (\bibinfo {year} {1939})}\BibitemShut {NoStop}%
\bibitem [{\citenamefont {Fishman}\ \emph {et~al.}(2022)\citenamefont
  {Fishman}, \citenamefont {White},\ and\ \citenamefont
  {Stoudenmire}}]{ITensor}%
  \BibitemOpen
  \bibfield  {author} {\bibinfo {author} {\bibfnamefont {M.}~\bibnamefont
  {Fishman}}, \bibinfo {author} {\bibfnamefont {S.~R.}\ \bibnamefont {White}},\
  and\ \bibinfo {author} {\bibfnamefont {E.~M.}\ \bibnamefont {Stoudenmire}},\
  }\bibfield  {title} {\bibinfo {title} {{The ITensor Software Library for
  Tensor Network Calculations}},\ }\href
  {https://doi.org/10.21468/SciPostPhysCodeb.4} {\bibfield  {journal} {\bibinfo
   {journal} {SciPost Phys. Codebases}\ ,\ \bibinfo {pages} {4}} (\bibinfo
  {year} {2022})}\BibitemShut {NoStop}%
\bibitem [{\citenamefont {Bridgeman}\ \emph {et~al.}(2023)\citenamefont
  {Bridgeman}, \citenamefont {Lootens},\ and\ \citenamefont
  {Verstraete}}]{Bridgeman_2023}%
  \BibitemOpen
  \bibfield  {author} {\bibinfo {author} {\bibfnamefont {J.~C.}\ \bibnamefont
  {Bridgeman}}, \bibinfo {author} {\bibfnamefont {L.}~\bibnamefont {Lootens}},\
  and\ \bibinfo {author} {\bibfnamefont {F.}~\bibnamefont {Verstraete}},\
  }\bibfield  {title} {\bibinfo {title} {Invertible bimodule categories and
  generalized schur orthogonality},\ }\href
  {https://doi.org/10.1007/s00220-023-04781-y} {\bibfield  {journal} {\bibinfo
  {journal} {Communications in Mathematical Physics}\ }\textbf {\bibinfo
  {volume} {402}},\ \bibinfo {pages} {2691} (\bibinfo {year}
  {2023})}\BibitemShut {NoStop}%
\bibitem [{\citenamefont {Levin}\ and\ \citenamefont
  {Wen}(2006)}]{levin_detecting_2006}%
  \BibitemOpen
  \bibfield  {author} {\bibinfo {author} {\bibfnamefont {M.}~\bibnamefont
  {Levin}}\ and\ \bibinfo {author} {\bibfnamefont {X.-G.}\ \bibnamefont
  {Wen}},\ }\bibfield  {title} {\bibinfo {title} {Detecting topological order
  in a ground state wave function},\ }\href
  {https://doi.org/10.1103/PhysRevLett.96.110405} {\bibfield  {journal}
  {\bibinfo  {journal} {Phys. Rev. Lett.}\ }\textbf {\bibinfo {volume} {96}},\
  \bibinfo {pages} {110405} (\bibinfo {year} {2006})}\BibitemShut {NoStop}%
\bibitem [{\citenamefont {Pollmann}\ \emph
  {et~al.}(2010{\natexlab{b}})\citenamefont {Pollmann}, \citenamefont {Turner},
  \citenamefont {Berg},\ and\ \citenamefont
  {Oshikawa}}]{pollmann_entanglement_2010}%
  \BibitemOpen
  \bibfield  {author} {\bibinfo {author} {\bibfnamefont {F.}~\bibnamefont
  {Pollmann}}, \bibinfo {author} {\bibfnamefont {A.~M.}\ \bibnamefont
  {Turner}}, \bibinfo {author} {\bibfnamefont {E.}~\bibnamefont {Berg}},\ and\
  \bibinfo {author} {\bibfnamefont {M.}~\bibnamefont {Oshikawa}},\ }\bibfield
  {title} {\bibinfo {title} {Entanglement spectrum of a topological phase in
  one dimension},\ }\href {https://doi.org/10.1103/PhysRevB.81.064439}
  {\bibfield  {journal} {\bibinfo  {journal} {Phys. Rev. B}\ }\textbf {\bibinfo
  {volume} {81}},\ \bibinfo {pages} {064439} (\bibinfo {year}
  {2010}{\natexlab{b}})}\BibitemShut {NoStop}%
\bibitem [{\citenamefont {Wong}\ \emph {et~al.}(2022)\citenamefont {Wong},
  \citenamefont {Raussendorf},\ and\ \citenamefont
  {Czech}}]{wong_gauge_2022-1}%
  \BibitemOpen
  \bibfield  {author} {\bibinfo {author} {\bibfnamefont {G.}~\bibnamefont
  {Wong}}, \bibinfo {author} {\bibfnamefont {R.}~\bibnamefont {Raussendorf}},\
  and\ \bibinfo {author} {\bibfnamefont {B.}~\bibnamefont {Czech}},\ }\href
  {https://doi.org/10.48550/arXiv.2207.10098} {\bibinfo {title} {The {{Gauge
  Theory}} of {{Measurement-Based Quantum Computation}}}} (\bibinfo {year}
  {2022}),\ \Eprint {https://arxiv.org/abs/2207.10098} {arXiv:2207.10098
  [cond-mat, physics:hep-th, physics:quant-ph]} \BibitemShut {NoStop}%
\bibitem [{\citenamefont {Raussendorf}\ \emph {et~al.}(2017)\citenamefont
  {Raussendorf}, \citenamefont {Wang}, \citenamefont {Prakash}, \citenamefont
  {Wei},\ and\ \citenamefont {Stephen}}]{Raussendorf17}%
  \BibitemOpen
  \bibfield  {author} {\bibinfo {author} {\bibfnamefont {R.}~\bibnamefont
  {Raussendorf}}, \bibinfo {author} {\bibfnamefont {D.-S.}\ \bibnamefont
  {Wang}}, \bibinfo {author} {\bibfnamefont {A.}~\bibnamefont {Prakash}},
  \bibinfo {author} {\bibfnamefont {T.-C.}\ \bibnamefont {Wei}},\ and\ \bibinfo
  {author} {\bibfnamefont {D.~T.}\ \bibnamefont {Stephen}},\ }\bibfield
  {title} {\bibinfo {title} {Symmetry-protected topological phases with uniform
  computational power in one dimension},\ }\href
  {https://doi.org/10.1103/PhysRevA.96.012302} {\bibfield  {journal} {\bibinfo
  {journal} {Phys. Rev. A}\ }\textbf {\bibinfo {volume} {96}},\ \bibinfo
  {pages} {012302} (\bibinfo {year} {2017})}\BibitemShut {NoStop}%
\bibitem [{\citenamefont {Raussendorf}\ \emph {et~al.}(2019)\citenamefont
  {Raussendorf}, \citenamefont {Okay}, \citenamefont {Wang}, \citenamefont
  {Stephen},\ and\ \citenamefont {Nautrup}}]{Raussendorf19}%
  \BibitemOpen
  \bibfield  {author} {\bibinfo {author} {\bibfnamefont {R.}~\bibnamefont
  {Raussendorf}}, \bibinfo {author} {\bibfnamefont {C.}~\bibnamefont {Okay}},
  \bibinfo {author} {\bibfnamefont {D.-S.}\ \bibnamefont {Wang}}, \bibinfo
  {author} {\bibfnamefont {D.~T.}\ \bibnamefont {Stephen}},\ and\ \bibinfo
  {author} {\bibfnamefont {H.~P.}\ \bibnamefont {Nautrup}},\ }\bibfield
  {title} {\bibinfo {title} {Computationally universal phase of quantum
  matter},\ }\href {https://doi.org/10.1103/PhysRevLett.122.090501} {\bibfield
  {journal} {\bibinfo  {journal} {Phys. Rev. Lett.}\ }\textbf {\bibinfo
  {volume} {122}},\ \bibinfo {pages} {090501} (\bibinfo {year}
  {2019})}\BibitemShut {NoStop}%
\bibitem [{\citenamefont {Clark}(2006)}]{clark_valence_2006}%
  \BibitemOpen
  \bibfield  {author} {\bibinfo {author} {\bibfnamefont {S.}~\bibnamefont
  {Clark}},\ }\bibfield  {title} {\bibinfo {title} {Valence bond solid
  formalism for d-level one-way quantum computation*},\ }\href
  {https://doi.org/10.1088/0305-4470/39/11/010} {\bibfield  {journal} {\bibinfo
   {journal} {Journal of Physics A: Mathematical and General}\ }\textbf
  {\bibinfo {volume} {39}},\ \bibinfo {pages} {2701} (\bibinfo {year}
  {2006})}\BibitemShut {NoStop}%
\bibitem [{\citenamefont {Brennen}\ \emph {et~al.}(2009)\citenamefont
  {Brennen}, \citenamefont {Aguado},\ and\ \citenamefont
  {Cirac}}]{brennen_simulations_2009-1}%
  \BibitemOpen
  \bibfield  {author} {\bibinfo {author} {\bibfnamefont {G.~K.}\ \bibnamefont
  {Brennen}}, \bibinfo {author} {\bibfnamefont {M.}~\bibnamefont {Aguado}},\
  and\ \bibinfo {author} {\bibfnamefont {J.~I.}\ \bibnamefont {Cirac}},\
  }\bibfield  {title} {\bibinfo {title} {Simulations of quantum double
  models},\ }\href {https://doi.org/10.1088/1367-2630/11/5/053009} {\bibfield
  {journal} {\bibinfo  {journal} {New Journal of Physics}\ }\textbf {\bibinfo
  {volume} {11}},\ \bibinfo {pages} {053009} (\bibinfo {year}
  {2009})}\BibitemShut {NoStop}%
\bibitem [{\citenamefont {Garre-Rubio}\ \emph {et~al.}(2023)\citenamefont
  {Garre-Rubio}, \citenamefont {Lootens},\ and\ \citenamefont
  {Moln{\'{a}}r}}]{garre-rubio_classifying_2022}%
  \BibitemOpen
  \bibfield  {author} {\bibinfo {author} {\bibfnamefont {J.}~\bibnamefont
  {Garre-Rubio}}, \bibinfo {author} {\bibfnamefont {L.}~\bibnamefont
  {Lootens}},\ and\ \bibinfo {author} {\bibfnamefont {A.}~\bibnamefont
  {Moln{\'{a}}r}},\ }\bibfield  {title} {\bibinfo {title} {Classifying phases
  protected by matrix product operator symmetries using matrix product
  states},\ }\href {https://doi.org/10.22331/q-2023-02-21-927} {\bibfield
  {journal} {\bibinfo  {journal} {{Quantum}}\ }\textbf {\bibinfo {volume}
  {7}},\ \bibinfo {pages} {927} (\bibinfo {year} {2023})}\BibitemShut {NoStop}%
\bibitem [{\citenamefont {Lichtman}\ \emph {et~al.}(2021)\citenamefont
  {Lichtman}, \citenamefont {Thorngren}, \citenamefont {Lindner}, \citenamefont
  {Stern},\ and\ \citenamefont {Berg}}]{lichtman_bulk_2021}%
  \BibitemOpen
  \bibfield  {author} {\bibinfo {author} {\bibfnamefont {T.}~\bibnamefont
  {Lichtman}}, \bibinfo {author} {\bibfnamefont {R.}~\bibnamefont {Thorngren}},
  \bibinfo {author} {\bibfnamefont {N.~H.}\ \bibnamefont {Lindner}}, \bibinfo
  {author} {\bibfnamefont {A.}~\bibnamefont {Stern}},\ and\ \bibinfo {author}
  {\bibfnamefont {E.}~\bibnamefont {Berg}},\ }\bibfield  {title} {\bibinfo
  {title} {Bulk {{Anyons}} as {{Edge Symmetries}}: {{Boundary Phase Diagrams}}
  of {{Topologically Ordered States}}},\ }\href
  {https://doi.org/10.1103/PhysRevB.104.075141} {\bibfield  {journal} {\bibinfo
   {journal} {Physical Review B}\ }\textbf {\bibinfo {volume} {104}},\ \bibinfo
  {pages} {075141} (\bibinfo {year} {2021})},\ \Eprint
  {https://arxiv.org/abs/2003.04328} {arXiv:2003.04328} \BibitemShut {NoStop}%
\bibitem [{\citenamefont {Kitaev}\ and\ \citenamefont
  {Kong}(2012)}]{kitaev_models_2012}%
  \BibitemOpen
  \bibfield  {author} {\bibinfo {author} {\bibfnamefont {A.}~\bibnamefont
  {Kitaev}}\ and\ \bibinfo {author} {\bibfnamefont {L.}~\bibnamefont {Kong}},\
  }\bibfield  {title} {\bibinfo {title} {Models for {{Gapped Boundaries}} and
  {{Domain Walls}}},\ }\href {https://doi.org/10.1007/s00220-012-1500-5}
  {\bibfield  {journal} {\bibinfo  {journal} {Communications in Mathematical
  Physics}\ }\textbf {\bibinfo {volume} {313}},\ \bibinfo {pages} {351}
  (\bibinfo {year} {2012})}\BibitemShut {NoStop}%
\bibitem [{\citenamefont {Verresen}\ \emph {et~al.}(2022)\citenamefont
  {Verresen}, \citenamefont {Borla}, \citenamefont {Vishwanath}, \citenamefont
  {Moroz},\ and\ \citenamefont {Thorngren}}]{verresen_higgs_2022}%
  \BibitemOpen
  \bibfield  {author} {\bibinfo {author} {\bibfnamefont {R.}~\bibnamefont
  {Verresen}}, \bibinfo {author} {\bibfnamefont {U.}~\bibnamefont {Borla}},
  \bibinfo {author} {\bibfnamefont {A.}~\bibnamefont {Vishwanath}}, \bibinfo
  {author} {\bibfnamefont {S.}~\bibnamefont {Moroz}},\ and\ \bibinfo {author}
  {\bibfnamefont {R.}~\bibnamefont {Thorngren}},\ }\href
  {https://doi.org/10.48550/arXiv.2211.01376} {\bibinfo {title} {Higgs
  {{Condensates}} are {{Symmetry-Protected Topological Phases}}: {{I}}.
  {{Discrete Symmetries}}}} (\bibinfo {year} {2022}),\ \Eprint
  {https://arxiv.org/abs/2211.01376} {arXiv:2211.01376 [cond-mat,
  physics:hep-th, physics:quant-ph]} \BibitemShut {NoStop}%
\bibitem [{\citenamefont {Feiguin}\ \emph {et~al.}(2007)\citenamefont
  {Feiguin}, \citenamefont {Trebst}, \citenamefont {Ludwig}, \citenamefont
  {Troyer}, \citenamefont {Kitaev}, \citenamefont {Wang},\ and\ \citenamefont
  {Freedman}}]{Feiguin07}%
  \BibitemOpen
  \bibfield  {author} {\bibinfo {author} {\bibfnamefont {A.}~\bibnamefont
  {Feiguin}}, \bibinfo {author} {\bibfnamefont {S.}~\bibnamefont {Trebst}},
  \bibinfo {author} {\bibfnamefont {A.~W.~W.}\ \bibnamefont {Ludwig}}, \bibinfo
  {author} {\bibfnamefont {M.}~\bibnamefont {Troyer}}, \bibinfo {author}
  {\bibfnamefont {A.}~\bibnamefont {Kitaev}}, \bibinfo {author} {\bibfnamefont
  {Z.}~\bibnamefont {Wang}},\ and\ \bibinfo {author} {\bibfnamefont {M.~H.}\
  \bibnamefont {Freedman}},\ }\bibfield  {title} {\bibinfo {title} {Interacting
  anyons in topological quantum liquids: The golden chain},\ }\href
  {https://doi.org/10.1103/PhysRevLett.98.160409} {\bibfield  {journal}
  {\bibinfo  {journal} {Phys. Rev. Lett.}\ }\textbf {\bibinfo {volume} {98}},\
  \bibinfo {pages} {160409} (\bibinfo {year} {2007})}\BibitemShut {NoStop}%
\bibitem [{\citenamefont {Affleck}\ \emph {et~al.}(1987)\citenamefont
  {Affleck}, \citenamefont {Kennedy}, \citenamefont {Lieb},\ and\ \citenamefont
  {Tasaki}}]{AKLT}%
  \BibitemOpen
  \bibfield  {author} {\bibinfo {author} {\bibfnamefont {I.}~\bibnamefont
  {Affleck}}, \bibinfo {author} {\bibfnamefont {T.}~\bibnamefont {Kennedy}},
  \bibinfo {author} {\bibfnamefont {E.~H.}\ \bibnamefont {Lieb}},\ and\
  \bibinfo {author} {\bibfnamefont {H.}~\bibnamefont {Tasaki}},\ }\bibfield
  {title} {\bibinfo {title} {Rigorous results on valence-bond ground states in
  antiferromagnets},\ }\href {https://doi.org/10.1103/PhysRevLett.59.799}
  {\bibfield  {journal} {\bibinfo  {journal} {Phys. Rev. Lett.}\ }\textbf
  {\bibinfo {volume} {59}},\ \bibinfo {pages} {799} (\bibinfo {year}
  {1987})}\BibitemShut {NoStop}%
\bibitem [{\citenamefont {Munk}\ \emph {et~al.}(2018)\citenamefont {Munk},
  \citenamefont {Rasmussen},\ and\ \citenamefont
  {Burrello}}]{munk_dyonic_2018_2}%
  \BibitemOpen
  \bibfield  {author} {\bibinfo {author} {\bibfnamefont {M.~I.~K.}\
  \bibnamefont {Munk}}, \bibinfo {author} {\bibfnamefont {A.}~\bibnamefont
  {Rasmussen}},\ and\ \bibinfo {author} {\bibfnamefont {M.}~\bibnamefont
  {Burrello}},\ }\bibfield  {title} {\bibinfo {title} {Dyonic zero-energy
  modes},\ }\href {https://doi.org/10.1103/PhysRevB.98.245135} {\bibfield
  {journal} {\bibinfo  {journal} {Physical Review B}\ }\textbf {\bibinfo
  {volume} {98}},\ \bibinfo {pages} {245135} (\bibinfo {year}
  {2018})}\BibitemShut {NoStop}%
\bibitem [{\citenamefont {Moradi}\ \emph {et~al.}(2022)\citenamefont {Moradi},
  \citenamefont {Moosavian},\ and\ \citenamefont
  {Tiwari}}]{moradi2022topological}%
  \BibitemOpen
  \bibfield  {author} {\bibinfo {author} {\bibfnamefont {H.}~\bibnamefont
  {Moradi}}, \bibinfo {author} {\bibfnamefont {S.~F.}\ \bibnamefont
  {Moosavian}},\ and\ \bibinfo {author} {\bibfnamefont {A.}~\bibnamefont
  {Tiwari}},\ }\bibfield  {title} {\bibinfo {title} {Topological holography:
  Towards a unification of landau and beyond-landau physics},\ }\href@noop {}
  {\bibfield  {journal} {\bibinfo  {journal} {arXiv preprint arXiv:2207.10712}\
  } (\bibinfo {year} {2022})}\BibitemShut {NoStop}%
\bibitem [{\citenamefont {Apruzzi}\ \emph {et~al.}(2023)\citenamefont
  {Apruzzi}, \citenamefont {Bonetti}, \citenamefont {Garc{\'{i}}a~Etxebarria},
  \citenamefont {Hosseini},\ and\ \citenamefont
  {Sch{\"{a}}fer-Nameki}}]{apruzzi2021symmetry}%
  \BibitemOpen
  \bibfield  {author} {\bibinfo {author} {\bibfnamefont {F.}~\bibnamefont
  {Apruzzi}}, \bibinfo {author} {\bibfnamefont {F.}~\bibnamefont {Bonetti}},
  \bibinfo {author} {\bibfnamefont {I.}~\bibnamefont
  {Garc{\'{i}}a~Etxebarria}}, \bibinfo {author} {\bibfnamefont {S.~S.}\
  \bibnamefont {Hosseini}},\ and\ \bibinfo {author} {\bibfnamefont
  {S.}~\bibnamefont {Sch{\"{a}}fer-Nameki}},\ }\bibfield  {title} {\bibinfo
  {title} {Symmetry tfts from string theory},\ }\href
  {https://doi.org/10.1007/s00220-023-04737-2} {\bibfield  {journal} {\bibinfo
  {journal} {Communications in Mathematical Physics}\ }\textbf {\bibinfo
  {volume} {402}},\ \bibinfo {pages} {895} (\bibinfo {year}
  {2023})}\BibitemShut {NoStop}%
\bibitem [{\citenamefont {Freed}\ \emph {et~al.}(2022)\citenamefont {Freed},
  \citenamefont {Moore},\ and\ \citenamefont {Teleman}}]{Freed2022topological}%
  \BibitemOpen
  \bibfield  {author} {\bibinfo {author} {\bibfnamefont {D.~S.}\ \bibnamefont
  {Freed}}, \bibinfo {author} {\bibfnamefont {G.~W.}\ \bibnamefont {Moore}},\
  and\ \bibinfo {author} {\bibfnamefont {C.}~\bibnamefont {Teleman}},\
  }\bibfield  {title} {\bibinfo {title} {Topological symmetry in quantum field
  theory},\ }\href@noop {} {\bibfield  {journal} {\bibinfo  {journal} {arXiv
  preprint arXiv:2209.07471}\ } (\bibinfo {year} {2022})}\BibitemShut {NoStop}%
\bibitem [{\citenamefont {Bhardwaj}\ and\ \citenamefont
  {Schafer-Nameki}(2023{\natexlab{b}})}]{Bhardwaj23}%
  \BibitemOpen
  \bibfield  {author} {\bibinfo {author} {\bibfnamefont {L.}~\bibnamefont
  {Bhardwaj}}\ and\ \bibinfo {author} {\bibfnamefont {S.}~\bibnamefont
  {Schafer-Nameki}},\ }\href@noop {} {\bibinfo {title} {Generalized charges,
  part ii: Non-invertible symmetries and the symmetry tft}} (\bibinfo {year}
  {2023}{\natexlab{b}}),\ \Eprint {https://arxiv.org/abs/2305.17159}
  {arXiv:2305.17159 [hep-th]} \BibitemShut {NoStop}%
\bibitem [{\citenamefont {Inamura}\ and\ \citenamefont
  {Wen}(2023)}]{Inamura2023}%
  \BibitemOpen
  \bibfield  {author} {\bibinfo {author} {\bibfnamefont {K.}~\bibnamefont
  {Inamura}}\ and\ \bibinfo {author} {\bibfnamefont {X.-G.}\ \bibnamefont
  {Wen}},\ }\href@noop {} {\bibinfo {title} {2+1d symmetry-topological-order
  from local symmetric operators in 1+1d}} (\bibinfo {year} {2023}),\ \Eprint
  {https://arxiv.org/abs/2310.05790} {arXiv:2310.05790 [cond-mat.str-el]}
  \BibitemShut {NoStop}%
\bibitem [{\citenamefont {Ostrik}(2003{\natexlab{b}})}]{Ostrik03}%
  \BibitemOpen
  \bibfield  {author} {\bibinfo {author} {\bibfnamefont {V.}~\bibnamefont
  {Ostrik}},\ }\bibfield  {title} {\bibinfo {title} {{Module categories over
  the Drinfeld double of a finite group}},\ }\href
  {https://doi.org/10.1155/S1073792803205079} {\bibfield  {journal} {\bibinfo
  {journal} {International Mathematics Research Notices}\ }\textbf {\bibinfo
  {volume} {2003}},\ \bibinfo {pages} {1507} (\bibinfo {year}
  {2003}{\natexlab{b}})},\ \Eprint
  {https://arxiv.org/abs/https://academic.oup.com/imrn/article-pdf/2003/27/1507/1931507/2003-27-1507.pdf}
  {https://academic.oup.com/imrn/article-pdf/2003/27/1507/1931507/2003-27-1507.pdf}
  \BibitemShut {NoStop}%
\end{thebibliography}%
\onecolumngrid
\clearpage
\raggedbottom

\appendix

\section{Qubit CSS Cluster State}\label{ap:CSS_cluster}
In this section, we briefly review the CSS cluster state defined on qubits and derive several results which are generalized in the main text.

\subsection{State and MPS}
In the case $G=\mathbb{Z}_2$, the $G$ cluster state reduces to the qubit CSS cluster state:
\begin{equation}
    |\text{CSS}\rangle=2^{-N/2}\sum_{\{g_i\}\in \mathbb{Z}_2}|g_1\rangle|g_1\oplus g_2\rangle|g_2\rangle\cdots|g_{N-1}\oplus{g_N}\rangle|g_N\rangle,
\end{equation}
where $\oplus$ denotes addition modulo 2. This state can be written as an MPS with tensors
\begin{equation}
  \begin{tikzpicture}
    % \node[irrep] at (0.2,0) {};
    \draw (-0.5,0)--(0.5,0);
    \draw (0,0) -- (0,0.5);
    \node[oddZ2] (t) at (0,0) {};
  \end{tikzpicture}
\ =\ketbra{0}{0}\otimes\ket{0}+\ketbra{1}{1}\otimes\ket{1},\quad
  \begin{tikzpicture}
    % \node[irrep] at (0.2,0) {};
    \draw (-0.5,0)--(0.5,0);
    \draw (0,0) -- (0,0.5);
    \node[evenZ2] (t) at (0,0) {};

  \end{tikzpicture}
\ =\mathbbm{1}\otimes\ket{0}+X\otimes\ket{1}.
\end{equation}
These tensors satisfy the following identities:
\begin{equation*}
  \begin{tikzpicture}
    \draw (-1,0)--(1,0);
    \draw (0,0) -- (0,1);
    \node[oddZ2] (t) at (0,0) {};
    % \node[] at (0.2,0.2) {o};
    \node[mpo] at (0,0.5) {$X$};
    % \node[] at (0.2,0.7) {$g$};

  \end{tikzpicture}
\ = \ketbra{0}{0}\otimes\ket{1}+\ketbra{1}{1}\otimes\ket{0} = \ \begin{tikzpicture}
    \draw (-1,0)--(1,0);
    \draw (0,0) -- (0,1);
    \node[oddZ2] (t) at (0,0) {};
    % \node[] at (0.2,0.2) {o};
    \node[mpo] at (-0.5,0) {$X$};
     \node[mpo] at (0.5,0) {$X$};
    % \node[] at (-0.3,0.2) {$g$};
  \end{tikzpicture} \ ,
\end{equation*}

\vspace{-1 em}
\begin{equation*}
  \begin{tikzpicture}
    \draw (-1,0)--(1,0);
    \draw (0,0) -- (0,1);
    \node[evenZ2] (t) at (0,0) {};
    % \node[] at (0.2,0.2) {o};
    \node[mpo] at (0,0.5) {$X$};
    % \node[] at (0.2,0.7) {$g$};

  \end{tikzpicture}
\ = \mathbbm{1}\otimes\ket{1}+X\otimes\ket{0} = \ \begin{tikzpicture}
    \draw (-1,0)--(1,0);
    \draw (0,0) -- (0,1);
    \node[evenZ2] (t) at (0,0) {};
    % \node[] at (0.2,0.2) {o};
    \node[mpo] at (-0.5,0) {$X$};
    % \node[] at (-0.3,0.2) {$g$};

  \end{tikzpicture} \ = \
  \begin{tikzpicture}
    \draw (-1,0)--(1,0);
    \draw (0,0) -- (0,1);
    \node[evenZ2] (t) at (0,0) {};
    % \node[] at (0.2,0.2) {o};
    \node[mpo] at (0.5,0) {$X$};
    % \node[] at (-0.3,0.2) {$g$};

  \end{tikzpicture} \ ,
\end{equation*}

\vspace{-1 em}
\begin{equation*}
  \begin{tikzpicture}
    \draw (-1,0)--(1,0);
    \draw (0,0) -- (0,1);
    \node[oddZ2] (t) at (0,0) {};
    % \node[] at (0.2,0.2) {o};
    \node[mpo] at (0,0.5) {$Z$};
    % \node[] at (0.2,0.7) {$g$};

  \end{tikzpicture}
\ = \ketbra{0}{0}\otimes\ket{0}-\ketbra{1}{1}\otimes\ket{1} = \ \begin{tikzpicture}
    \draw (-1,0)--(1,0);
    \draw (0,0) -- (0,1);
    \node[oddZ2] (t) at (0,0) {};
    % \node[] at (0.2,0.2) {o};
    \node[mpo] at (-0.5,0) {$Z$};
    % \node[] at (-0.3,0.2) {$g$};

  \end{tikzpicture} \ = \
  \begin{tikzpicture}
    \draw (-1,0)--(1,0);
    \draw (0,0) -- (0,1);
    \node[oddZ2] (t) at (0,0) {};
    % \node[] at (0.2,0.2) {o};
    \node[mpo] at (0.5,0) {$Z$};
    % \node[] at (-0.3,0.2) {$g$};

  \end{tikzpicture} \ ,
\end{equation*}

\vspace{-1 em}
\begin{equation*}
  \begin{tikzpicture}
    \draw (-1,0)--(1,0);
    \draw (0,0) -- (0,1);
    \node[evenZ2] (t) at (0,0) {};
    % \node[] at (0.2,0.2) {o};
    \node[mpo] at (0,0.5) {$Z$};
    % \node[] at (0.2,0.7) {$g$};

  \end{tikzpicture}
\ = \mathbbm{1}\otimes\ket{0}-X\otimes\ket{1} = \ \begin{tikzpicture}
    \draw (-1,0)--(1,0);
    \draw (0,0) -- (0,1);
    \node[evenZ2] (t) at (0,0) {};
    % \node[] at (0.2,0.2) {o};
    \node[mpo] at (0.5,0) {$Z$};
    \node[mpo] at (-0.5,0) {$Z$};
    % \node[] at (-0.3,0.2) {$g$};
  \end{tikzpicture} \ ,
\end{equation*}

\vspace{-1 em}
\begin{equation*}
  \begin{tikzpicture}
    \draw (-1,0)--(1,0);
    \draw (0,0) -- (0,1);
    \node[oddZ2] (t) at (0,0) {};
    % \node[] at (0.2,0.2) {o};
    \node[mpo] at (-0.5,0) {$Z$};
     \node[mpo] at (0.5,0) {$Z$};
    % \node[] at (-0.3,0.2) {$g$};
  \end{tikzpicture} \ = \ketbra{0}{0}\otimes\ket{0}+\ketbra{1}{1}\otimes\ket{1} = \
  \begin{tikzpicture}
    \draw (-1,0)--(1,0);
    \draw (0,0) -- (0,1);
    \node[oddZ2] (t) at (0,0) {};
    % \node[] at (0.2,0.2) {o};
    % \node[] at (0.2,0.7) {$g$};
  \end{tikzpicture}
\ ,
\end{equation*}

\vspace{-1 em}
\begin{equation*}
  \begin{tikzpicture}
    \draw (-1,0)--(1,0);
    \draw (0,0) -- (0,1);
    \node[evenZ2] (t) at (0,0) {};
    % \node[] at (0.2,0.2) {o};
    \node[mpo] at (-0.5,0) {$X$};
     \node[mpo] at (0.5,0) {$X$};
    % \node[] at (-0.3,0.2) {$g$};
  \end{tikzpicture} \ = \mathbbm{1}\otimes\ket{0}+X\otimes\ket{1} = \
  \begin{tikzpicture}
    \draw (-1,0)--(1,0);
    \draw (0,0) -- (0,1);
    \node[evenZ2] (t) at (0,0) {};
    % \node[] at (0.2,0.2) {o};
    % \node[] at (0.2,0.7) {$g$};
  \end{tikzpicture}
\ .
\end{equation*}

The corresponding product state is $\ket{\psi_0}=\ket{+,0,+,0,\ldots}$.

\subsection{Hamiltonian and Symmetries}
The qubit CSS cluster state is the ground state of the commuting stabilizer Hamiltonian
\begin{equation}
    H = -\sum_{i\text{ odd}} X^{(i+1)}X^{(i+2)}X^{(i+3)} + Z^{(i)}Z^{(i+1)}Z^{(i+2)}.
\end{equation}
This Hamiltonian is obtained from the usual cluster state stabilizers by performing a Hadamard gate on the middle qubit of each string. This Hamiltonian has global symmetries
\begin{equation}
    \hat A= \prod_{i\text{ odd}}X^{(i)},\quad\hat B = \prod_{i\text{ even}}Z^{(i)}.
\end{equation}
These symmetries commute with one another and each square to 1. Together, they form a representation of the symmetry group $\ztwo\times\ztwo$.

\subsection{Gauging the Ising Model}
Consider the quantum Ising model without transverse field,
\begin{equation}
   H=-\sum_{i}Z^{(i)}Z^{(i+1)}.
\end{equation}
 $X=\prod_iX^{(i)}$ is a global symmetry, but $X^{(i)}$ is not a local symmetry. However, we can promote $X^{(i)}$ to a local symmetry through minimal coupling to a background $\mathbb{Z}_2$ gauge field. Define Pauli operators for the gauge sites $\tilde{X}^{(i\pm 1/2)},\tilde{Z}^{(i\pm 1/2)}$.
 The local symmetry action is given by the Gauss law $\tilde{X}^{(i-1/2)}X^{(i)}\tilde{X}^{(i+1/2)}$
 and in order for the kinetic term $ZZ$ to commute with the local symmetry action, it must be minimally-coupled resulting in $Z^{(i)}\tilde{Z}^{(i+1/2)}Z^{(i+1)}$. Indeed, we see that $[Z^{(i)}\tilde{Z}^{(i+1/2)}Z^{(i+1)},\tilde{X}^{(i-1/2)}X^{(i)}\tilde{X}^{(i+1/2)}]=0$. In general, the gauged Hamiltonian includes the gauged ZZ term, a term energetically enforcing the gauge symmetry, and a term enforcing zero flux. In an open 1d chain there is no flux term and the Hamiltonian becomes
\begin{equation}
   H_\text{gauged}=-\sum_{i}Z^{(i)}\tilde{Z}^{(i+1/2)}Z^{(i+1)}+\tilde{X}^{(i-1/2)}X^{(i)}\tilde{X}^{(i+1/2)}.
\end{equation}
This is precisely the Hamiltonian for the qubit CSS cluster state, up to a trivial rescaling of position coordinates.

\subsection{String Order Parameter}
In the qubit setting, the string order parameter is simply a string of Pauli $Z$ operators which acts on two odd sites spaced by $2k$ as well as every intermediary even site:
\begin{equation}
     \hat{\mathcal{S}}^{(i,k)}=Z^{(i)}\left(\prod_{j=0}^{k-1}Z^{(i+1+2j)}\right)Z^{(i+2k)}.
\end{equation}

We can see Diagrammatically that $\hat{\mathcal{S}}^{(i,k)}$ acts trivially on the CSS cluster state:
\begin{equation}
\hat{\mathcal{S}}^{(i,k)}\ket{\text{CSS}}=\begin{tikzpicture}[baseline={(0,.4)}]
    \draw (0,0)--(5.5,0);
    \node[fill=white] at (-0.25,0) {$\dots$};
    \node[fill=white] at (5.75,0) {$\dots$};
    \foreach \x in {5}{
	    \draw (\x,0) -- (\x,1);
		\node[oddZ2] at (\x,0) {};
		\node[mpo] (t\x) at (\x,0.5) {$Z$};
		  }
    \foreach \x in {2}{
		 \draw (\x,0) -- (\x,1);
		 \node[oddZ2] at (\x,0) {};
		  }
		  \foreach \x in {1.25,2.75,4.25}{
		     \draw (\x,0) -- (\x,1);
		    \node[evenZ2] at (\x,0) {};
		    \node[mpo] (t\x) at (\x,0.5) {$Z$};
		  }
	\foreach \x in {.5}{
		     \draw (\x,0) -- (\x,1);
		    \node[oddZ2] at (\x,0) {};
		    \node[mpo] (t\x) at (\x,0.5) {$Z$};}

        \node[fill=white] at (3.5,0) {$\dots$};
  \end{tikzpicture} \ =
  \begin{tikzpicture}[baseline={(0,.4)}]
    \draw (-0.25,0)--(5.75,0);
    \foreach \x in {0.5,2,5}{
		     \draw (\x,0) -- (\x,1);
		    \node[oddZ2] at (\x,0) {};
		  }
		  \foreach \x in {1.25,2.75,4.25}{
		     \draw (\x,0) -- (\x,1);
		    \node[evenZ2] at (\x,0) {};
		  }
		  \node[edge] at (0,0) {};
		  \node[edge] at (5.5,0) {};
    \node[fill=white] at (3.5,0) {$\dots$};
        \node[fill=white] at (-0.25,0) {$\dots$};
        \node[fill=white] at (5.75,0) {$\dots$};
  \end{tikzpicture} \ = \ket{\text{CSS}},
\end{equation}
so that $\bra{\text{CSS}}\hat{\mathcal{S}}^{(i,k)}\ket{\text{CSS}}=1$, as expected.

We can see that the product state has vanishing expectation value of the string order parameter because it has no effect on the even sites, and flips the odd endpoints from $\ket{+}$ to $\ket{-}$:
\begin{align}
\begin{split}
    \hat{\mathcal{S}}^{(i,k)}\ket{\psi_0}&=\hat{\mathcal{S}}^{(i,k)}\ket{+,0,+,0,\ldots}\\
    &=\ket{+,0,\ldots,-_{i},\ldots,-_{i+2k},\ldots}\\
    \Rightarrow \bra{\psi_0}\hat{\mathcal{S}}^{(i,k)}\ket{\psi_0}&=\braket{+,0,\ldots,+_{i},\ldots,+_{i+2k},\ldots|+,0,\ldots,-_{i},\ldots,-_{i+2k},\ldots}\\
    &=0.
\end{split}
\end{align}

\subsection{Edge Modes}
Consider the action of the global symmetries $\hat A$ and $\hat B$ on the qubit CSS cluster state:
\begin{equation*}
\hat A\ket{\text{CSS}} = \begin{tikzpicture}[baseline={(0,.4)}]
    \draw (0,0)--(5.5,0);
    \foreach \x in {0.5,2,3.5,5}{

		     \draw (\x,0) -- (\x,1);
		    \node[mpo] (t\x) at (\x,0.5) {$X$};
		    \node[oddZ2] at (\x,0) {};
		  }
		  \foreach \x in {1.25,4.25}{
		     \draw (\x,0) -- (\x,1);
		    \node[evenZ2] at (\x,0) {};
		  }
		  \node[fill=white] at (2.75,0) {$\dots$};
		  \node[edge] at (0,0) {};
		  \node[edge] at (5.5,0) {};

  \end{tikzpicture} \ =
  \begin{tikzpicture}[baseline={(0,.4)}]
    \draw (-0.5,0)--(6,0);
    \foreach \x in {0.5,2,3.5,5}{

		     \draw (\x,0) -- (\x,1);
		    \node[oddZ2] at (\x,0) {};
		  }
		  \foreach \x in {1.25,4.25}{
		     \draw (\x,0) -- (\x,1);
		    \node[evenZ2] at (\x,0) {};
		  }
		  \node[fill=white] at (2.75,0) {$\dots$};
		  \node[edge] at (-0.5,0) {};
		  \node[edge] at (6,0) {};
		  \node[mpo] at (0,0) {$X$};
		  \node[mpo] at (5.5,0) {$X$};

  \end{tikzpicture} \
\end{equation*}

\begin{equation*}
\hat B\ket{\text{CSS}}=\begin{tikzpicture}[baseline={(0,.4)}]
    \draw (0,0)--(5.5,0);
    \foreach \x in {0.5,2,3.5,5}{

		     \draw (\x,0) -- (\x,1);

		    \node[oddZ2] at (\x,0) {};
		  }
		  \foreach \x in {1.25,4.25}{
		     \draw (\x,0) -- (\x,1);
		    \node[evenZ2] at (\x,0) {};
		    \node[mpo] (t\x) at (\x,0.5) {$Z$};
		  }
		  \node[fill=white] at (2.75,0) {$\dots$};
		  \node[edge] at (0,0){};
		  \node[edge] at (5.5,0){};

  \end{tikzpicture} \ =
  \begin{tikzpicture}[baseline={(0,.4)}]
    \draw (-0.5,0)--(6,0);
    \foreach \x in {0.5,2,3.5,5}{
		     \draw (\x,0) -- (\x,1);
		    \node[oddZ2] at (\x,0) {};
		  }
		  \foreach \x in {1.25,4.25}{
		     \draw (\x,0) -- (\x,1);
		    \node[evenZ2] at (\x,0) {};
		  }
		  \node[fill=white] at (2.75,0) {$\dots$};
		  \node[edge] at (-0.5,0) {};
		  \node[edge] at (6,0) {};
		  \node[mpo] at (0,0) {$Z$};
		  \node[mpo] at (5.5,0) {$Z$};
  \end{tikzpicture} \ ,
\end{equation*}

From this diagramtic derivation, we can read off the edge modes
 \begin{equation}
 \hat A: X^{(L)}X^{(R)},\quad
 \hat B:Z^{(L)}Z^{(R)}
\end{equation}
 acting on the edge Hilbert space $\mathcal{H}_L\otimes\mathcal{H}_R$.

Notice that \begin{equation}
    \{X^{(L)},Z^{(L)}\}=\{X^{(R)},Z^{(R)}\}=0,
\end{equation}
so that the left and right operators generate the single-qubit Pauli algebra on the left and right ends of the chain, respectively. Recognizing that the Pauli matrices are a projective representation of $\ztwo\times\ztwo$, we have found edge modes transforming as a projective representation of the protecting symmetry, as is expected in a 1d SPT phase with on-site group symmetry.

\subsection{Topological Response}\label{app:responseZ2}
There is only a single nontrivial $\mathbb{Z}_2$-flux state, which corresponds to $g=1$:
\begin{equation}
    |\text{CSS}_1\rangle:=
    2^{-(N-1)/2}\sum_{{g_i}\in \mathbb{Z}_2}|g_1\rangle|g_1\oplus g_2\rangle|g_2\rangle\cdots|g_{N-1}\oplus(g_1\oplus{1})\rangle.
\end{equation}
This state is obtained from $\ket{\text{CSS}}$ by twisting the boundary condition from one end of the chain to the other. It acquires a nontrivial charge under $\hat B$:
\begin{equation}
    \hat B |\text{CSS}_1\rangle=(-1)^{g_1+g_2+g_2+\cdots+g_{N-1}+g_{N-1}+g_1+1}|\text{CSS}_1\rangle=-|\text{CSS}_1\rangle,
\end{equation}
where we have used the fact that $2g_i$ is even for all $g_i$ in $\mathbb{Z}_2$. Similarly, there is only a single nontrivial $\rep \ztwo$-flux state corresponding to the sign irrep $\Gamma=\Gamma_-$:
\begin{equation}
|\text{CSS}_{\Gamma_-}\rangle=2^{-(N-1)/2}\sum_{\{g_i\}\in \mathbb{Z}_2}\Gamma_-(g_1)|g_1\rangle|g_1\oplus g_2\rangle|g_2\rangle\cdots|g_{N-1}\oplus{g_1}\rangle.
\end{equation}
We find that this state is nontrivially charged under $\hat A$:
\begin{equation}
\begin{split}
    \hat A|\text{CSS}_{\Gamma_-}\rangle=2^{-(N-1)/2}&\sum_{\{g_i\}\in \mathbb{Z}_2}\Gamma_-(g_1)|g_1\oplus1\rangle|g_1\oplus g_2\rangle|g_2\oplus1\rangle\cdots|g_{N-1}\oplus{g_1}\rangle\\&=2^{-(N-1)/2}\sum_{\{g_i\}\in \mathbb{Z}_2}\Gamma_-(g_1\oplus 1)|g_1\rangle|g_1\oplus g_2\rangle|g_2\rangle\cdots|g_{N-1}\oplus{g_1}\rangle=-|\text{CSS}_{\Gamma_-}\rangle.
\end{split}
\end{equation}
Because $\ztwo$ is abelian, it is easy to see that the response is symmetric. That is, we find a $\ztwo$ charge of -1 in response to $\rep \ztwo$ flux, and a $\rep\ztwo$ charge of -1 in response to $\ztwo$ flux.

The response for the cluster state can be understood from the perspective of charges of one $\ztwo$ symmetry decorated on the domain wall of the other $\ztwo$ symmetry. Inserting a flux generates a single domain wall from which we can see the charge response~\cite{Chen14DDW}.

\section{Additional Symmetry}\label{ap:additional_sym}
In this section, we seek to better understand the full symmetry group of the $G$ cluster state. There are four symmetries acting independently, so that the total symmetry group is $G_L\times G_R\times \rep G\times \inn G$. We would like to understand which symmetries are actually necessary for protection of the SPT phase. For example, it is insightful to recall that the AKLT model~\cite{AKLT} is a non-trivial SPT phase when protected separately either under SO$(3)$ rotations or time reversal, meaning that weak local perturbations which respect one subgroup while breaking the other will not lift the degeneracy in the thermodynamic limit. We want to understand whether a similar mechanism is at play here for the $G$ cluster state.

As alluded to in the main text, this is indeed the case. If we wish to protect the ground state degeneracy, one sufficient subgroup to preserve is $G_R\times \rep G$. This can be understood from the fact that the full algebra
\begin{equation}
    \left\{\rx_g\zt_\Gamma:g\in G,\Gamma\in\rep G\right\}
\end{equation}
of a $G$-qudit is present at each edge, as can be seen in \cref{eq:edge_modes}. This was also confirmed numerically in the case $G=D_3$ in Sec.~\ref{sec:D3_degen} by forming perturbations which respect only $G_R\times \rep G$ and observing the stability of the phase.

However, $G_R\times \rep G$ may not be the only choice of preserved subgroup which protects the SPT. We can repeat the edge mode analysis for the symmetry $G_L$:
\begin{equation*}
\overrightarrow A_g\ket{\Cl} = \begin{tikzpicture}[baseline={(0,.4)}]
    \draw (0,0)--(5.5,0);
    \foreach \x in {0.5,2,3.5,5}{

		     \draw (\x,0) -- (\x,1);
		    \node[mpo] (t\x) at (\x,0.5) {$\lx_g$};
		    \node[odd] at (\x,0) {};
		  }
		  \foreach \x in {1.25,4.25}{
		     \draw (\x,0) -- (\x,1);
		    \node[even] at (\x,0) {};
            \node[mpo] (t\x) at (\x,0.5) {$\cx_g$};
		  }
		  \node[fill=white] at (2.75,0) {$\dots$};
		  \node[edge] at (0,0) {};
		  \node[edge] at (5.5,0) {};

  \end{tikzpicture} \ =
  \begin{tikzpicture}[baseline={(0,.4)}]
    \draw (-0.5,0)--(6,0);
    \foreach \x in {0.5,2,3.5,5}{

		     \draw (\x,0) -- (\x,1);
		    \node[odd] at (\x,0) {};
		  }
		  \foreach \x in {1.25,4.25}{
		     \draw (\x,0) -- (\x,1);
		    \node[even] at (\x,0) {};
		  }
		  \node[fill=white] at (2.75,0) {$\dots$};
		  \node[edge] at (-0.5,0) {};
		  \node[edge] at (6,0) {};
		  \node[mpo] at (0,0) {$\lx_g^\dagger$};
		  \node[mpo] at (5.5,0) {$\lx_g$};
  \end{tikzpicture} \ ,
\end{equation*}
and we see that we also have an algebra of $X$-type operators at each edge. Because $Z$-type operators can be combined with either left or right $X$-type operators to span the full algebra of a single $G$-qudit, this suggests that $G_L$ may be able to protect the SPT in the case where $G_R$ is broken. We leave further exploration of this alternative to future work. We have also not considered the role of spatial symmetries, as our perturbations were translation invariant with the periodicity of the unit cell. We leave this question to future work as well.

\section{Gauging the Flux Ladder}
We can proceed in much the same was as above to gauge the group-based classical Ising model, introduced in~\cite{munk_dyonic_2018_2} as the flux ladder. The Hamiltonian of this model is given by
\begin{equation}
    H=-\sum_{i}\sum_{\Gamma\in\rep{G}}\Tr\left[\zt^{\dagger(i)}_\Gamma.\zt^{(i+1)}_\Gamma\right].
\end{equation}
Notice that $H$ has a global symmetry $\lx_g=\prod_i\lx_g^{(i)}$, but $\lx_g^{(i)}$ is not a local symmetry. We can gauge this symmetry in the usual way, resulting in the Hamiltonian,
\begin{equation}
   H_\text{gauged}=-\sum_{i}\left(\sum_{\Gamma\in\rep{G}}\Tr\left[\zt^{\dagger(i)}_\Gamma.\zt_\Gamma^{(i+1/2)}.\zt^{(i+1)}_\Gamma\right]+\sum_{g\in G}\rx_g^{(i+1/2)}\lx_g^{(i+1)}{\lx_g^{(i+3/2)}}\right),
\end{equation}
where the half-integer sites correspond to gauge sites inserted between the original matter sites. $\lx_g^{(i)}$ is a symmetry of $H_\text{gauged}$, so we say that the gauging procedure was successful. Notice that a trivial relabeling of indices shows that $H_\text{gauged}$ is equivalent to the $G$ cluster state Hamiltonian.

\section{Deriving the Fusion Category Symmetry}\label{algebraic_deriv}
Consider the product of two $\hat B$ operators labeled by irreps $\Gamma_1$ and $\Gamma_2$,
\begin{equation}
    \hat B_{\Gamma_1}\hat B_{\Gamma_2}=\Tr\left[\prod_{i\text{ even}}\zt^{(i)}_{\Gamma_1}\right]\Tr\left[\prod_{i\text{ even}}\zt^{(i)}_{\Gamma_2}\right].
\end{equation}
Recall that $\Tr[A]\Tr[B]=\Tr[A\otimes B]$, so
\begin{equation}
    =\Tr\left[\prod_{i\text{ even}}\zt^{(i)}_{\Gamma_1}\otimes\prod_{i\text{ even}}\zt^{(i)}_{\Gamma_2}\right].
\end{equation}
We also know that $(AB)\otimes(CD)=(A\otimes C)(B\otimes D)$, so
\begin{equation}
    =\Tr\left[\prod_{i\text{ even}}\zt^{(i)}_{\Gamma_1}\otimes\zt^{(i)}_{\Gamma_2}\right].
\end{equation}
The tensor product of irreps (of any linear representations, in fact) can be decomposed into a direct sum of irreps according to the fusion rules  $\Gamma_i\otimes\Gamma_j=\bigoplus_k \Gamma_k^{\oplus N^{\Gamma_k}_{\Gamma_m,\Gamma_j}}$. We then have
\begin{equation}
   =\Tr\left[\prod_{i\text{ even}}\bigoplus_k \left(\zt^{(i)}_{\Gamma_k}\right)^{\oplus N^{\Gamma_k}_{\Gamma_1,\Gamma_2}}\right]=\Tr\left[\bigoplus_k\prod_{i\text{ even}} \left(\zt^{(i)}_{\Gamma_k}\right)^{\oplus N^{\Gamma_k}_{\Gamma_1,\Gamma_2}}\right].
\end{equation}
We finally use the property that the trace of a direct sum is the sum of the traces, so
\begin{equation}
    =\sum_k N^{\Gamma_k}_{\Gamma_1,\Gamma_2}\Tr\left[\prod_{i\text{ even}} \zt^{(i)}_{\Gamma_k}\right]=\sum_k N^{\Gamma_k}_{\Gamma_1,\Gamma_2}\hat B_{\Gamma_k}.
\end{equation}
We now see that
\begin{equation}\label{algebraic}
    \hat B_{\Gamma_1}\hat B_{\Gamma_2}=\sum_k N^{\Gamma_k}_{\Gamma_1,\Gamma_2}\hat B_{\Gamma_k}.
\end{equation}
This means that $\hat B$ is a fusion category symmetry.

Alternatively, this can be shown in the MPS language:
\begin{equation*}
    \begin{tikzpicture}[baseline={(0,0.8)}]
    \draw[red] (0,0.5)--(1,0.5);
    \draw[red] (0,1.25)--(1,1.25);
		  \foreach \x in {0.5}{
		     \draw (\x,0) -- (\x,1.75);
		    \node[mpo] (t\x) at (\x,0.5) {$\zt_{\Gamma_i}$};
            \node[mpo] (t\x) at (\x,1.25) {$\zt_{\Gamma_j}$};
		  }
  \end{tikzpicture} \ = \sum_g\zt_{\Gamma_i}(g)\otimes\zt_{\Gamma_j}(g)\otimes \ketbra{g}{g}= \sum_kN_{i,j}^k\sum_g\zt_{\Gamma_k}(g)\otimes \ketbra{g}{g}=\sum_k N_{i,j}^k\begin{tikzpicture}[baseline={(0,.4)}]
    \draw[red] (0,0.5)--(1,0.5);
		  \foreach \x in {0.5}{
		     \draw (\x,0) -- (\x,1);
		    \node[mpo] (t\x) at (\x,0.5) {$\zt_{\Gamma_k}$};
		  }
  \end{tikzpicture}.
\end{equation*}

In the case where $G$ is Abelian, the algebraic relation (\ref{algebraic}) reduces to a group-like relation. One way to see that this must be true is to recall that
\begin{equation}
    d_{\Gamma_i}d_{\Gamma_j}=\sum_kN^{\Gamma_k}_{\Gamma_i,\Gamma_j}d_{\Gamma_k}.
\end{equation}
For abelian groups, all irreps are one-dimensional, so for each pair $(\Gamma_i,\Gamma_j)$, some $N^{\Gamma_k}_{\Gamma_i,\Gamma_j}=1$ while the rest are zero. This means that for $G$ abelian, we have
\begin{equation}
    \hat B_{\Gamma_i}\hat B_{\Gamma_j}=\hat B_{\Gamma_k}.
\end{equation}

\section{Review of Relationship Between Cluster State and KW Duality}\label{ap:KW}
The Kramers-Wannier duality is a mapping which enacts
\begin{equation}
    X\leftrightarrow XX, \quad ZZ\leftrightarrow Z, \quad Z\leftrightarrow ZZ\cdots,\quad XX\cdots \leftrightarrow X.
\end{equation}
It is also occasionally referred to as the gauging map because it maps a pure matter theory to the pure gauge theory obtained by gauging the pure matter theory then discarding the matter degrees of freedom. Note that our convention differs from the more common convention which exchanges $ZZ$ and $X$.

It was observed in~\cite{tantivasadakarn_long-range_2022} that the qubit cluster state MPS tensor, with odd physical legs flipped to form an operator, realizes this duality on the symmetric sector of the theory. Namely,
\begin{subequations}
\begin{equation}\label{eq:Z2_X_to_xx}
 \begin{tikzpicture}[baseline={(0,-.09)}]
    \draw (0,0)--(2.5,0);
    \foreach \x in {0.5,2}{
		     \draw (\x,0) -- (\x,0.5);

		    \node[evenZ2] at (\x,0) {};
		  }
		  \foreach \x in {1.25}{
		     \draw (\x,0) -- (\x,-1);
		    \node[oddZ2] at (\x,0) {};
		    \node[mpo] (t\x) at (\x,-0.5) {$X$};
		  }
  \end{tikzpicture} \ = \
\begin{tikzpicture}[baseline={(0,-.09)}]
    \draw (0,0)--(2.5,0);
    \foreach \x in {0.5,2}{
		     \draw (\x,0) -- (\x,1);
		    \node[evenZ2] at (\x,0) {};
		  }\
		  \node[mpo] at (0.5,0.5) {$X$};
		  \node[mpo] at (2,0.5) {$X$};
		  \foreach \x in {1.25}{
		     \draw (\x,0) -- (\x,-0.5);
		    \node[oddZ2] at (\x,0) {};
		  }
  \end{tikzpicture} \ ,
\end{equation}
\begin{equation}\label{eq:Z2_ZZ_to_Z}
 \begin{tikzpicture}[baseline={(0,-.09)}]
    \draw (0,0)--(2.5,0);
    \foreach \x in {0.5,2}{
		     \draw (\x,0) -- (\x,-1);
		    \node[oddZ2] at (\x,0) {};
		  }
		  \foreach \x in {1.25}{
		     \draw (\x,0) -- (\x,0.5);
		    \node[evenZ2] at (\x,0) {};
		  }
  \node[mpo] at (0.5,-0.5) {$Z$};
  \node[mpo] at (2,-0.5) {$Z$};
  \end{tikzpicture} \ = \
\begin{tikzpicture}[baseline={(0,-.09)}]
    \draw (0,0)--(2.5,0);
    \foreach \x in {0.5,2}{
		     \draw (\x,0) -- (\x,-0.5);
		    \node[oddZ2] at (\x,0) {};
		  }
		  \foreach \x in {1.25}{
		     \draw (\x,0) -- (\x,1);
		    \node[evenZ2] at (\x,0) {};
		  }
  \node[mpo] at (1.25,0.5) {$Z$};
  \end{tikzpicture} \ ,
\end{equation}
\begin{equation}
 \begin{tikzpicture}[baseline={(0,-.09)}]
    \draw (0,0)--(6.25,0);
    \draw (2,0) -- (2,-1);
    \foreach \x in {0.5,2,3.5,5}{
		     \draw (\x,0) -- (\x,-0.5);
		    \node[oddZ2] at (\x,0) {};
		  }
		  \foreach \x in {1.25,2.75,4.25,5.75}{
		     \draw (\x,0) -- (\x,0.5);
		    \node[evenZ2] at (\x,0) {};
		  }
  % \node[mpo] at (0.5,-0.5) {$Z$};
  \node[mpo] at (2,-0.5) {$Z$};
  \node[fill=white] at (6.5,0) {$\dots$};
  \end{tikzpicture} \ = \
\begin{tikzpicture}[baseline={(0,-.09)}]
    \draw (0,0)--(6.25,0);
		  \foreach \x in {1.25}{
		     \draw (\x,0) -- (\x,0.5);
		    \node[evenZ2] at (\x,0) {};
		  }
   \foreach \x in {2.75,4.25,5.75}{
        \draw (\x,0) -- (\x,1);
		     \node[mpo] at (\x,0.5) {$Z$};
       \node[evenZ2] at (\x,0) {};
		  }
    \foreach \x in {0.5,2,3.5,5}{
		     \draw (\x,0) -- (\x,-0.5);
		    \node[oddZ2] at (\x,0) {};
		  }
    \node[fill=white] at (6.5,0) {$\dots$};
  % \node[mpo] at (0.5,-0.5) {$Z$};
  \end{tikzpicture}  \ ,
\end{equation}
\begin{equation}
 \begin{tikzpicture}[baseline={(0,-.09)}]
    \draw (0,0)--(6.25,0);
    \foreach \x in {0.5}{
		     \draw (\x,0) -- (\x,-0.5);
		    \node[oddZ2] at (\x,0) {};
		  }
    \foreach \x in {2,3.5,5}{
        \draw (\x,0) -- (\x,-1);
		     \node[mpo] at (\x,-0.5) {$X$};
		    \node[oddZ2] at (\x,0) {};
		  }
		  \foreach \x in {1.25,2.75,4.25,5.75}{
		     \draw (\x,0) -- (\x,0.5);
		    \node[evenZ2] at (\x,0) {};
		  }
  % \node[mpo] at (0.5,-0.5) {$Z$};
  \node[fill=white] at (6.5,0) {$\dots$};
  \end{tikzpicture} \ = \
\begin{tikzpicture}[baseline={(0,-.09)}]
    \draw (0,0)--(6.25,0);
		  \foreach \x in {2.75,4.25,5.75}{
		     \draw (\x,0) -- (\x,0.5);
		    \node[evenZ2] at (\x,0) {};
		  }
   \foreach \x in {1.25}{
        \draw (\x,0) -- (\x,1);
		     \node[mpo] at (\x,0.5) {$X$};
       \node[evenZ2] at (\x,0) {};
		  }
    \foreach \x in {0.5,2,3.5,5}{
		     \draw (\x,0) -- (\x,-0.5);
		    \node[oddZ2] at (\x,0) {};
		  }
    \node[fill=white] at (6.5,0) {$\dots$};
  \end{tikzpicture}  \ .
\end{equation}
\end{subequations}
If we generalize the tensors in \cref{eq:Z2_X_to_xx} and \cref{eq:Z2_ZZ_to_Z} from $G=\ztwo$ to arbitrary non-abelian $G$, we find
\begin{subequations}
\begin{equation}
 \begin{tikzpicture}[baseline={(0,-.09)}]
    \draw (0,0)--(2.5,0);
    \foreach \x in {0.5,2}{
		     \draw (\x,0) -- (\x,0.5);

		    \node[even] at (\x,0) {};
		  }
		  \foreach \x in {1.25}{
		     \draw (\x,0) -- (\x,-1);
		    \node[odd] at (\x,0) {};
		    \node[mpo] (t\x) at (\x,-0.5) {$\lx_g$};
		  }
  \end{tikzpicture} \ = \
\begin{tikzpicture}[baseline={(0,-.09)}]
    \draw (0,0)--(2.5,0);
    \foreach \x in {0.5,2}{
		     \draw (\x,0) -- (\x,1);
		    \node[even] at (\x,0) {};
		  }\
		  \node[mpo] at (0.5,0.5) {$\rx_g$};
		  \node[mpo] at (2,0.5) {$\lx_g$};
		  \foreach \x in {1.25}{
		     \draw (\x,0) -- (\x,-0.5);
		    \node[odd] at (\x,0) {};
		  }
  \end{tikzpicture} \ ,
\end{equation}
\begin{equation}
 \begin{tikzpicture}[baseline={(0,-.09)}]
    \draw (0,0)--(2.5,0);
    \draw[red] (0,-0.5) -- (2.5,-0.5);
    \foreach \x in {0.5,2}{
		     \draw (\x,0) -- (\x,-1);
		    \node[odd] at (\x,0) {};
		  }
		  \foreach \x in {1.25}{
		     \draw (\x,0) -- (\x,0.5);
		    \node[even] at (\x,0) {};
		  }
  \node[mpo] at (0.5,-0.5) {$\zt_\Gamma$};
  \node[mpo] at (2,-0.5) {$\zt^\dagger_\Gamma$};
  \end{tikzpicture} \ = \
\begin{tikzpicture}[baseline={(0,-.09)}]
    \draw (0,0)--(2.5,0);
    \draw[red] (0,0.5) -- (2.5,0.5);
    \foreach \x in {0.5,2}{
		     \draw (\x,0) -- (\x,-0.5);
		    \node[odd] at (\x,0) {};
		  }
		  \foreach \x in {1.25}{
		     \draw (\x,0) -- (\x,1);
		    \node[even] at (\x,0) {};
		  }
  \node[mpo] at (1.25,0.5) {$\zt_\Gamma$};
  \end{tikzpicture} \ .
\end{equation}
\end{subequations}
which are used to implement the duality in the main text.

\section{Edge Modes from Hamiltonian}\label{ap:edge_modes}
In the main text, we used the MPS form of the $G$ cluster state to derive the edge modes. However, it is also possible to do so without refering to the explicit form of the ground state, using only the stabilizer Hamiltonian. We can rewrite the action of the global symmetries on the $G$ cluster state using the fact that the terms of the Hamiltonian act stabilize the $G$ cluster state, so that they can be inserted to the right of the symmetry without changing its effect on the $G$ cluster state. Let $\hat X|_{\ket{\Cl}}$ denote the effective action of an arbitrary operator $\hat X$ on the $G$ cluster state. For the $X$-type symmetry $\overleftarrow{A}_g$, we then have
\begin{equation}
    \begin{split}
        \overleftarrow{A}_g|_{\ket{\Cl}}=\left(\prod_{i\text{ odd}} \rx_g^{(i)}\right)\left(\rx_g^{(2)}\lx_g^{(3)}\lx_g^{(4)}\right)\left(\rx_g^{(4)}\lx_g^{(5)}\lx_g^{(6)}\right)\cdots\\
        =\left\{\begin{array}{lr}
             \rx_g^{(1)}\rx_g^{(2)}\mathbbm{1}^{(3)}\cdots \mathbbm{1}^{(N-1)}\lx_g^{(N)},& N\text{ even}\\
             \rx_g^{(1)}\rx_g^{(2)}\mathbbm{1}^{(3)}\cdots\mathbbm{1}^{(N-2)} \lx_g^{(N-1)}\rx_g^{(N)},& N\text{ odd}
        \end{array}\right..
    \end{split}
\end{equation}
We can apply a similar procedure to analyze the $Z$-type symmetry $\hat B_\Gamma$:
\begin{equation}
    \begin{split}
        \hat{B_\Gamma}|_{\ket{\Cl}}&=\left.\Tr\left[\prod_{i\text{ even}}\zt^{(i)}_\Gamma\right]\right|_{\ket{\Cl}}\\
        &=\left\{\begin{array}{lr}
             \Tr\left[\zt_\Gamma^{(2)}.\zt_\Gamma^{(4)}.\cdots.\zt_\Gamma^{(N-2)}\zt_\Gamma^{(N)}\right],& N\text{ even}\\
             \Tr\left[\zt_\Gamma^{(2)}.\zt_\Gamma^{(4)}.\cdots.\zt_\Gamma^{(N-3)}\zt_\Gamma^{(N-1)}\right],& N\text{ odd}
        \end{array}\right.\\
        &=\left\{\begin{array}{lr}
             \Tr\left[\left(\zt_\Gamma^{(1)}.\zt_\Gamma^{\dagger(3)}\right).\left(\zt_\Gamma^{(3)}.\zt_\Gamma^{\dagger(5)}\right).\cdots.\left(\zt_\Gamma^{(N-3)}.\zt_\Gamma^{\dagger(N-1)}\right).\zt_\Gamma^{(N)}\right],& N\text{ even}\\
             \Tr\left[\left(\zt_\Gamma^{(1)}.\zt_\Gamma^{\dagger(3)}\right).\left(\zt_\Gamma^{(3)}.\zt_\Gamma^{\dagger(5)}\right).\cdots.\left(\zt_\Gamma^{(N-4)}.\zt_\Gamma^{\dagger(N-2)}\right).\left(\zt_\Gamma^{(N-2)}.\zt_\Gamma^{\dagger(N)}\right)\right],& N\text{ odd}
        \end{array}\right.\\
        &=\left\{\begin{array}{lr}
             \Tr\left[\zt_\Gamma^{(1)}.\zt_\Gamma^{\dagger(N-1)}.\zt_\Gamma^{(N)}\right],& N\text{ even}\\
             \Tr\left[\zt_\Gamma^{(1)}.\zt_\Gamma^{\dagger(N)}\right],& N\text{ odd}
        \end{array}\right.
    \end{split},
\end{equation}
where we have used the fact that $\zt_\Gamma^{(i)}.\zt_\Gamma^{\dagger(i)}=\zt_\Gamma^{\dagger(i)}.\zt_\Gamma^{(i)}=\mathbbm{1}_{d_\Gamma}$ because $\zt$ is unitary as well as the fact that for even $i$ we have $\zt_\Gamma^{(i)}=\zt_\Gamma^{(i-1)}.\zt_\Gamma^{\dagger(i+1)}$ on the code space because $\zt_\Gamma^{\dagger(i-1)}.\zt_\Gamma^{(i)}.\zt_\Gamma^{(i+1)}$ is a stabilizer for even $i$.

It is important to note that this approach has the benefit of not relying on the explicit form of the ground state, but it also has the drawback of not clearly illustrating how the edge operators reduce to an action on the edge Hilbert space, as described in the main text.

\section{String Order Parameter}\label{ap:string_order}
In this appendix, we will prove that the string order parameter introduced in \cref{eq:string_order_param} of the main text has zero expectation value in the $G\times\rep G$-symmetric product state. We begin by proving \cref{eq:string_delta} which relates the string order parameter to the delta function on the group.

\begin{proof}
\begin{align}
\begin{split}
    \delta^G_{e,\bar{g}_i\left(\prod_{j=0}^{k-1}g_{i+1+2j}\right)g_{i+2k}}&=\sum_{\Gamma}\frac{d_\Gamma}{|G|}\Tr\left[\zt_\Gamma^{\dagger(i)}.\left(\prod_{j=0}^{k-1}\zt_\Gamma^{(i+1+2j)}\right).\zt_\Gamma^{(i+2k)}\right]\\
    &=\frac{|G|-1}{|G|}\hat{\mathcal{S}}^{(i,i+2k)}+\frac{d_\mathbf{1}}{|G|}\Tr\left[\zt_\mathbf{1}^{\dagger(i)}\left(\prod_{j=0}^{k-1}\zt_\mathbf{1}^{(i+1+2j)}\right).\zt_\mathbf{1}^{(i+2k)}\right]\\
    &=\frac{|G|-1}{|G|}\hat{\mathcal{S}}^{(i,i+2k)}+\frac{1}{|G|}\\
    \Rightarrow \hat{\mathcal{S}}^{(i,k)}&=\frac{|G|}{|G|-1}\delta^G_{e,\bar{g}_i\left(\prod_{j=0}^{k-1}g_{i+1+2j}\right)g_{i+2k}}-\frac{1}{|G|-1}
\end{split}
\end{align}
\end{proof}

We are now prepared to prove that the string order parameter has vanishing expectation value in the product state $\ket{\psi_0}$.
\begin{proof}
\begin{align}
\begin{split}
    \hat{\mathcal{S}}^{(i,k)}\ket{\psi_0}&=\frac{|G|}{|G|-1}\delta_{g_i,g_{i+2k}}\ket{\psi_0}-\frac{1}{|G|-1}\ket{\psi_0}\\
    &=\frac{1}{|G|-1}\sum_{g_i,g_{i+2k}}\delta_{g_i,g_{i+2k}}\ket{1,e,\ldots,g_i,\ldots,g_{i+2k},\ldots}-\frac{1}{|G|-1}\ket{\psi_0}\\
    &=\frac{1}{|G|-1}\sum_{g_i}\ket{1,e,\ldots,g_i,\ldots,g_{i},\ldots}-\frac{1}{|G|-1}\ket{\psi_0}\\
    \Rightarrow\bra{\psi_0}\hat{\mathcal{S}}^{(i,k)}\ket{\psi_0}&=\frac{1}{|G|-1}\bra{\psi_0}\sum_{g_i}\ket{1,e,\ldots,g_i,\ldots,g_{i},\ldots}-\frac{1}{|G|-1}\braket{\psi_0|\psi_0}\\
    &=\frac{1}{|G|-1}\bra{\mathbf{1},\mathbf{1}}\sum_{g_i}\ket{g_i,g_{i}}-\frac{1}{|G|-1}\\
    &=\frac{1}{|G|-1}\left(\frac{|G|}{\sqrt{|G|}^2}\right)-\frac{1}{|G|-1}\\
    &=0.
\end{split}
\end{align}
\end{proof}

\section{MBQC}
In this appendix, we will prove two claims which we used in our discussion of universal MBQC in the main text.
\subsection{Proof that ${\protect\overrightarrow\Lambda}$ is Imprimitive}\label{ap:imprimitive}
Consider a generic separable state in $\mathbb{C}[Q]\otimes \mathbb{C}[N]$:
\begin{equation}
   \ket{\psi_{QN}}:= \left(\sum_{q\in Q}c_q\ket{q}\right)\left(\sum_{n\in N}d_n\ket{n}\right)=\sum_{q\in Q, n\in N}c_qd_n\ket{q,n}.
\end{equation}
Acting on this state with $\overrightarrow\Lambda$, we find:
\begin{equation}
    \overrightarrow\Lambda \ket{\psi_{QN}}=\sum_{q\in Q, n\in N}c_qd_n\ket{q,\lambda(q)n}.
\end{equation}
As long as $\lambda(q)$ depends non-trivially on $q$ -- that is, there exists at least one pair $q_1,q_2\in Q$ such that $\lambda(q_1)\neq\lambda(q_2)$ -- then this state is not separable, and the gate $\overrightarrow\Lambda$ is imprimitive.

\subsection{Proof of Single-Qudit Universality}\label{ap:universality}
In this section we will prove that the gate set $R_G$ is universal for single $G$-qudit quantum computation.

    \noindent \textit{Proof.} Let $G$ be a finite abelian group and consider the gate set $R_G$. We want to show that the lie algebra $\mathfrak{a}$ generated by the generators of the rotations in $R_G$ is equivalent to $\mathfrak{su}(|G|)$. We can do this by calculating commutators of elements of $\mathfrak{a}$ and showing that this procedure produces $|G|^2$ independent Hermitian operators. We will show that there exists one independent Hermitian operator per group element besides the identity, each irrep besides the trivial irrep, and each product thereof. Along with the identity operator, this results in a total of $(|G|-1)+(|G|-1)+(|G|-1)^2+1=|G|^2$ independent Hermitian operators. It is clear from the form of the rotations in \cref{eq:left_rot} that we already have one Hermitian operator per nontrivial group element and irrep, so we will now proceed to construct the products of these operators by evaluating commutators.\\

    \textbf{Case 1)} Let $\Gamma(g)\neq 1$. This also implies that $\Gamma(\bar g)\neq 1$.\\
        \begin{adjustwidth}{2 em}{0 em}
        \textbf{Case 1.1)} Let $g=\bar g$, $\Gamma=\Gamma^\dagger$.\\
        We have $\lx_g$, $\zt_\Gamma\in\mathfrak{a}$. We can then evaluate $[\lx_g,\zt_\Gamma]=(1-\Gamma(g))\lx_g\zt_\Gamma$, so that $\lx_g\zt_\Gamma\in \mathfrak{a}$. This commutator is nonzero because we have taken $\Gamma(g)\neq 1$.\\

        \noindent\textbf{Case 1.2)} Let $g=\bar g$, $\Gamma\neq\Gamma^\dagger$.\\
        We have $\lx_g$, $(\zt_\Gamma+\zt_\Gamma^\dagger)$, $i(\zt_\Gamma-\zt_\Gamma^\dagger)\in\mathfrak{a}$, so we can construct the following elements of $\mathfrak{a}$:
        \begin{align*}
            \begin{split}
                \left[\lx_g,\zt_\Gamma+\zt_\Gamma^\dagger\right]&=(1-\Gamma(g))\lx_g(\zt_\Gamma+\zt_\Gamma^\dagger)\\
                \left[\lx_g,i(\zt_\Gamma-\zt_\Gamma^\dagger)\right]&=(1-\Gamma(g))i\lx_g(\zt_\Gamma-\zt_\Gamma^\dagger).
            \end{split}
        \end{align*}

        \noindent This means that $\lx_g(\zt_\Gamma+\zt_\Gamma^\dagger)$, $i\lx_g(\zt_\Gamma-\zt_\Gamma^\dagger)\in\mathfrak{a}$. We see that we generate two new basis elements, which accounts for the fact that we have used both $\Gamma$ and $\Gamma^\dagger$.\\

        \noindent\textbf{Case 1.3)} Let $g\neq\bar g$, $\Gamma=\Gamma^\dagger$.\\
        We have $(\lx_g+\lx_g^\dagger)$, $i(\lx_g-\lx_g^\dagger)$, $\zt_\Gamma\in\mathfrak{a}$, so we can construct the following elements of $\mathfrak{a}$:
        \begin{align*}
            \begin{split}
                \left[\lx_g+\lx_g^\dagger,\zt_\Gamma\right]&=(1-\Gamma(g))(\lx_g+\lx_g^\dagger)\zt_\Gamma\\
                \left[i(\lx_g-\lx_g^\dagger),\zt_\Gamma\right]&=(1-\Gamma(g))i(\lx_g-\lx_g^\dagger)\zt_\Gamma.
            \end{split}
        \end{align*} Therefore, $(\lx_g+\lx_g^\dagger)\zt_\Gamma$, $i(\lx_g-\lx_g^\dagger)\zt_\Gamma\in\mathfrak{a}$. As above, we generate two new basis elements, which accounts for the fact that we have used both $g$ and $\bar g$.\\

        \noindent\textbf{Case 1.4)} Let $g\neq\bar g$, $\Gamma\neq\Gamma^\dagger$.\\
        \noindent We have $(\lx_g+\lx_g^\dagger)$, $i(\lx_g-\lx_g^\dagger)$, $(\zt_\Gamma+\zt_\Gamma^\dagger)$, $i(\zt_\Gamma-\zt_\Gamma^\dagger)\in\mathfrak{a}$, so we can construct the following elements of $\mathfrak{a}$:
        \begin{align*}
            \begin{split}
                \left[(\lx_g+\lx_g^\dagger),(\zt_\Gamma+\zt_\Gamma^\dagger)\right]&=(1-\Gamma(g))\left(\lx_g\zt_\Gamma+\lx_g^\dagger\zt_\Gamma^\dagger\right)+(1-\Gamma(\bar g))\left(\lx_g^\dagger\zt_\Gamma+\lx_g\zt_\Gamma^\dagger\right),\\
                \left[i(\lx_g-\lx_g^\dagger),(\zt_\Gamma+\zt_\Gamma^\dagger)\right]&=i(1-\Gamma(g))\left(\lx_g\zt_\Gamma-\lx_g^\dagger\zt_\Gamma^\dagger\right)+i(1-\Gamma(\bar g))\left(-\lx_g^\dagger\zt_\Gamma+\lx_g\zt_\Gamma^\dagger\right),\\
                \left[(\lx_g+\lx_g^\dagger),i(\zt_\Gamma-\zt_\Gamma^\dagger)\right]&=i(1-\Gamma(g))\left(\lx_g\zt_\Gamma-\lx_g^\dagger\zt_\Gamma^\dagger\right)+i(1-\Gamma(\bar g))\left(\lx_g^\dagger\zt_\Gamma-\lx_g\zt_\Gamma^\dagger\right),\\
                \left[i(\lx_g-\lx_g^\dagger),i(\zt_\Gamma-\zt_\Gamma^\dagger)\right]&=-(1-\Gamma(g))\left(\lx_g\zt_\Gamma+\lx_g^\dagger\zt_\Gamma^\dagger\right)-(1-\Gamma(\bar g))\left(-\lx_g^\dagger\zt_\Gamma-\lx_g\zt_\Gamma^\dagger\right).
            \end{split}
        \end{align*} By summing these elements with various sets of coefficients, we can generate:
        \begingroup
        \renewcommand*{\arraystretch}{2}
        $$\begin{array}{c|c}
        \text{Coefficients} & \text{Sums To} \\ \hline
             \left(\frac{1-\text{Re}[\Gamma(g)]}{(\Gamma(g)-1)(\Gamma(\bar g)-1)},0,0,\frac{-i\text{Im}[\Gamma(g)]}{(\Gamma(g)-1)(\Gamma(\bar g)-1)}\right) & (\lx_g+\lx_g^\dagger)(\zt_\Gamma+\zt_\Gamma^\dagger) \\
             \left(0,\frac{i\text{Im}[\Gamma(g)]}{(\Gamma(g)-1)(\Gamma(\bar g)-1)},\frac{1-\text{Re}[\Gamma(g)]}{(\Gamma(g)-1)(\Gamma(\bar g)-1)},0\right) & i(\lx_g+\lx_g^\dagger)(\zt_\Gamma-\zt_\Gamma^\dagger)\\
             \left(0,\frac{1-\text{Re}[\Gamma(g)]}{(\Gamma(g)-1)(\Gamma(\bar g)-1)},\frac{i\text{Im}[\Gamma(g)]}{(\Gamma(g)-1)(\Gamma(\bar g)-1)},0\right) & i(\lx_g-\lx_g^\dagger)(\zt_\Gamma+\zt_\Gamma^\dagger)\\
             \left(\frac{i\text{Im}[\Gamma(g)]}{(\Gamma(g)-1)(\Gamma(\bar g)-1)},0,0,\frac{1}{2} \left(\frac{1}{\Gamma(g) -1}+\frac{1}{\Gamma (\bar g)-1}\right)\right) & (\lx_g-\lx_g^\dagger)(\zt_\Gamma-\zt_\Gamma^\dagger)\\
        \end{array}$$
        \endgroup
        We can therefore conclude that $(\lx_g+\lx_g^\dagger)(\zt_\Gamma+\zt_\Gamma^\dagger)$, $i(\lx_g+\lx_g^\dagger)(\zt_\Gamma-\zt_\Gamma^\dagger)$, $i(\lx_g-\lx_g^\dagger)(\zt_\Gamma+\zt_\Gamma^\dagger)$, $(\lx_g-\lx_g^\dagger)(\zt_\Gamma-\zt_\Gamma^\dagger)\in\mathfrak{a}$.\\
        \end{adjustwidth}

\noindent Notice that when $\Gamma(g)=1$, the commutators in \textbf{Case 1} evaluate to 0. In order to generate basis elements corresponding to these combinations, we will have to circumvent this issue by taking further commutators.\\

\noindent\textbf{Case 2)} Let $\Gamma(g)=1$.\\
    \noindent By the well-known fact that the intersection of the kernels of all irreducible characters of a finite group is trivial (the only $g\in G$ such that $\Gamma(g)=1$ for all $\Gamma\in\irr G$ is $e$), there exists some irrep $\Gamma'$ such that $\Gamma'(g)\neq1$ since $g\neq e$. Because in the case of abelian $G$ the irreps form an abelian group, there also exists some $\Gamma''$ such that $\Gamma'\Gamma''=\Gamma$. Furthermore, because $\Gamma'(g)\neq 1$ and $\Gamma'(g)\Gamma''(g)=\Gamma(g)=1$, we know that $\Gamma''(g)\neq 1$.\\
    \begin{adjustwidth}{2 em}{0 em}
    \textbf{Case 2.1)} Let $g=\bar g$, $\Gamma=\Gamma^\dagger$.\\
    Because $\Gamma$ is real, $\Gamma'$ and $\Gamma''$ must either both be real or both be complex. We will consider these cases separately:\\
    \begin{adjustwidth}{2 em}{0 em}
    \textbf{Case 2.1.1)} Let $\Gamma',\Gamma''$ real.\\
    \noindent We know that $\zt_{\Gamma''}\in\mathfrak{a}$, and we see from \textbf{Case 1} that $\lx_g\zt_{\Gamma'}\in\mathfrak{a}$. We can then evaluate $[\lx_g\zt_{\Gamma'},\zt_{\Gamma''}]=(1-\Gamma''(g))\lx_g\zt_\Gamma$, so that $\lx_g\zt_\Gamma\in\mathfrak{a}$. This is nonzero because $\Gamma''(g)\neq0$, and we have used the fact that $\zt_{\Gamma'}\zt_{\Gamma''}=\zt_{\Gamma}$. \\

    \noindent\textbf{Case 2.1.2)} Let $\Gamma',\Gamma''$ complex.\\
    \noindent We know that $\lx_g(\zt_{\Gamma'}+\zt_{\Gamma'}^\dagger)$, $i\lx_g(\zt_{\Gamma'}-\zt_{\Gamma'}^\dagger)$, $(\zt_{\Gamma''}+\zt_{\Gamma''}^\dagger)$, $i(\zt_{\Gamma''}-\zt_{\Gamma''}^\dagger)\in \mathfrak{a}$. This means that we can construct the following elements of $\mathfrak{a}$:\begin{align*}
    \begin{split}
        \left[\lx_g(\zt_{\Gamma'}+\zt_{\Gamma'}^\dagger),(\zt_{\Gamma''}+\zt_{\Gamma''}^\dagger)\right]&=(1-\Gamma''(g))\left(\lx_g\zt_\Gamma+\lx_g\zt_{\Gamma'^\dagger\Gamma''}
        \right)+(1-\Gamma''(\bar g))\left(\lx_g\zt_\Gamma+\lx_g\zt_{\Gamma'\Gamma''^\dagger}
        \right),\\
        \left[\lx_g(\zt_{\Gamma'}+\zt_{\Gamma'}^\dagger),i(\zt_{\Gamma''}-\zt_{\Gamma''}^\dagger)\right]&=i(1-\Gamma''(g))\left(\lx_g\zt_\Gamma+\lx_g\zt_{\Gamma'^\dagger\Gamma''}
        \right)+i(1-\Gamma''(\bar g))\left(-\lx_g\zt_\Gamma-\lx_g\zt_{\Gamma'\Gamma''^\dagger}
        \right),\\
        \left[i\lx_g(\zt_{\Gamma'}-\zt_{\Gamma'}^\dagger),(\zt_{\Gamma''}+\zt_{\Gamma''}^\dagger)\right]&=i(1-\Gamma''(g))\left(\lx_g\zt_\Gamma-\lx_g\zt_{\Gamma'^\dagger\Gamma''}
        \right)+i(1-\Gamma''(\bar g))\left(-\lx_g\zt_\Gamma+\lx_g\zt_{\Gamma'\Gamma''^\dagger}
        \right),\\
        \left[i\lx_g(\zt_{\Gamma'}-\zt_{\Gamma'}^\dagger),i(\zt_{\Gamma''}-\zt_{\Gamma''}^\dagger)\right]&=-(1-\Gamma''(g))\left(\lx_g\zt_\Gamma-\lx_g\zt_{\Gamma'^\dagger\Gamma''}
        \right)-(1-\Gamma''(\bar g))\left(\lx_g\zt_\Gamma-\lx_g\zt_{\Gamma'\Gamma''^\dagger}
        \right).
    \end{split}
    \end{align*} \\
    By summing these four elements with appropriate coefficients $\frac{1}{4}(1,-i,-i,-1)$, we find $(1-\Gamma''(g))\lx_g\zt_\Gamma$, so that $\lx_g\zt_\Gamma\in\mathfrak{a}$. \\
    \end{adjustwidth}

    \noindent\textbf{Case 2.2)} Let $g=\bar g$, $\Gamma\neq\Gamma^\dagger$.\\
    \noindent Because $\Gamma$ is complex, we know that $\Gamma'$ and $\Gamma''$ cannot both be real. This leads to two distinct cases:\\

    \begin{adjustwidth}{2 em}{0 em}
    \textbf{Case 2.2.1)} Without loss of generality, let $\Gamma'$ be real, $\Gamma''$ be complex.\\
    We know that $\lx_g\zt_{\Gamma'}$, $(\zt_{\Gamma''}+\zt_{\Gamma''}^\dagger)$, $i(\zt_{\Gamma''}-\zt_{\Gamma''}^\dagger)\in \mathfrak{a}$. This means that we can construct the following elements of $\mathfrak{a}$:\begin{align*}
    \begin{split}
        \left[\lx_g\zt_{\Gamma'},(\zt_{\Gamma''}+\zt_{\Gamma''}^\dagger)\right]&=(1-\Gamma''(g))\left(\lx_g\zt_\Gamma+\lx_g\zt_{\Gamma}^\dagger
        \right),\\
        \left[\lx_g\zt_{\Gamma'},i(\zt_{\Gamma''}-\zt_{\Gamma''}^\dagger)\right]&=i(1-\Gamma''(g))\left(\lx_g\zt_\Gamma-\lx_g\zt_{\Gamma}^\dagger
        \right).\\
    \end{split}
    \end{align*} \\
    This means that $\lx_g(\zt_{\Gamma}+\zt_{\Gamma}^\dagger)$, $i\lx_g(\zt_{\Gamma}-\zt_{\Gamma}^\dagger)\in\mathfrak{a}$.\\

    \noindent\textbf{Case 2.2.2)} Let $\Gamma',\Gamma''$ complex.\\
    \noindent We know that $\lx_g(\zt_{\Gamma'}+\zt_{\Gamma'}^\dagger)$, $i\lx_g(\zt_{\Gamma'}-\zt_{\Gamma'}^\dagger)$, $(\zt_{\Gamma''}+\zt_{\Gamma''}^\dagger)$, $i(\zt_{\Gamma''}-\zt_{\Gamma''}^\dagger)\in \mathfrak{a}$. This means that we can construct the following elements of $\mathfrak{a}$:
    {\allowdisplaybreaks
    \begin{align*}
        \left[\lx_g(\zt_{\Gamma'}+\zt_{\Gamma'}^\dagger),(\zt_{\Gamma''}+\zt_{\Gamma''}^\dagger)\right]&=(1-\Gamma''(g))\lx_g\left(\zt_\Gamma+\zt_{\Gamma'^\dagger\Gamma''}
        +\zt_\Gamma^\dagger+\zt_{\Gamma'\Gamma''^\dagger}
        \right),\\
        \left[\lx_g(\zt_{\Gamma'}+\zt_{\Gamma'}^\dagger),i(\zt_{\Gamma''}-\zt_{\Gamma''}^\dagger)\right]&=i(1-\Gamma''(g))\lx_g\left(\zt_\Gamma+\zt_{\Gamma'^\dagger\Gamma''}
        -\zt_\Gamma^\dagger-\zt_{\Gamma'\Gamma''^\dagger}
        \right),\\
        \left[i\lx_g(\zt_{\Gamma'}-\zt_{\Gamma'}^\dagger),(\zt_{\Gamma''}+\zt_{\Gamma''}^\dagger)\right]&=i(1-\Gamma''(g))\lx_g\left(\zt_\Gamma-\zt_{\Gamma'^\dagger\Gamma''}
        -\zt_\Gamma^\dagger+\zt_{\Gamma'\Gamma''^\dagger}
        \right),\\
        \left[i\lx_g(\zt_{\Gamma'}-\zt_{\Gamma'}^\dagger),i(\zt_{\Gamma''}-\zt_{\Gamma''}^\dagger)\right]&=-(1-\Gamma''(g))\lx_g\left(\zt_\Gamma-\zt_{\Gamma'^\dagger\Gamma''}
        +\zt_\Gamma^\dagger-\zt_{\Gamma'\Gamma''^\dagger}
        \right).
    \end{align*}}\\
    By summing the first and last elements with the coefficients $\frac{1}{2}(1,-1)$, we find $(1-\Gamma''(g))\lx_g(\zt_\Gamma+\zt_\Gamma^\dagger)$. By summing the second and third elements with the coefficients $\frac{1}{2}(1,1)$, we find $(1-\Gamma''(g))i\lx_g(\zt_\Gamma-\zt_\Gamma^\dagger)$. We can therefore conclude that $\lx_g(\zt_\Gamma+\zt_\Gamma^\dagger)$, $i\lx_g(\zt_\Gamma-\zt_\Gamma^\dagger)\in\mathfrak{a}$.\\
    \end{adjustwidth}

    \noindent\textbf{Case 2.3)} Let $g\neq\bar g$, $\Gamma=\Gamma^\dagger$.\\
    Because $\Gamma$ is complex, we know that $\Gamma'$ and $\Gamma''$ cannot both be real. This leads to two distinct cases:\\
    \begin{adjustwidth}{2 em}{0 em}
    \textbf{Case 2.3.1)} Let $\Gamma',\Gamma''$ be real.\\
    We know that $(\lx_g+\lx_g^\dagger)\zt_{\Gamma'}$, $i(\lx_g-\lx_g^\dagger)\zt_{\Gamma'}$, $\zt_{\Gamma''}\in \mathfrak{a}$. This means that we can construct the following elements of $\mathfrak{a}$:\begin{align*}
    \begin{split}
        \left[(\lx_g+\lx_g^\dagger)\zt_{\Gamma'},\zt_{\Gamma''}\right]&=(1-\Gamma''(g))\left(\lx_g\zt_\Gamma+\lx_g^\dagger\zt_{\Gamma}
        \right),\\
        \left[i(\lx_g-\lx_g^\dagger)\zt_{\Gamma'},\zt_{\Gamma''}\right]&=i(1-\Gamma''(g))\left(\lx_g\zt_\Gamma-\lx_g^\dagger\zt_{\Gamma}
        \right).
    \end{split}
    \end{align*} \\
    This means that $(\lx_g+\lx_g^\dagger)\zt_{\Gamma}$, $i(\lx_g-\lx_g^\dagger)\zt_{\Gamma}\in\mathfrak{a}$.\\

    \noindent\textbf{Case 2.3.2)} Let $\Gamma',\Gamma''$ be complex.\\
    \noindent We know that $(\lx_g+\lx_g^\dagger)(\zt_{\Gamma'}+\zt_{\Gamma'}^\dagger)$, $i(\lx_g-\lx_g^\dagger)(\zt_{\Gamma'}+\zt_{\Gamma'}^\dagger)$, $i(\lx_g+\lx_g^\dagger)(\zt_{\Gamma'}-\zt_{\Gamma'}^\dagger)$, $(\lx_g-\lx_g^\dagger)(\zt_{\Gamma'}-\zt_{\Gamma'}^\dagger)$, $(\zt_{\Gamma''}+\zt_{\Gamma''}^\dagger)$, $i(\zt_{\Gamma''}-\zt_{\Gamma''}^\dagger)\in \mathfrak{a}$. This means that we can construct the following elements of $\mathfrak{a}$:
    {\allowdisplaybreaks
    \begin{align*}
        \left[(\lx_g+\lx_g^\dagger)(\zt_{\Gamma'}+\zt_{\Gamma'}^\dagger),(\zt_{\Gamma''}+\zt_{\Gamma''}^\dagger)\right]&=(1-\Gamma''(g))\left((\lx_g+\lx_g^\dagger)\zt_\Gamma+\lx_g\zt_{\Gamma'^\dagger\Gamma''}
        +\lx_g^\dagger\zt_{\Gamma'\Gamma''^\dagger}
        \right)\\&\quad+(1-\Gamma''(\bar{g}))\left((\lx_g+\lx_g^\dagger)\zt_\Gamma+\lx_g^\dagger\zt_{\Gamma'^\dagger\Gamma''}
        +\lx_g\zt_{\Gamma'\Gamma''^\dagger}
        \right),\\
        \left[(\lx_g+\lx_g^\dagger)(\zt_{\Gamma'}+\zt_{\Gamma'}^\dagger),i(\zt_{\Gamma''}-\zt_{\Gamma''}^\dagger)\right]&=i(1-\Gamma''(g))\left((\lx_g-\lx_g^\dagger)\zt_\Gamma+\lx_g\zt_{\Gamma'^\dagger\Gamma''}
        -\lx_g^\dagger\zt_{\Gamma'\Gamma''^\dagger}
        \right)\\&\quad+i(1-\Gamma''(\bar{g}))\left((-\lx_g+\lx_g^\dagger)\zt_\Gamma+\lx_g^\dagger\zt_{\Gamma'^\dagger\Gamma''}
        -\lx_g\zt_{\Gamma'\Gamma''^\dagger}
        \right),\\
        \left[i(\lx_g+\lx_g^\dagger)(\zt_{\Gamma'}-\zt_{\Gamma'}^\dagger),(\zt_{\Gamma''}+\zt_{\Gamma''}^\dagger)\right]&=i(1-\Gamma''(g))\left((\lx_g-\lx_g^\dagger)\zt_\Gamma-\lx_g\zt_{\Gamma'^\dagger\Gamma''}
        +\lx_g^\dagger\zt_{\Gamma'\Gamma''^\dagger}
        \right)\\&\quad+i(1-\Gamma''(\bar{g}))\left((-\lx_g+\lx_g^\dagger)\zt_\Gamma-\lx_g^\dagger\zt_{\Gamma'^\dagger\Gamma''}
        +\lx_g\zt_{\Gamma'\Gamma''^\dagger}
        \right),\\
        \left[i(\lx_g+\lx_g^\dagger)(\zt_{\Gamma'}-\zt_{\Gamma'}^\dagger),i(\zt_{\Gamma''}-\zt_{\Gamma''}^\dagger)\right]&=-(1-\Gamma''(g))\left((\lx_g+\lx_g^\dagger)\zt_\Gamma-\lx_g\zt_{\Gamma'^\dagger\Gamma''}
        -\lx_g^\dagger\zt_{\Gamma'\Gamma''^\dagger}
        \right)\\&\quad-(1-\Gamma''(\bar{g}))\left((\lx_g+\lx_g^\dagger)\zt_\Gamma-\lx_g^\dagger\zt_{\Gamma'^\dagger\Gamma''}
        -\lx_g\zt_{\Gamma'\Gamma''^\dagger}
        \right),\\
        \left[i(\lx_g-\lx_g^\dagger)(\zt_{\Gamma'}+\zt_{\Gamma'}^\dagger),(\zt_{\Gamma''}+\zt_{\Gamma''}^\dagger)\right]&=i(1-\Gamma''(g))\left((\lx_g-\lx_g^\dagger)\zt_\Gamma+\lx_g\zt_{\Gamma'^\dagger\Gamma''}
        -\lx_g^\dagger\zt_{\Gamma'\Gamma''^\dagger}
        \right)\\&\quad+i(1-\Gamma''(\bar{g}))\left((\lx_g-\lx_g^\dagger)\zt_\Gamma-\lx_g^\dagger\zt_{\Gamma'^\dagger\Gamma''}
        +\lx_g\zt_{\Gamma'\Gamma''^\dagger}
        \right),\\
        \left[i(\lx_g-\lx_g^\dagger)(\zt_{\Gamma'}+\zt_{\Gamma'}^\dagger),i(\zt_{\Gamma''}-\zt_{\Gamma''}^\dagger)\right]&=-(1-\Gamma''(g))\left((\lx_g+\lx_g^\dagger)\zt_\Gamma+\lx_g\zt_{\Gamma'^\dagger\Gamma''}
        +\lx_g^\dagger\zt_{\Gamma'\Gamma''^\dagger}
        \right)\\&\quad-(1-\Gamma''(\bar{g}))\left((-\lx_g-\lx_g^\dagger)\zt_\Gamma-\lx_g^\dagger\zt_{\Gamma'^\dagger\Gamma''}
        -\lx_g\zt_{\Gamma'\Gamma''^\dagger}
        \right),\\
        \left[(\lx_g-\lx_g^\dagger)(\zt_{\Gamma'}-\zt_{\Gamma'}^\dagger),(\zt_{\Gamma''}+\zt_{\Gamma''}^\dagger)\right]&=(1-\Gamma''(g))\left((\lx_g+\lx_g^\dagger)\zt_\Gamma-\lx_g\zt_{\Gamma'^\dagger\Gamma''}
        -\lx_g^\dagger\zt_{\Gamma'\Gamma''^\dagger}
        \right)\\&\quad+(1-\Gamma''(\bar{g}))\left((-\lx_g-\lx_g^\dagger)\zt_\Gamma+\lx_g^\dagger\zt_{\Gamma'^\dagger\Gamma''}
        +\lx_g\zt_{\Gamma'\Gamma''^\dagger}
        \right),\\
        \left[(\lx_g-\lx_g^\dagger)(\zt_{\Gamma'}-\zt_{\Gamma'}^\dagger),i(\zt_{\Gamma''}-\zt_{\Gamma''}^\dagger)\right]&=i(1-\Gamma''(g))\left((\lx_g-\lx_g^\dagger)\zt_\Gamma-\lx_g\zt_{\Gamma'^\dagger\Gamma''}
        +\lx_g^\dagger\zt_{\Gamma'\Gamma''^\dagger}
        \right)\\&\quad+i(1-\Gamma''(\bar{g}))\left((\lx_g-\lx_g^\dagger)\zt_\Gamma+\lx_g^\dagger\zt_{\Gamma'^\dagger\Gamma''}
        -\lx_g\zt_{\Gamma'\Gamma''^\dagger}
        \right).\\
    \end{align*}}
    By summing these elements with the coefficients $\frac{1}{4}(1,0,0,-1,0,-1,1,0)$, we find $(1-\Gamma''(g))(\lx_g+\lx_g^\dagger)\zt_\Gamma$. By summing these elements with the coefficients $\frac{1}{4}(0,1,1,0,1,0,0,1)$, we find $(1-\Gamma''(g))i(\lx_g-\lx_g^\dagger)\zt_\Gamma$. We can therefore conclude that $(\lx_g+\lx_g^\dagger)\zt_\Gamma$, $i(\lx_g-\lx_g^\dagger)\zt_\Gamma\in\mathfrak{a}$.\\
    \end{adjustwidth}

    \noindent\textbf{Case 2.4)} Let $g\neq\bar g$, $\Gamma\neq\Gamma^\dagger$.\\
    Because $\Gamma$ is complex, $\Gamma'$ and $\Gamma''$ must either both be real or both be complex. We will consider these cases separately:\\
    \begin{adjustwidth}{2 em}{0 em}
    \textbf{Case 2.4.1)} Without loss of generality, let $\Gamma'$ be complex, $\Gamma''$ be real.\\
    We know that $(\lx_g+\lx_g^\dagger)(\zt_{\Gamma'}+\zt_{\Gamma'}^\dagger)$, $i(\lx_g-\lx_g^\dagger)(\zt_{\Gamma'}+\zt_{\Gamma'}^\dagger)$, $i(\lx_g+\lx_g^\dagger)(\zt_{\Gamma'}-\zt_{\Gamma'}^\dagger)$, $(\lx_g-\lx_g^\dagger)(\zt_{\Gamma'}-\zt_{\Gamma'}^\dagger)$, $\zt_{\Gamma''}\in \mathfrak{a}$. This means that we can construct the following elements of $\mathfrak{a}$:
    \begin{align*}
    \begin{split}
        \left[(\lx_g+\lx_g^\dagger)(\zt_{\Gamma'}+\zt_{\Gamma'}^\dagger),\zt_{\Gamma''}\right]&=(1-\Gamma''(g))\left(\lx_g\zt_\Gamma+\lx_g^\dagger\zt_{\Gamma}^\dagger
        +\lx_g^\dagger\zt_\Gamma+\lx_g\zt_{\Gamma}^\dagger
        \right),\\
        \left[i(\lx_g-\lx_g^\dagger)(\zt_{\Gamma'}+\zt_{\Gamma'}^\dagger),\zt_{\Gamma''}\right]&=i(1-\Gamma''(g))\left(\lx_g\zt_\Gamma-\lx_g^\dagger\zt_{\Gamma}^\dagger
        -\lx_g^\dagger\zt_\Gamma+\lx_g\zt_{\Gamma}^\dagger
        \right),\\
        \left[i(\lx_g+\lx_g^\dagger)(\zt_{\Gamma'}-\zt_{\Gamma'}^\dagger),\zt_{\Gamma''}\right]&=i(1-\Gamma''(g))\left(\lx_g\zt_\Gamma-\lx_g^\dagger\zt_{\Gamma}^\dagger
        +\lx_g^\dagger\zt_\Gamma-\lx_g\zt_{\Gamma}^\dagger
        \right),\\
        \left[(\lx_g-\lx_g^\dagger)(\zt_{\Gamma'}-\zt_{\Gamma'}^\dagger),\zt_{\Gamma''}\right]&=(1-\Gamma''(g))\left(\lx_g\zt_\Gamma+\lx_g^\dagger\zt_{\Gamma}^\dagger
        -\lx_g^\dagger\zt_\Gamma-\lx_g\zt_{\Gamma}^\dagger
        \right).\\
    \end{split}
    \end{align*} \\
    This means that $(\lx_g+\lx_g^\dagger)(\zt_{\Gamma}+\zt_{\Gamma}^\dagger)$, $i(\lx_g-\lx_g^\dagger)(\zt_{\Gamma}+\zt_{\Gamma}^\dagger)$, $i(\lx_g+\lx_g^\dagger)(\zt_{\Gamma}-\zt_{\Gamma}^\dagger)$, $(\lx_g-\lx_g^\dagger)(\zt_{\Gamma}-\zt_{\Gamma}^\dagger)\in \mathfrak{a}$.\\

    \noindent\textbf{Case 2.4.2)} Let $\Gamma'$, $\Gamma''$ be complex.\\
    \noindent We know that $(\lx_g+\lx_g^\dagger)(\zt_{\Gamma'}+\zt_{\Gamma'}^\dagger)$, $i(\lx_g-\lx_g^\dagger)(\zt_{\Gamma'}+\zt_{\Gamma'}^\dagger)$, $i(\lx_g+\lx_g^\dagger)(\zt_{\Gamma'}-\zt_{\Gamma'}^\dagger)$, $(\lx_g-\lx_g^\dagger)(\zt_{\Gamma'}-\zt_{\Gamma'}^\dagger)$, $(\zt_{\Gamma''}+\zt_{\Gamma''}^\dagger)$, $i(\zt_{\Gamma''}-\zt_{\Gamma''}^\dagger)\in \mathfrak{a}$. This means that we can construct the following elements of $\mathfrak{a}$:{\allowdisplaybreaks
    \begin{align*}
        \left[(\lx_g+\lx_g^\dagger)(\zt_{\Gamma'}+\zt_{\Gamma'}^\dagger),(\zt_{\Gamma''}+\zt_{\Gamma''}^\dagger)\right]&=(1-\Gamma''(g))\left(\lx_g\zt_\Gamma+\lx_g^\dagger\zt_\Gamma^\dagger+\lx_g\zt_{\Gamma'^\dagger\Gamma''}
        +\lx_g^\dagger\zt_{\Gamma'\Gamma''^\dagger}
        \right)\\&\quad+(1-\Gamma''(\bar{g}))\left(\lx_g\zt_\Gamma^\dagger+\lx_g^\dagger\zt_\Gamma+\lx_g^\dagger\zt_{\Gamma'^\dagger\Gamma''}
        +\lx_g\zt_{\Gamma'\Gamma''^\dagger}
        \right),\\
        \left[(\lx_g+\lx_g^\dagger)(\zt_{\Gamma'}+\zt_{\Gamma'}^\dagger),i(\zt_{\Gamma''}-\zt_{\Gamma''}^\dagger)\right]&=i(1-\Gamma''(g))\left(\lx_g\zt_\Gamma-\lx_g^\dagger\zt_\Gamma^\dagger+\lx_g\zt_{\Gamma'^\dagger\Gamma''}
        -\lx_g^\dagger\zt_{\Gamma'\Gamma''^\dagger}
        \right)\\&\quad+i(1-\Gamma''(\bar{g}))\left(-\lx_g\zt_\Gamma^\dagger+\lx_g^\dagger\zt_\Gamma+\lx_g^\dagger\zt_{\Gamma'^\dagger\Gamma''}
        -\lx_g\zt_{\Gamma'\Gamma''^\dagger}
        \right),\\
        \left[i(\lx_g+\lx_g^\dagger)(\zt_{\Gamma'}-\zt_{\Gamma'}^\dagger),(\zt_{\Gamma''}+\zt_{\Gamma''}^\dagger)\right]&=i(1-\Gamma''(g))\left(\lx_g\zt_\Gamma-\lx_g^\dagger\zt_\Gamma^\dagger-\lx_g\zt_{\Gamma'^\dagger\Gamma''}
        +\lx_g^\dagger\zt_{\Gamma'\Gamma''^\dagger}
        \right)\\&\quad+i(1-\Gamma''(\bar{g}))\left(-\lx_g\zt_\Gamma^\dagger+\lx_g^\dagger\zt_\Gamma-\lx_g^\dagger\zt_{\Gamma'^\dagger\Gamma''}
        +\lx_g\zt_{\Gamma'\Gamma''^\dagger}
        \right),\\
        \left[i(\lx_g+\lx_g^\dagger)(\zt_{\Gamma'}-\zt_{\Gamma'}^\dagger),i(\zt_{\Gamma''}-\zt_{\Gamma''}^\dagger)\right]&=-(1-\Gamma''(g))\left(\lx_g\zt_\Gamma+\lx_g^\dagger\zt_\Gamma^\dagger-\lx_g\zt_{\Gamma'^\dagger\Gamma''}
        -\lx_g^\dagger\zt_{\Gamma'\Gamma''^\dagger}
        \right)\\&\quad-(1-\Gamma''(\bar{g}))\left(\lx_g\zt_\Gamma^\dagger+\lx_g^\dagger\zt_\Gamma-\lx_g^\dagger\zt_{\Gamma'^\dagger\Gamma''}
        -\lx_g\zt_{\Gamma'\Gamma''^\dagger}
        \right),\\
        \left[i(\lx_g-\lx_g^\dagger)(\zt_{\Gamma'}+\zt_{\Gamma'}^\dagger),(\zt_{\Gamma''}+\zt_{\Gamma''}^\dagger)\right]&=i(1-\Gamma''(g))\left(\lx_g\zt_\Gamma-\lx_g^\dagger\zt_\Gamma^\dagger+\lx_g\zt_{\Gamma'^\dagger\Gamma''}
        -\lx_g^\dagger\zt_{\Gamma'\Gamma''^\dagger}
        \right)\\&\quad+i(1-\Gamma''(\bar{g}))\left(\lx_g\zt_\Gamma^\dagger-\lx_g^\dagger\zt_\Gamma-\lx_g^\dagger\zt_{\Gamma'^\dagger\Gamma''}
        +\lx_g\zt_{\Gamma'\Gamma''^\dagger}
        \right),\\
        \left[i(\lx_g-\lx_g^\dagger)(\zt_{\Gamma'}+\zt_{\Gamma'}^\dagger),i(\zt_{\Gamma''}-\zt_{\Gamma''}^\dagger)\right]&=-(1-\Gamma''(g))\left(\lx_g\zt_\Gamma+\lx_g^\dagger\zt_\Gamma^\dagger+\lx_g\zt_{\Gamma'^\dagger\Gamma''}
        +\lx_g^\dagger\zt_{\Gamma'\Gamma''^\dagger}
        \right)\\&\quad-(1-\Gamma''(\bar{g}))\left(-\lx_g\zt_\Gamma^\dagger-\lx_g^\dagger\zt_\Gamma-\lx_g^\dagger\zt_{\Gamma'^\dagger\Gamma''}
        -\lx_g\zt_{\Gamma'\Gamma''^\dagger}
        \right),\\
        \left[(\lx_g-\lx_g^\dagger)(\zt_{\Gamma'}-\zt_{\Gamma'}^\dagger),(\zt_{\Gamma''}+\zt_{\Gamma''}^\dagger)\right]&=(1-\Gamma''(g))\left(\lx_g\zt_\Gamma+\lx_g^\dagger\zt_\Gamma^\dagger-\lx_g\zt_{\Gamma'^\dagger\Gamma''}
        -\lx_g^\dagger\zt_{\Gamma'\Gamma''^\dagger}
        \right)\\&\quad+(1-\Gamma''(\bar{g}))\left(-\lx_g\zt_\Gamma^\dagger-\lx_g^\dagger\zt_\Gamma+\lx_g^\dagger\zt_{\Gamma'^\dagger\Gamma''}
        +\lx_g\zt_{\Gamma'\Gamma''^\dagger}
        \right),\\
        \left[(\lx_g-\lx_g^\dagger)(\zt_{\Gamma'}-\zt_{\Gamma'}^\dagger),i(\zt_{\Gamma''}-\zt_{\Gamma''}^\dagger)\right]&=i(1-\Gamma''(g))\left(\lx_g\zt_\Gamma-\lx_g^\dagger\zt_\Gamma^\dagger-\lx_g\zt_{\Gamma'^\dagger\Gamma''}
        +\lx_g^\dagger\zt_{\Gamma'\Gamma''^\dagger}
        \right)\\&\quad+i(1-\Gamma''(\bar{g}))\left(\lx_g\zt_\Gamma^\dagger-\lx_g^\dagger\zt_\Gamma+\lx_g^\dagger\zt_{\Gamma'^\dagger\Gamma''}
        -\lx_g\zt_{\Gamma'\Gamma''^\dagger}
        \right).\\
    \end{align*}}
    By summing these elements with various sets of coefficients, we can generate:
        \begingroup
        \renewcommand*{\arraystretch}{2}
        $$\begin{array}{c|c}
        \text{Coefficients} & \text{Sums To} \\ \hline
             \left(\frac{1-\Re(\Gamma''(g) )}{2 (\Gamma''(g) -1) \left(\Gamma''(\bar g)-1\right)},0,0,\frac{1}{4} \left(\frac{1}{\Gamma''(g) -1}+\frac{1}{\Gamma''(\bar g)-1}\right),0,\frac{-i \Im(\Gamma''(g) )}{2 (\Gamma''(g) -1) \left(\Gamma''(\bar g)-1\right)},\frac{i \Im(\Gamma''(g) )}{2 (\Gamma''(g) -1) \left(\Gamma''(\bar g)-1\right)},0\right) & (\lx_g+\lx_g^\dagger)(\zt_\Gamma+\zt_\Gamma^\dagger) \\
             \left(0,\frac{1-\Re(\Gamma''(g) )}{2 (\Gamma''(g) -1) \left(\Gamma''(\bar g)-1\right)},\frac{1-\Re(\Gamma''(g) )}{2 (\Gamma''(g) -1) \left(\Gamma''(\bar g)-1\right)},0,\frac{i \Im(\Gamma''(g) )}{2 (\Gamma''(g) -1) \left(\Gamma''(\bar g)-1\right)},0,0,\frac{i \Im(\Gamma''(g) )}{2 (\Gamma''(g) -1) \left(\Gamma''(\bar g)-1\right)}\right) & i(\lx_g+\lx_g^\dagger)(\zt_\Gamma-\zt_\Gamma^\dagger)\\
             \left(0,\frac{i \Im(\Gamma''(g) )}{2 (\Gamma''(g) -1) \left(\Gamma''(\bar g)-1\right)},\frac{i \Im(\Gamma''(g) )}{2 (\Gamma''(g) -1) \left(\Gamma''(\bar g)-1\right)},0,\frac{1-\Re(\Gamma''(g) )}{2 (\Gamma''(g) -1) \left(\Gamma''(\bar g)-1\right)},0,0,\frac{1-\Re(\Gamma''(g) )}{2 (\Gamma''(g) -1) \left(\Gamma''(\bar g)-1\right)}\right) & i(\lx_g-\lx_g^\dagger)(\zt_\Gamma+\zt_\Gamma^\dagger)\\
             \left(\frac{i \Im(\Gamma''(g) )}{2 (\Gamma''(g) -1) \left(\Gamma''(\bar g)-1\right)},0,0,\frac{-i \Im(\Gamma''(g) )}{2 (\Gamma''(g) -1) \left(\Gamma''(\bar g)-1\right)},0,\frac{1}{4} \left(\frac{1}{\Gamma''(g) -1}+\frac{1}{\Gamma''(\bar g)-1}\right),\frac{1-\Re(\Gamma''(g) )}{2 (\Gamma''(g) -1) \left(\Gamma''(\bar g)-1\right)},0\right) & (\lx_g-\lx_g^\dagger)(\zt_\Gamma-\zt_\Gamma^\dagger)\\
        \end{array}$$
        \endgroup
        We can therefore conclude that $(\lx_g+\lx_g^\dagger)(\zt_\Gamma+\zt_\Gamma^\dagger)$, $i(\lx_g+\lx_g^\dagger)(\zt_\Gamma-\zt_\Gamma^\dagger)$, $i(\lx_g-\lx_g^\dagger)(\zt_\Gamma+\zt_\Gamma^\dagger)$, $(\lx_g-\lx_g^\dagger)(\zt_\Gamma-\zt_\Gamma^\dagger)\in\mathfrak{a}$.\\
    \end{adjustwidth}

    \end{adjustwidth}
    \noindent Having demonstrated that we can construct $(|G|-1)^2$ independent Hermitian operators, in addition to the identity and the $2|G|-2$ operators with which we began, we can conclude that  $\mathfrak{a}=\mathfrak{su}(|G|)$.

\section{SymTFT construction of group-based cluster state}\label{app:SymTFT}

In this appendix, we provide a SymTFT construction of the group-based cluster state, as well as relating the abstract construction to the explicit lattice model discussed in the paper.

A 1+1D lattice model with $G\times \rep{G}$ symmetry can be realized via the SymTFT~\cite{lichtman_bulk_2021,moradi2022topological,apruzzi2021symmetry,Freed2022topological,Bhardwaj23,Inamura2023} construction using a topological order in one-higher dimension, namely $\mathcal D(G\times G)$. The reference boundary is the gapped boundary where all charges of the first copy is condensed, and all fluxes of the second copy is condensed.

In this paper, we considered two types of condensations on the physical boundary to obtain the lattice model of interest. The first is where all fluxes of the first copy and all charges of the second copy is condensed. This results in the product state SPT. The second is where we condense all anyons according to the algebra $\mathcal A =\bigoplus_{\boldsymbol a \in \mathcal D(G)} d_{\boldsymbol a} \boldsymbol a \otimes \bar {\boldsymbol a}$. We claim that this gives rise to the cluster state SPT.

\begin{figure}[h]
    \centering
    \includegraphics{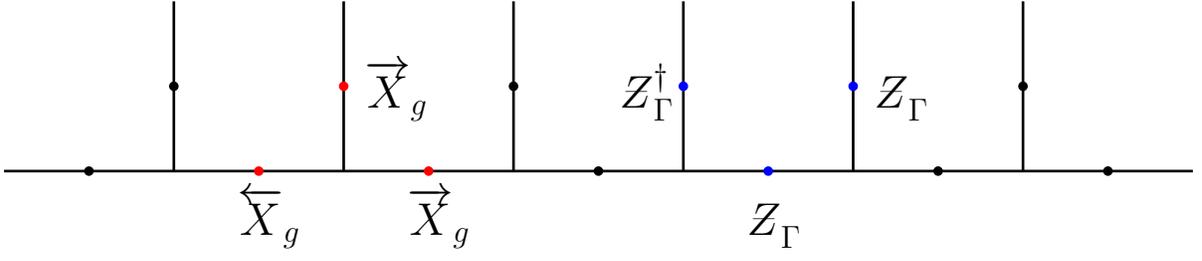}
    \caption{The group based cluster state can be obtained from a thin strip of the quantum double model with a rough boundary on side and a smooth boundary on the other. The vertex term (red) and plaquette term (blue) gives rise to the stabilizers of the cluster Hamiltonian.}
    \label{fig:clusterfromstrip}
\end{figure}

First, we notice that the boundary $\mathcal A$ corresponds to a trivial defect after unfolding. Thus the construction of the cluster state SPT corresponds to a thin model of $ D(G)$ where all fluxes are condensed on one side and all charges are condensed on the other after unfolding into a single copy. We may now construct an explicit lattice model for this SPT using the Kitaev's lattice model for the quantum double\cite{kitaev_fault-tolerant_2003-2}. Consider the thinnest slab of $\mathcal D(G)$ where we choose the smooth boundary on one side (corresponding to condensing all fluxes) and the rough boundary on the other side (corresponding to condensing all fluxes)\footnote{We thank Sahand Seifnashri for pointing out this construction to us} (see Fig.~\ref{fig:clusterfromstrip}). We can explicitly see that the Hamiltonian for the quantum double model defined on this lattice is exactly the cluster Hamiltonian. Morever, the $G$ and $\rep{G}$ symmetry correspond to the Wilson lines of the fluxes and charges, respectively.

\subsection{SPTs for $G=D_3$}\label{ap:D3_SymTFT}

To count the number of possible SPTs one may instantiate all the gapped boundaries and compute the overlap of the condensate of the reference boundary and the physical boundary. SPTs then correspond to the case where the only overlap is the trivial anyon.

We may perform this calculation explicitly for the group $G=D_3$. The bulk corresponds to $\mathcal D(D_3 \times D_3)$. The anyons in $\mathcal D(D_3)$ are labeled by the conjugacy class and an irreducible representation of the centralizer subgroup as follows

\begin{center}
    \begin{tabular}{|l|c|c|c|c|c|c|c|c|}
    \hline
        Anyon $(\boldsymbol a)$ & $\boldsymbol A$ & $\boldsymbol B$ & $\boldsymbol C$ &$\boldsymbol D$ & $\boldsymbol E$ & $\boldsymbol F$ & $\boldsymbol G$ & $\boldsymbol H$  \\
        $d_{\boldsymbol a}$ & 1 & 1 & 2 & 3 & 3 & 2 & 2 & 2\\
        \hline
        Conjugacy class & $\{e\}$ & $\{e\}$ & $\{e\}$ & $\{S\}$ & $\{S\}$& $\{R\}$& $\{R\}$&$\{R\}$\\
        Irrep & $\Gamma_1$ & $\Gamma_s$ & $\Gamma_{2d}$ & $\Gamma_1$ & $\Gamma_s$ & $\Gamma_1$ & $\Gamma_{\omega}$ & $\Gamma_{\bar \omega}$ \\
        \hline
    \end{tabular}
\end{center}
where $\Gamma_\omega$ and $\Gamma_{\bar\omega}$ are the two non-trivial irreps of $\mathbb Z_3$. We also note that each anyon is its own conjugate except $\bar {\boldsymbol G} = \boldsymbol H$ and $\bar {\boldsymbol H} = \boldsymbol G$.

The gapped boundaries of $\mathcal D(D_3 \times D_3)$ are labeled by subgroups of $\mathcal D(D_3 \times D_3)$ and the corresponding cocycles of that subgroup. There are 28 such gapped boundaries (see Sec.4 of Ref.~\onlinecite{Ostrik03} for a complete enumeration.)

The reference boundary corresponds to the algebra $\mathcal A_\text{ref} = (\boldsymbol A\oplus \boldsymbol B \oplus 2\boldsymbol C) \otimes (\boldsymbol A\oplus \boldsymbol D\oplus \boldsymbol F)$, corresponding to the subgroup $ \{e\} \times G$ (i.e. $G^\text{even}$  following the discussion in Sec.~\ref{sec:dualityargument}.) We find three algebras (all corresponding to $D_3$ subgroups of $G \times G$) whose only overlap with $\mathcal A_\text{ref}$ is $\boldsymbol A\otimes \boldsymbol A$:

\begin{enumerate}
    \item $\mathcal A_1 =  (\boldsymbol A\oplus \boldsymbol D\oplus \boldsymbol F) \otimes  (\boldsymbol A\oplus \boldsymbol B \oplus 2\boldsymbol C) $ corresponding to the subgroup $G \times \{e\}$ (i.e. $G^\text{odd}$).
    \item $\displaystyle \mathcal A_2 =\bigoplus_{\boldsymbol a \in \mathcal D(D_3)} d_{\boldsymbol a} \boldsymbol a \otimes \bar {\boldsymbol a}$ corresponding to the diagonal subgroup of $G\times G$.
    \item $\mathcal A_3 =(\boldsymbol A \otimes \boldsymbol A) \oplus(\boldsymbol B \otimes \boldsymbol B)  \oplus((\boldsymbol A \oplus \boldsymbol B)  \otimes \boldsymbol C) \oplus (\boldsymbol D \otimes \boldsymbol D)  \oplus(\boldsymbol E \otimes \boldsymbol E) \oplus (\boldsymbol F \otimes (\boldsymbol A\oplus \boldsymbol B \oplus 2\boldsymbol C)) $ corresponding to the subgroup generated by $\{R\} \times \{e\}$ and $\{S\} \times \{S\}$.
\end{enumerate}
The first and second choice correspond to the product state and the cluster state respectively. It would be interesting to see whether the third SPT can be constructed from a repeated action of $U_{\mathcal C}$ as in Sec.~\ref{sec:repeated_U_C} or if it is something beyond.

\end{document}